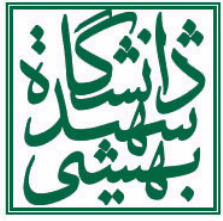

Laser and Plasma Research Institute

Shahid Beheshti Uiniversity

Graduate Thesis of
Photonics and Photonic Materials

# Investigation of light conduction in single-wall and double-wall carbon nanotubes

Supervisor:
Prof. Babak Shokri

By:
Clara Javaherian

Date:
31-July- 2007

# Abstract

Carbon Nanotubes (CNs) are made of tubular graphite layers. These macromolecules have interesting features which are generally introduced in the first chapter of the thesis. In the rest of the thesis, the light conduction in metallic CNs is investigated. In chapter 2, by modeling metallic single-wall CNs as 2D free electron gas layers, we find the density of electrons on a 2D electron gas and solve the Maxwell's equations for these cylindrical waveguides. We assume the same time-dependency or frequency for the electrons oscillations on the 2D gas layers and the propagating electromagnetic waves along the CNs. This assumption is the necessary condition for the light conduction with surface plasmons, that is, the only mechanism to conduct laser beams on nanoscale and below the diffraction limit of light. In chapters 3 and 4, we apply boundary conditions to the solutions of Maxwell's equations, and find the dispersion relations of single-wall and double-wall CNs for the transverse electric (TE) and transverse magnetic (TM) modes. In chapter 5, we investigate the effect of applying relativistic electron beams on single-wall and double-wall CNs to create surface plasmons, and calculate the dispersion relation ghraphs of CNs in the presence of fast electron beams. This yield the frequencies, wavelengths and group velocities of surface plasmon waves created by relativistic beams along CN metallic waveguides.

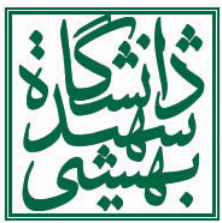

پژوهشکده لیزر و پلاسما
پایان‌نامه کارشناسی ارشد فوتونیک
گرایش مواد فوتونیکی

عنوان پایان نامه:

# بررسی هدایت نوری
# در نانولوله‌های کربنی تک‌جداره و دوجداره

استاد راهنما: دکتر بابک شکری

دانشجو: کلارا جواهریان

مرداد ماه ۱۳۸۶

تقدیم

به :

پدر و مادر عزیزم

# فهرست

# چکیده

نانوتیوپ‌های کربنی از ورقه‌های لوله‌شده‌ی گرافیت تشکیل شده‌اند. این درشت مولکول‌ها خواص جالبی دارند که در فصل اول این پایان‌نامه به طور عمومی به آن‌ها پرداخته می‌شود. در ادامه‌ی پایان‌نامه، هدایت نوری در نانویتوپ‌های فلزی بررسی خواهد شد. در فصل دوم، با در نظر گرفتن مدل گاز الکترون آزاد دوبعدی برای نانوتیوپ‌های کربنی فلزی، چگالی الکترونی آن‌ها را محاسبه می‌کنیم. همچنین در این فصل جواب‌های معادلات ماکسول برای موجبرهای استوانه‌ای به دست آورده می‌شوند. تغییرات زمانی چگالی الکترونی و امواج الکترومغناطیسی را یکسان و با یک فرکانس در نظر می‌گیریم. زیرا این فرض شرط لازمی برای هدایت نوری با مکانیزم پلزمون‌های سطحی است که در ابعاد نانومتری جز آن امکان‌پذیر نیست. با ترکیب معادلات شرط مرزی و جواب‌های فرضی برای میدان‌های الکترومغناطیسی، معادلات پاشندگی برای نانوتیوپ‌های تک‌جداره و دوجداره به ترتیب در فصل‌های سوم و چهارم برای دو مد TE,TM محاسبه خواهند شد. در فصل پنجم به رسم نمودارهای پاشندگی نانوتیوپ‌های تک‌جداره و دوجداره در حضور باریکه‌ی الکترونی خواهیم پرداخت. با نمودارهای موجود در فصل پنجم فرکانس، طول موج و سرعت گروه امواج پلزمونی منتشر شده در نانوتیوپ‌های کربنی تک‌جداره و دوجداره قابل محاسبه هستند.

# فصل اول- نانوتیوپ های کربنی

## ۱-۱- معرفی نانوتیوپ های کربنی

در سال ۱۹۸۰ تنها سه حالت برای اتم‌های کربن متصور بود: الماس، گرافیت و آمورف (کربن غیر کریستالی). امروزه خانواده‌ی بزرگی از دیگر حالات کربن به وجود آمده است. اولین آن‌ها مولکول $C60$ (Fullerene – buckyball) بوده و در حال حاضر بیش از ۳۰ نوع فولرین به وجود دارد. این خانواده‌ی بزرگ به خواهران و برادرهای آن (نانوتیوپ‌ها) گسترش یافته است. اولین نانوتیوپ تک‌جداره در سال ۱۹۹۱ توسط شخصی به نام Sumio Iijima ساخته شد.[۳] نانوتیوپ‌های کربنی ورقه‌های گرافیتی هستند که به شکل یک استوانه پیچیده شده‌اند. قبل از معرفی نانوتیوپ‌های کربنی، ابتدا به معرفی ساختار گرافیت می‌پردازیم.

گرافیت متشکل از لایه‌های اتم‌های کربن است. اتم‌های کربنِ لایه‌های گرافیت در گوشه‌های شش‌ضلعی‌های منتظمی قرار گرفته و با پیوندهای قوی کووالانسی به همسایه‌های خود متصل هستند. کمترین فاصله‌ی دو اتم کربن (طول هر ضلع شش ضلعی منتظم)، $0.14nm$ است. لایه‌های کربنی در گرافیت با پیوندهای ضعیف واندر والس به یکدیگر متصل بوده و فواصل بین صفحات $0.34nm$ هستند. به علت وجود چنین پیوندهای ضعیفی در بین صفحات است که از گرافیت می‌توان به عنوان ماده‌ای نرم در نوشت افزار استفاده کرد. در شکل (۱-۱) نسبت فواصل بین اتم‌ها و لایه‌ها در ورقه‌های گرافیت مشخص است.

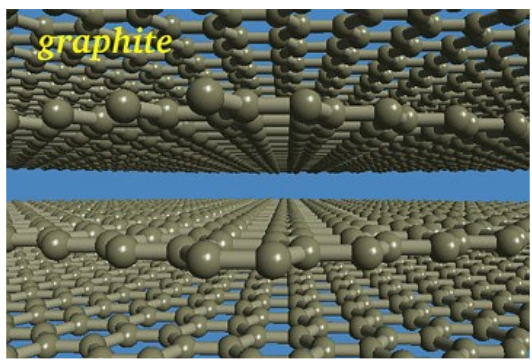

شکل (۱-۱)- ورقه‌های گرافیت که با پیوندهای بسیار ضعیف واندر والس به یکدیگر متصلند.

به طور کلی از لحاظ تعداد جداره‌ها، نانوتیوپ‌ها را می‌توان به دو دسته‌ی عمده تحت عناوین تانوتیوپ‌های تک‌جداره و چند جداره تقسیم‌بندی نمود.

نانوتیوپ‌های تک‌جداره (Single Walled Carbon Nanotubes) که مختصرا به آن‌ها (SWCNT) گفته می‌شود، از پیچیده شدن یک لایه‌ی کربنی گرافیت به شکل یک استوانه تشکیل می‌شوند. نانوتیوپ‌های کربنی تک‌جداره با شعاع‌های یک تا ۱۵ نانومتر ساخته شده‌اند[۱]. در شکل (۱-۲) نماهایی از نانوتیوپ‌های کربنی تک‌جداره نشان داده شده است. دو سر این نانوتیوپ‌ها می‌توانند سربسته (end-capped) و یا سرباز باشند.

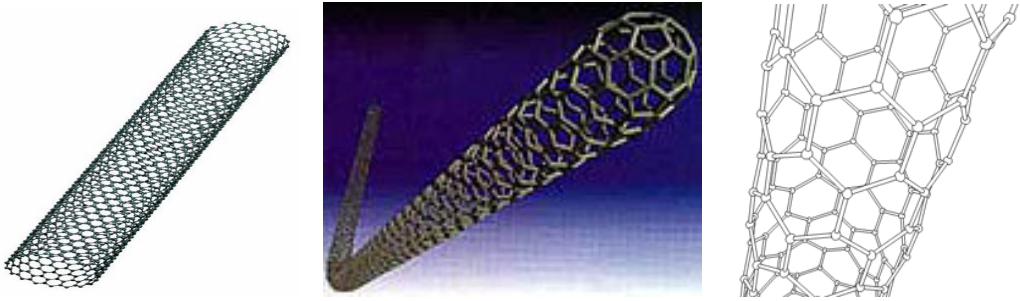

شکل (1-2)- نانوتیوپ‌های کربنی تک‌جداره

نانوتیوپ‌های چندجداره از به هم پیچیده شدن چند لایه‌ی کربنی گرافیت تشکیل می‌شوند. به این نانوتیوپ‌ها مختصرا MWCNT (Multiwalled Carbon NanoTubes) گفته می‌شود. چون نانوتیوپ‌های چند جداره لوله‌های هم‌مرکزی هستند، به آن‌ها نانوتیوپ‌های هم‌محور (Coaxial Carbon NanoTubes) نیز گفته می‌شود. یکی از انواع معروف نانوتیوپ‌های چندجداره، نانوتیوپ دوجداره است که مختصرا به آن DWCNT (Double Walled Carbon NanoTube) می‌گویند. در شکل (1-3) نمایی از نانوتیوپ‌های دوجداره و چند جداره آورده شده است.

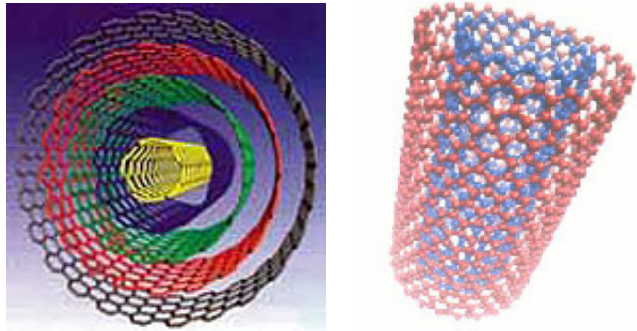

شکل (1-3)- نانوتیوپ‌های کربنی دو جداره و چند جداره

لازم به ذکر است که در بعضی از روش‌های ساخت نانوتیوپ‌ها، مثل (Chemical Vapor Diposition) CVD ، بیش از ۷۰٪ از نانوتیوپ‌های تولیدی دوجداره هستند و بقیه یک جداره و یا بیش از دو جداره دارند [2]. به علت راحت‌تر درست شدن و بیشتر در دسترس بودن نانوتیوپ‌های دوجداره، بررسی خواص آن‌ها امری ضروری به نظر می‌رسد.

امروزه در بیشتر کاربردها از نانوتیوپ‌هایی با تعداد جداره‌های دو عدد به بالا استفاده می‌شود. زیرا از این نانوتیوپ‌ها می‌توان در فواصل طولانی‌تری استفاده نمود و نیز بیشتر کاربردها، مربوط به اثرات کوانتومی یک بعدی که در نانوتیوپ‌های تک‌جداره بیشتر وجود دارد نیستند. در زیر به چند کاربرد مهم از نانوتیوپ‌ها اشاره شده است.

- موجبرها در مدارات نوری نانومتری که در زمینه‌ی نانوفوتونیک (پلزمونیک) مورد استفاده قرار می‌گیرند.
- طناب‌های نانومتری
- خنک کردن تراشه‌ها
- میکروسکوپ‌های روبشی با نوک نانوتیوپ کربنی که به طور تجاری ساخته شده است. [4]
- ترانزیستورهای اثرمیدان (Field Effect Transistors (FET's با استفاده از نانوتیوپ‌های کربنی تک‌جداره [6] و دوجداره[7]

## ۱-۲- نام گذاری نانوتیوپ های کربنی

به علت این‌که ورقه‌ی گرافیت را می‌توان در راستاهای مختلفی پیچید، ساختارهای مختلفی برای نانوتیوپ‌ها موجود است. نام‌گذاری ساختارهای مختلف نانوتیوپ‌ها توسط یکی از روش‌های معمول شناسایی به صورت زیر است.
یک اتم کربن دلخواه در یک لایه‌ی کربنی تخت را به عنوان مبدا فرض می‌کنیم. بردارهای $\vec{a}, \vec{b}$ را با اندازه‌ی قطر کوچک (فاصله تا دومین همسایه) در یک شش ضلعی شبکه‌ی گرافیت، در نظر می‌گیریم. حال هر بردار در شبکه‌ی گرافیت که از مبدا به هر اتم دیگر متصل شود را می‌توان یک ترکیب خطی از دو بردار $\vec{a}, \vec{b}$ در نظر گرفت.

$$\vec{R} = n\vec{a} + m\vec{b} \qquad ; n, m \in Z$$

برای ساخت یک نانوتیوپ، اگر ورقه‌ی گرافیت در راستای عمود بر بردار $\vec{R}$ پیچیده شود، به طوری‌که اندازه‌ی بردار $\vec{R}$ برابر با محیط نانوتیوپ گردد، نانوتیوپ فوق با عنوان $(n,m)$ خوانده می‌شود. در شکل (۱-۴) یک بردار دلخواه $\vec{R}$ به ازای $m = 1, n = 4$ نشان داده شده است.[۸]

زاویه‌ی پیچش هر نانوتیوپ، زاویه‌ی بین بردار هادی $\vec{R} = n\vec{a} + m\vec{b}$ در آن نانوتیوپ و بردار اصلی $\vec{a}$ است. [۹]

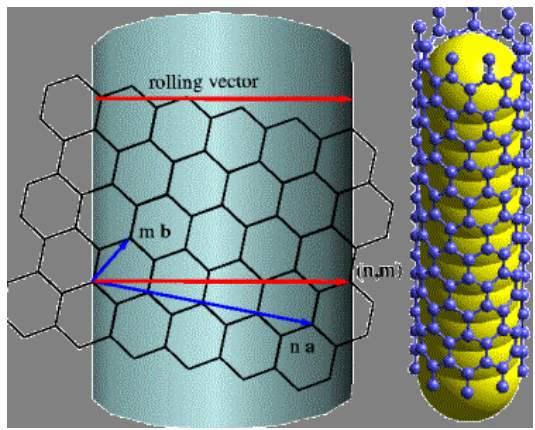

شکل (۱-۴) – یک بردار دلخواه که ترکیب خطی دو بردار اصلی $\vec{a}, \vec{b}$ در نانوتیوپ است.

با روش نام‌گذاری فوق، نانوتیوپ‌ها را به سه دسته تقسیم می‌کنند: نانوتیوپ‌های Zigzag، Armchair و Chiral. این نانوتیوپ‌ها در شکل (۱-۵) به نمایش گذاشته شده‌اند.

نانوتیوپ‌های Zigzag با صفر در نظر گرفتن یکی از شاخص‌های m یا n ایجاد خواهند شد. برای این نوع از نانوتیوپ‌ها زاویه‌ی پیچش صفر درجه فرض می‌شود.

نانوتیوپ‌های Armchair وقتی تولید می‌شوند که دو شاخص n,m با یکدیگر برابر باشند. در این حالت، زاویه‌ی پیچش نانوتیوپ نسبت به حالت Zigzag (که زاویه‌ی مبدا است) برابر با ۳۰ درجه خواهد بود. بقیه‌ی نانوتیوپ‌ها که زاویه‌ی پیچشی بین صفر تا ۳۰ درجه دارند، نانوتیوپ‌های Chiral هستند.

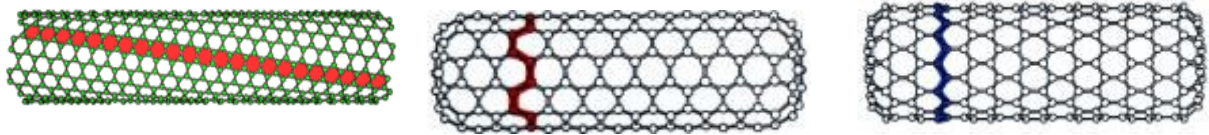

الف          ب          ج

شکل (۱-۵)- الف) zigzag (n,۰)   ب) armchair (n,n)   ج) chiral (n,m)

خواص نانوتیوپ‌ها به قطر و زاویه‌ی پیچش آن‌ها بستگی دارد. قابل ذکر است که هم قطر و هم زاویه‌ی پیچش در یک نانوتیوپ به دو شاخص m,n در آن مربوط است.

روابط قطر و زاویه‌ی پیچش در نانوتیوپ بر حسب دو شاخص m,n به صورت زیر هستند.

$$d_t = (\sqrt{3}/\pi) a_{c-c} (m^2 + mn + n^2)^{1/2}$$

$d_t$ – قطر نانوتیوپ $(n,m)$

$a_{c-c}$ – فاصله‌ی دو اتم کربن همسایه در ورقه‌ی گرافیت

$$\theta = \tan^{-1}(\sqrt{3}n/(2m+n)) \quad ; m > n$$

$\theta$ - زاویه‌ی پیچش نانوتیوپ [۹]

اثبات:

محیط نانوتیوپی با بردار هادی $\vec{R} = n\vec{a}_1 + m\vec{a}_2$ برابر با طول بردار $|\vec{R}|$ است.

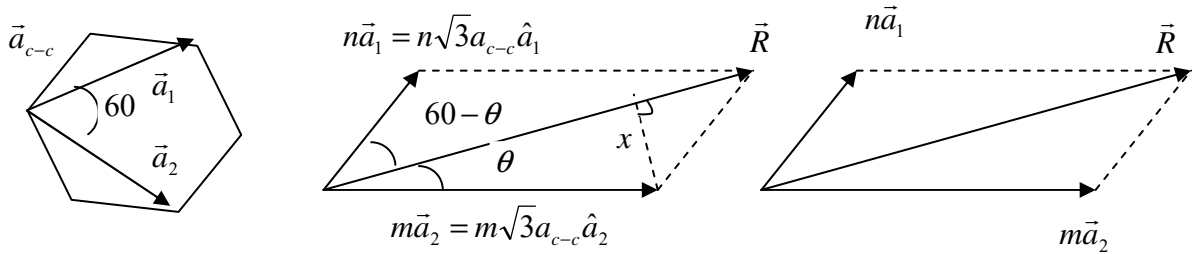

شکل (۱-۶) – بردار $\vec{R}$ هادی یک نانوتیوپ دلخواه بر حسب مولفه‌های آن و بردارهای اصلی $\vec{a}_1, \vec{a}_2$ در شبکه‌ی شش‌ضلعی کربن

به این ترتیب مطابق با تصاویر اول و سوم در شکل (۱-۶) روابط زیر را خواهیم داشت.

$$\vec{R} = n\vec{a}_1 + m\vec{a}_2 \quad \Rightarrow |\vec{R}| = \sqrt{|n\vec{a}_1|^2 + |m\vec{a}_2|^2 + 2|n\vec{a}_1||m\vec{a}_2|\cos(\vec{a}_1, \vec{a}_2)}$$

$$|\vec{a}_1|, |\vec{a}_2| = \sqrt{3} a_{c-c}$$

$$\cos(\vec{a}_1, \vec{a}_2) = \cos(60^o) = \sqrt{3}/2$$

$$|\vec{R}| = \sqrt{3} a_{c-c} \sqrt{n^2 + m^2 + mn}$$

رابطه‌ی قطر نانوتیوپ و محیط آن ($|\vec{R}|$) به صورت زیر است.

$$|\vec{R}| = 2\pi r = \pi d_t$$
$$\Rightarrow d_t = (\sqrt{3}/\pi) a_{c-c} \sqrt{n^2 + m^2 + mn}$$

برای اثبات رابطه‌ی زاویه‌ی پیچش مطابق تصویر میانی در شکل (۱-۶) روابط زیر را می‌نویسیم.

$$\sin(\theta) = x/(m\sqrt{3}a) \quad , \sin(60-\theta) = x/(n\sqrt{3}a)$$
$$\Rightarrow n\sin(60-\theta) = m\sin(\theta)$$
$$n\sqrt{3}/2 \cos(\theta) - (n/2)\sin(\theta) = m\sin(\theta)$$
$$\tan(\theta) = (n\sqrt{3}/2)/(m+n/2) = \sqrt{3}n/(2m+n)$$

تاکنون اندازه‌گیری‌های قطر و زاویه‌ی پیچش نانوتیوپ‌ها با روش‌های STM (Scanning Tunnelling Microscopy) و TEM (Transmission Electron Microscopy) انجام شده‌اند. ولی به دست آوردن قطر و زاویه‌ی پیچش از روی خواص نانوتیوپ‌ها مثل مقاومت الکتریکی، به علت کوچک بودن نانوتیوپ‌ها و حرکت گرمایی اتم‌های کربن هنوز به راحتی امکان‌پذیر نیست. همچنین نانوتیوپ‌ها توسط باریکه‌ی الکترونیِ میکروسکوپ‌ها نیز، مورد آسیب قرار می‌گیرند.[۹]

## ۱-۳- خواص مختلف در نانوتیوپ های کربنی

از جمله خواص جالب نانوتیوپ‌های کربنی می‌توان به خواص مکانیکی، الکتریکی و نوری آن‌ها اشاره کرد. در ادامه توضیحاتی در مورد هر یک از این خواص آورده شده است.

### ۱-۳-۱- استحکام مکانیکی

نانوتیوپ‌های کربنی بسیار قوی بوده و مدول الاستیکی بالایی دارند. در سال ۱۹۹۶ یک گروه تحقیقاتی در دانشگاه پرینستون، مدول یانگ را برای نانوتیوپ‌های کربنی تخمین زدند. این کار توسط مرتعش کردن یک سر نانوتیوپی انجام شد که انتهای دیگرش ثابت نگه داشته شده بود. برآورد این گروه با مقادیر مدول‌های یانگ بالایی که برای ورقه‌ی گرافیت اندازه‌گیری شده، همخوانی داشته است.

شبیه‌سازی‌های دینامیک مولکولی نشان می‌دهند که در بیشتر حالات، وقتی فشارهایی که باعث تغییر شکل نانوتیوپ‌ها می‌شوند از روی آن‌ها برداشته شوند نانوتیوپ‌ها به شکل اولیه‌ی خود باز می‌گردند.

یک گروه تحقیقاتی در دانشگاه Carolina نشان داده‌اند که خواص مکانیکی نانوتیوپ‌ها آن‌ها را به وسایل ایده‌آلی برای دستکاری دیگر مواد نانومتری تبدیل کرده است.

در شکل (۱-۷) پیچش و خمش نانوتیوپ‌ها نشان داده شده است. این تصویر توسط میکروسکوپ TEM در دانشگاه Shinshu ژاپن عکس‌برداری شده و نمایان‌گر استحکام بالای مکانیکی نانوتیوپ‌های کربنی است.[۹]

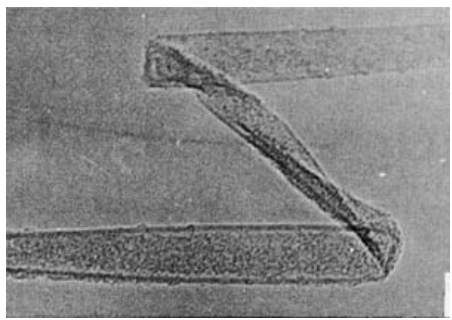

شکل (۱-۷) - تصویر پیچش و خمش در یک نانوتیوپ

### ۱-۳-۲- هدایت الکتریکی

گرافیت ماده‌ای نیمه‌رسانا (نیمه‌فلز) با BandGap صفر است. به این معنی که ماگزیمم انرژیِ نوار ظرفیت و مینیمم انرژی نوار هدایت هم‌مقدار هستند. نانوتیوپ‌ها با این‌که از به هم پیچیده شدن ورقه‌های گرافیت تشکیل شده‌اند، ولی خواص متفاوتی با آن‌ها داشته و مثلا بر خلاف گرافیت می‌توانند فلزی باشند.

همه‌ی نانوتیوپ‌ها را با توجه به میزان گذردهی جریان الکتریکی می‌توان به دو نوع تقسیم نمود: نانوتیوپ‌های فلزی و نیمه‌رسانا. نانوتیوپ‌های فلزی، جریان الکتریکی را به راحتی از خود عبور می‌دهند و نانوتیوپ‌های نیمه‌رسانا به علت وجود گاف‌های انرژی در نوارهای انرژی آن‌ها، جریان الکتریکی را اندکی از خود عبور می‌دهند.

Noriaki Hamada و همکارانش با محاسبات نوارهای انرژی برای نانوتیوپ‌های با قطرهای کوچک نشان داده‌اند که مثلا برای یک نانوتیوپ Armchair با شاخصِ (5,5) و یک نانوتیوپ Zigzag (9,0)، مقدار بسیار ناچیزی انرژی لازم است تا

الکترون از نوار ظرفیت به نوار هدایت انتقال یابد. ولی برای یک نانوتیوپ Zigzag (10,0) فاصله‌ای بین نوارهای هدایت و ظرفیت انرژی ایجاد شده و نانوتیوپ خاصیت فلزی خود را از دست می‌دهد و نیمه‌رسانا می‌شود. این امر بدان معنی است که تغییراتی جزئی در شکل نانوتیوپ می‌تواند سبب تغییر میزان هدایت الکتریکی در آن شود. در نتیجه فلزی و نیمه‌رسانایی نانوتیوپ‌ها به قطر و زاویه‌ی پیچش در آن‌ها بستگی دارد. به طور کل، برای هر نانوتیوپ اگر رابطه‌ی زیر بین شاخص‌های $(n,m)$ آن برقرار باشد نانوتیوپ رفتار فلزی از خود نشان خواهد داد.[۹]

$$n - m = 3q \quad ; q \in Z$$

چون قطر و زاویه‌ی پیچش نانوتیوپ را می‌توان بر حسب شاخص‌های $(n,m)$ نوشت، فرمول بالا رابطه‌ای بین قطر و زاویه‌ی پیچش در نانوتیوپ نیز به دست خواهد داد که درصورت برقرار بودن، نانوتیوپ خاصیت فلزی پیدا می‌کند.

مطابق با رابطه‌ی بالا، یک‌سوم نانوتیوپ‌ها فلزی بوده و بقیه‌ی آن‌ها نیمه‌رسانا خواهند بود. همچنین همه‌ی نانوتیوپ‌های از نوع Armchair و یک‌سوم نانوتیوپ‌های Zigzag نیز، فلزی هستند.[۹]

نانوتیوپ‌هایی که رابطه‌ی بالا بین شاخص‌های $(n,m)$ در آن‌ها برقرار نیست، خاصیت نیمه‌رسانایی دارند. در نوارهای انرژی این نانوتیوپ‌ها BandGap هایی وجود دارد که اندازه‌ی آن‌ها با قطر نانوتیوپ رابطه‌ی عکس داشته و از مرتبه‌ی $0.5 ev$ هستند.[۵] در چنین نانوتیوپ‌هایی با زیاد شدن شعاع و به بی‌نهایت میل کردن آن، اندازه‌ی BandGap به صفر میل کرده و شبیه به BandGap در ورقه‌ی گرافیت می‌شود. [۹]

نانوتیوپ‌های دوجداره بر حسب فلزی یا نیمه‌رسانا بودن هرکدام از جداره‌های تشکیل‌دهنده‌ی آن‌ها می‌توانند به چهار دسته تقسیم شوند. نانوتیوپ‌هایی با جداره‌های داخلی و خارجی فلز- فلز، فلز- نیمه‌رسانا، نیمه‌رسانا- فلز و نیمه‌رسانا- نیمه‌رسانا. خاصیت هدایت الکتریکی نانوتیوپ‌های دوجداره، بر حسب فاصله‌ی بین جداره‌های آن‌ها معین می‌شود. به کمک محاسبات ساختار نواری و چگالی حالات می‌توان نشان داد که در نانوتیوپ‌های دوجداره‌ی فلز- فلز، در فواصل مختلف بین جداره‌ای، نانوتیوپ دوجداره‌ی حاصل‌شده فلزی باقی می‌ماند. برای نانوتیوپ‌های فلز- نیمه‌رسانا، کم‌شدن فاصله‌ی بین‌جداره‌ای باعث وقوع گذار فازی شده که در آن هر دو جداره‌ی نانوتیوپ خاصیت فلزی پیدا می‌کنند. برای نانوتیوپ‌های دوجداره‌ی نیمه‌رسانا- فلز، در بعضی فواصل بین‌جداره‌ای دیواره‌ی داخلی خاصیت فلزی و دیواره‌ی خارجی خاصیت نیمه‌رسانایی پیدا می‌کند. یعنی دو جداره‌ی نانوتیوپ خواص الکتریکی خود را با یکدیگر تعویض می‌نمایند. با کاهش مجدد فاصله‌ی جداره‌ها، هر دو دیواره خاصیت فلزی پیدا خواهند کرد. برای حالت نیمه‌رسانا- نیمه‌رسانا وقتی دوجداره فاصله‌ی بسیار زیادی از هم داشته باشند، نانوتیوپ دوجداره‌ی حاصل شده نیمه‌رسانا باقی می‌ماند. اما با کاهش فاصله‌ی دوجداره، ابتدا جداره‌ی داخلی و سپس جداره‌ی خارجی فلزی خواهند شد. در این حالت، زمانی‌که گاف‌های نواریِ نانوتیوپ‌های تک‌جداره‌ی نیمه‌رسانا هم‌پوشانی ندارند، نانوتیوپ دوجداره‌ی تشکیل‌شده فلزی است. ولی اگر این گاف‌های نواری هم‌پوشانی داشته باشند، فاصله‌ی بین‌جداره‌ای تعیین‌کننده‌ی آن است که هر یک از جداره‌ها خاصیت نیمه‌رسانایی داشته باشند یا فلزی. اگر فاصله‌ی بین جداره‌ها بزرگ باشد، هر دو نانوتیوپ نیمه‌رسانا باقی می‌مانند. اما اگر فاصله‌ی آن‌ها کم باشد نانوتیوپ‌ها فلزی خواهند شد. گذار فاز از نیمه‌رسانا به فلز برای فواصل بین‌جداره‌ای کم، به علت هم‌پوشانی اوربیتال‌های $2p_z$ اتم‌های کربن در دیواره‌های داخلی و خارجی است. [۵]

### ۱-۳-۳- هدایت نوری

در نانوتیوپ‌های کربنی خاصیت هدایت نوری وجود دارد و می‌توان از آن‌ها به عنوان موجبر استفاده نمود. لازم به ذکر است که هدایت نوری در ابعاد نانومتر تنها با موجبرهای پلزمونی امکان‌پذیر است و با موجبرهای معمولی و یا موجبرهای کریستال فوتونی امکان‌پذیر نخواهد بود. این امر بدان دلیل است که نمی‌توان نور را در پرتویی که قطر آن کمتر از حد پراش ($\frac{\lambda}{2n}$) باشد متمرکز و کانونی کرد.

پلزمون به معنای کوانتای انرژی در امواج الکترونی است و پلزمون‌های سطحی یا حجمی، کوانتای انرژی در امواج الکترونی سطحی یا حجمی هستند.

موجبرهای پلزمونی ساختارهای فلزی هستند که نور تابیده شده به آن‌ها به پلزمون‌های سطحی تبدیل می‌شود و پس از عبور از موجبر، مجددا فوتون‌هایی با مشخصات نور اولیه تشکیل خواهند شد. به عنوان مثال‌هایی از این موجبرها می‌توان موجبر پلزمون- گاف[۱۲]، موجبر نانوذره‌ای[۱۱] و موجبرهای کربنی (نانوتیوپ‌های کربنی) را نام برد. اساس خاصیت هدایت نوری در این موجبرها، جفت‌شدگی میدان الکترومغناطیسی با پلزمون‌های سطحی است. به این معنی که فوتون‌های موج الکترومغناطیسی فرودی می‌توانند در شرایطی خاص که به خصوصیات موجبر بستگی دارد، تمامی انرژی خود را به پلزمون‌های سطحی در موجبر منتقل کنند. این پلزمون‌ها پس از عبور کردن از موجبر به فوتون‌هایی با انرژی فوتون‌های اولیه تبدیل می‌شوند. پلزمون‌های سطحی به سرعت میرا می‌شوند ولی طول انتشار آن‌ها برای انتقال اطلاعات در مدارات مجتمع نوری نانومتری مناسب است.

نانوتیوپ‌های کربنی که توانایی هدایت الکتریکی بالایی دارند و فلز محسوب می‌شوند، می‌توانند به عنوان موجبرهای پلزمونی برای هدایت نور در مدارات نانومتری مورد استفاده قرار گیرند.

برای بررسی هدایت نوری در موجبرهای پلزمونی به حل معادلات ماکسول در آن‌ها می‌پردازند و امواجی را که اجازه‌ی انتشار دارند می‌یابند. بسامد این امواج برحسب بردار موجشان در قالب معادلات پاشندگی قابل دستیابی هستند. معادله‌ی پاشندگی، علاوه بر تعیین فرکانس‌های امواج مجاز سیستم، سرعت گروه امواج مختلف، فرکانس قطع و فرکانس امواج ایستا (با سرعت گروه صفر) را در موجبر مشخص می‌کند. همچنین رسم معادلات پاشندگی برای مدهای عرضی مختلف تعیین می‌کند که چه مدهای عرضی برای نانوتیوپی با هندسه‌ی خاص امکان ایجاد شدن را دارند. معادلات پاشندگی می‌توانند برای هندسه‌های مختلف از موجبرها به دست آیند.

در این پایان‌نامه خاصیت هدایت نوری برای موجبرهای کربنی تک‌جداره و دوجداره به ترتیب در فصل‌های سوم و چهارم مورد بررسی قرار گرفته و معادلات پاشندگی برای آن‌ها در دو مد TE,TM محاسبه و رسم شده‌اند.

# فصل دوم – محاسبات پایه برای بررسی هدایت نوری در نانوتیوپ های کربنی

### مقدمه

در این فصل بعضی روابطی که برای محاسبه‌ی معادلات پاشندگی نانوتیوپ‌های تک‌جداره و دوجداره در فصول ۳و۴ مورد استفاده قرار خواهند گرفت به دست می‌آیند.

## ۲-۱- چگالی الکترونی در گاز الکترونی دوبعدی (نانوتیوپ تک جداره)

در این پایان‌نامه برای بررسی هدایت نوری در نانوتیوپ‌های کربنی آن‌ها را با یک گاز الکترونی دوبعدی مدل‌سازی می‌کنیم. مشخصه‌ی گاز فوق، چگالی الکترونی آن است که با $n$ نشان می‌دهیم. فرض می‌کنیم که توسط میدان الکترومغناطیسی اختلالی به یک گاز الکترونی دوبعدی وارد می‌شود. با در نظر گرفتن اختلال مرتبه‌ی اول (اختلال خطی)، خواهیم داشت:

$$n = n_0 + \delta n$$

$n_0$ - چگالی الکترونی گاز دوبعدی در حالت تعادل (زمانی که هیچ اختلال یا میدانی به الکترون‌های وارد نشده است)
$\delta n$ - چگالی الکترونی اختلالی که نسبت به چگالی الکترونی اولیه کوچک فرض می‌شود.

حال برای جمله‌ی اختلالی، بسط موج تخت را در نظر می‌گیریم.

$$\delta n(a,\varphi,z,t) = \sum_{m=-\infty}^{+\infty} \int_{-\infty}^{+\infty} dq N_m e^{i(m\varphi+qz-\omega t)} \qquad (2-1)$$

برای به دست آوردن $\delta n$ و یا به طور معادل $N_m$، معادلات پیوستگی و حرکت را برای گاز الکترونی دوبعدی در نظر گرفته و با تلفیق آن‌ها، $N_m$ را بر حسب میدان الکتریکیِ در راستای سطح و دیگر ضرایب به دست می‌آوریم. برای میدان‌های الکتریکی و مغناطیسی اعمال شده بر گاز الکترونی دوبعدی نیز بسط موج تخت را به صورت زیر در نظر می‌گیریم.

$$E(r,\varphi,z,t) = \sum_{m=-\infty}^{+\infty} \int_{-\infty}^{+\infty} dq E_m(r,q) e^{i(m\varphi+qz-\omega t)}, B(r,\varphi,z,t) = \sum_{m=-\infty}^{+\infty} \int_{-\infty}^{+\infty} dq B_m(r,q) e^{i(m\varphi+qz-\omega t)} \qquad (2-2)$$

بسط فوریه‌ی گسسته‌ی توابع فوق بر حسب زاویه‌ی سمتی، به آن دلیل است که این توابع بر حسب $\varphi$ متناوب هستند و در بازه‌ی $(0, 2\pi)$ تکرار می‌شوند. ولی برای این توابع نسبت به $z$، بسط انتگرال فوریه را در نظر گرفته‌ایم. زیرا این توابع در راستای $z$ می‌توانند متناوب نباشند.

### نکته:

همان‌طور که در فصل اول اشاره شد، تنها مکانیزم عبور موج نوری از یک موجبر نانومتری (به عنوان مثال نانوتیوپ کربنی) تبدیل فوتون‌ها به پلزمون‌های سطحی است. پلزمون‌های سطحی امواج الکترونی هستند که با امواج

الکترومغناطیسی فرودی جفت شده و با یکدیگر در روی سطح به جلو می‌روند. به همین دلیل در معادلات (۲-۱) و (۲-۲) تحول زمانی میدان‌ها (یا مدهای سیستم) و تابع چگالی الکترونی را با فرکانس‌های برابر در نظر گرفته‌ایم.

**معادلات پیوستگی و حرکت برای گاز الکترونی دوبعدی**

معادله‌ی پیوستگی با استفاده از معادله‌ی شرودینگر و بعضی تعاریف به دست می‌آید.
معادله شرودینگر برای الکترون‌های گاز الکترونی دوبعدی واقع در میدان الکترومغناطیسی به شکل زیر است.

$$i\hbar \frac{\partial \psi}{\partial t} = \hat{H}\psi = \frac{1}{2m}[(\hat{P} - \frac{e}{c}A)^2 + e\varphi]\psi$$

که در آن m جرم تعدیل‌یافته‌ی الکترون و $\psi$ تابع حالت الکترون در گاز دوبعدی است.
با توجه به اتحاد زیر و در نظر گرفتن پیمانه‌ی کولن ($\nabla \cdot A = 0$) برای جواب‌ها، معادله شرودینگر به شکل ساده‌تری نوشته می‌شود.

$$\nabla.(\vec{A}\psi) = \psi \nabla.A + A.\nabla \psi \qquad \nabla.A = 0 \rightarrow \nabla.(A\psi) = A.\nabla \psi$$

$$i\hbar \frac{\partial \psi}{\partial t} = (-\frac{\hbar^2}{2m}\nabla^2 + i\hbar \frac{e}{mc}A\nabla + \frac{e^2}{2mc^2}A^2 + e\varphi)\psi$$

با استفاده از تعریف‌های زیر و به کمک معادله‌ی شرودینگر، معادله‌ی پیوستگی به دست می‌آید.

چگالی بار الکترونی(شار احتمال) $\qquad \rho = ne = e|\psi|^2$

چگالی (سطحی) جریان الکترونی(بردار جریان احتمال) $\qquad j = env = \frac{ie\hbar}{2m}(\psi \nabla \psi^* - \psi^* \nabla \psi) - \frac{e^2}{mc}A\psi\psi^*$

معادله پیوستگی $\qquad \frac{\partial \rho}{\partial t} + \nabla \cdot J = 0$

با تقسیم معادله‌ی پیوستگی بر $e$ فرم دیگر آن به شکل روبه‌رو به دست می‌آید. $\qquad \frac{\partial n}{\partial t} + \nabla \cdot (nv) = 0$

**اثبات معادله‌ی پیوستگی**

معادله‌ی شرودینگر $\qquad i\hbar \frac{\partial \psi}{\partial t} = \hat{H}\psi = \frac{1}{2m}[(\hat{P} - \frac{e}{c}A)^2 + e\varphi]\psi$

$P = -i\hbar \nabla$

۱۴

$$i\hbar\frac{\partial \psi}{\partial t} = (-\frac{\hbar^2}{2m}\nabla^2 + i\hbar\frac{e}{2mc}A.\nabla + i\hbar\frac{e}{2mc}\nabla.A + \frac{e^2}{2mc^2}A^2 + e\varphi)\psi$$

معادله‌ی فوق را در مزدوج تابع حالت $\psi$ ضرب می‌کنیم.

$$\psi^*(i\hbar\frac{\partial \psi}{\partial t} = (-\frac{\hbar^2}{2m}\nabla^2 + i\hbar\frac{e}{mc}A.\nabla + \frac{e^2}{2mc^2}A^2 + e\varphi)\psi)$$

بار دیگر از معادله‌ی شرودینگر مزدوج گرفته و نتیجه‌ی آن‌را در تابع حالت $\psi$ ضرب می‌نماییم.

$$\psi(-i\hbar\frac{\partial \psi^*}{\partial t} = (-\frac{\hbar^2}{2m}\nabla^2 - i\hbar\frac{e}{mc}A.\nabla + \frac{e^2}{2mc^2}A^2 + e\varphi)\psi^*)$$

با جمع کردن دو معادله‌ی به دست آمده خواهیم داشت:

$$i\hbar\frac{\partial(\psi\psi^*)}{\partial t} = \frac{\hbar^2}{2m}(\psi\nabla^2\psi^* - \psi^*\nabla^2\psi) + i\hbar\frac{e}{mc}A.\nabla(\psi\psi^*)$$

$$\frac{\partial(\psi\psi^*)}{\partial t} = \frac{-i\hbar}{2m}(\psi\nabla^2\psi^* - \psi^*\nabla^2\psi) + \frac{e}{mc}A.\nabla(\psi\psi^*) \qquad (2-3)$$

احتمال کل (=۱) / ( احتمال حضور الکترون در یک مکان) = (تعداد کل الکترون‌ها) / (تعدادالکترون‌ها در یک مکان)

پس احتمال حضور الکترون در یک مکان، همان چگالی الکترونی(n) است: $n = |\psi|^2$

$$j = env = \frac{ie\hbar}{2m}(\psi\nabla\psi^* - \psi^*\nabla\psi) - \frac{e^2}{mc}A\psi\psi^*$$

$$\to nv = \frac{i\hbar}{2m}(\psi\nabla\psi^* - \psi^*\nabla\psi) - \frac{e}{mc}A\psi\psi^*$$

$$\begin{cases}\nabla.(\psi\nabla\psi^*) = \nabla\psi^*.\nabla\psi + \psi\nabla.(\nabla\psi^*) \\ \nabla.(A\psi\psi^*) = A.\nabla(\psi\psi^*) + \psi\psi^*\nabla.A\end{cases}$$

$$\nabla.(nv) = \frac{i\hbar}{2m}(\nabla\psi.\nabla\psi^* + \psi\nabla^2\psi^* - \nabla\psi.\nabla\psi^* - \psi^*\nabla^2\psi) - \frac{e}{mc}A.\nabla(\psi\psi^*) - \frac{e}{mc}\psi\psi^*\nabla.A$$

$$\nabla.(nv) = \frac{i\hbar}{2m}(\psi\nabla^2\psi^* - \psi^*\nabla^2\psi) - \frac{e}{mc}A.\nabla(\psi\psi^*) \qquad (2-4)$$

$(2-3)$ و $(2-4)$ $\to \frac{\partial n}{\partial t} + \nabla.(nv) = 0$

فصل ۲ - چگالی الکترونی در گاز الکترونی دوبعدی

با ضرب کردن طرفین رابطه در e معادله‌ی پیوستگی به دست می‌آید.

$$\rho = ne, \; j = nev \rightarrow \frac{\partial \rho}{\partial t} + \nabla \cdot j = 0$$

$$\frac{\partial \delta n}{\partial t} + n_0 \nabla \cdot v = 0 \tag{۲-۵}$$

حال معادله‌ی حرکت را برای الکترونی در گاز الکترونی دوبعدی به صورت زیر در نظر می‌گیریم. [۱]

$$\frac{\partial v}{\partial t} = -\frac{e}{m} E_{ll} - \frac{\alpha}{n_0} \nabla_{ll} \delta n + \frac{\beta}{n_0} \nabla_{ll} (\nabla_{ll}^2 \delta n) \tag{۲-۶}$$

در این معادله (معادله‌ی هیدرودینامیکی) $E_{ll}$ مولفه‌ی میدان در راستای سطح استوانه است.

$$E_{ll} = E_z \hat{e}_z + E_\varphi \hat{e}_\varphi$$

همچنین $v$ نمایان‌گر میدان سرعت است که در زمان و مکان روی سطح استوانه تغییر می‌کند.
$$v(r,t) \quad ; r = (\varphi, z)$$

گرادیان موازی نیز، مشتق بر حسب مولفه‌های در راستای سطح است.

$$\nabla_{ll} = \hat{e}_z \frac{\partial}{\partial z} + \frac{1}{a} \hat{e}_\varphi \frac{\partial}{\partial \varphi}$$

جمله‌ی دوم در معادله حرکت (۲-۶) به نوعی معرف فشار وارد شده از طرف دیگر الکترون‌های گاز الکترون آزاد فرمی به یک الکترون خاص است که معادله‌ی حرکت برای آن نوشته شده است. ضریب $\alpha$ در این جمله معرف سرعت انتشار اغتشاشات چگالی الکترون‌ها در گاز الکترونی است که به صورت زیر تعریف می‌شود.

$$\alpha = v_F^2 / 2 = \pi n_0 a_B^2 v_B^2 = \frac{e^2 n_0 a_B}{4 \varepsilon_0 m_e}$$

$v_F$ - سرعت فرمی گاز الکترونی دوبعدی
$a_B$ - شعاع بور

جمله‌ی سوم در معادله‌ی (۲-۶)، تصحیح کوانتومی انرژی جنبشی است و ضریب $\beta$ در آن به صورت زیر تعریف می‌شود.

$$\beta = \frac{a_B^2 v_B^2}{4}$$

$a_B$ - شعاع بور
$v_B$ - سرعت بور

$a_B$ شعاع الکترون در اتم هیدروژن در مدل اتمی بور است. البته در این مدل جرم کاهش‌یافته‌ی الکترون، وارد نشده و به همین دلیل $a_B$، مقدار واقعی شعاع الکترون در هسته‌ی اتم هیدروژن نیست. ولی امروزه از این عدد به عنوان مقیاسی برای اندازه‌ی ابر الکترونی استفاده می‌شود.

$$a_B = \frac{4\pi\varepsilon_0 \hbar^2}{m_e e^2} = 5.29 \times 10^{-11} m \approx 0.53 A^0 \approx 0.05 nm$$

**محاسبه‌ی سرعت بور ($v_B$)**

سرعت بور، سرعتِ الکترون هنگام چرخش به دور هسته‌ی اتم هیدروژن در مدل اتمی بور است. می‌دانیم که نیروی کولنی باعث حرکت یک تک‌الکترون به دورهسته‌ای با یک پروتون در اتم هیدروژن می‌شود. با مساوی قرار دادن نیروی کولنی و نیروی جانب مرکز، سرعت حرکت تک‌الکترون به دور هسته در اتم هیدروژن به صورت زیر به‌دست می‌آید.

$$\frac{m_e v_B^2}{a_B} = \frac{1}{4\pi\varepsilon_0}\frac{e^2}{a_B^2} \Rightarrow v_B = (\frac{e^2}{4\pi\varepsilon_0 m_e a_B})^{1/2}$$

و ضریب $\beta$ به صورت زیر خواهد بود.

$$\beta = \frac{a_B^2 v_B^2}{4} = \frac{e^2 a_B}{16\pi\varepsilon_0 m_e}$$

با ضرب طرفین معادله‌ی (۲-۶) در $\delta n$، دیورژانس گرفتن از معادله و جایگذاری معادله‌ی (۲-۵) در آن به رابطه‌ی زیر می‌رسیم.

$$\frac{\partial^2 \delta n}{\partial t^2} = \frac{en_0}{m_e}\nabla_{ll}\cdot E_{ll} + \alpha\nabla_{ll}^2 \delta n - \beta\nabla_{ll}^2(\nabla_{ll}^2 \delta n)$$

بسط‌های فوریه‌ی میدان $\delta n, E_\varphi, E_z$ که در معادلات (۲-۱) و (۲-۲) معرفی شده‌اند را در معادله دیفرانسیل بالا جایگذاری می‌کنیم.

$$-\omega^2 \sum_{m=-\infty}^{+\infty}\int_{-\infty}^{+\infty}dq N_m e^{i(m\varphi+qz-\omega t)} = \frac{en_0}{m_e}(\frac{\partial E_z}{\partial z}+\frac{1}{a}\frac{\partial E_\varphi}{\partial \varphi}) + \alpha\nabla_{ll}^2(\sum_{m=-\infty}^{+\infty}\int_{-\infty}^{+\infty}dq N_m e^{i(m\varphi+qz-\omega t)})$$
$$-\beta\nabla_{ll}^2(\nabla_{ll}^2 \sum_{m=-\infty}^{+\infty}\int_{-\infty}^{+\infty}dq N_m e^{i(m\varphi+qz-\omega t)})$$

$$\nabla_{ll} = \hat{e}_z\frac{\partial}{\partial z}+\frac{1}{a}\hat{e}_\varphi\frac{\partial}{\partial \varphi}$$

$$\nabla^2_{ll} = \nabla_{ll}.\nabla_{ll} = (\hat{e}_z\frac{\partial}{\partial z}+\frac{1}{a}\hat{e}_\varphi\frac{\partial}{\partial \varphi}).(\hat{e}_z\frac{\partial}{\partial z}+\frac{1}{a}\hat{e}_\varphi\frac{\partial}{\partial \varphi}) = \frac{\partial^2}{\partial z^2}+\frac{1}{a^2}\frac{\partial^2}{\partial \varphi^2}$$

$$-\omega^2 \sum_{m=-\infty}^{+\infty} \int_{-\infty}^{+\infty} dq N_m e^{i(m\varphi+qz-\omega t)} = \frac{en_0}{m_e}(iqE_z + i\frac{m}{a}E_\varphi) + \alpha(\frac{\partial^2}{\partial z^2} + \frac{1}{a^2}\frac{\partial^2}{\partial \varphi^2})(\sum_{m=-\infty}^{+\infty}\int_{-\infty}^{+\infty} dq N_m e^{i(m\varphi+qz-\omega t)})$$

$$-\beta(\frac{\partial^2}{\partial z^2} + \frac{1}{a^2}\frac{\partial^2}{\partial \varphi^2})[(\frac{\partial^2}{\partial z^2} + \frac{1}{a^2}\frac{\partial^2}{\partial \varphi^2})(\sum_{m=-\infty}^{+\infty}\int_{-\infty}^{+\infty} dq N_m e^{i(m\varphi+qz-\omega t)})]$$

$$-\omega^2 \sum_{m=-\infty}^{+\infty} \int_{-\infty}^{+\infty} dq N_m e^{i(m\varphi+qz-\omega t)} = \frac{ien_0}{m_e}(qE_z + \frac{m}{a}E_\varphi) + \alpha(-q^2 - \frac{m^2}{a^2})(\sum_{m=-\infty}^{+\infty}\int_{-\infty}^{+\infty} dq N_m e^{i(m\varphi+qz-\omega t)})$$

$$-\beta(\frac{\partial^2}{\partial z^2} + \frac{1}{a^2}\frac{\partial^2}{\partial \varphi^2})(-q^2 - \frac{m^2}{a^2})(\sum_{m=-\infty}^{+\infty}\int_{-\infty}^{+\infty} dq N_m e^{i(m\varphi+qz-\omega t)})$$

$$-\omega^2 \sum_{m=-\infty}^{+\infty} \int_{-\infty}^{+\infty} dq N_m e^{i(m\varphi+qz-\omega t)} = \frac{ien_0}{m_e}(qE_z + \frac{m}{a}E_\varphi) + \alpha(-q^2 - \frac{m^2}{a^2})(\sum_{m=-\infty}^{+\infty}\int_{-\infty}^{+\infty} dq N_m e^{i(m\varphi+qz-\omega t)})$$

$$-\beta(-q^2 - \frac{m^2}{a^2})^2(\sum_{m=-\infty}^{+\infty}\int_{-\infty}^{+\infty} dq N_m e^{i(m\varphi+qz-\omega t)})$$

$$-\omega^2 \sum_{m=-\infty}^{+\infty} \int_{-\infty}^{+\infty} dq N_m e^{i(m\varphi+qz-\omega t)} + \alpha(q^2 + \frac{m^2}{a^2})(\sum_{m=-\infty}^{+\infty}\int_{-\infty}^{+\infty} dq N_m e^{i(m\varphi+qz-\omega t)})$$

$$+\beta(q^2 + \frac{m^2}{a^2})^2(\sum_{m=-\infty}^{+\infty}\int_{-\infty}^{+\infty} dq N_m e^{i(m\varphi+qz-\omega t)})$$

$$= \frac{ien_0}{m_e}(qE_z + \frac{m}{a}E_\varphi)$$

$$E_z = \sum_{m=-\infty}^{+\infty}\int_{-\infty}^{+\infty} dq E_{zm}(r,q) e^{i(m\varphi+qz-\omega t)} \quad , E_\varphi(r,\varphi,z,t) = \sum_{m=-\infty}^{+\infty}\int_{-\infty}^{+\infty} dq E_{\varphi m}(r,q) e^{i(m\varphi+qz-\omega t)}$$

$$\sum_{m=-\infty}^{+\infty} \int_{-\infty}^{+\infty} dq [\omega^2 N_m] e^{i(m\varphi + qz - \omega t)} + (\sum_{m=-\infty}^{+\infty} \int_{-\infty}^{+\infty} dq [-\alpha(q^2 + \frac{m^2}{a^2}) N_m] e^{i(m\varphi + qz - \omega t)})$$

$$+ (\sum_{m=-\infty}^{+\infty} \int_{-\infty}^{+\infty} dq [-\beta(q^2 + \frac{m^2}{a^2})^2 N_m] e^{i(m\varphi + qz - \omega t)})$$

$$= -\frac{ien_0}{m_e} [\sum_{m=-\infty}^{+\infty} \int_{-\infty}^{+\infty} dq [q E_{zm}(r,q)] e^{i(m\varphi + qz - \omega t)} + \sum_{m=-\infty}^{+\infty} \int_{-\infty}^{+\infty} dq [\frac{m}{a} E_{\varphi m}(r,q)] e^{i(m\varphi + qz - \omega t)}]$$

$$\sum_{m=-\infty}^{+\infty} \int_{-\infty}^{+\infty} dq [\omega^2 - \alpha(q^2 + \frac{m^2}{a^2}) - \beta(q^2 + \frac{m^2}{a^2})^2] N_m e^{i(m\varphi + qz - \omega t)}$$

$$= -\frac{ien_0}{m_e} [\sum_{m=-\infty}^{+\infty} \int_{-\infty}^{+\infty} dq [q E_{zm}(r,q) + \frac{m}{a} E_{\varphi m}(r,q)] e^{i(m\varphi + qz - \omega t)}$$

با مساوی قرار دادن آرگومان‌های سیگما و انتگرال به رابطه‌ی زیر می‌رسیم.

$$N_m = -i \frac{en_0}{m_e} \frac{(qE_{zm} + m/a E_{\varphi m})}{\omega^2 - \alpha(q^2 + m^2/a^2) - \beta(q^2 + m^2/a^2)^2} \qquad (2\text{-}7)$$

$$\delta n = \sum_{m=-\infty}^{+\infty} \int_{-\infty}^{+\infty} dq [-i \frac{en_0}{m_e} \frac{(qE_{zm} + m/a E_{\varphi m})}{\omega^2 - \alpha(q^2 + m^2/a^2) - \beta(q^2 + m^2/a^2)^2}] e^{i(m\varphi + qz - \omega t)}$$

## ۲-۲- معادلات دیفرانسیل میدان های الکترومغناطیسی در موجبرهای استوانه ای

برای به دست آوردن معادلات دیفرانسیل حاکم بر میدان‌های الکترومغناطیسی در موجبرهای استوانه‌ای بسط فوریه را مطابق معادله‌ی (۲-۸) در زیر، در نظر می‌گیریم و آن‌ها را در معادلات چرخشی ماکسول با مختصات استوانه‌ای قرار می‌دهیم.

$$\begin{cases} E(r,\varphi,z,t) = \sum_{m=-\infty}^{+\infty} \int_{-\infty}^{+\infty} dq\, E_m(r,q) e^{i(m\varphi+qz-\omega t)} \Rightarrow E_z = \sum_{m=-\infty}^{+\infty} \int_{-\infty}^{+\infty} dq\, E_{zm}(r,q) e^{i(m\varphi+qz-\omega t)} \quad E_\varphi, E_r \cdots \\ B(r,\varphi,z,t) = \sum_{m=-\infty}^{+\infty} \int_{-\infty}^{+\infty} dq\, B_m(r,q) e^{i(m\varphi+qz-\omega t)} \Rightarrow B_z = \sum_{m=-\infty}^{+\infty} \int_{-\infty}^{+\infty} dq\, B_{zm}(r,q) e^{i(m\varphi+qz-\omega t)} \quad B_\varphi, B_r \cdots \end{cases} \quad (۲-۸)$$

معادلات چرخشی ماکسول

$$\nabla \times E = -\frac{\partial B}{\partial t}$$

$$\nabla \times B = \frac{1}{c^2}\frac{\partial E}{\partial t} \quad ; J = 0$$

شرط $J = 0$ بدان معنی است که در موجبرهای استوانه‌ایِ نانولوله‌ی کربنی، اختلاف ولتاژی به دو سر نانولوله اعمال نشده است و جریان خالصی روی سطح نانولوله‌ی کربنی موجود نیست.

در ادامه می‌بینیم که مولفه‌های عرضی میدان‌ها $(E_\varphi, E_r, B_\varphi, B_r)$ را می‌توان بر حسب دو مولفه‌ی $E_z, B_z$ که در راستای محور موجبر استوانه‌ای هستند به دست آورد (معادلات (۲-۹) ). دو مولفه‌ی $E_z, B_z$ نیز خود در معادلات بسل تعمیم‌یافته صدق می‌کنند (معادلات (۲-۱۰) ).

ارتباط مولفه‌های عرضی میدان با مولفه‌های در راستای میدان به صورت زیر محاسبه خواهند شد.

$$E_{\varphi m} = \frac{i\omega}{\kappa^2}\frac{\partial B_{zm}}{\partial r} + \frac{qm}{\kappa^2 r} E_{zm} \qquad \kappa^2 = q^2 - k^2 \quad , k = \frac{\omega}{c} \qquad (I)$$

$$E_{rm} = -\frac{iq}{\kappa^2}\frac{\partial E_{zm}}{\partial r} + \frac{m\omega}{\kappa^2 r} B_{zm} \qquad\qquad (II)$$

$$B_{\varphi m} = \frac{i}{\omega}(1-\frac{q^2}{\kappa^2})\frac{\partial E_{zm}}{\partial r} - \frac{qm}{\kappa^2 r}(1-\frac{q^2}{\kappa^2}) B_{zm} \qquad (III)$$

(۲-۹)

$$B_{rm} = -\frac{iq}{\kappa^2}\frac{\partial B_{zm}}{\partial r} + \frac{m}{r\omega}(1-\frac{q^2}{\kappa^2}) E_{zm} \qquad (IV)$$

معادلات بسل تعمیم‌یافته

$$\frac{d^2 E_{zm}}{dr^2} + \frac{1}{r}\frac{dE_{zm}}{dr} - (\kappa^2 + \frac{m^2}{r^2})E_{zm} = 0$$

$$\frac{d^2 B_{zm}}{dr^2} + \frac{1}{r}\frac{dB_{zm}}{dr} - (\kappa^2 + \frac{m^2}{r^2})B_{zm} = 0$$

(۲-۱۰)

**اثبات روابط (۲-۹) :**

معادلات چرخشی ماکسول در مختصات استوانه‌ای به صورت زیر در می‌آیند.

$$\frac{1}{r}\begin{vmatrix} \hat{r} & r\hat{\varphi} & \hat{z} \\ \frac{\partial}{\partial r} & \frac{\partial}{\partial \varphi} & \frac{\partial}{\partial z} \\ E_r & rE_\varphi & E_z \end{vmatrix} = \frac{\hat{r}}{r}(\frac{\partial E_z}{\partial \varphi} - r\frac{\partial E_\varphi}{\partial z}) - \hat{\varphi}(\frac{\partial E_r}{\partial z} - r\frac{\partial E_z}{\partial r}) + \frac{\hat{z}}{r}(\frac{\partial (rE_\varphi)}{\partial r} - \frac{\partial E_r}{\partial \varphi}) = -\frac{\partial B}{\partial t}$$

$$\frac{1}{r}\begin{vmatrix} \hat{r} & r\hat{\varphi} & \hat{z} \\ \frac{\partial}{\partial r} & \frac{\partial}{\partial \varphi} & \frac{\partial}{\partial z} \\ B_r & rB_\varphi & B_z \end{vmatrix} = \frac{\hat{r}}{r}(\frac{\partial B_z}{\partial \varphi} - r\frac{\partial B_\varphi}{\partial z}) - \hat{\varphi}(\frac{\partial B_r}{\partial z} - r\frac{\partial B_z}{\partial r}) + \frac{\hat{z}}{r}(\frac{\partial (rB_\varphi)}{\partial r} - \frac{\partial B_r}{\partial \varphi}) = \frac{1}{c^2}\frac{\partial E}{\partial t}$$

با برابر قرار دادن مولفه‌های مشابه در دو طرف تساوی‌ها خواهیم داشت:

$$\frac{1}{r}\frac{\partial E_z}{\partial \varphi} - \frac{\partial E_\varphi}{\partial z} = -\frac{\partial B_r}{\partial t} \qquad\qquad \frac{1}{r}\frac{\partial B_z}{\partial \varphi} - \frac{\partial B_\varphi}{\partial z} = \frac{1}{c^2}\frac{\partial E_r}{\partial t}$$

$$\frac{\partial E_r}{\partial z} - \frac{\partial E_z}{\partial r} = -\frac{\partial B_\varphi}{\partial t} \qquad \text{و} \qquad \frac{\partial B_r}{\partial z} - \frac{\partial B_z}{\partial r} = \frac{1}{c^2}\frac{\partial E_\varphi}{\partial t}$$

$$\frac{1}{r}E_\varphi + \frac{\partial E_\varphi}{\partial r} - \frac{1}{r}\frac{\partial E_r}{\partial \varphi} = -\frac{\partial B_z}{\partial t} \qquad\qquad \frac{1}{r}B_\varphi + \frac{\partial B_\varphi}{\partial r} - \frac{1}{r}\frac{\partial B_r}{\partial \varphi} = \frac{1}{c^2}\frac{\partial E_z}{\partial t}$$

روابط (۲-۸) را در معادلات بالا جایگزین می‌کنیم. آرگومان‌های جلوی $\sum \int dq e^{i(m\varphi+qz-\omega t)}$ در دو طرف معادله را مساوی یکدیگر در نظر می‌گیریم. به این ترتیب سری معادلات زیر را خواهیم داشت.

$$\frac{1}{r}(im)E_{zm} - (iq)E_{\varphi m} = (i\omega)B_{rm} \quad (a) \qquad \frac{1}{r}(im)B_{zm} - (iq)B_{\varphi m} = \frac{-i\omega}{c^2}E_{rm} \quad (d)$$

$$(iq)E_{rm} - \frac{\partial E_{zm}}{\partial r} = (i\omega)B_{\varphi m} \quad (b) \quad \text{و} \quad (iq)B_{rm} - \frac{\partial B_{zm}}{\partial r} = \frac{-i\omega}{c^2}E_{\varphi m} \quad (e)$$

$$\frac{1}{r}E_{\varphi m} + \frac{\partial E_{\varphi m}}{\partial r} - \frac{1}{r}(im)E_{rm} = (i\omega)B_{zm} \quad (c) \qquad \frac{1}{r}B_{\varphi m} + \frac{\partial B_{\varphi m}}{\partial r} - \frac{1}{r}(im)B_{rm} = \frac{-i\omega}{c^2}E_{zm} \quad (f)$$

ترکیب روابط $(a),(e)$:

$$\frac{-i\omega}{c^2}E_{\varphi m} = -\frac{\partial B_{zm}}{\partial r} + \frac{iq}{i\omega}(\frac{1}{r}imE_{zm} - iqE_{\varphi m}) = -\frac{\partial B_{zm}}{\partial r} + \frac{iqm}{\omega r}E_{zm} - \frac{iq^2}{\omega}E_{\varphi m}$$

$$(\frac{-i\omega}{c^2} + \frac{iq^2}{\omega})E_{\varphi m} = -\frac{\partial B_{zm}}{\partial r} + \frac{iqm}{\omega r}E_{zm}$$

$$E_{\varphi m} = \frac{1}{-i/\omega(\omega^2/c^2 - q^2)}\frac{-\partial B_{zm}}{\partial r} + \frac{iqm/\omega r}{-i/\omega(\omega^2/c^2 - q^2)}E_{zm}$$

$$E_{\varphi m} = \frac{i\omega}{\kappa^2}\frac{\partial B_{zm}}{\partial r} + \frac{qm}{\kappa^2 r}E_{zm} \qquad ; \kappa^2 = q^2 - k^2, k^2 = \omega^2/c^2 \qquad \text{(۲-۹) -(I)}$$

ترکیب روابط $(b),(d)$:

$$\frac{-i\omega}{c^2}E_{rm} = \frac{im}{r}B_{zm} - \frac{iq}{i\omega}(iqE_{rm} - \frac{\partial E_{zm}}{\partial r}) = \frac{im}{r}B_{zm} - \frac{iq^2}{\omega}E_{rm} + \frac{q}{\omega}\frac{\partial E_{zm}}{\partial r}$$

$$(\frac{-i\omega}{c^2} + \frac{iq^2}{\omega})E_{rm} = \frac{im}{r}B_{zm} + \frac{q}{\omega}\frac{\partial E_{zm}}{\partial r}$$

$$E_{rm} = \frac{1}{-i/\omega(\omega^2/c^2 - q^2)}\frac{im}{r}B_{zm} + \frac{q/\omega}{-i/\omega(\omega^2/c^2 - q^2)}\frac{\partial E_{zm}}{\partial r}$$

$$E_{rm} = \frac{m\omega}{r\kappa^2}B_{zm} - \frac{iq}{\kappa^2}\frac{\partial E_{zm}}{\partial r} \qquad ;\kappa^2 = q^2 - k^2, k^2 = \omega^2/c^2 \qquad \text{(۲-۹) -(II)}$$

ترکیب روابط $(e),(a)$:

$$i\omega B_{rm} = \frac{im}{r}E_{zm} - \frac{iq}{-i\omega/c^2}(iqB_{rm} - \frac{\partial B_{zm}}{\partial r}) = \frac{im}{r}E_{zm} - \frac{iq^2c^2}{\omega}B_{rm} - \frac{qc^2}{\omega}\frac{\partial B_{zm}}{\partial r}$$

$$(i\omega - \frac{iq^2c^2}{\omega})E_{rm} = \frac{im}{r}E_{zm} - \frac{qc^2}{\omega}\frac{\partial B_{zm}}{\partial r}$$

$$B_{rm} = \frac{im/r}{i\omega(1 - q^2/k^2)}E_{zm} + \frac{-qc^2/\omega}{-ic^2/\omega(-\omega^2/c^2 + q^2)}\frac{\partial B_{zm}}{\partial r}$$

$$1/(1 - q^2/k^2) = k^2/(k^2 - q^2) = (k^2 + q^2 - q^2)/(k^2 - q^2) = 1 + q^2/(k^2 - q^2) = 1 - q^2/(q^2 - k^2)$$

$$B_{rm} = \frac{m}{r\omega}(1 - q^2/\kappa^2)E_{zm} - \frac{iq}{\kappa^2}\frac{\partial B_{zm}}{\partial r} \qquad ;\kappa^2 = q^2 - k^2, k^2 = \omega^2/c^2 \qquad \text{(۲-۹) -(III)}$$

۲۲

ترکیب روابط $(d),(b)$:

$$i\omega B_{\varphi m} = -\frac{\partial E_{zm}}{\partial r} + \frac{iqc^2}{-i\omega}(\frac{im}{r}B_{zm} - iqB_{\varphi m}) = -\frac{\partial E_{zm}}{\partial r} - \frac{iqc^2 m}{r\omega}B_{zm} + \frac{iq^2c^2}{\omega}B_{\varphi m}$$

$$(i\omega - \frac{iq^2c^2}{\omega})B_{\varphi m} = -\frac{\partial E_{zm}}{\partial r} - \frac{iqc^2 m}{r\omega}B_{zm}$$

$$B_{\varphi m} = \frac{-1}{i\omega(1-q^2/k^2)}\frac{\partial E_{zm}}{\partial r} + \frac{-iqc^2 m/r\omega}{-ic^2/\omega(-\omega^2/c^2 + q^2)}B_{zm}$$

$$1/(1-q^2/k^2) = k^2/(k^2-q^2) = (k^2+q^2-q^2)/(k^2-q^2) = 1 + q^2/(k^2-q^2) = 1 - q^2/(q^2-k^2)$$

$$B_{\varphi m} = \frac{i}{\omega}(1-q^2/\kappa^2)\frac{\partial E_{zm}}{\partial r} + \frac{qm}{r\kappa^2}B_{zm} \quad ; \kappa^2 = q^2 - k^2, k^2 = \omega^2/c^2 \qquad (VI)-(۲-۹)$$

**اثبات روابط (۲-۱۰) :**

با کرل گرفتن از روابط ماکسول خواهیم داشت:

$$\nabla \times (\nabla \times E) = -\frac{\partial}{\partial t}\nabla \times B, \nabla \times (\nabla \times B) = (1/c^2)\frac{\partial}{\partial t}\nabla \times E$$

$$A \times (B \times C) = B(A.C) - C(A.B) = B(A.C) - (A.B)C$$
$$\nabla \times (\nabla \times E) = \nabla(\nabla.E) - (\nabla.\nabla)E$$
$$= -\frac{\partial}{\partial t}(\nabla \times B)$$

محیطی که در آن‌ها میدان‌های الکترومغناطیسی موجود هستند (در مسئله‌ی ما، محیط اطراف نانولوله‌های کربنی تک‌جداره و دوجداره) از لحاظ الکتریکی همسانگرد بوده و گذردهی آن یک عدد ثابت است. $\varepsilon = \varepsilon_{cte}$. (در مسئله‌ی ما محیط فوق را هوا در نظر گرفته‌ایم.)
$$\nabla.D = \nabla.\varepsilon E = \rho$$

بار خالص در ناحیه‌ای که میدان‌ها در آن محاسبه می‌شوند (هوای داخل و خارج موجبر) صفر است. $\rho = 0 \Rightarrow \nabla.E = 0$.

$$\xrightarrow{\nabla.E=0, \nabla \times B = \frac{1}{c^2}\frac{\partial E}{\partial t}} \nabla^2 E = \frac{1}{c^2}\frac{\partial^2 E}{\partial t^2}$$

$$\nabla \times (\nabla \times B) = \nabla(\nabla.B) - (\nabla.\nabla)B$$
$$= (1/c^2)\frac{\partial}{\partial t}(\nabla \times E)$$
$$\xrightarrow{\nabla.B=0, \nabla \times E = -\frac{\partial B}{\partial t}} \nabla^2 B = \frac{1}{c^2}\frac{\partial^2 B}{\partial t^2}$$

با استفاده از معادلات (۲-۲) روابط زیر به دست می‌آیند.

$$\nabla^2 E + (\omega^2/c^2)E = 0 \to \begin{cases} \nabla^2 E_r + (\omega^2/c^2)E_r = 0 \\ \nabla^2 E_\varphi + (\omega^2/c^2)E_\varphi = 0 \\ \nabla^2 E_z + (\omega^2/c^2)E_z = 0 \end{cases}$$

$$\nabla^2 B + (\omega^2/c^2)B = 0 \to \begin{cases} \nabla^2 B_r + (\omega^2/c^2)B_r = 0 \\ \nabla^2 B_\varphi + (\omega^2/c^2)B_\varphi = 0 \\ \nabla^2 B_z + (\omega^2/c^2)B_z = 0 \end{cases}$$

در فرمول (۲-۹) همه‌ی مولفه‌های $r, \varphi$ میدان‌ها موازی سطح را بر حسب مولفه‌ی z به دست آورده‌ایم. حال اگر مولفه‌های z میدان الکتریکی و مغناطیسی را از معادله‌ی هلمهولتز زیر بیابیم تمام مولفه‌ها در دسترس خواهند بود.

$$\nabla^2 E_z + (\omega^2/c^2)E_z = 0$$
$$\nabla^2 B_z + (\omega^2/c^2)B_z = 0$$

در مختصات استوانه‌ای معادله‌ی هلمهولتز برای میدان الکتریکی به صورت است. (معادلات از این پس برای هر دو میدان الکتریکی و مغناطیسی مشابه‌اند.)

$$\frac{1}{h_1 h_2 h_3}\left[\frac{\partial}{\partial q_1}\left(\frac{h_2 h_3}{h_1}\frac{\partial E_z}{\partial q_1}\right) + \frac{\partial}{\partial q_2}\left(\frac{h_1 h_3}{h_2}\frac{\partial E_z}{\partial q_2}\right) + \frac{\partial}{\partial q_3}\left(\frac{h_1 h_2}{h_3}\frac{\partial E_z}{\partial q_3}\right)\right] + (\omega^2/c^2)E_z = 0$$

$$\frac{1}{r}\left[\frac{\partial}{\partial r}\left(r\frac{\partial E_z}{\partial r}\right) + \frac{\partial}{\partial \varphi}\left(\frac{1}{r}\frac{\partial E_z}{\partial \varphi}\right) + \frac{\partial}{\partial z}\left(r\frac{\partial E_z}{\partial z}\right)\right] + (\omega^2/c^2)E_z = 0$$

بسط فوریه‌ی مولفه‌ی z میدان الکتریکی را در معادله‌ی هلمهولتز جایگذاری می‌کنیم.

$$\frac{1}{r}\left[\frac{\partial}{\partial r}\left(r\frac{\partial E_z}{\partial r}\right) + \frac{\partial}{\partial \varphi}\left(\frac{1}{r}imE_z\right) + \frac{\partial}{\partial z}(r[iqE_z])\right] + (\omega^2/c^2)E_z = 0$$

$$\frac{1}{r}\frac{\partial E_z}{\partial r} + \frac{\partial^2 E_z}{\partial r^2} - \frac{m^2}{r^2}E_z - q^2 E_z + (\omega^2/c^2)E_z = 0$$

$$\frac{\partial^2 E_z}{\partial r^2} + \frac{1}{r}\frac{\partial E_z}{\partial r} - \left(q^2 - \omega^2/c^2 + \frac{m^2}{r^2}\right)E_z = 0$$

معادلات بسل تعمیم‌یافته برای مولفه‌های z میدان الکتریکی و مغناطیسی به صورت معادلات (2-10) به دست می‌آیند.

$$\frac{\partial^2 E_{zm}}{\partial r^2} + \frac{1}{r}\frac{\partial E_{zm}}{\partial r} - (\kappa^2 + \frac{m^2}{r^2})E_{zm} = 0 \quad ; \kappa^2 = q^2 - \omega^2/c^2$$

$$\frac{\partial^2 B_{zm}}{\partial r^2} + \frac{1}{r}\frac{\partial B_{zm}}{\partial r} - (\kappa^2 + \frac{m^2}{r^2})B_{zm} = 0 \quad ; \kappa^2 = q^2 - \omega^2/c^2$$

(2-10)

# فصل سوم – نانوتیوپ کربنی تک جداره

## ۳-۱- مد TM

### ۳-۱-۱- معادله پاشندگی مد TM برای نانوتیوپ کربنی تک جداره

معادلات شرط مرزی برای میدان الکتریکی در یک نانولوله‌ی کربنی تک‌جداره با در نظر گرفتن بسط موج تخت برای میدان‌ها و چگالی بار الکترونی (معادلات (۲-۱) و (۲-۲)) به صورت زیر هستند.

$$E_{rm}(a)\big|_{r>a} - E_{rm}(a)\big|_{r<a} = -\frac{eN_m}{\varepsilon_0} \tag{۳-۱}$$

$$E_{llm}(a)\big|_{r>a} - E_{llm}(a)\big|_{r<a} = 0 \rightarrow \begin{cases} E_{zm}(a)\big|_{r>a} - E_{zm}(a)\big|_{r<a} = 0 \\ E_{\varphi m}(a)\big|_{r>a} - E_{\varphi m}(a)\big|_{r<a} = 0 \end{cases} \tag{۳-۲}$$

که در آن $N_m$ مطابق رابطه‌ی (۲-۷) قبلا معرفی شده است.

با توجه به معادلات (۲-۱۰) در فصل قبل و نمودارهای توابع بسل تعمیم‌یافته‌ی نوع اول و دوم که در شکل (۳-۱) نشان داده شده‌اند، جواب‌های فرضی میدان در اطراف نانولوله‌ی کربنی تک‌جداره به صورت زیر خواهند بود.

$$\begin{aligned} E_{zm}(r) &= C_m I_m(\kappa r) \qquad r < a \\ E_{zm}(r) &= D_m K_m(\kappa r) \qquad r > a \end{aligned} \tag{۳-۳}$$

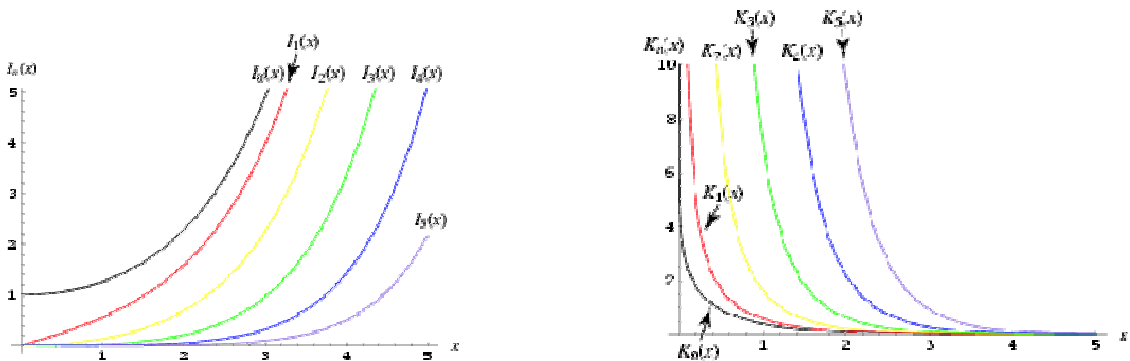

شکل (۳-۱) – نمودارهای توابع بسل تعمیم‌یافته نوع اول و دوم

با در نظر گرفتن مد TM ($B_z = 0$)، روابط بین مولفه‌های $r, \varphi$ میدان الکتریکی و مولفه‌ی z آن که قبلا به صورت رابطه‌ی (۲-۹) بوده به شکل زیر تبدیل می‌شوند.

$$E_{\varphi m} = \frac{qm}{\kappa^2 r} E_{zm} \qquad \kappa^2 = q^2 - k^2 \quad , k = \frac{\omega}{c}$$

$$E_{rm} = -\frac{iq}{\kappa^2} \frac{\partial E_{zm}}{\partial r} \tag{۳-۴}$$

معادلات (۳-۴) را در معادلات شرط مرزی (۳-۱) و (۳-۲) جایگزین می‌کنیم.

$$-\frac{iq}{\kappa^2}(\frac{\partial E_{zm}(a)}{\partial r}\Big|_{r>a} - \frac{\partial E_{zm}(a)}{\partial r}\Big|_{r<a}) = -\frac{e}{\varepsilon_0} N_m$$

$$\begin{cases} E_{zm}(a)\big|_{r>a} - E_{zm}(a)\big|_{r<a} = 0 \\ \frac{qm}{\kappa^2 r}(E_{zm}(a)\big|_{r>a} - E_{zm}(a)\big|_{r<a}) = 0 \end{cases}$$

دو معادله‌ی شرط مرزی که در کروشه به دست آمده‌اند مشابه بوده و از هم مستقل نیست.
حال معادلات (۳-۳) را در دو معادله‌ی شرط مرزی بالا جایگزین می‌کنیم.

$$-\frac{iq}{\kappa^2}[D_m \frac{\partial K_m(\kappa r)}{\partial r}\Big|_{r=a} - C_m \frac{\partial I_m(\kappa r)}{\partial r}\Big|_{r=a}] = -\frac{e}{\varepsilon_0}(-i\frac{en_0}{m_e} \frac{(qE_{zm} + (m/a)[(qm/\kappa^2 r)E_{zm}])}{\omega^2 - \alpha(q^2 + m^2/a^2) - \beta(q^2 + m^2/a^2)^2})$$

$$-\frac{iq}{\kappa}[D_m \frac{\partial K_m(\kappa r)}{\partial(\kappa r)}\Big|_{r=a} - C_m \frac{\partial I_m(\kappa r)}{\partial(\kappa r)}\Big|_{r=a}] = -\frac{e}{\varepsilon_0}(-i\frac{en_0}{m_e} \frac{(qE_{zm} + (m/a)[(qm/\kappa^2 r)E_{zm}])}{\omega^2 - \alpha(q^2 + m^2/a^2) - \beta(q^2 + m^2/a^2)^2})$$

$$\frac{qm}{\kappa^2 r}[D_m K_m(\kappa a) - C_m I_m(\kappa a)] = 0 \rightarrow D_m = \frac{I_m(\kappa a)}{K_m(\kappa a)} C_m$$

با حذف ضریب $C_m$ در اولین معادله‌ی شرط مرزی بالا معادله‌ی زیر به دست می‌آید.

$$-\frac{q}{\kappa}[\frac{I_m(\kappa a)}{K_m(\kappa a)} K'_m(\kappa a) - I'_m(\kappa a)][\omega^2 - \alpha(q^2 + \frac{m^2}{a^2}) - \beta(q^2 + \frac{m^2}{a^2})^2]$$

$$= \frac{e^2 n_0}{\varepsilon_0 m_e}(q + \frac{qm^2}{a^2 \kappa^2}) I_m(\kappa a)$$

$$\frac{1}{\kappa}[-I_m(\kappa a) K'_m(\kappa a) + I'_m(\kappa a) K_m(\kappa a)][\omega^2 - \alpha(q^2 + \frac{m^2}{a^2}) - \beta(q^2 + \frac{m^2}{a^2})^2]$$

$$= \frac{e^2 n_0}{\varepsilon_0 m_e}(1 + \frac{m^2}{a^2 \kappa^2}) I_m(\kappa a) K_m(\kappa a)$$

با استفاده از تعریف رونسکین ($I'_m(x) K_m(x) - I_m(x) K'_m(x) = 1/x$) رابطه به شکل زیر ساده خواهد شد.

۲۷

$$\frac{1}{\kappa(\kappa a)}[\omega^2 - \alpha(q^2 + \frac{m^2}{a^2}) - \beta(q^2 + \frac{m^2}{a^2})^2] = \frac{e^2 n_0}{\varepsilon_0 m_e} \frac{(a^2\kappa^2 + m^2)}{(\kappa a)^2} I_m(\kappa a) K_m(\kappa a)$$

$$\omega^2 - \alpha(q^2 + \frac{m^2}{a^2}) - \beta(q^2 + \frac{m^2}{a^2})^2 = \frac{e^2 n_0}{\varepsilon_0 m_e} \frac{\kappa(m^2 + a^2\kappa^2)}{(\kappa a)} I_m(\kappa a) K_m(\kappa a)$$

**معادله پاشندگی $\omega(\kappa)$**

$$\omega^2 - \alpha(q^2 + \frac{m^2}{a^2}) - \beta(q^2 + \frac{m^2}{a^2})^2 = \Omega_P^2 (m^2 + \kappa^2 a^2) I_m(\kappa a) K_m(\kappa a)$$

$$\Omega_P = (\frac{e^2 n_0}{\varepsilon_0 m_e a})^{1/2}, \kappa^2 = q^2 - \omega^2/c^2$$

$\Omega_P$ تنها به متغیرهای ساختاری و هندسه‌ی نانوتیوپ کربنی تک‌جداره وابسته بوده و دارای بعد فرکانس است.

## ۳-۱-۲- نمودارهای معادله‌ی پاشندگی در شرایط مختلف

در این بخش قبل از رسم معادله‌ی پاشندگی مد TM در شرایط مختلف، به بی‌بعد کردن معادله‌ی پاشندگی می‌پردازیم.

$$\omega^2 - \alpha(\kappa^2 + \omega^2/c^2 + m^2/a^2) - \beta(\kappa^2 + \omega^2/c^2 + m^2/a^2)^2 = \Omega_P^2(m^2 + \kappa^2 a^2)I_m(\kappa a)K_m(\kappa a)$$

$$\Omega_p = (e^2 n_0/\varepsilon_0 m_e a)^{1/2}$$

$$\kappa = (q^2 - k^2); k = \omega/c$$

برای تبدیل $\omega, \kappa$ در معادله‌ی پاشندگی بالا، به متغیرهای بی‌بعد $x = \kappa a, y = \omega/\Omega_P$ طرفین رابطه را بر $\Omega_P^2$ تقسیم می‌کنیم.

$$y^2 - \frac{\alpha}{\Omega_p^2}(\kappa^2 + \omega^2/c^2 + m^2/a^2) - \frac{\beta}{\Omega_p^2}(\kappa^2 + \omega^2/c^2 + m^2/a^2)^2 = (m^2 + x^2)I_m(\kappa a)K_m(\kappa a)$$

$$y^2 - \frac{\alpha}{\Omega_p^2 a^2}(x^2 + \frac{\Omega_P^2 a^2}{c^2}y^2 + m^2) - \frac{\beta}{\Omega_p^2 a^4}(x^2 + \frac{\Omega_P^2 a^2}{c^2}y^2 + m^2)^2 = (m^2 + x^2)I_m(\kappa a)K_m(\kappa a)$$

با تعریف ضرایب بی‌بعد $\alpha_1 = \alpha/(\Omega_p a)^2, \beta_1 = \beta/(\Omega_p^2 a^4), \sigma = \Omega_P^2 a^2/c^2$، معادله به صورت زیر ساده می‌شود.

$$y^2 = \alpha_1(x^2 + \sigma y^2 + m^2) + \beta_1(x^2 + \sigma y^2 + m^2)^2 + (x^2 + m^2)I_m(x)K_m(x)$$

[اثبات بی‌بعد بودن ضرایب در معادله‌ی بالا:

$$\Omega_P = (\frac{e^2 n_0}{\varepsilon_0 m_e a})^{1/2}, \alpha = \frac{e^2 n_0 a_B}{4\varepsilon_0 m_e}, \beta = \frac{e^2 a_B}{16\pi\varepsilon_0 m_e}$$

$$\alpha_1 = \frac{\alpha}{(\Omega_p a)^2} = \frac{e^2 n_0 a_B}{4\varepsilon_0 m_e} \cdot \frac{1}{(\frac{e^2 n_0}{\varepsilon_0 m_e a})a^2} = \frac{a_B}{4a} \to [\alpha_1] = 1$$

$$\beta_1 = \frac{\beta}{(\Omega_p^2 a^4)} = \frac{e^2 a_B}{16\pi\varepsilon_0 m_e} \cdot \frac{1}{(\frac{e^2 n_0}{\varepsilon_0 m_e a})a^4} = \frac{a_B}{16\pi n_0 a^3} \xrightarrow{[n_0]=1/L^3} [\beta_1] = 1$$

$\Omega_P$ دارای بعد فرکانس است.

$$[\sigma] = [\Omega_p^2 a^2/c^2] = \frac{(1/T^2)(L^2)}{L^2/T^2} = 1$$

[

[مقدار $n_0$ :

چگالی اتمی یا تعداد اتم‌ها در واحد سطح یک ورقه گرافیت $38nm^{-2}$ است[۱]. چون هر اتم کربن دارای چهار الکترون در لایه‌ی ظرفیت خود است، چگالی الکترونی نانوتیوپ کربنی برابر با $n_0 = 4\times 38 nm^{-2} = 152 nm^{-2}$ خواهد بود.
]

حال که مقدار $n_0$ مشخص است، پارامترهای $\Omega_P$ و $\sigma$ قابل محاسبه خواهند بود.

$$\sigma = \Omega_P^2 \frac{a^2}{c^2} = \frac{e^2 n_0}{\varepsilon_0 m_e} \frac{a}{c^2} = 5.5\times 10^{-3} \approx 0.05$$

به علت کوچک بودن $\sigma$، از جمله‌ی $\sigma y^2$ در برابر جمله‌ی $y^2$ در معادله‌ی پاشندگی می‌توان صرفنظر کرد. با حذف $\sigma y^2$ معادله‌ی پاشندگی به صورت زیر تبدیل می‌شود.

$$y^2 = \alpha_1(x^2+m^2) + \beta_1(x^2+m^2)^2 + (x^2+m^2)I_m(x)K_m(x)$$
$$y = \pm[\alpha_1(x^2+m^2) + \beta_1(x^2+m^2)^2 + (x^2+m^2)I_m(x)K_m(x)]^{1/2}$$

جواب‌های منفی $y = \omega/\Omega_P$ ، غیر فیزیکی هستند و از آن‌ها صرفنظر می‌کنیم.

**نکته:**

معادله‌ی پاشندگی بالا رابطه‌ای بین فرکانس سیستم و متغیر $\kappa$ تعیین می‌کند. چون جواب‌های فرضی مولفه‌های $z$ میدان بر حسب توابع بسل تعمیم‌یافته با آرگومان‌های $\kappa r$ به دست آمد (معادلات (۳-۳) )، $\kappa$ می‌تواند نمایانگر معکوس عمق نفوذ در موجبر کربنی باشد. یعنی $\kappa$ معکوس فاصله‌ای را تعیین می‌کند که در آن میدان در اطراف پوسته‌ی گاز الکترونی به $1/e$ مقدار ماگزیمم خود (مقدار میدان روی پوسته) می‌رسد. در ادامه نشان می‌دهیم که اگر معادله‌ی پاشندگی را برحسب عدد موج q به دست آوریم، باز هم با حذف متغیر $\sigma y^2$، به معادله‌ای شبیه به معادله‌ی پاشندگی در بالا می‌رسیم.

معادله‌ی پاشندگی به دست آمده بر حسب عکس عمق نفوذ را به معادله‌ای برحسب عدد موج تبدیل می‌کنیم.

$$\omega^2 - \alpha(\kappa^2 + \omega^2/c^2 + m^2/a^2) - \beta(\kappa^2 + \omega^2/c^2 + m^2/a^2)^2 = \Omega_p^2(m^2 + \kappa^2 a^2)I_m(\kappa a)K_m(\kappa a)$$

$$\frac{\omega^2}{\Omega_P^2} - \frac{\alpha}{\Omega_P^2 a^2}([\kappa^2 + \omega^2/c^2]a^2 + m^2) - \frac{\beta}{\Omega_P^2 a^4}([\kappa^2 + \omega^2/c^2]a^2 + m^2)^2 = (m^2 + \kappa^2 a^2)I_m(\kappa a)K_m(\kappa a)$$

متغیرهای بی‌بعد را به صورت زیر تعریف می‌نماییم.

$$y = \frac{\omega}{\Omega_p}, x = qa \Rightarrow \begin{cases} \kappa^2 + \omega^2/c^2 = q^2, \frac{\alpha}{\Omega_P^2 a^2} = \alpha_1, \frac{\beta}{\Omega_P^2 a^4} = \beta_1 \\ \kappa = (q^2 - \omega^2/c^2)^{1/2}, \kappa a = [(qa)^2 - \frac{\Omega_P^2 a^2}{c^2}\frac{\omega^2}{\Omega_P^2}]^{1/2} = (x^2 - \sigma y^2)^{1/2}; \sigma = \Omega_P^2 a^2/c^2 \end{cases}$$

$$y^2 - \alpha_1(x^2+m^2) - \beta_1(x^2+m^2)^2 = (x^2 - \sigma y^2 + m^2)I_m([x^2-\sigma y^2]^{1/2})K_m([x^2-\sigma y^2]^{1/2})$$

۳۰

ملاحظه می‌شود که با حذف $\sigma y^2$ معادله به شکل معادله‌ی فرکانس برحسب معکوس عمق نفوذ تبدیل می‌شود.

$$y^2 - \alpha_1(x^2+m^2) - \beta_1(x^2+m^2)^2 = (x^2+m^2)I_m(x)K_m(x)$$

حال به محاسبه‌ی ضرایب $\alpha_1, \beta_1$ و رسم معادله‌ی پاشندگی می‌پردازیم.
با توجه به مقدار شعاع بور ($a_B = 5.29\times10^{-11}m \approx 0.53 A^0 \approx 0.05 nm$)، ضرایب $\alpha_1, \beta_1$ در معادله برحسب شعاع نانوتیوپ و چگالی الکترونی سطحی نوشته می‌شوند.

$$\alpha_1 = \frac{a_B}{4a} = \frac{1.32\times10^{-11}m}{a(m)} = \frac{1.32\times10^{-2}nm}{a(nm)}$$

$$n_0 = 152 nm^{-2}(for\ \ CNT)$$

$$\beta_1 = \frac{a_B}{16\pi n_0 a^3} = \frac{5.29\times10^{-11}m}{16\times3.14\times4\times38\times10^{18}m^{-2}\times a^3(m^3)} = \frac{6.92\times10^{-4}\times10^{-29}}{[a(m)]^3} = \frac{6.92\times10^{-6}}{[a(nm)]^3}$$

شکل (۳-۲) نمودار معادله پاشندگی مد TM را به ازای مدهای عرضی m=۰،۱،۲،۳،۴ برای نانوتیوپ تک‌جداره به شعاع 5nm نشان می‌دهد. معادلات و ضرایب آن‌ها در زیر آورده شده است.

$$\beta_1 = \frac{6.92\times10^{-6}}{[a(nm)]^3} = \frac{6.92\times10^{-6}}{[5(nm)]^3} = 5.54\times10^{-8}$$

$$\alpha_1 = \frac{1.32\times10^{-2}nm}{a(nm)} = \frac{1.32\times10^{-2}nm}{5(nm)} = 2.64\times10^{-3}$$

$$m=0 \quad y = [2.64\times10^{-3}x^2 + 5.54\times10^{-8}x^4 + x^2 I_0(x)K_0(x)]^{1/2}$$

$$m=1 \quad y = [2.64\times10^{-3}(x^2+1) + 5.54\times10^{-8}(x^2+1)^2 + (x^2+1)I_1(x)K_1(x)]^{1/2}$$

$$m=2 \quad y = [2.64\times10^{-3}(x^2+4) + 5.54\times10^{-8}(x^2+4)^2 + (x^2+4)I_2(x)K_2(x)]^{1/2}$$

$$m=3 \quad y = [2.64\times10^{-3}(x^2+9) + 5.54\times10^{-8}(x^2+9)^2 + (x^2+9)I_3(x)K_3(x)]^{1/2}$$

$$m=4 \quad y = [2.64\times10^{-3}(x^2+16) + 5.54\times10^{-8}(x^2+16)^2 + (x^2+16)I_4(x)K_4(x)]^{1/2}$$

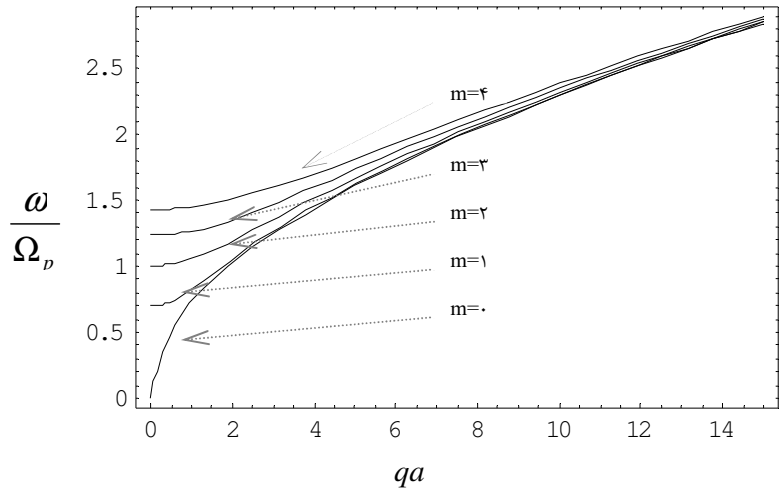

شکل (۳-۲)- نمودارهای پاشندگی با مدهای عرضی مختلف در نانوتیوپ کربنی تک‌جداره

همان‌طور که پیش‌تر در نکته اشاره شد، این نمودار هم نشانگر رابطه‌ی فرکانس و عکس عمق نفوذ در نانوتیوپ بوده و هم نمایان‌گر رابطه‌ی فرکانس و عدد موج q است.

مطابق نمودار، سرعت گروه برای مدهای عرضی غیر صفر در یک فرکانس خاص برابر با صفر می‌شود. در این فرکانس موج ایستا در نانوتیوپ تک‌جداره تشکیل خواهد شد.

همچنین فرکانس قطع در نانوتیوپ فوق برای مد عرضی صفر برابر با صفر بوده و با افزایش مدهای عرضی مقدار فرکانس قطع بیشتر می‌شود.

شیب مجانبی نمودار پاشندگی برای همه‌ی مدهای عرضی مشابه است و معادلات پاشندگی تنها در $qa$ های کمتر از ۱۲ با یکدیگر تفاوت دارند. از این مقدار به بالا نمودارهای پاشندگی مدهای مختلف تقریبا بر هم منطبق هستند.

با در نظر گرفتن نمودار فوق برای فرکانس و عکس عمق نفوذ به این نتیجه می‌رسیم که برای نانوتیوپ کربنی تک‌جداره با هر مد عرضی، افزایش فرکانس‌ها باعث کاهش عمق نفوذ شعاعی می‌شود. به این معنی که هرچه فرکانس عبوری از نانوتیوپ تک‌جداره بیشتر باشد، عمق نفوذ (و یا میزان گستردگی میدان **الکتریکی** در جهت شعاعی) کمتر خواهد شد.

در شکل (۳-۳) نمودارهای معادلات پاشندگی در مد عرضی یک و شعاع‌های $a = 2, 5, 10, 15 nm$ رسم شده‌اند. در این شکل وابستگی معادله‌ی پاشندگی به شعاع نانوتیوپ تک‌جداره نشان داده می‌شود.

$$\beta_1 = \frac{6.92 \times 10^{-6}}{[a(nm)]^3}$$

$$\alpha_1 = \frac{1.32 \times 10^{-2} nm}{a(nm)}$$

$m=1,$
$a=2nm \quad \alpha_1=6.61\times10^{-3}, \beta_1=8.65\times10^{-7} \quad y=[6.61\times10^{-3}(x^2+1)+8.65\times10^{-7}(x^2+1)^2+(x^2+1)I_1(x)K_1(x)]^{1/2}$
$a=5nm \quad \alpha_1=2.64\times10^{-3}, \beta_1=5.54\times10^{-8} \quad y=[2.64\times10^{-3}(x^2+1)+5.54\times10^{-8}(x^2+1)^2+(x^2+1)I_1(x)K_1(x)]^{1/2}$
$a=10nm \quad \alpha_1=1.32\times10^{-3}, \beta_1=5.54\times10^{-11} \quad y=[1.32\times10^{-3}(x^2+1)+5.54\times10^{-11}(x^2+1)^2+(x^2+1)I_1(x)K_1(x)]^{1/2}$
$a=15nm \quad \alpha_1=8.81\times10^{-4}, \beta_1=1.64\times10^{-11} \quad y=[8.81\times10^{-4}(x^2+1)+1.64\times10^{-11}(x^2+1)^2+(x^2+1)I_1(x)K_1(x)]^{1/2}$

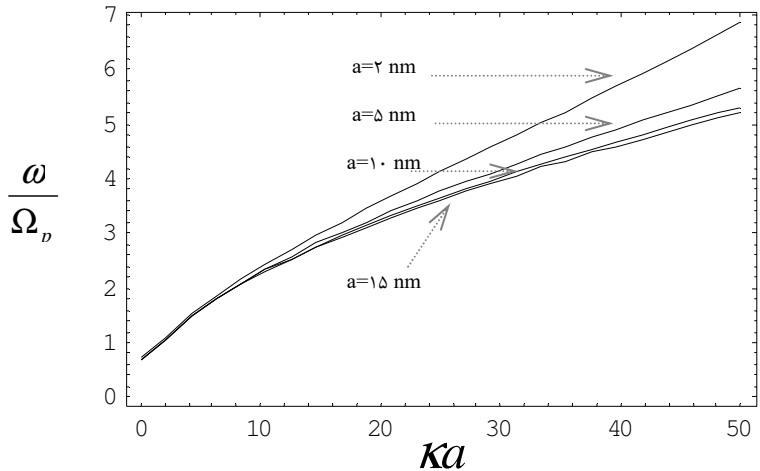

شکل(۳-۳)- نمودارهای پاشندگی با شعاع‌های مختلف نانوتیوپ کربنی تک‌جداره

این نمودار نشان می‌دهد که شعاع نانوتیوپ روی فرکانس قطع آن تاثیری ندارد و برای تمام شعاع‌های بررسی شده در مد عرضی یک، فرکانس قطع ثابت است. این امر نشان می‌دهد که فرکانس قطع تنها در مدهای عرضی متفاوت تغییر می‌کند.

مطابق این نمودار با افزایش شعاع نانوتیوپ شیب منحنی پاشندگی که نمایان‌گر سرعت گروه امواج ایجاد شده در نانوتیوپ است کاهش می‌یابد و با زیاد شدن شعاع نانوتیوپ‌ها، اختلاف شیب در نمودارهای پاشندگی کاهش می‌یابد.

همچنین این نمودار نشان می‌دهد که با افزایش فرکانس‌ها، عمق نفوذ شعاعی در نانوتیوپ کم خواهد شد. به این معنی که هرچه فرکانس عبوری از نانوتیوپ تک‌جداره بیشتر باشد، عمق نفوذ (و یا میزان گستردگی میدان **الکتریکی** در جهت شعاعی) کمتر می‌شود.

برای بررسی اثر نیروهای برهم‌کنش داخلی در گاز الکترونی دوبعدی، برای معادله پاشندگی مد TM ، به صورت زیر عمل می‌کنیم.

ضرایب $\alpha, \beta$ در معادله‌ی حرکت حاکم بر الکترون‌های گاز الکترون آزاد دوبعدی، مربوط به نیروهای برهم‌کنشی داخلی هستند.

$$\frac{\partial v}{\partial t}=-\frac{e}{m}E_{ll}-\frac{\alpha}{n_0}\nabla\delta n+\frac{\beta}{n_0}\nabla(\nabla^2\delta n)$$

معادلات پاشندگی را با در نظر گرفتن یا صرف‌نظرکردن از بعضی یا همه‌ی این ضرایب ($\alpha, \beta$)، در مـد عرضـی m=۰ و شعاع $a = 5nm$ به دست می‌آوریم. نمودارهای این معادلات در شکل(۳-۴) رسم شده‌اند.

$$m=0, a=5nm \rightarrow \alpha_1 = \alpha/(\Omega_p a)^2 = \frac{1.32 \times 10^{-2} nm}{a(nm)} = 2.64 \times 10^{-3}$$

$$\beta_1 = \beta/(\Omega_p^2 a^4) = \frac{6.92 \times 10^{-6}}{[a(nm)]^3} = 5.54 \times 10^{-8}$$

$$\alpha_1, \beta_1 \neq 0 \rightarrow y = [2.64 \times 10^{-3} x^2 + 5.54 \times 10^{-8} x^4 + x^2 I_0(x) K_0(x)]^{1/2}$$
$$\alpha_1 = 0 \rightarrow y = [5.54 \times 10^{-8} x^4 + x^2 I_0(x) K_0(x)]^{1/2}$$
$$\alpha_1 = \beta_1 = 0 \rightarrow y = [x^2 I_0(x) K_0(x)]^{1/2}$$

**نکته:**

شرط‌های $\alpha_1 or \beta_1 = 0$ تنها هنگامی رخ می‌دهند که:
۱) $\alpha, \beta$ صفر باشند (نیروهای داخلی در گاز الکترونی ناچیز فرض شوند.)
۲) شعاع نانولوله‌ی کربنی بی‌نهایت باشد.

در این قسمت با توجه به این‌که شعاع نانوتیوپ کربنی را ثابت و برابر با $5nm$ در نظر گرفته‌ایـم، صـفر فـرض کردن ضرایب $\alpha_1, \beta_1$ تنها از ناچیز انگاشتن نیروهای داخلی در گاز الکترونی نتیجه می‌شود.

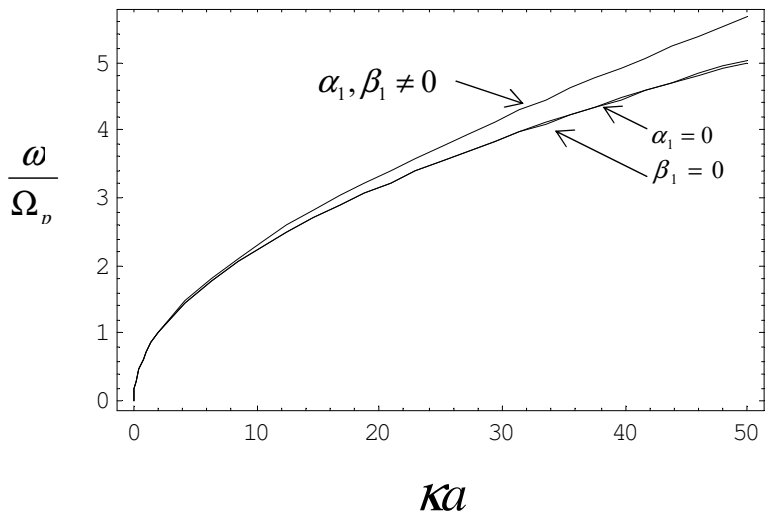

شکل (۳-۴)- نمودارهای پاشندگی با در نظر گرفتن نیروهای داخلی و بدون درنظر گرفتن آن‌ها و در نظر گرفتن بعضی از آن‌ها

با مقایسه نمودارها در شکل (۳-۴) درمی‌یابیم که نیروهای برهم‌کنش داخلی اثر زیادی روی معادله پاشندگی مد TM نداشته و تنها شیب مجانبی آن‌را اندکی کاهش می‌دهند.

## 3-2- مد TE

### 3-2-1- معادله پاشندگی مد TE برای نانوتیوپ کربنی تک جداره

شرایط مرزی برای میدان الکتریکی نانوتیوپ تک‌جداره در بخش مد TM و در قالب معادلات (3-1) و (3-2) نشان داده شد.

با توجه به معادله‌ی تعمیم‌یافته‌ی بسل برای میدان مغناطیسی (معادله‌ی (2-10)) و نمودارهای توابع بسل تعمیم‌یافته‌ی نوع اول و دوم که در شکل (3-1) نشان داده شده‌اند، جواب‌های فرضی میدان مغناطیسی در اطراف نانولوله‌ی کربنی تک‌جداره را به صورت زیر در نظر می‌گیریم.

$$B_{zm}(r) = C_m I_m(\kappa r) \quad r < a$$
$$B_{zm}(r) = D_m K_m(\kappa r) \quad r > a$$
(3-5)

با در نظر گرفتن مد TE ($E_z = 0$)، مولفه‌های $r, \varphi$ میدان الکتریکی بر حسب مولفه‌ی z که قبلا به صورت رابطه‌ی (2-9) بوده به شکل زیر تبدیل می‌شوند.

$$E_{\varphi m} = \frac{i\omega}{\kappa^2}\frac{\partial B_{zm}}{\partial r} \qquad \kappa^2 = q^2 - k^2 \quad , k = \frac{\omega}{c}$$
$$E_{rm} = \frac{m\omega}{\kappa^2 r} B_{zm}$$
(3-6)

معادلات (3-6) را در معادلات شرط مرزی (3-1) و (3-2) جایگزین می‌کنیم.

$$\frac{m\omega}{\kappa^2 r}(B_{zm}|_{r>a} - B_{zm}|_{r<a}) = -\frac{e}{\varepsilon_0} N_m$$

$$\begin{cases} (\frac{\partial B_{zm}}{\partial r}|_{r>a} - \frac{\partial B_{zm}}{\partial r}|_{r<a}) = 0 \\ \frac{i\omega}{\kappa^2}(\frac{\partial B_{zm}}{\partial r}|_{r>a} - \frac{\partial B_{zm}}{\partial r}|_{r<a}) = 0 \end{cases}$$

دو معادله‌ی شرط مرزی که در کروشه به دست آمده‌اند مشابه بوده و از هم مستقل نیست. حال معادلات (3-5) را در دو معادله‌ی شرط مرزی بالا جایگزین می‌کنیم.

$$\frac{m\omega}{\kappa^2 a}(D_m K_m(\kappa a) - C_m I_m(\kappa a)) = -\frac{e}{\varepsilon_0}(-i\frac{en_0}{m_e}\frac{(qE_{zm} + \frac{m}{a}[\frac{i\omega}{\kappa}\frac{\partial B_{zm}}{\partial(\kappa r)}])}{\omega^2 - \alpha(q^2 + \frac{m^2}{a^2}) - \beta(q^2 + \frac{m^2}{a^2})^2})$$

$$\frac{i\omega}{\kappa}(D_m \frac{\partial K_m(\kappa a)}{\partial(\kappa r)} - C_m \frac{\partial I_m(\kappa a)}{\partial(\kappa r)}) = 0 \rightarrow D_m = \frac{I'_m(\kappa a)}{K'_m(\kappa a)} C_m$$

$$\frac{m\omega}{\kappa^2 a}[(\frac{I'_m(\kappa a)}{K'_m(\kappa a)}C_m)K_m(\kappa a) - C_m I_m(\kappa a)] = \frac{ie^2 n_0}{\varepsilon_0 m_e} \frac{\frac{im\omega}{\kappa a}C_m I'_m(\kappa a)}{\omega^2 - \alpha(q^2 + \frac{m^2}{a^2}) - \beta(q^2 + \frac{m^2}{a^2})^2}$$

با حذف $C_m$ از طرفین و ساده کردن معادله خواهیم داشت:

$$\frac{1}{\kappa}[\frac{I'_m(\kappa a)}{K'_m(\kappa a)}K_m(\kappa a) - I_m(\kappa a)] = -\frac{e^2 n_0}{\varepsilon_0 m_e} \frac{I'_m(\kappa a)}{\omega^2 - \alpha(q^2 + \frac{m^2}{a^2}) - \beta(q^2 + \frac{m^2}{a^2})^2}$$

$$\frac{1}{\kappa}[I'_m(\kappa a)K_m(\kappa a) - I_m(\kappa a)K'_m(\kappa a)][\omega^2 - \alpha(q^2 + \frac{m^2}{a^2}) - \beta(q^2 + \frac{m^2}{a^2})^2]$$

$$= -\frac{e^2 n_0}{\varepsilon_0 m_e} I'_m(\kappa a)K'_m(\kappa a)$$

با استفاده از تعریف رونسکین ($I'_m(x)K_m(x) - I_m(x)K'_m(x) = 1/x$) رابطه به شکل زیر ساده می‌شود.

$$[1/(\kappa a)][\omega^2 - \alpha(q^2 + \frac{m^2}{a^2}) - \beta(q^2 + \frac{m^2}{a^2})^2] = -\frac{e^2 n_0 \kappa}{\varepsilon_0 m_e} I'_m(\kappa a)K'_m(\kappa a)$$

$$[\omega^2 - \alpha(q^2 + \frac{m^2}{a^2}) - \beta(q^2 + \frac{m^2}{a^2})^2] = -\frac{e^2 n_0}{\varepsilon_0 m_e a}(\kappa a)^2 I'_m(\kappa a)K'_m(\kappa a)$$

معادله پاشندگی $\omega(\kappa)$

$$\omega^2 - \alpha(\kappa^2 + \omega^2/c^2 + \frac{m^2}{a^2}) - \beta(\kappa^2 + \omega^2/c^2 + \frac{m^2}{a^2})^2 = -\Omega_P^2(\kappa a)^2 I'_m(\kappa a)K'_m(\kappa a)$$

$$\Omega_P = (\frac{e^2 n_0}{\varepsilon_0 m_e a})^{1/2}, \kappa^2 = q^2 - \omega^2/c^2$$

## ۳-۲-۲- نمودارهای معادله‌ی پاشندگی در شرایط مختلف

در این بخش معادله‌ی پاشندگی نانوتیوپ تک‌جداره در مد TE را در حالات مختلف رسم خواهیم کرد.

$$\omega^2 - \alpha(\kappa^2 + \omega^2/c^2 + \frac{m^2}{a^2}) - \beta(\kappa^2 + \omega^2/c^2 + \frac{m^2}{a^2})^2 = -\Omega_P^2(\kappa a)^2 I'_m(\kappa a) K'_m(\kappa a)$$

$$\Omega_p = (e^2 n_0 / \varepsilon_0 m_e a)^{1/2}$$

$$\kappa = (q^2 - k^2); k = \omega/c$$

به منظور بی‌بعد کردن معادله‌ی پاشندگی پارامترهای بی‌بعد زیر را تعریف می‌نماییم.

$$x = \kappa a$$
$$y = \omega/\Omega_p$$

طرفین رابطه‌ی پاشندگی را بر $\Omega_P^2$ تقسیم می‌کنیم.

$$y^2 - \frac{\alpha}{\Omega_p^2}(\kappa^2 + \omega^2/c^2 + m^2/a^2) - \frac{\beta}{\Omega_p^2}(\kappa^2 + \omega^2/c^2 + m^2/a^2)^2 = -x^2 I'_m(\kappa a) K'_m(\kappa a)$$

$$y^2 - \frac{\alpha}{\Omega_P^2 a^2}(x^2 + \frac{\Omega_P^2 a^2}{c^2}\frac{\omega^2}{\Omega_P^2} + m^2) - \frac{\beta}{\Omega_P^2 a^4}(x^2 + \frac{\Omega_P^2 a^2}{c^2}\frac{\omega^2}{\Omega_P^2} + m^2)^2 = -x^2 I'_m(\kappa a) K'_m(\kappa a)$$

$$y^2 - \frac{\alpha}{\Omega_P^2 a^2}(x^2 + \frac{\Omega_P^2 a^2}{c^2}y^2 + m^2) - \frac{\beta}{\Omega_P^2 a^4}(x^2 + \frac{\Omega_P^2 a^2}{c^2}y^2 + m^2)^2 = -x^2 I'_m(\kappa a) K'_m(\kappa a)$$

با تعریف ضرایب بی‌بعد $\alpha_1 = \alpha/(\Omega_p a)^2, \beta_1 = \beta/(\Omega_p^2 a^4), \sigma = \Omega_p^2 a^2/c^2$، معادله به صورت زیر ساده می‌شود.

$$y^2 - \alpha_1(x^2 + \sigma y^2 + m^2) - \beta_1(x^2 + \sigma y^2 + m^2)^2 = -x^2 I'_m(\kappa a) K'_m(\kappa a)$$

اثبات بی‌بعد بودن ضرایب معادله‌ی بالا در بخش ۲-۱-۳ آورده شده است.
همان‌طور که در بخش ۲-۱-۲ اشاره شد مقدار $\sigma$ نسبت به ضرایب دیگر $y^2$ در معادله پاشندگی کوچک است و می‌توان از جمله $\sigma y^2$ در معادله پاشندگی صرف‌نظر کرد.
با حذف $\sigma y^2$ معادله به شکل زیر تبدیل می‌شود.

$$y^2 = \alpha_1(x^2 + m^2) + \beta_1(x^2 + m^2)^2 - x^2 I'_m(x) K'_m(x)$$
$$y = \pm[\alpha_1(x^2 + m^2) + \beta_1(x^2 + m^2)^2 - x^2 I'_m(x) K'_m(x)]^{1/2}$$

مقادیر منفی برای $y$ یا فرکانس، جواب‌های غیر فیزیکی مسئله هستند و از آن‌ها صرف‌نظر می‌کنیم.

**نکته:**

همان‌طور که در بخش ۳-۱-۲ نیز اشاره شد، فرم معادله‌ی فرکانس بر حسب عکس عمق نفوذ و عدد موج یکی است. در این بخش نیز به تایید مجدد این امر می‌پردازیم.

$$\omega^2 - \alpha(\kappa^2 + \omega^2/c^2 + \frac{m^2}{a^2}) - \beta(\kappa^2 + \omega^2/c^2 + \frac{m^2}{a^2})^2 = -\Omega_P^2(\kappa a)^2 I'_m(\kappa a) K'_m(\kappa a)$$

$$\frac{\omega^2}{\Omega_P^2} - \frac{\alpha}{\Omega_P^2 a^2}([\kappa^2 + \frac{\omega^2}{c^2}]a^2 + m^2) - \frac{\beta}{\Omega_P^2 a^4}([\kappa^2 + \frac{\omega^2}{c^2}]a^2 + m^2)^2 = -(\kappa a)^2 I'_m(\kappa a) K'_m(\kappa a)$$

$$y = \frac{\omega}{\Omega_p}, x = qa \Rightarrow \begin{cases} \kappa^2 + \omega^2/c^2 = q^2, \frac{\alpha}{\Omega_P^2 a^2} = \alpha_1, \frac{\beta}{\Omega_P^2 a^4} = \beta_1 \\ \kappa = (q^2 - \omega^2/c^2)^{1/2}, \kappa a = [(qa)^2 - \frac{\Omega_P^2 a^2}{c^2}\frac{\omega^2}{\Omega_P^2}]^{1/2} = (x^2 - \sigma y^2)^{1/2}; \sigma = \Omega_P^2 a^2/c^2 \end{cases}$$

$$y^2 - \alpha_1(x^2 + m^2) - \beta_1(x^2 + m^2)^2 = -(x^2 - \sigma y^2) I'_m([x^2 - \sigma y^2]^{1/2}) K'_m([x^2 - \sigma y^2]^{1/2})$$

ملاحظه می‌شود که با حذف $\sigma y^2$ معادله به شکل معادله‌ی فرکانس برحسب معکوس عمق نفوذ در می‌آید.

$$y^2 - \alpha_1(x^2 + m^2) - \beta_1(x^2 + m^2)^2 = -x^2 I'_m(x) K'_m(x)$$

با توجه به مقدار شعاع بور $a_B = 5.29 \times 10^{-11} m \approx 0.53 A^0 \approx 0.05 nm$، ضرایب $\alpha_1, \beta_1$ برحسب شعاع نانوتیوپ و چگالی الکترونی سطحی در آن نوشته می‌شوند.

$$\alpha_1 = \frac{\alpha}{(\Omega_p a)^2} = \frac{e^2 n_0 a_B}{4\varepsilon_0 m_e} \cdot \frac{1}{(\frac{e^2 n_0}{\varepsilon_0 m_e a})a^2} = \frac{a_B}{4a} = \frac{1.32 \times 10^{-11} m}{a(m)} = \frac{1.32 \times 10^{-2} nm}{a(nm)}$$

$$n_0 = 152 nm^{-2} (for \quad CNT)$$

$$\beta_1 = \frac{\beta}{(\Omega_P^2 a^4)} = \frac{e^2 a_B}{16\pi\varepsilon_0 m_e} \cdot \frac{1}{(\frac{e^2 n_0}{\varepsilon_0 m_e a})a^4} = \frac{a_B}{16\pi n_0 a^3}$$

$$= \frac{5.29 \times 10^{-11} m}{16 \times 3.14 \times 4 \times 38 \times 10^{18} m^{-2} \times a^3(m^3)} = \frac{6.92 \times 10^{-4} \times 10^{-29}}{[a(m)]^3} = \frac{6.92 \times 10^{-6}}{[a(nm)]^3}$$

معادلات پاشندگی مد TE به ازای مدهای عرضی m=۰،۱،۲،۳،۴ برای نانوتیوپ تک‌جداره به شعاع ۵nm در شکل (۳-۷) نشان داده شده‌اند.

$$\alpha_1 = \frac{1.32 \times 10^{-2} nm}{a(nm)} = \frac{1.32 \times 10^{-2} nm}{5(nm)} = 2.64 \times 10^{-3}$$

$$\beta_1 = \frac{6.92 \times 10^{-6}}{[a(nm)]^3} = \frac{6.92 \times 10^{-6}}{[5(nm)]^3} = 5.54 \times 10^{-8}$$

$m=0 \quad y = [2.64 \times 10^{-3} x^2 + 5.54 \times 10^{-8} x^4 - x^2 I_0'(x) K_0'(x)]^{1/2}$

$m=1 \quad y = [2.64 \times 10^{-3}(x^2+1) + 5.54 \times 10^{-8}(x^2+1)^2 - x^2 I_1'(x) K_1'(x)]^{1/2}$

$m=2 \quad y = [2.64 \times 10^{-3}(x^2+4) + 5.54 \times 10^{-8}(x^2+4)^2 - x^2 I_2'(x) K_2'(x)]^{1/2}$

$m=3 \quad y = [2.64 \times 10^{-3}(x^2+9) + 5.54 \times 10^{-8}(x^2+9)^2 - x^2 I_3'(x) K_3'(x)]^{1/2}$

$m=4 \quad y = [2.64 \times 10^{-3}(x^2+16) + 5.54 \times 10^{-8}(x^2+16)^2 - x^2 I_4'(x) K_4'(x)]^{1/2}$

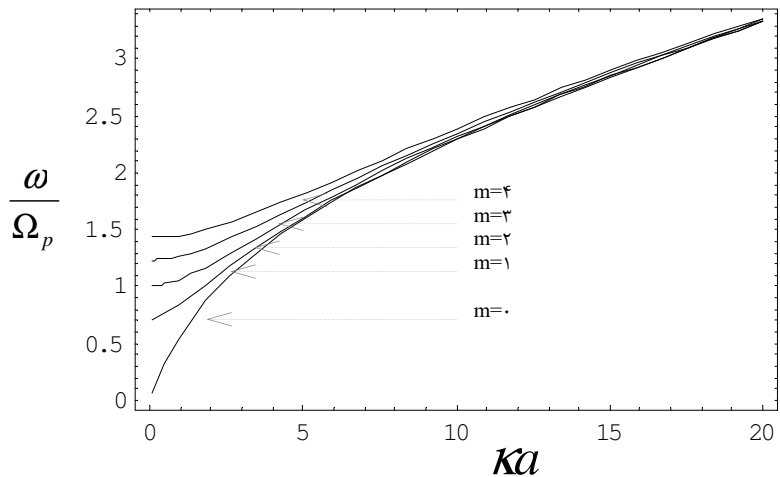

شکل (۳-۷) – نمودارهای پاشندگی نانوتیوپ تک‌جداره در مد TE برای m=۰،۱،۲،۳،۴

برای نشان دادن وابستگی معادله پاشندگی به شعاع نانوتیوپ، معادله‌ی پاشندگی را در مد عرضی یـک ($m=1$) و بـرای شعاع‌های مختلف ($a = 2,5,10 nm$) رسم می‌کنیم. نمودارهای مربوطه در شکل (۳-۸) آورده شده‌اند.

$a = 2nm \quad \alpha_1 = 6.61 \times 10^{-3}, \beta_1 = 8.65 \times 10^{-7} \quad y = [6.61 \times 10^{-3}(x^2+1) + 8.65 \times 10^{-7}(x^2+1)^2 - x^2 I_1'(x) K_1'(x)]^{1/2}$

$a = 5nm \quad \alpha_1 = 2.64 \times 10^{-3}, \beta_1 = 5.54 \times 10^{-8} \quad y = [2.64 \times 10^{-3}(x^2+1) + 5.54 \times 10^{-8}(x^2+1)^2 - x^2 I_1'(x) K_1'(x)]^{1/2}$

$a = 10nm \quad \alpha_1 = 1.32 \times 10^{-3}, \beta_1 = 5.54 \times 10^{-11} \quad y = [1.32 \times 10^{-3}(x^2+1) + 5.54 \times 10^{-11}(x^2+1)^2 - x^2 I_1'(x) K_1'(x)]^{1/2}$

$a = 15nm \quad \alpha_1 = 8.81 \times 10^{-4}, \beta_1 = 1.64 \times 10^{-11} \quad y = [8.81 \times 10^{-4}(x^2+1) + 1.64 \times 10^{-11}(x^2+1)^2 - x^2 I_1'(x) K_1'(x)]^{1/2}$

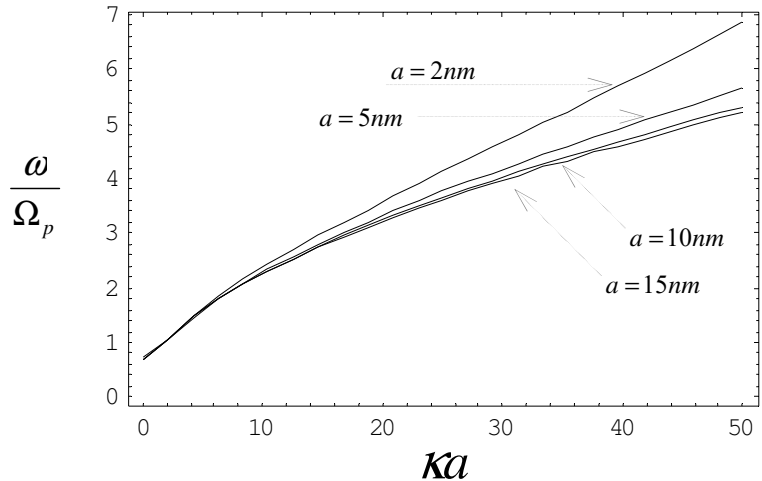

شکل (۳-۸) – نمودارهای پاشندگی نانوتیوپ تک‌جداره در مد TE برای شعاع‌های $a = 2, 5, 10, 15\ nm$

معادلات پاشندگی را با در نظر گرفتن یا صرف‌نظر کردن از بعضی یا همه‌ی این ضرایب، در مد عرضی m=۰ و شعاع $a = 5nm$ به دست می‌آوریم. نمودارهای این معادلات در شکل(۳-۹) رسم شده‌اند. همان‌طور که می‌بینیم حذف نیروهای برهم‌کنش داخلی، تاثیر محسوسی بر معادله‌ی پاشندگی نمی‌گذارد.

$$m = 0, a = 5nm \rightarrow \alpha_1 = \alpha/(\Omega_p a)^2 = \frac{1.32 \times 10^{-2}\ nm}{a(nm)} = 2.64 \times 10^{-3}$$

$$\beta_1 = \beta/(\Omega_p^2 a^4) = \frac{6.92 \times 10^{-6}}{[a(nm)]^3} = 5.54 \times 10^{-8}$$

$$\alpha_1, \beta_1 \neq 0 \quad \rightarrow y = [2.64 \times 10^{-3} x^2 + 5.54 \times 10^{-8} x^4 - x^2 I_0'(x) K_0'(x)]^{1/2}$$
$$\alpha_1 = 0 \quad \quad \rightarrow y = [5.54 \times 10^{-8} x^4 - x^2 I_0'(x) K_0'(x)]^{1/2}$$
$$\alpha_1 = \beta_1 = 0 \quad \rightarrow y = [-x^2 I_0'(x) K_0'(x)]^{1/2}$$

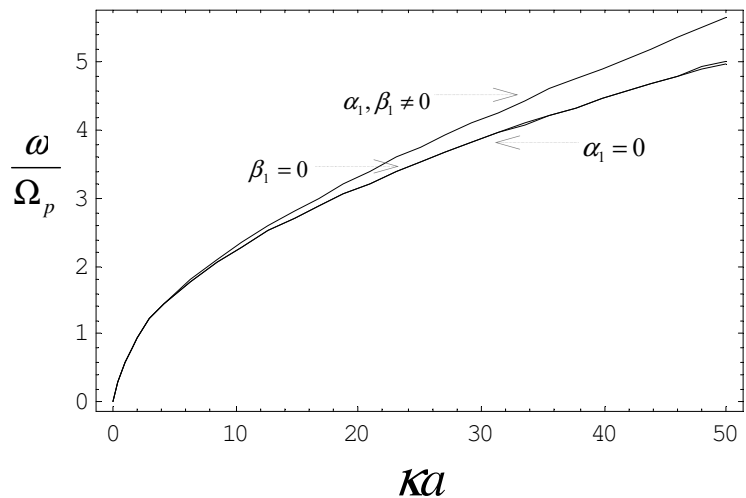

شکل (۳-۹) – نمودارهای پاشندگی با در نظر گرفتن نیروهای داخلی یا بدون در نظر گرفتن آن‌ها

نمودارهای پاشندگی مد TE مشابه با نمودارهای پاشندگی مد TM هستند که در بخش قبل رسم شده‌اند. تشابه این نمودارها بدان دلیل است که دو تابع $(x^2+m^2)I_m(x)K_m(x)$ و $-x^2 I'_m(x)K'_m(x)$ که تنها تفاوتِ معادلات پاشندگی مدهای TE,TM هستند، در x های کمتر از ۵ اختلاف بسیار ناچیزی با یکدیگر داشته و برای x های بزرگتر از ۵ کاملا بر هم منطبق می‌شوند.

لازم به ذکر است که اگر مد TE در سیستم ایجاد شود، چون میدان الکتریکی مولفه‌ای در راستای انتشار موج ندارد، پلزمون‌های سطحی عرضی در گاز الکترونی عامل انتقال انرژی خواهند بود.

معانی برگرفته شده از نمودارها در این بخش کاملا مشابه با بخش مد TM بوده و تنها این تفاوت در تعریف وجود دارد که عمق نفوذ در این مد مربوط به میزان گستردگی میدان **مغناطیسی** در جهت شعاعی نانوتیوپ خواهد بود.

# فصل چهارم – نانوتیوپ کربنی دو جداره

## ۴-۱- مد TM

### ۴-۱-۱- معادله پاشندگی مد TM برای نانوتیوپ کربنی دوجداره

معادلات شرط مرزی برای میدان الکتریکی در اطراف یک نانولوله‌ی کربنی دوجداره به شعاع‌های درونی و بیرونی $a_1$ و $a_2$، با در نظر گرفتن بسط موج تخت برای میدان و چگالی بار الکترونی نانولوله (معادلات (۱-۲) و (۲-۲) ) به صورت زیر هستند.

پوسته‌ی درونی:

$$E_{rm}(a_1)\big|_{r>a_1} - E_{rm}(a_1)\big|_{r<a_1} = -\frac{eN_{m1}}{\varepsilon_0}$$

$$E_{llm}(a_1)\big|_{r>a_1} - E_{llm}(a_1)\big|_{r<a_1} = 0 \rightarrow \begin{cases} E_{zm}(a_1)\big|_{r>a_1} - E_{zm}(a_1)\big|_{r<a_1} = 0 \\ E_{\varphi m}(a_1)\big|_{r>a_1} - E_{\varphi m}(a_1)\big|_{r<a_1} = 0 \end{cases}$$

$$N_{m1} = -i\frac{en_0}{m_e}\frac{(qE_{zm} + \frac{m}{a_1}E_{\varphi m})}{\omega^2 - \alpha(q^2 + \frac{m^2}{a_1^2}) - \beta(q^2 + \frac{m^2}{a_1^2})^2}$$

پوسته بیرونی:

$$E_{rm}(a_2)\big|_{r>a_2} - E_{rm}(a_2)\big|_{r<a_2} = -\frac{eN_{m2}}{\varepsilon_0}$$

$$E_{llm}(a_2)\big|_{r>a_2} - E_{llm}(a_2)\big|_{r<a_2} = 0 \rightarrow \begin{cases} E_{zm}(a_2)\big|_{r>a_2} - E_{zm}(a_2)\big|_{r<a_2} = 0 \\ E_{\varphi m}(a_2)\big|_{r>a_2} - E_{\varphi m}(a_2)\big|_{r<a_2} = 0 \end{cases}$$

$$N_{m2} = -i\frac{en_0}{m_e}\frac{(qE_{zm} + \frac{m}{a_2}E_{\varphi m})}{\omega^2 - \alpha(q^2 + \frac{m^2}{a_2^2}) - \beta(q^2 + \frac{m^2}{a_2^2})^2}$$

با توجه به معادله‌ی (۹-۲) مولفه‌های $r, \varphi$ میدان الکتریکی در مد TM ($B_z = 0$) بر حسب مولفه‌ی z میدان الکتریکی، به شکل زیر خواهند بود.

$$E_{\varphi m} = \frac{qm}{\kappa^2 r}E_{zm} \quad ; \kappa^2 = q^2 - k^2 \quad , k = \frac{\omega}{c}$$

$$E_{rm} = -\frac{iq}{\kappa^2}\frac{\partial E_{zm}}{\partial r}$$

روابط بالا را در معادلات شرط مرزی دو پوسته‌ی بیرونی و درونی قرار می‌دهیم.

$$-\frac{iq}{\kappa^2}(\frac{\partial E_{zm}(a_1)}{\partial r}\Big|_{r>a_1} - \frac{\partial E_{zm}(a_1)}{\partial r}\Big|_{r<a_1}) = -\frac{e}{\varepsilon_0}N_{m1} \tag{4-1}$$

$$\begin{cases} E_{zm}(a_1)\Big|_{r>a_1} - E_{zm}(a_1)\Big|_{r<a_1} = 0 \\ \frac{qm}{\kappa^2 r}(E_{zm}(a_1)\Big|_{r>a_1} - E_{zm}(a_1)\Big|_{r<a_1}) = 0 \end{cases} \tag{4-2}$$

$$-\frac{iq}{\kappa^2}(\frac{\partial E_{zm}(a_2)}{\partial r}\Big|_{r>a_2} - \frac{\partial E_{zm}(a_2)}{\partial r}\Big|_{r<a_2}) = -\frac{e}{\varepsilon_0}N_{m2} \tag{4-3}$$

$$\begin{cases} E_{zm}(a_2)\Big|_{r>a_2} - E_{zm}(a_2)\Big|_{r<a_2} = 0 \\ \frac{qm}{\kappa^2 r}(E_{zm}(a_2)\Big|_{r>a_2} - E_{zm}(a_2)\Big|_{r<a_2}) = 0 \end{cases} \tag{4-4}$$

معادلات شرط مرزی در کروشه‌ها یکسان هستند.

مولفه‌های z میدان الکتریکی در معادله‌ی بسل تعمیم‌یافته صدق می‌کنند (معادلات (۲-۱۰)) و جواب فرضی میدان در نواحی مختلف نانوتیوپ بر حسب فاصله‌ی شعاعی تا محور آن به شکل زیر است.

$$\begin{aligned} E_{zm}(r) &= C_{1m} I_m(\kappa r) & r < a_1 \\ E_{zm}(r) &= C_{2m} I_m(\kappa r) + C_{3m} K_m(\kappa r) & a_1 < r < a_2 \\ E_{zm}(r) &= C_{4m} K_m(\kappa r) & r > a_2 \end{aligned}$$

مقدار $N_{m1}$ را در معادله‌ی (۴-۱) قرار می‌دهیم.

$$-\frac{iq}{\kappa^2}(\frac{\partial E_{zm}(a_1)}{\partial r}\Big|_{r>a_1} - \frac{\partial E_{zm}(a_1)}{\partial r}\Big|_{r<a_1}) = -\frac{e}{\varepsilon_0}(-i\frac{en_0}{m_e}\frac{(qE_{zm}+\frac{m}{a_1}E_{\varphi m})}{\omega^2 - \alpha(q^2+\frac{m^2}{a_1^2}) - \beta(q^2+\frac{m^2}{a_1^2})^2})$$

$$\frac{q}{\kappa^2}(\frac{\partial E_{zm}(a_1)}{\partial r}\Big|_{r>a_1} - \frac{\partial E_{zm}(a_1)}{\partial r}\Big|_{r<a_1}) = -\frac{e^2 n_0}{m_e \varepsilon_0}\frac{(qE_{zm}+\frac{m}{a_1}E_{\varphi m})}{\omega^2 - \alpha q_1^2 - \beta q_1^4} \quad ; q_1^2 = q^2 + \frac{m^2}{a_1^2}$$

$$E_{\varphi m} = \frac{qm}{\kappa^2 r}E_{zm} \Rightarrow$$

$$\frac{q}{\kappa^2}(\frac{\partial E_{zm}(a_1)}{\partial r}\Big|_{r>a_1} - \frac{\partial E_{zm}(a_1)}{\partial r}\Big|_{r<a_1}) = -\frac{e^2 n_0}{m_e \varepsilon_0}\frac{(qE_{zm}+\frac{qm^2}{\kappa^2 a_1^2}E_{zm})}{\omega^2 - \alpha q_1^2 - \beta q_1^4}$$

با استفاده از جواب‌های فرضی مولفه‌ی $z$ میدان الکتریکی، معادله‌ی بالا به شکل زیر در می‌آید.

$$\frac{1}{\kappa^2}\left(\frac{\partial E_{zm}(a_1)}{\partial r}\bigg|_{r>a_1} - \frac{\partial E_{zm}(a_1)}{\partial r}\bigg|_{r<a_1}\right)(\omega^2 - \alpha q_1^2 - \beta q_1^4) = -\frac{e^2 n_0}{m_e \varepsilon_0}(1+\frac{m^2}{\kappa^2 a_1^2})E_{zm}$$

$$\frac{1}{\kappa}\left(C_{2m}\frac{\partial I(\kappa r)}{\partial (\kappa r)}\bigg|_{r>a_1} + C_{3m}\frac{\partial K(\kappa r)}{\partial (\kappa r)}\bigg|_{r>a_1} - C_{1m}\frac{\partial I(\kappa r)}{\partial (\kappa r)}\bigg|_{r<a_1}\right)(\omega^2 - \alpha q_1^2 - \beta q_1^4) = -\frac{e^2 n_0}{m_e \varepsilon_0}(1+\frac{m^2}{\kappa^2 a_1^2})C_{1m}I(\kappa a_1)$$

$$(\omega^2 - \alpha q_1^2 - \beta q_1^4)[C_{2m}I'_m(\kappa a_1) + C_{3m}K'_m(\kappa a_1) - C_{1m}I'_m(\kappa a_1)] = -\frac{e^2 n_0}{m_e \varepsilon_0}\kappa(1+\frac{m^2}{\kappa^2 a_1^2})C_{1m}I_m(\kappa a_1)$$

$$(\omega^2 - \alpha q_1^2 - \beta q_1^4)[C_{2m}I'_m(\kappa a_1) + C_{3m}K'_m(\kappa a_1) - C_{1m}I'_m(\kappa a_1)] = -\frac{e^2 n_0}{m_e \varepsilon_0}\frac{1}{\kappa}(\kappa^2 + m^2/a_1^2)C_{1m}I_m(\kappa a_1)$$

با استفاده از تعریف دو پارامتر جدید $N, R$ معادله‌ی بالا را به شکل ساده‌تری می‌نویسیم.

$$R = (\omega^2 - \alpha q_1^2 - \beta q_1^4) \qquad , N = -\frac{e^2 n_0}{m_e \varepsilon_0}\frac{1}{\kappa}(\kappa^2 + m^2/a_1^2)$$

$$RC_{2m}I'_m(\kappa a_1) + RC_{3m}K'_m(\kappa a_1) - RC_{1m}I'_m(\kappa a_1) = NC_{1m}I_m(\kappa a_1) \tag{4-5}$$

همان‌طور که انتظار داریم بُعد متغیرهای N,R برابر با توان دو فرکانس است و رابطه‌ی (4-5) نیز از لحاظ ابعادی درست خواهد بود. (متغیرهای $C_{im}I_{im}, C_{im}K_{im}; i=1,2$ دارای بعد میدان الکتریکی هستند. در متغیرهای $C_{im}I'_{im}, C_{im}K'_{im}; i=1,2$ نیز چون مشتق‌گیری نسبت به متغیر بی‌بعد $\kappa r$ انجام شده، بعد این متغیرها نیز برابر با بعد میدان الکتریکی است.)

برای ساده‌سازی معادله‌ی(4-3) نیز مطابق روند بالا عمل کرده و معادله‌ی زیر را به دست می‌آوریم.

$$HC_{4m}I'_m(\kappa a_2) - HC_{2m}I'_m(\kappa a_2) - HC_{3m}K'_m(\kappa a_2) = FC_{4m}K_m(\kappa a_2) \tag{4-6}$$

$$H = (\omega^2 - \alpha q_2^2 - \beta q_2^4), F = -\frac{e^2 n_0}{m_e \varepsilon_0}\frac{1}{\kappa}(\kappa^2 + m^2/a_2^2)$$

با قرار دادن جواب‌های فرضی میدان در معادلات (4-2) و (4-4)، معادلات زیر به دست می‌آیند.

$$C_{2m}I_m(\kappa a_1) + C_{3m}K_m(\kappa a_1) = C_{1m}I_m(\kappa a_1) \tag{4-7}$$

$$C_{2m}I_m(\kappa a_2) + C_{3m}K_m(\kappa a_2) = C_{4m}K_m(\kappa a_2) \tag{۴-۸}$$

در پیوست الف، ضرایب $C_{4m}, C_{3m}, C_{21m}, C_{1m}$ را از معادلات (۴-۵)، (۴-۶)، (۴-۷) و (۴-۸) حذف می‌کنیم و معادله‌ی زیر را به‌دست می‌آوریم.

معادله پاشندگی $\omega(\kappa)$:

$$\omega^4 - [\omega_1^2 + \omega_2^2 + \alpha(q_1^2 + q_2^2) + \beta(q_1^4 + q_2^4)]\omega^2$$
$$+ \omega_1^2\omega_2^2 - \omega_1^2\omega_2^2 F_{12} + (\alpha q_2^2 + \beta q_2^4)\omega_1^2 + (\alpha q_1^2 + \beta q_1^4)\omega_2^2 + (\alpha q_1^2 + \beta q_1^4)(\alpha q_2^2 + \beta q_2^4) = 0$$

$$\begin{cases} q_1^2 = (q^2 + m^2/a_1^2) \\ q_2^2 = (q^2 + m^2/a_2^2) \\ \omega_1^2 = \Omega_{p1}^2(\kappa^2 a_1^2 + m^2)K_m(\kappa a_1)I_m(\kappa a_1); \Omega_{p1} = (e^2 n_0/m_e\varepsilon_0 a_1)^{1/2} \\ \omega_2^2 = \Omega_{p2}^2(\kappa^2 a_2^2 + m^2)K_m(\kappa a_2)I_m(\kappa a_2); \Omega_{p2} = (e^2 n_0/m_e\varepsilon_0 a_2)^{1/2} \\ F_{12} = [I_m(\kappa a_1)/K_m(\kappa a_1)][K_m(\kappa a_2)/I_m(\kappa a_2)] \end{cases}$$

$\Omega_{p1}, \Omega_{p2}$ دارای بعد فرکانس بوده و تنها به هندسه و ساختار جداره‌های داخلی و خارجی نانوتیوب وابسته‌اند.

## بررسی صحت معادله پاشندگی محاسبه شده

۱. با توجه به این‌که $\Omega_{p2}^2, \Omega_{p1}^2, \omega_1^2, \omega_2^2$ دارای بعد توان دو فرکانس هستند، تمام جملات معادله پاشندگی دارای بعد توان چهارم فرکانس خواهند بود. پس معادله‌ی به دست آمده از لحاظ ابعادی درست است.

۲. وقتی نیروهای برهم‌کنش داخلی در گاز الکترونی دوبعدی را در نظر نمی‌گیریم، ($\alpha, \beta = 0$)، معادله‌ی پاشندگی به‌دست آمده به معادله‌ی (۲۰) در مرجع [۱۰] تبدیل می‌شود.

$$\alpha, \beta = 0 \Rightarrow \omega^2 = \frac{1}{2}(\omega_1^2 + \omega_2^2) \pm \sqrt{\frac{1}{4}(\omega_1^2 - \omega_2^2)^2 + \omega_1^2\omega_2^2 F_{12}} \qquad \text{رابطه‌ی (۲۰) مرجع [۱۰]}$$

در مرجع [۱۰] معادلات پاشندگی نانوتیوپ‌های چندجداره را بدون در نظر گرفتن نیروهای برهم‌کنش داخلی در یک گاز دوبعدی الکترونی به‌دست آورده است.

۳. روش دیگر رسیدن به معادله‌ی پاشندگی به صورت زیر است. چهار معادله‌ی شرط مرزی میدان الکتریکی، دستگاه همگنی از معادلات تشکیل می‌دهند که مجهولات آن ضرایب $C_{1m}, C_{2m}, C_{3m}, C_{4m}$ هستند. شرط وجود جواب غیربدیهی برای این دستگاه و در نتیجه وجود میدان غیرصفر در نواحی اطراف و داخل نانوتیوپ دوجداره، آن است که دترمینان ماتریس ضرایب صفر باشد. جواب دترمینان برابر با صفر، در حقیقت معادل با محاسبات موجود در

پیوست الف است و معادله‌ی پاشندگی را نتیجه می‌دهد. قابل ذکر است که این راه، به یک فرمول میانی در پیوست الف ختم می‌شود و به نوعی چک کردن محاسبات پیوست الف تا فرمول فوق خواهد بود.

$$RC_{2m}I'_m(\kappa a_1) + RC_{3m}K'_m(\kappa a_1) - RC_{1m}I'_m(\kappa a_1) = NC_{1m}I_m(\kappa a_1) \tag{۴-۵}$$

$$HC_{4m}K'_m(\kappa a_2) - HC_{2m}I'_m(\kappa a_2) - HC_{3m}K'_m(\kappa a_2) = FC_{4m}K_m(\kappa a_2) \tag{۴-۶}$$

$$C_{2m}I_m(\kappa a_1) + C_{3m}K_m(\kappa a_1) = C_{1m}I_m(\kappa a_1) \tag{۴-۷}$$

$$C_{2m}I_m(\kappa a_2) + C_{3m}K_m(\kappa a_2) = C_{4m}K_m(\kappa a_2) \tag{۴-۸}$$

$$\begin{pmatrix} -RI'_m(\kappa a_1)-NI_m(\kappa a_1) & RI'_m(\kappa a_1) & RK'_m(\kappa a_1) & 0 \\ 0 & HI'_m(\kappa a_2) & HK'_m(\kappa a_2) & FK_m(\kappa a_2)-HK'_m(\kappa a_2) \\ -I_m(\kappa a_1) & I_m(\kappa a_1) & K_m(\kappa a_1) & 0 \\ 0 & I_m(\kappa a_2) & K_m(\kappa a_2) & -K_m(\kappa a_2) \end{pmatrix} \begin{pmatrix} C_{1m} \\ C_{2m} \\ C_{3m} \\ C_{4m} \end{pmatrix} = \begin{pmatrix} 0 \\ 0 \\ 0 \\ 0 \end{pmatrix}$$

شرط وجود جواب غیربدیهی آن است که رابطه‌ی زیر برقرار باشد.

$$\begin{vmatrix} -RI'_m(\kappa a_1)-NI_m(\kappa a_1) & RI'_m(\kappa a_1) & RK'_m(\kappa a_1) & 0 \\ 0 & HI'_m(\kappa a_2) & HK'_m(\kappa a_2) & FK_m(\kappa a_2)-HK'_m(\kappa a_2) \\ -I_m(\kappa a_1) & I_m(\kappa a_1) & K_m(\kappa a_1) & 0 \\ 0 & I_m(\kappa a_2) & K_m(\kappa a_2) & -K_m(\kappa a_2) \end{vmatrix} = 0$$

بسط دترمینان حول ستون اول به صورت زیر است.

$$\begin{aligned} & -RI'_m(\kappa a_1)-NI_m(\kappa a_1) \begin{vmatrix} HI'_m(\kappa a_2) & HK'_m(\kappa a_2) & FK_m(\kappa a_2)-HK'_m(\kappa a_2) \\ I_m(\kappa a_1) & K_m(\kappa a_1) & 0 \\ I_m(\kappa a_2) & K_m(\kappa a_2) & -K_m(\kappa a_2) \end{vmatrix} \\ & -I_m(\kappa a_1) \begin{vmatrix} RI'_m(\kappa a_1) & RK'_m(\kappa a_1) & 0 \\ HI'_m(\kappa a_2) & HK'_m(\kappa a_2) & FK_m(\kappa a_2)-HK'_m(\kappa a_2) \\ I_m(\kappa a_2) & K_m(\kappa a_2) & -K_m(\kappa a_2) \end{vmatrix} = 0 \end{aligned}$$

هر دو دترمینان فوق را حول سطرهای اول آن‌ها بسط می‌دهیم.

$$[-RI'_m(\kappa a_1) - NI_m(\kappa a_1)][HI'_m(\kappa a_2)][-K_m(\kappa a_1)K_m(\kappa a_2))]$$
$$[RI'_m(\kappa a_1) + NI_m(\kappa a_1)][HK'_m(\kappa a_2)][-I_m(\kappa a_1)K_m(\kappa a_2)]$$
$$[-RI'_m(\kappa a_1) - NI_m(\kappa a_1)][FK_m(\kappa a_2) - HK'_m(\kappa a_2)][I_m(\kappa a_1)K_m(\kappa a_2) - K_m(\kappa a_1)I_m(\kappa a_2)]$$
$$-RI'_m(\kappa a_1)I_m(\kappa a_1)[-HK_m(\kappa a_2)K'_m(\kappa a_2) + K_m(\kappa a_2)[HK'_m(\kappa a_2) - FK_m(\kappa a_2)]]$$
$$+RK'_m(\kappa a_1)I_m(\kappa a_1)[-HI'_m(\kappa a_2)K_m(\kappa a_2) + I_m(\kappa a_2)[HK'_m(\kappa a_2) - FK_m(\kappa a_2)]]$$
$$= 0$$

$$[-RI'_m(\kappa a_1)][HI'_m(\kappa a_2)][-K_m(\kappa a_1)K_m(\kappa a_2))]$$
$$[-NI_m(\kappa a_1)][HI'_m(\kappa a_2)][-K_m(\kappa a_1)K_m(\kappa a_2))]$$
$$[RI'_m(\kappa a_1)][HK'_m(\kappa a_2)][-I_m(\kappa a_1)K_m(\kappa a_2)]$$
$$[NI_m(\kappa a_1)][HK'_m(\kappa a_2)][-I_m(\kappa a_1)K_m(\kappa a_2)]$$
$$[-RI'_m(\kappa a_1)][FK_m(\kappa a_2) - HK'_m(\kappa a_2)][I_m(\kappa a_1)K_m(\kappa a_2) - K_m(\kappa a_1)I_m(\kappa a_2)]$$
$$[-NI_m(\kappa a_1)][FK_m(\kappa a_2) - HK'_m(\kappa a_2)][I_m(\kappa a_1)K_m(\kappa a_2) - K_m(\kappa a_1)I_m(\kappa a_2)]$$
$$-RI'_m(\kappa a_1)I_m(\kappa a_1)[-HK_m(\kappa a_2)K'_m(\kappa a_2)]$$
$$-RI'_m(\kappa a_1)I_m(\kappa a_1)K_m(\kappa a_2)[HK'_m(\kappa a_2) - FK_m(\kappa a_2)]]$$
$$+RK'_m(\kappa a_1)I_m(\kappa a_1)[-HK'_m(\kappa a_2)K_m(\kappa a_2)]$$
$$+RK'_m(\kappa a_1)I_m(\kappa a_1)I_m(\kappa a_2)[HK'_m(\kappa a_2) - FK_m(\kappa a_2)]$$
$$= 0$$

$+ RHI'_m(\kappa a_1)I'_m(\kappa a_2)K_m(\kappa a_1)K_m(\kappa a_2)$

$+ NHI_m(\kappa a_1)I'_m(\kappa a_2)K_m(\kappa a_1)K_m(\kappa a_2)$

$- RHI'_m(\kappa a_1)K'_m(\kappa a_2)I_m(\kappa a_1)K_m(\kappa a_2)$

$- NHI_m(\kappa a_1)K'_m(\kappa a_2)I_m(\kappa a_1)K_m(\kappa a_2)$

$- RFI'_m(\kappa a_1)K_m(\kappa a_2)I_m(\kappa a_1)K_m(\kappa a_2)$

$+ RFI'_m(\kappa a_1)K_m(\kappa a_2)K_m(\kappa a_1)I_m(\kappa a_2)$

$+ RHI'_m(\kappa a_1)K'_m(\kappa a_2)I_m(\kappa a_1)K_m(\kappa a_2)$

$- RHI'_m(\kappa a_1)K'_m(\kappa a_2)K_m(\kappa a_1)I_m(\kappa a_2)$

$- NFI_m(\kappa a_1)K_m(\kappa a_2)I_m(\kappa a_1)K_m(\kappa a_2)$

$+ NFI_m(\kappa a_1)K_m(\kappa a_2)K_m(\kappa a_1)I_m(\kappa a_2)$

$+ NHI_m(\kappa a_1)K'_m(\kappa a_2)I_m(\kappa a_1)K_m(\kappa a_2)$

$- NHI_m(\kappa a_1)K'_m(\kappa a_2)K_m(\kappa a_1)I_m(\kappa a_2)$

$+ RHI'_m(\kappa a_1)I_m(\kappa a_1)K_m(\kappa a_2)K'_m(\kappa a_2)$

$- RHI'_m(\kappa a_1)I_m(\kappa a_1)K_m(\kappa a_2)K'_m(\kappa a_2)$

$+ RFI'_m(\kappa a_1)I_m(\kappa a_1)K_m(\kappa a_2)K_m(\kappa a_2)$

$- RHK'_m(\kappa a_1)I_m(\kappa a_1)K'_m(\kappa a_2)K_m(\kappa a_2)$

$+ RHK'_m(\kappa a_1)I_m(\kappa a_1)I_m(\kappa a_2)K'_m(\kappa a_2)$

$- RFK'_m(\kappa a_1)I_m(\kappa a_1)I_m(\kappa a_2)K_m(\kappa a_2)$

$= 0$

با ساده کردن بعضی جملات مشابه و با علامت مخالف، معادله‌ی زیر به دست می‌آید.

$+ RHI'_m(\kappa a_1)I'_m(\kappa a_2)K_m(\kappa a_1)K_m(\kappa a_2)$

$+ NHI_m(\kappa a_1)I'_m(\kappa a_2)K_m(\kappa a_1)K_m(\kappa a_2)$

$+ RFI'_m(\kappa a_1)K_m(\kappa a_2)K_m(\kappa a_1)I_m(\kappa a_2)$

$- RHI'_m(\kappa a_1)K'_m(\kappa a_2)K_m(\kappa a_1)I_m(\kappa a_2)$

$- NFI_m(\kappa a_1)K_m(\kappa a_2)I_m(\kappa a_1)K_m(\kappa a_2)$

$+ NFI_m(\kappa a_1)K_m(\kappa a_2)K_m(\kappa a_1)I_m(\kappa a_2)$

$- NHI_m(\kappa a_1)K'_m(\kappa a_2)K_m(\kappa a_1)I_m(\kappa a_2)$

$- RHK'_m(\kappa a_1)I_m(\kappa a_1)K'_m(\kappa a_2)K_m(\kappa a_2)$

$+ RHK'_m(\kappa a_1)I_m(\kappa a_1)I_m(\kappa a_2)K'_m(\kappa a_2)$

$- RFK'_m(\kappa a_1)I_m(\kappa a_1)I_m(\kappa a_2)K_m(\kappa a_2)$

$= 0$

فرمول (A) از پیوست الف :

-R* H*K'(ka۱) * K'(ka۲) *I (ka۲)/ K (ka۲) * I (ka۱)/ K (ka۱)

+ R* H* I'(ka۱)* K'(ka۲) * I (ka۲)/ K (ka۲)

+N* H* K'(ka۲) *I (ka۲)/ K (ka۲) *I (ka۱)

+ R* H*I'(ka۲) *K'(ka۱) * I (ka۱)/ K (ka۱)

 - R*H*I'(ka۲) *I'(ka۱)

 - N* H*I'(ka۲) * I (ka۱)

+ R* F*K'(ka۱)* I (ka۲) * I (ka۱)/ K (ka۱)

- R* F* I'(ka۱) *I (ka۲)

- N* F* K (ka۲) *I (ka۱) *I (ka۲)/ K (ka۲)

+ N* F* K (ka۲) * I (ka۱)*I (ka۱)/ K (ka۱)

= ۰

طرفین معادله‌ی (A) را در (ka۲) K (ka۱) K ضرب می‌کنیم.

-R * H * K'(ka۱) * K'(ka۲) *I (ka۲) * I (ka۱)

+ R * H * I'(ka۱) * K'(ka۲) * I (ka۲)* K (ka۱)

+N * H* K'(ka۲) * I (ka۲)* K (ka۱) *I (ka۱)

+ R* H*I'(ka۲) * K'(ka۱) * I (ka۱)* K (ka۲)

 - R * H * I'(ka۲) * I'(ka۱)* K (ka۱) * K (ka۲)

 - N * H*I'(ka۲) * I (ka۱)* K (ka۱) * K (ka۲)

+ R * F * K'(ka۱)* I (ka۲) * I (ka۱) * K (ka۲)

- R * F * I'(ka۱) * I (ka۲) * K (ka۱) * K (ka۲)

- N * F * K (ka۲) * I (ka۱) * I (ka۲) * K (ka۱)

+ N * F * K (ka۲) * I (ka۱) * I (ka۱)* K (ka۲)

= ۰

معادله‌ی فوق که از پیوست الف به دست آمد منفی معادله‌ی حاصل از حل دترمینان است. این امر درستی محاسبات پیوست الف تا معادله‌ی (A) را نشان می‌دهد.

## ۴-۱-۲- نمودارهای معادله‌ی پاشندگی در شرایط مختلف

معادله‌ی پاشندگی نانوتیوپ دوجداره در مد TM به صورت زیر است.

$$\omega^4 - [\omega_1^2 + \omega_2^2 + \alpha(q_1^2 + q_2^2) + \beta(q_1^4 + q_2^4)]\omega^2$$
$$+ \omega_1^2\omega_2^2 - \omega_1^2\omega_2^2 F_{12} + (\alpha q_2^2 + \beta q_2^4)\omega_1^2 + (\alpha q_1^2 + \beta q_1^4)\omega_2^2 + (\alpha q_1^2 + \beta q_1^4)(\alpha q_2^2 + \beta q_2^4) = 0$$

$$\begin{cases} q_1^2 = (q^2 + m^2/a_1^2) \\ q_2^2 = (q^2 + m^2/a_2^2) \\ \omega_1^2 = \Omega_{p1}^2(\kappa^2 a_1^2 + m^2)K_m(\kappa a_1)I_m(\kappa a_1); \Omega_{p1} = (e^2 n_0/m_e \varepsilon_0 a_1)^{1/2} \\ \omega_2^2 = \Omega_{p2}^2(\kappa^2 a_2^2 + m^2)K_m(\kappa a_2)I_m(\kappa a_2); \Omega_{p2} = (e^2 n_0/m_e \varepsilon_0 a_2)^{1/2} \\ F_{12} = [I_m(\kappa a_1)/K_m(\kappa a_1)][K_m(\kappa a_2)/I_m(\kappa a_2)] \end{cases}$$

قبل از رسم معادله‌ی پاشندگی، به بی‌بعد کردن آن می‌پردازیم.
کمیت‌های بی‌بعد $\omega, \kappa$ را به شکل زیر تعریف می‌کنیم.

$$x = \kappa\sqrt{a_1 a_2}, \quad y = \frac{\omega}{\sqrt{\Omega_{p1}\Omega_{p2}}} \quad ; \Omega_{p1} = (e^2 n_0/\varepsilon_0 m_e a_1)^{1/2}, \Omega_{p2} = (e^2 n_0/\varepsilon_0 m_e a_2)^{1/2}$$

[ برای بی‌بعد کردن کمیت‌های $\omega, \kappa$ در معادله‌ی پاشندگی نانوتیوپ دوجداره، از میانگین هندسی فرکانس‌های ویژه‌ی هر کدام از جداره‌ها و دو شعاعِ جداره‌ی داخلی و خارجی استفاده کرده‌ایم. بدین منظور از میانگین معمولی یا حسابی این متغیرها نیز می‌توان استفاده نمود، ولی به دلیل پیچیده‌تر شدن محاسباتِ بی‌بعد کردن، میانگین هندسی را برگزیدیم.
پارامترهای بی‌بعد $\omega, \kappa$ با استفاده از میانگین‌های حسابی به شکل زیر هستند.

$$x = \kappa(\frac{a_1 + a_2}{2}), \quad y = \frac{2\omega}{\Omega_{p1} + \Omega_{p2}}$$

از خصوصیات میانگین هندسی دو کمیت آن است که از میانگین حسابی کم‌تر بوده و به نوعی گرایش به مرکز را نشان می‌دهد. میانگین هندسی دقیقا در بین فاصله‌ی بین دو داده واقع نمی‌شود و به عدد کوچک‌تر نزدیک‌تر است.
[ $\frac{a_1 + a_2}{2} \succ \sqrt{a_1 a_2} \Leftrightarrow (\frac{a_1 + a_2}{2})^2 \succ a_1 a_2 \Leftrightarrow \frac{1}{4}(a_1 + a_2)^2 \succ a_1 a_2 \Leftrightarrow (a_1 + a_2)^2 \succ 4a_1 a_2 \Leftrightarrow (a_1 - a_2)^2 \succ 0$

به منظور بی‌بعد کردن معادله‌ی پاشندگی طرفین معادله را بر $\Omega_{p1}^2\Omega_{p2}^2$ تقسیم می‌کنیم.

$$\frac{\omega^4}{\Omega_{p1}^2\Omega_{p2}^2} - [\frac{\omega_1^2}{\Omega_{p1}\Omega_{p2}} + \frac{\omega_2^2}{\Omega_{p1}\Omega_{p2}} + \frac{\alpha}{\Omega_{p1}\Omega_{p2}}(q_1^2+q_2^2) + \frac{\beta}{\Omega_{p1}\Omega_{p2}}(q_1^4+q_2^4)]\frac{\omega^2}{\Omega_{p1}\Omega_{p2}}$$

$$-\frac{\omega_1^2\omega_2^2 F_{12}}{\Omega_{p1}^2\Omega_{p2}^2} + \frac{\omega_1^2\omega_2^2}{\Omega_{p1}^2\Omega_{p2}^2} + \frac{(\alpha q_2^2+\beta q_2^4)\omega_1^2}{\Omega_{p1}^2\Omega_{p2}^2} + \frac{(\alpha q_1^2+\beta q_1^4)\omega_2^2}{\Omega_{p1}^2\Omega_{p2}^2} + \frac{(\alpha q_1^2+\beta q_1^4)(\alpha q_2^2+\beta q_2^4)}{\Omega_{p1}^2\Omega_{p2}^2} = 0$$

برای تبدیل پارامترهای $q_1,q_2$ به $\kappa$ و پس از آن به x، عبارت $\sqrt{a_1 a_2}$ را در $q_1,q_2$ های معادله ضرب و تقسیم می‌کنیم.

$$y^4 - [\frac{\omega_1^2}{\Omega_{p1}\Omega_{p2}} + \frac{\omega_2^2}{\Omega_{p1}\Omega_{p2}} + \frac{\alpha}{\Omega_{p1}\Omega_{p2}a_1a_2}(a_1a_2q_1^2+a_1a_2q_2^2)$$

$$+\frac{\beta}{\Omega_{p1}\Omega_{p2}a_1^2a_2^2}(a_1^2a_2^2q_1^4+a_1^2a_2^2q_2^4)]y^2$$

$$-\frac{\omega_1^2\omega_2^2 F_{12}}{\Omega_{p1}^2\Omega_{p2}^2} + \frac{\omega_1^2\omega_2^2}{\Omega_{p1}^2\Omega_{p2}^2}$$

$$+(\frac{\alpha}{\Omega_{p1}\Omega_{p2}a_1a_2}a_1a_2q_2^2)\frac{\omega_1^2}{\Omega_{p1}\Omega_{p2}} + (\frac{\beta}{\Omega_{p1}\Omega_{p2}a_1^2a_2^2}a_1^2a_2^2q_2^4)\frac{\omega_1^2}{\Omega_{p1}\Omega_{p2}}$$

$$+(\frac{\alpha}{\Omega_{p1}\Omega_{p2}a_1a_2}a_1a_2q_1^2)\frac{\omega_2^2}{\Omega_{p1}\Omega_{p2}} + (\frac{\beta}{\Omega_{p1}\Omega_{p2}a_1^2a_2^2}a_1^2a_2^2q_1^4)\frac{\omega_2^2}{\Omega_{p1}\Omega_{p2}}$$

$$+(\frac{\alpha}{\Omega_{p1}\Omega_{p2}a_1a_2}a_1a_2q_1^2 + \frac{\beta}{\Omega_{p1}\Omega_{p2}a_1^2a_2^2}a_1^2a_2^2q_1^4)(\frac{\alpha}{\Omega_{p1}\Omega_{p2}a_1a_2}a_1a_2q_2^2 + \frac{\beta}{\Omega_{p1}\Omega_{p2}a_1^2a_2^2}a_1^2a_2^2q_2^4)$$

$$= 0$$

متغیرهای بی‌بعد $\sigma, \alpha_1, \beta_1$ را به صورت زیر تعریف می‌کنیم.

$$\sigma = \Omega_p^2 a^2/c^2 : SingleWalled \qquad , \sigma = \Omega_{p1}\Omega_{p2}a_1a_2/c^2 : DoubleWalled$$

$$\alpha_1 = \frac{\alpha}{\Omega_p^2 a^2} : SingleWalled \qquad , \alpha_1 = \frac{\alpha}{\Omega_{p1}\Omega_{p2}a_1a_2} : DoubleWalled$$

$$\beta_1 = \frac{\beta}{\Omega_P^2 a^2} ; SingleWalled \qquad , \beta_1 = \frac{\beta}{\Omega_{p1}\Omega_{p2}a_1^2a_2^2} ; DoubleWalled$$

معادله‌ی پاشندگی با استفاده از متغیرهای بالا به صورت زیر است.

$$y^4 - [\frac{\omega_1^2}{\Omega_{p1}\Omega_{p2}} + \frac{\omega_2^2}{\Omega_{p1}\Omega_{p2}} + \alpha_1(a_1a_2q_1^2 + a_1a_2q_2^2)$$
$$+ \beta_1(a_1^2a_2^2q_1^4 + a_1^2a_2^2q_2^4)]y^2$$
$$- \frac{\omega_1^2\omega_2^2 F_{12}}{\Omega_{p1}^2\Omega_{p2}^2} + \frac{\omega_1^2\omega_2^2}{\Omega_{p1}^2\Omega_{p2}^2}$$
$$+ \alpha_1(a_1a_2q_2^2)\frac{\omega_1^2}{\Omega_{p1}\Omega_{p2}} + \beta_1(a_1^2a_2^2q_2^4)\frac{\omega_1^2}{\Omega_{p1}\Omega_{p2}}$$
$$+ \alpha_1(a_1a_2q_1^2)\frac{\omega_2^2}{\Omega_{p1}\Omega_{p2}} + \beta_1(a_1^2a_2^2q_1^4)\frac{\omega_2^2}{\Omega_{p1}\Omega_{p2}}$$
$$+ (\alpha_1a_1a_2q_1^2 + \beta_1a_1^2a_2^2q_1^4)(\alpha_1a_1a_2q_2^2 + \beta_1a_1^2a_2^2q_2^4)$$
$$= 0$$

بقیه‌ی پارامترها را در معادله پاشندگی می‌توان به صورت زیر ساده نمود.

$$q_1^2 = (q^2 + m^2/a_1^2) = [\kappa^2 + \frac{\omega^2}{c^2} + m^2/a_1^2]$$
$$a_1a_2q_1^2 = [\kappa^2a_1a_2 + \frac{\Omega_{P1}\Omega_{P2}a_1a_2}{c^2}\frac{\omega^2}{\Omega_{P1}\Omega_{P2}} + m^2a_1a_2/a_1^2] = [x^2 + \sigma y^2 + m^2a_2/a_1]$$

$$q_2^2 = (q^2 + m^2/a_2^2) = [(\kappa^2 + \frac{\omega^2}{c^2}) + m^2/a_2^2]$$
$$a_1a_2q_2^2 = [a_1a_2\kappa^2 + \frac{\Omega_{P1}\Omega_{P2}a_1a_2}{c^2}\frac{\omega^2}{\Omega_{P1}\Omega_{P2}} + m^2a_1a_2/a_2^2] = [x^2 + \sigma y^2 + m^2a_1/a_2]$$

$$\omega_1^2 = \Omega_{p1}^2(\kappa^2a_1^2 + m^2)K_m(\kappa a_1)I_m(\kappa a_1)$$
$$\kappa a_1 = \kappa\sqrt{a_1/a_2}\sqrt{a_1a_2} \rightarrow$$
$$\omega_1^2 = \Omega_{P1}^2[(a_1/a_2)\kappa^2(a_1a_2) + m^2]K_m(\kappa a_1)I_m(\kappa a_1)$$
$$= \Omega_{P1}^2[(a_1/a_2)x^2 + m^2]K_m(\sqrt{a_1/a_2}\kappa\sqrt{a_1a_2})I_m(\sqrt{a_1/a_2}\kappa\sqrt{a_1a_2})$$
$$= \Omega_{P1}^2[(a_1/a_2)x^2 + m^2]K_m(\sqrt{a_1/a_2}x)I_m(\sqrt{a_1/a_2}x)$$

$$\omega_2^2 = \Omega_{p2}^2(\kappa^2 a_2^2 + m^2)K_m(\kappa a_2)I_m(\kappa a_2)$$
$$\kappa a_2 = \kappa\sqrt{a_1 a_2}\sqrt{a_2/a_1} \rightarrow$$
$$\omega_2^2 = \Omega_{P2}^2[(a_2/a_1)\kappa^2(a_1 a_2) + m^2]K_m(\kappa a_2)I_m(\kappa a_2)$$
$$= \Omega_{P2}^2[(a_2/a_1)x^2 + m^2]K_m(\kappa\sqrt{a_1 a_2}\sqrt{a_2/a_1})I_m(\kappa\sqrt{a_1 a_2}\sqrt{a_2/a_1})$$
$$= \Omega_{P2}^2[(a_2/a_1)x^2 + m^2]K_m(\sqrt{a_2/a_1}x)I_m(\sqrt{a_2/a_1}x)$$

$$\Omega_{p1} = (e^2 n_0/\varepsilon_0 m_e a_1)^{1/2}, \Omega_{p2} = (e^2 n_0/\varepsilon_0 m_e a_2)^{1/2} \rightarrow \Omega_{p1}/\Omega_{p2} = \sqrt{a_2/a_1}$$

$$\frac{\omega_1^2}{\Omega_{p1}\Omega_{p2}} = \frac{\Omega_{p1}}{\Omega_{p2}}[(a_1/a_2)x^2 + m^2]K_m(\sqrt{a_1/a_2}x)I_m(\sqrt{a_1/a_2}x)$$
$$= \sqrt{a_2/a_1}[(a_1/a_2)x^2 + m^2]K_m(\sqrt{a_1/a_2}x)I_m(\sqrt{a_1/a_2}x)$$

$$\frac{\omega_2^2}{\Omega_{p1}\Omega_{p2}} = \frac{\Omega_{p2}}{\Omega_{p1}}[(a_2/a_1)x^2 + m^2]K_m(\sqrt{a_2/a_1}x)I_m(\sqrt{a_2/a_1}x)$$
$$= \sqrt{a_1/a_2}[(a_2/a_1)x^2 + m^2]K_m(\sqrt{a_2/a_1}x)I_m(\sqrt{a_2/a_1}x)$$

$$\frac{\omega_1^2 \omega_2^2}{\Omega_{p1}^2 \Omega_{p2}^2} = [(a_1/a_2)x^2 + m^2]K_m(\sqrt{a_1/a_2}x)I_m(\sqrt{a_1/a_2}x)$$
$$\times [(a_2/a_1)x^2 + m^2]K_m(\sqrt{a_2/a_1}x)I_m(\sqrt{a_2/a_1}x)$$
$$= [(a_1/a_2)x^2 + m^2][(a_2/a_1)x^2 + m^2]$$
$$\times I_m(\sqrt{a_1/a_2}x)I_m(\sqrt{a_2/a_1}x)K_m(\sqrt{a_1/a_2}x)K_m(\sqrt{a_2/a_1}x)$$

$$\frac{\omega_1^2 \omega_2^2 F_{12}}{\Omega_{p1}^2 \Omega_{p2}^2} = [(a_1/a_2)x^2 + m^2]K_m(\sqrt{a_1/a_2}x)I_m(\sqrt{a_1/a_2}x)$$
$$\times [(a_2/a_1)x^2 + m^2]K_m(\sqrt{a_2/a_1}x)I_m(\sqrt{a_2/a_1}x)$$
$$\times I_m(\sqrt{a_1/a_2}x)K_m(\sqrt{a_2/a_1}x)/(K_m(\sqrt{a_1/a_2}x)I_m(\sqrt{a_2/a_1}x))$$
$$= [(a_1/a_2)x^2 + m^2][(a_2/a_1)x^2 + m^2]I_m^2(\sqrt{a_1/a_2}x)K_m^2(\sqrt{a_2/a_1}x)$$

با جایگذاری مقادیر ثابت، معادله به شکل زیر تبدیل می‌شود.

$$y^4 - y^2[\sqrt{a_2/a_1}[(a_1/a_2)x^2 + m^2]K_m(\sqrt{a_1/a_2}x)I_m(\sqrt{a_1/a_2}x)$$
$$+ \sqrt{a_1/a_2}[(a_2/a_1)x^2 + m^2]K_m(\sqrt{a_2/a_1}x)I_m(\sqrt{a_2/a_1}x)$$
$$+ \alpha_1([x^2 + \sigma y^2 + m^2 a_2/a_1] + [x^2 + \sigma y^2 + m^2 a_1/a_2])$$
$$+ \beta_1([x^2 + \sigma y^2 + m^2 a_2/a_1]^2 + [x^2 + \sigma y^2 + m^2 a_1/a_2]^2)]$$
$$- [(a_1/a_2)x^2 + m^2][(a_2/a_1)x^2 + m^2]I_m^2(\sqrt{a_1/a_2}x)K_m^2(\sqrt{a_2/a_1}x)$$
$$+ [(a_1/a_2)x^2 + m^2][(a_2/a_1)x^2 + m^2]I_m(\sqrt{a_1/a_2}x)I_m(\sqrt{a_2/a_1}x)K_m(\sqrt{a_1/a_2}x)K_m(\sqrt{a_2/a_1}x)$$
$$+ \alpha_1[x^2 + \sigma y^2 + m^2 a_1/a_2]\sqrt{a_2/a_1}[(a_1/a_2)x^2 + m^2]K_m(\sqrt{a_1/a_2}x)I_m(\sqrt{a_1/a_2}x)$$
$$+ \beta_1[x^2 + \sigma y^2 + m^2 a_1/a_2]^2\sqrt{a_2/a_1}[(a_1/a_2)x^2 + m^2]K_m(\sqrt{a_1/a_2}x)I_m(\sqrt{a_1/a_2}x)$$
$$+ \alpha_1[x^2 + \sigma y^2 + m^2 a_2/a_1]\sqrt{a_1/a_2}[(a_2/a_1)x^2 + m^2]K_m(\sqrt{a_2/a_1}x)I_m(\sqrt{a_2/a_1}x)$$
$$+ \beta_1[x^2 + \sigma y^2 + m^2 a_2/a_1]^2\sqrt{a_1/a_2}[(a_2/a_1)x^2 + m^2]K_m(\sqrt{a_2/a_1}x)I_m(\sqrt{a_2/a_1}x)$$
$$+ (\alpha_1[x^2 + \sigma y^2 + m^2 a_2/a_1] + \beta_1[x^2 + \sigma y^2 + m^2 a_2/a_1]^2) \times$$
$$(\alpha_1[x^2 + \sigma y^2 + m^2 a_1/a_2] + \beta_1[x^2 + \sigma y^2 + m^2 a_1/a_2]^2)$$
$$= 0$$

جملات $\sigma y^2$ معادله‌ی پاشندگیِ بی‌بعد شده را مانند حالت تک‌جداره حذف می‌کنیم.

$$y^4 - y^2[\sqrt{a_2/a_1}[(a_1/a_2)x^2 + m^2]K_m(\sqrt{a_1/a_2}x)I_m(\sqrt{a_1/a_2}x)$$
$$+ \sqrt{a_1/a_2}[(a_2/a_1)x^2 + m^2]K_m(\sqrt{a_2/a_1}x)I_m(\sqrt{a_2/a_1}x)$$
$$+ \alpha_1([x^2 + m^2 a_2/a_1] + [x^2 + m^2 a_1/a_2])$$
$$+ \beta_1([x^2 + m^2 a_2/a_1]^2 + [x^2 + m^2 a_1/a_2]^2)]$$
$$- [(a_1/a_2)x^2 + m^2][(a_2/a_1)x^2 + m^2]I_m^2(\sqrt{a_1/a_2}x)K_m^2(\sqrt{a_2/a_1}x)$$
$$+ [(a_1/a_2)x^2 + m^2][(a_2/a_1)x^2 + m^2]I_m(\sqrt{a_1/a_2}x)I_m(\sqrt{a_2/a_1}x)K_m(\sqrt{a_1/a_2}x)K_m(\sqrt{a_2/a_1}x)$$
$$+ \alpha_1[x^2 + m^2 a_1/a_2]\sqrt{a_2/a_1}[(a_1/a_2)x^2 + m^2]K_m(\sqrt{a_1/a_2}x)I_m(\sqrt{a_1/a_2}x)$$
$$+ \beta_1[x^2 + m^2 a_1/a_2]^2\sqrt{a_2/a_1}[(a_1/a_2)x^2 + m^2]K_m(\sqrt{a_1/a_2}x)I_m(\sqrt{a_1/a_2}x)$$
$$+ \alpha_1[x^2 + m^2 a_2/a_1]\sqrt{a_1/a_2}[(a_2/a_1)x^2 + m^2]K_m(\sqrt{a_2/a_1}x)I_m(\sqrt{a_2/a_1}x)$$
$$+ \beta_1[x^2 + m^2 a_2/a_1]^2\sqrt{a_1/a_2}[(a_2/a_1)x^2 + m^2]K_m(\sqrt{a_2/a_1}x)I_m(\sqrt{a_2/a_1}x)$$
$$+ (\alpha_1[x^2 + m^2 a_2/a_1] + \beta_1[x^2 + m^2 a_2/a_1]^2) \times (\alpha_1[x^2 + m^2 a_1/a_2] + \beta_1[x^2 + m^2 a_1/a_2]^2)$$
$$= 0$$

معادله‌ی فوق، معادله پاشندگی بی‌بعد شده‌ی نانوتیوپ کربنی دوجداره در مد TM است.

برای رسم این معادله‌ی درجه‌ی چهار، به حل آن می‌پردازیم.

$$Ay^4 + By^2 + C = 0$$

$$y^2 = \frac{-B \pm \sqrt{B^2 - 4AC}}{2A} = \frac{-B}{2A} \pm \sqrt{\frac{B^2}{4A^2} - \frac{C}{A}}$$

$$A = 1$$

$$B = -[\sqrt{a_2/a_1}[(a_1/a_2)(x^2) + m^2]K_m(\sqrt{a_1/a_2}\,x)I_m(\sqrt{a_1/a_2}\,x)$$
$$+ \sqrt{a_1/a_2}[(a_2/a_1)(x^2) + m^2]K_m(\sqrt{a_2/a_1}\,x)I_m(\sqrt{a_2/a_1}\,x)$$
$$+ \alpha_1([x^2 + m^2 a_2/a_1] + [x^2 + m^2 a_1/a_2])$$
$$+ \beta_1([x^2 + m^2 a_2/a_1]^2 + [x^2 + m^2 a_1/a_2]^2)]$$

$$C = -[(a_1/a_2)(x^2) + m^2][(a_2/a_1)(x^2) + m^2]I_m^2(\sqrt{a_1/a_2}\,x)K_m^2(\sqrt{a_2/a_1}\,x)$$
$$+ [(a_1/a_2)(x^2) + m^2][(a_2/a_1)(x^2) + m^2]I_m(\sqrt{a_1/a_2}\,x)I_m(\sqrt{a_2/a_1}\,x)K_m(\sqrt{a_1/a_2}\,x)K_m(\sqrt{a_2/a_1}\,x)$$
$$+ \alpha_1[x^2 + m^2 a_1/a_2]\sqrt{a_2/a_1}[(a_1/a_2)(x^2) + m^2]K_m(\sqrt{a_1/a_2}\,x)I_m(\sqrt{a_1/a_2}\,x)$$
$$+ \beta_1[x^2 + m^2 a_1/a_2]^2\sqrt{a_2/a_1}[(a_1/a_2)(x^2) + m^2]K_m(\sqrt{a_1/a_2}\,x)I_m(\sqrt{a_1/a_2}\,x)$$
$$+ \alpha_1[x^2 + m^2 a_2/a_1]\sqrt{a_1/a_2}[(a_2/a_1)(x^2) + m^2]K_m(\sqrt{a_2/a_1}\,x)I_m(\sqrt{a_2/a_1}\,x)$$
$$+ \beta_1[x^2 + m^2 a_2/a_1]^2\sqrt{a_1/a_2}[(a_2/a_1)(x^2) + m^2]K_m(\sqrt{a_2/a_1}\,x)I_m(\sqrt{a_2/a_1}\,x)$$
$$+ (\alpha_1[x^2 + m^2 a_2/a_1] + \beta_1[x^2 + m^2 a_2/a_1]^2) \times (\alpha_1[x^2 + m^2 a_1/a_2] + \beta_1[x^2 + m^2 a_1/a_2]^2)$$

$$\Rightarrow y^2 = \frac{-B}{2} \pm \sqrt{\frac{B^2}{4} - C}, \quad y = \pm\sqrt{\frac{-B}{2} \pm \sqrt{\frac{B^2}{4} - C}}$$

علامت منفی پشت رادیکال بیرونی، نمایانگر $y$ یا فرکانس منفی در سیستم است که جوابی غیر فیزیکی بوده و قابل صرف‌نظر کردن است.

$$\Rightarrow y = \sqrt{\frac{-B}{2} \pm \sqrt{\frac{B^2}{4} - C}} \quad ; (\frac{B^2}{4} - C) \geq 0 \quad , (\frac{-B}{2} \pm \sqrt{\frac{B^2}{4} - C}) \geq 0$$

ضرایب $\alpha_1, \beta_1$ را می‌توان برحسب شعاع‌های $a_1, a_2$ مشخص نمود.

$$\alpha_1 = \alpha/(\Omega_{p1}\Omega_{p2}a_1 a_2) \quad ; \alpha = \frac{e^2 n_0 a_B}{4\varepsilon_0 m_e} \quad , a_B = 5.29 \times 10^{-11} m$$

$$, \Omega_{P1} = (\frac{e^2 n_0}{\varepsilon_0 m_e a_1})^{1/2}, \Omega_{P2} = (\frac{e^2 n_0}{\varepsilon_0 m_e a_2})^{1/2}$$

$$\beta_1 = \beta/(\Omega_{p1}\Omega_{p2}a_1^2 a_2^2) \quad ; \beta = \frac{e^2 a_B}{16\pi\varepsilon_0 m_e}$$

$$\alpha_1 = \frac{\alpha}{\Omega_{p1}\Omega_{p2}a_1a_2} = \frac{e^2 n_0 a_B}{4\varepsilon_0 m_e} \cdot \frac{1}{(\frac{e^2 n_0}{\varepsilon_0 m_e a_1})^{1/2}(\frac{e^2 n_0}{\varepsilon_0 m_e a_2})^{1/2} a_1 a_2} = \frac{a_B}{4\sqrt{a_1 a_2}}$$

$$= \frac{1.32 \times 10^{-11} m}{\sqrt{a_1(m)a_2(m)}} = \frac{1.32 \times 10^{-2} nm}{\sqrt{a_1(nm)a_2(nm)}}$$

با توجه به مقدار $n_0 = 152 nm^{-2}$ که در فصل ۳ محاسبه شد، داریم:

$$\beta_1 = \frac{\beta}{\Omega_{p1}\Omega_{p2}a_1^2 a_2^2} = \frac{e^2 a_B}{16\pi\varepsilon_0 m_e} \cdot \frac{1}{(\frac{e^2 n_0}{\varepsilon_0 m_e a_1})^{1/2}(\frac{e^2 n_0}{\varepsilon_0 m_e a_2})^{1/2} a_1^2 a_2^2} = \frac{a_B}{16\pi n_0 \sqrt{(a_1 a_2)^3}}$$

$$= \frac{5.29 \times 10^{-11} m}{16 \times 3.14 \times 152 \times 10^{18} m^{-2} \times \sqrt{[a_1(m)a_2(m)]^3}} = \frac{6.92 \times 10^{-4} \times 10^{-29}}{\sqrt{[a_1(m)a_2(m)]^3}}$$

$$= \frac{6.92 \times 10^{-6}}{\sqrt{[a_1(nm)a_2(nm)]^3}}$$

نانوتیوپی به شعاع‌های داخلی و خارجی $\begin{cases} a_1 = 1nm \\ a_2 = 1.35nm \end{cases}$ را در نظر می‌گیریم. برای مدهای عرضی m=۰،۱،۲،۳،۴، نمودارهای معادله‌ی پاشندگی مربوط به آن را رسم می‌کنیم.

ضرایب به‌کار رفته در معادله‌ی پاشندگی فوق به صورت زیر هستند.

$$\alpha_1 = \frac{1.32 \times 10^{-2} nm}{\sqrt{a_1(nm)a_2(nm)}} = \frac{1.32 \times 10^{-2} nm}{\sqrt{1(nm)1.35(nm)}} = 1.14 \times 10^{-3}$$

$$\beta_1 = \frac{6.92 \times 10^{-6}}{\sqrt{[a_1(nm)a_2(nm)]^3}} = \frac{6.92 \times 10^{-6}}{\sqrt{[1(nm)1.35(nm)]^3}} = 4.41 \times 10^{-6}$$

$$a_1/a_2 = 0.74, \sqrt{a_1/a_2} = 0.86, a_2/a_1 = 1.35, \sqrt{a_2/a_1} = 1.16$$

در شکل (۴-۱) نمودارهای پاشندگی برای مدهای عرضی m=۰،۱،۲،۳،۴ درنانوتیوپ کربنی فوق به طور جداگانه نشان داده شده‌اند.

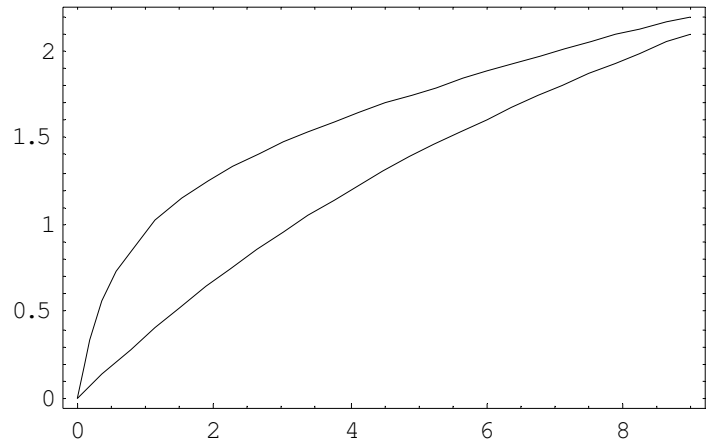

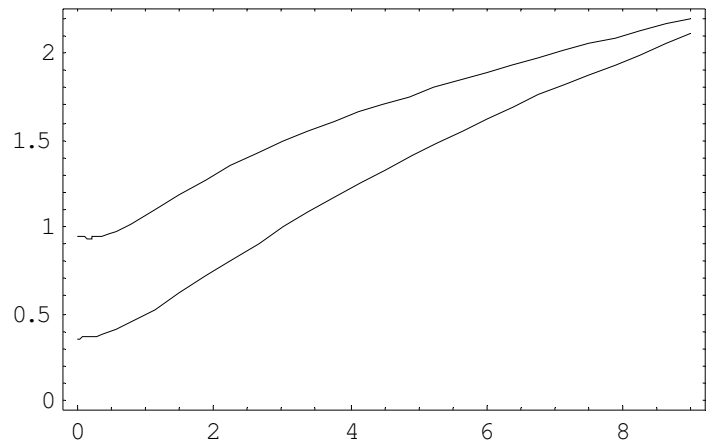

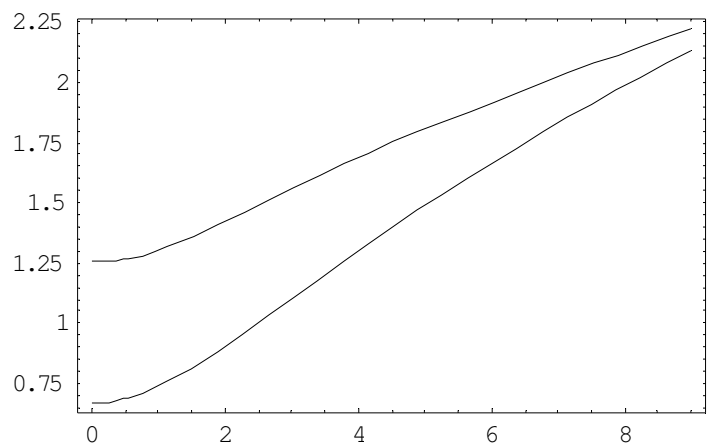

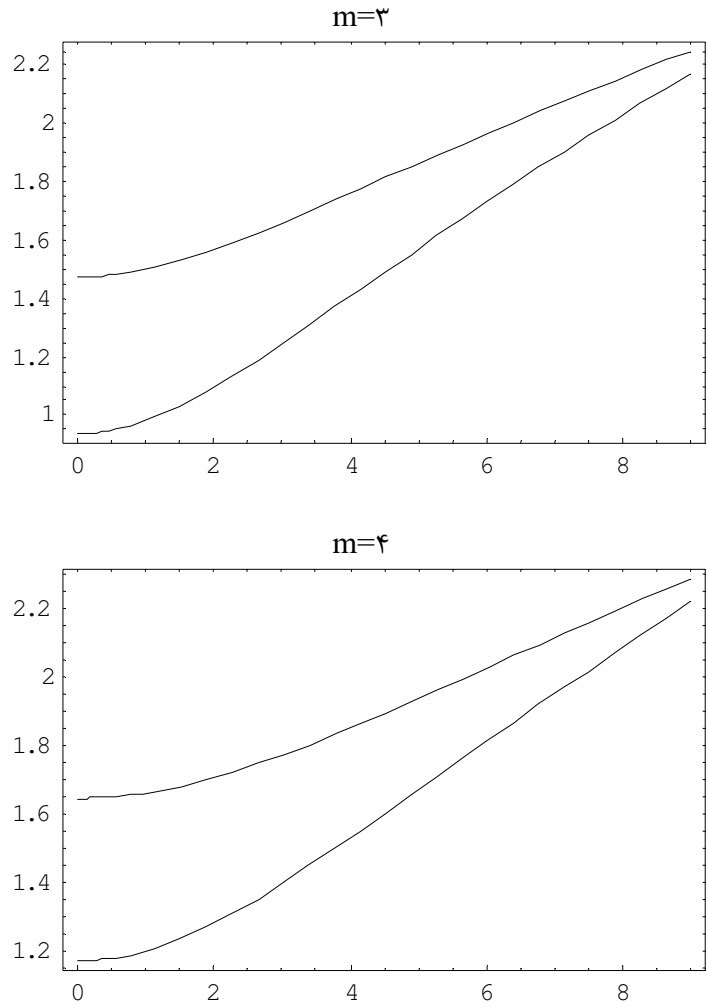

شکل (۴-۱)- نمودار پاشندگی با مدهای عرضی مختلف نانوتیوپ دوجداره. محورهای عمودی فرکانس بی‌بعد شده‌ی $\omega/\sqrt{\Omega_{p1}\Omega_{p2}}$ و محورهای افقی عدد موج بی‌بعد شده‌ی $q\sqrt{a_1 a_2}$ هستند.

همان‌طور که در شکل (۴-۱) مشخص است نمودار پاشندگی هر مد عرضی TM دارای دو شاخه است. این بدان معنی است که در یک فرکانس خاص، می‌توان امواجی با طول موج‌های مختلف در نانوتیوپ کربنی دوجداره تولید نمود.

در مدهای عرضی بالاتر از صفر، در شاخه‌ی بالایی، نقطه‌ی عطفی وجود دارد. این نقطه در مد صفر بیشتر از بقیه‌ی مدها نمایان است. در فرکانس فوق، پاشندگی سرعت گروه موج تشکیل شده در نانوتیوپ دوجداره صفر خواهد بود.

در شکل (۴-۲) نمودارهای مربوط به مدهای مختلف عرضی همگی در یک نمودار آورده شده‌اند و می‌توان آن‌ها را با یکدیگر مقایسه نمود.

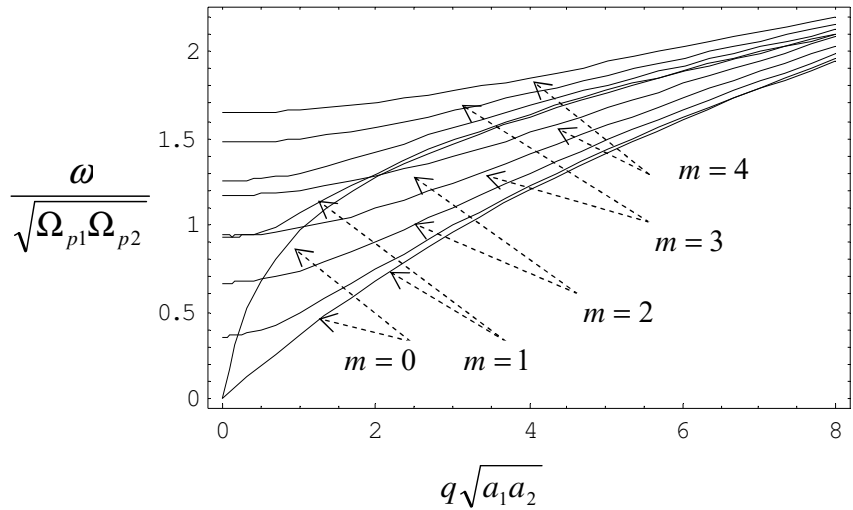

شکل (۴-۲)- نمودار پاشندگی با مدهای عرضی مختلف نانوتیوپ کربنی دوجداره

با افزایش مدهای عرضی در نانوتیوپ دوجداره، فرکانس قطع در هر دوشاخه که از صفر شروع می‌شود، افزایش می‌یابد. این رفتار در نمودارهای معادلات پاشندگی نانوتیوپ‌های تک‌جداره نیز برای مدهای عرضی مختلف وجود داشته است.

همان‌طور که در فصل ۳ در مورد معادلات پاشندگی نانوتیوپ‌های کربنی تک‌جداره اشاره شد معادلات پاشندگی در این فصل نیز، با در نظر گرفتن تقریب بکار رفته در محاسبات، برای روابط بین فرکانس- عدد موج و فرکانس- معکوس عمق نفوذ، هر دو برقرارند. به این ترتیب طبق شکل‌های بالا با افزایش فرکانس موج تشکیل شده در سیستم، عمق نفوذ شعاعی (اندازه‌ی میدان الکتریکی در راستای Z و در اطراف نانوتیوپ) کاهش می‌یابد.

حال به بررسی معادلات پاشندگی با شعاع‌های داخلی و خارجی متغیر می‌پردازیم. فاصله‌ی بین جداره‌ها را در نانوتیوپ دوجداره ثابت و برابر با $0.35nm$ در نظر می‌گیریم. لازم به ذکر است که به طور کلی این فاصله از $0.34nm$ تا $0.41nm$ متغیر است.

$$\begin{cases} a_1 = 0.5 nm \\ a_2 = 0.85 nm \end{cases}$$

$$a_1/a_2 = 0.59, \sqrt{a_1/a_2} = 0.77, a_2/a_1 = 1.7, \sqrt{a_2/a_1} = 1.3$$

$$\alpha_1 = \frac{1.32 \times 10^{-2} nm}{\sqrt{a_1(nm)a_2(nm)}} = \frac{1.32 \times 10^{-2} nm}{\sqrt{0.5(nm)0.85(nm)}} = 2.03 \times 10^{-3}$$

$$\beta_1 = \frac{6.92 \times 10^{-6}}{\sqrt{[a_1(nm)a_2(nm)]^3}} = \frac{6.92 \times 10^{-6}}{\sqrt{[0.5(nm)0.85(nm)]^3}} = 24.7 \times 10^{-6}$$

$$\begin{cases} a_1 = 1nm \\ a_2 = 1.35nm \end{cases}$$

$$a_1/a_2 = 0.74, \sqrt{a_1/a_2} = 0.86, a_2/a_1 = 1.35, \sqrt{a_2/a_1} = 1.16$$

$$\alpha_1 = \frac{1.32 \times 10^{-2} nm}{\sqrt{a_1(nm)a_2(nm)}} = \frac{1.32 \times 10^{-2} nm}{\sqrt{1(nm)1.35(nm)}} = 1.14 \times 10^{-3}$$

$$\beta_1 = \frac{6.92 \times 10^{-6}}{\sqrt{[a_1(nm)a_2(nm)]^3}} = \frac{6.92 \times 10^{-6}}{\sqrt{[1(nm)1.35(nm)]^3}} = 4.41 \times 10^{-6}$$

$$\begin{cases} a_1 = 2nm \\ a_2 = 2.35nm \end{cases}$$

$$a_1/a_2 = 0.85, \sqrt{a_1/a_2} = 0.92, a_2/a_1 = 1.175, \sqrt{a_2/a_1} = 1.08$$

$$\alpha_1 = \frac{1.32 \times 10^{-2} nm}{\sqrt{a_1(nm)a_2(nm)}} = \frac{1.32 \times 10^{-2} nm}{\sqrt{2(nm)2.35(nm)}} = 0.6 \times 10^{-3}$$

$$\beta_1 = \frac{6.92 \times 10^{-6}}{\sqrt{[a_1(nm)a_2(nm)]^3}} = \frac{6.92 \times 10^{-6}}{\sqrt{[2(nm)2.35(nm)]^3}} = 0.22 \times 10^{-6}$$

$$\begin{cases} a_1 = 4nm \\ a_2 = 4.35nm \end{cases}$$

$$a_1/a_2 = 0.92, \sqrt{a_1/a_2} = 0.96, a_2/a_1 = 1.09, \sqrt{a_2/a_1} = 1.04$$

$$\alpha_1 = \frac{1.32 \times 10^{-2} nm}{\sqrt{a_1(nm)a_2(nm)}} = \frac{1.32 \times 10^{-2} nm}{\sqrt{4(nm)4.35(nm)}} = 0.32 \times 10^{-3}$$

$$\beta_1 = \frac{6.92 \times 10^{-6}}{\sqrt{[a_1(nm)a_2(nm)]^3}} = \frac{6.92 \times 10^{-6}}{\sqrt{[2(nm)2.35(nm)]^3}} = 0.68 \times 10^{-6}$$

نمودارهای پاشندگی نانوتیوپ‌های بالا در مد عرضی صفر در شکل (۴-۳) نشان داده شده اند.

$\begin{cases} a_1 = 0.5nm \\ a_2 = 0.85nm \end{cases}$

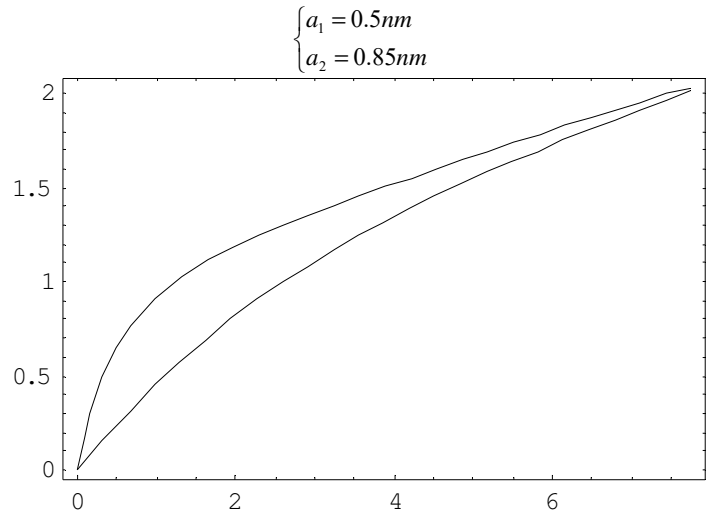

$\begin{cases} a_1 = 1nm \\ a_2 = 1.35nm \end{cases}$

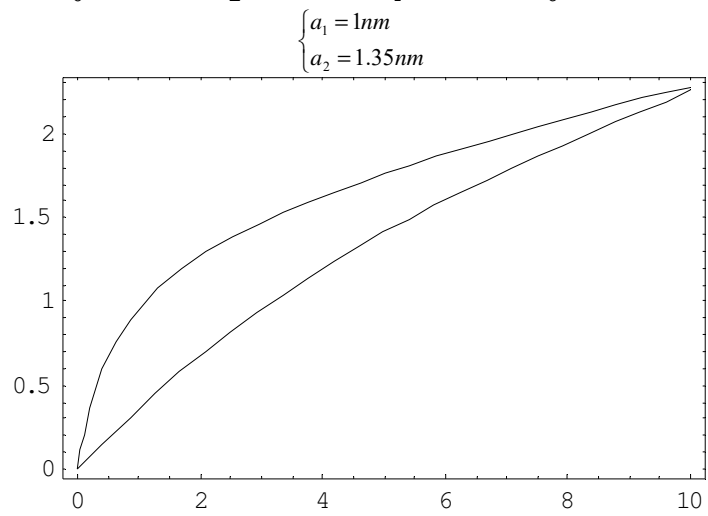

$\begin{cases} a_1 = 2nm \\ a_2 = 2.35nm \end{cases}$

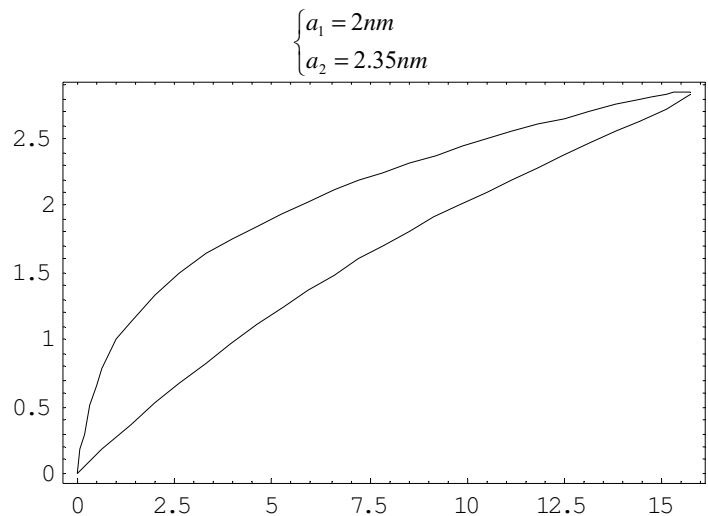

$\begin{cases} a_1 = 4nm \\ a_2 = 4.35nm \end{cases}$

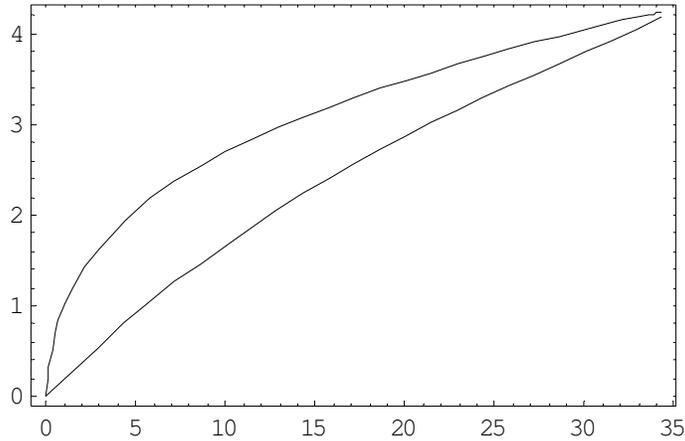

شکل (۴-۳)- نمودار پاشندگی با شعاع‌های مختلف داخلی و خارجی در نانوتیوپ کربنی دوجداره. محورهای عمودی فرکانس بی‌بعد شده‌ی $y = \omega/\sqrt{\Omega_{p1}\Omega_{p2}}$ ومحورهای افقی عدد موج بی‌بعد شده‌ی $x = q\sqrt{a_1 a_2}$ هستند.

در شکل (۴-۴) نمودارهای پاشندگی نانوتیوپ‌های دوجداره با شعاع‌های داخلی و خارجی مختلف همگی در یـک نمـودار آورده شده‌اند.

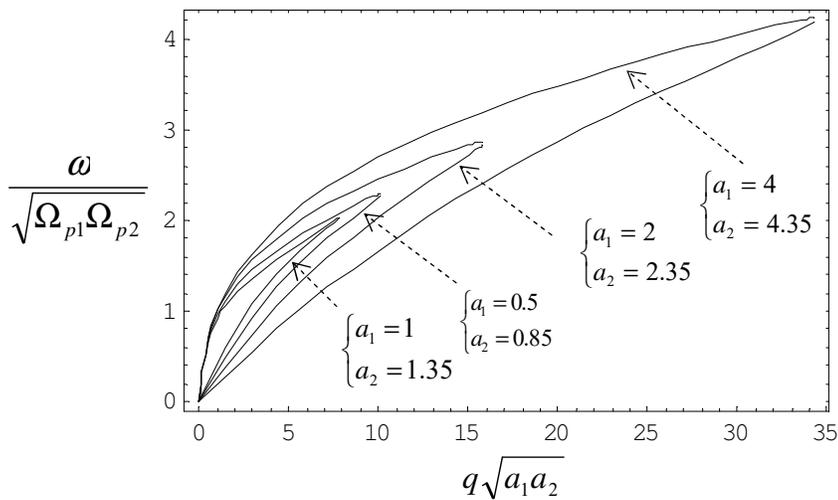

شکل (۴-۴)- نمودار پاشندگی با شعاع‌های مختلف داخلی و خارجی در نانوتیوپ کربنی دوجداره

در اینجا هم مانند نانوتیوپ تک‌جداره، با افزایش شعاع‌های جداره‌های داخلی و خارجی، فرکاس قطع سیسـتم تغییر نکرده و همان فرکانس قطع مربوط به مد صفر باقی خواهد ماند. با زیاد شدن اندازه‌ی شعاع‌های داخلی و خارجی، شیب شاخه‌ی بالایی مثل حالت تک‌جداره بیشتر می‌شود و دو شاخه‌ی معادله‌ی پاشندگی در فرکانس‌های بالاتری به یکدیگر می‌رسند.

حال اثر نیروهای برهم‌کنش داخلی در دو پوسته‌ی گاز الکترونی دوبعدی را بر روی معادله پاشندگی نانوتیوپ دوجداره در مد TM بررسی می‌نماییم.

با صفر در نظر گرفتن ضرایب $\alpha, \beta$ می‌توان اثر نیروهای برهم‌کنش داخلی گاز الکترونی را حذف نمود. معادلات پاشندگی در این شرایط برای مد عرضی صفر و شعاع‌های داخلی و خارجی $\begin{cases} a_1 = 1nm \\ a_2 = 1.35nm \end{cases}$ در شکل (۴-۵) رسم شده‌اند.

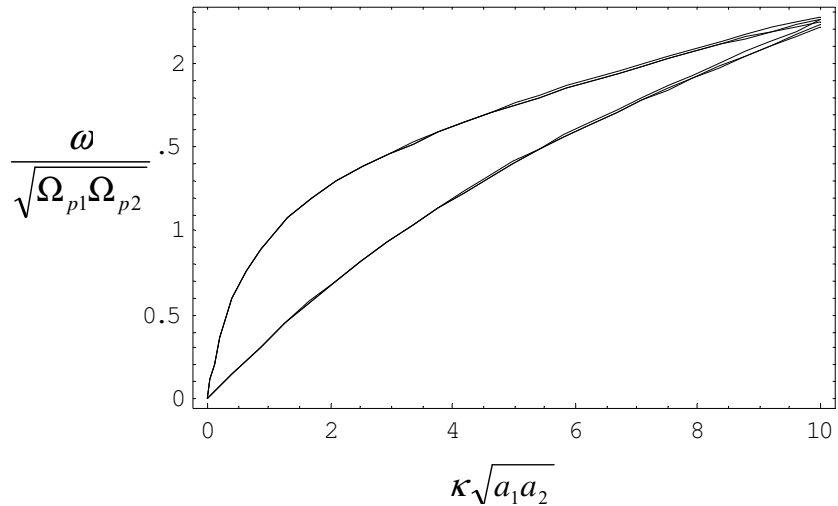

شکل (۴-۵)- نمودارهای پاشندگی با در نظر گرفتن نیروهای داخلی و بدون درنظر گرفتن آن‌ها و یا با در نظر گرفتن بعضی از آن‌ها که در سه بازه نشان داده شده است.

نمودارهای مربوط به حالات $\alpha_1 = 0$، $\beta_1 = 0$ و $\alpha_1, \beta_1 = 0$ تنها اندکی در قسمت انتهایی با نمودار $\alpha_1, \beta_1 \neq 0$ متفاوتند. این امر نشان می‌دهد که در نانوتیوپ دوجداره اثر نیروهای برهم‌کنش داخلی در گاز الکترونی تاثیر کمتری بر روی معادله‌ی پاشندگی نسبت به حالت تک‌جداره دارند.

## ۴-۲- مد TE

### ۴-۲-۱- معادله پاشندگی مد TE برای نانوتیوپ کربنی دوجداره

معادلات شرط مرزی برای میدان الکتریکی در اطراف یک نانولوله‌ی کربنی دوجداره به شعاع‌های درونی و بیرونی $a_1$ و $a_2$، با در نظر گرفتن بسط موج تخت برای میدان و چگالی بار الکترونی (معادلات (۲-۱) و (۲-۲)) به صورت زیر است.

پوسته‌ی درونی:

$$E_{rm}(a_1)\big|_{r>a_1} - E_{rm}(a_1)\big|_{r<a_1} = -\frac{eN_{m1}}{\varepsilon_0}$$

$$E_{llm}(a_1)\big|_{r>a_1} - E_{llm}(a_1)\big|_{r<a_1} = 0 \rightarrow \begin{cases} E_{zm}(a_1)\big|_{r>a_1} - E_{zm}(a_1)\big|_{r<a_1} = 0 \\ E_{\varphi m}(a_1)\big|_{r>a_1} - E_{\varphi m}(a_1)\big|_{r<a_1} = 0 \end{cases}$$

$$N_{m1} = -i\frac{en_0}{m_e} \frac{(qE_{zm} + \frac{m}{a_1}E_{\varphi m})}{\omega^2 - \alpha(q^2 + \frac{m^2}{a_1^2}) - \beta(q^2 + \frac{m^2}{a_1^2})^2}$$

پوسته بیرونی:

$$E_{rm}(a_2)\big|_{r>a_2} - E_{rm}(a_2)\big|_{r<a_2} = -\frac{eN_{m2}}{\varepsilon_0}$$

$$E_{llm}(a_2)\big|_{r>a_2} - E_{llm}(a_2)\big|_{r<a_2} = 0 \rightarrow \begin{cases} E_{zm}(a_2)\big|_{r>a_2} - E_{zm}(a_2)\big|_{r<a_2} = 0 \\ E_{\varphi m}(a_2)\big|_{r>a_2} - E_{\varphi m}(a_2)\big|_{r<a_2} = 0 \end{cases}$$

$$N_{m2} = -i\frac{en_0}{m_e} \frac{(qE_{zm} + \frac{m}{a_2}E_{\varphi m})}{\omega^2 - \alpha(q^2 + \frac{m^2}{a_2^2}) - \beta(q^2 + \frac{m^2}{a_2^2})^2}$$

با توجه بـه معادلـه‌ی (۲-۹) مولفه‌هـای $z, \varphi$ میـدان الکتریکـی در مـد TE ($E_z = 0$) برحسـب مولفـه‌ی z میدان الکتریکی، به شکل زیر خواهند بود.

$$E_{\varphi m} = \frac{i\omega}{\kappa^2}\frac{\partial B_{zm}}{\partial r} \qquad \kappa^2 = q^2 - k^2 \quad , k = \frac{\omega}{c}$$

$$E_{rm} = \frac{m\omega}{\kappa^2 r} B_{zm}$$

روابط بالا را در معادلات شرط مرزی دو پوسته‌ی بیرونی و درونی قرار می‌دهیم.

$$\frac{m\omega}{\kappa^2 r}(B_{zm}(a_1)\big|_{r>a_1} - B_{zm}(a_1)\big|_{r<a_1}) = -\frac{e}{\varepsilon_0}N_{m1} \tag{۴-۹}$$

$$\begin{cases} \dfrac{\partial B_{zm}(a_1)}{\partial r}\bigg|_{r>a_1} - \dfrac{\partial B_{zm}(a_1)}{\partial r}\bigg|_{r<a_1} = 0 \\ \dfrac{i\omega}{\kappa^2}(\dfrac{\partial B_{zm}(a_1)}{\partial r}\bigg|_{r>a_1} - \dfrac{\partial B_{zm}(a_1)}{\partial r}\bigg|_{r<a_1}) = 0 \end{cases} \tag{۴-۱۰}$$

$$\frac{m\omega}{\kappa^2 r}(B_{zm}(a_2)\big|_{r>a_2} - B_{zm}(a_2)\big|_{r<a_2}) = -\frac{e}{\varepsilon_0}N_{m2} \tag{۴-۱۱}$$

$$\begin{cases} \dfrac{\partial B_{zm}(a_2)}{\partial r}\bigg|_{r>a_2} - \dfrac{\partial B_{zm}(a_2)}{\partial r}\bigg|_{r<a_2} = 0 \\ \dfrac{i\omega}{\kappa^2}(\dfrac{\partial B_{zm}(a_2)}{\partial r}\bigg|_{r>a_2} - \dfrac{\partial B_{zm}(a_2)}{\partial r}\bigg|_{r<a_2}) = 0 \end{cases} \tag{۴-۱۲}$$

معادلات شرط مرزی درون کروشه‌ها یکسان هستند.
مولفه‌های z میدان مغناطیسی در معادله‌ی بسل تعمیم‌یافته صدق می‌کنند (معادلات (۲-۱۰)) و جواب فرضی میدان در نواحی مختلف نانوتیوپ بر حسب فاصله‌ی شعاعی تا محور آن به شکل زیر است.

$$\begin{aligned} B_{zm}(r) &= C_{1m}I_m(\kappa r) & r < a_1 \\ B_{zm}(r) &= C_{2m}I_m(\kappa r) + C_{3m}K_m(\kappa r) & a_1 < r < a_2 \\ B_{zm}(r) &= C_{4m}K_m(\kappa r) & r > a_2 \end{aligned}$$

مقدار $N_{m1}$ را در معادله‌ی (۴-۹) قرار می‌دهیم.

$$\frac{m\omega}{\kappa^2 r}(B_{zm}(a_1)\big|_{r>a_1} - B_{zm}(a_1)\big|_{r<a_1}) = -\frac{e}{\varepsilon_0}(-i\frac{en_0}{m_e}\frac{(qE_{zm}+\frac{m}{a_1}E_{\varphi m})}{\omega^2 - \alpha(q^2+\frac{m^2}{a_1^2}) - \beta(q^2+\frac{m^2}{a_1^2})^2})$$

$$\frac{m\omega}{\kappa^2 r}(B_{zm}(a_1)\big|_{r>a_1} - B_{zm}(a_1)\big|_{r<a_1}) = \frac{ie^2 n_0}{m_e \varepsilon_0}\frac{(qE_{zm}+\frac{m}{a_1}E_{\varphi m})}{\omega^2 - \alpha q_1^2 - \beta q_1^4} \qquad ; q_1^2 = q^2 + \frac{m^2}{a_1^2}$$

$$E_{zm} = 0, \quad E_{\varphi m} = \frac{i\omega}{\kappa^2}\frac{\partial B_{zm}}{\partial r}$$

$$\frac{m\omega}{\kappa^2 r}(B_{zm}(a_1)\big|_{r>a_1} - B_{zm}(a_1)\big|_{r<a_1}) = \frac{ie^2 n_0}{m_e \varepsilon_0} \frac{(\frac{m}{a_1}\frac{i\omega}{\kappa^2}\frac{\partial B_{zm}}{\partial r})}{\omega^2 - \alpha q_1^2 - \beta q_1^4}$$

با استفاده از جواب‌های فرضی برای مولفه‌ی $z$ میدان مغناطیسی در نواحی مختلف، معادله‌ی بالا به شکل زیر در می‌آید.

$$(\omega^2 - \alpha q_1^2 - \beta q_1^4)[C_{2m}I_m(\kappa a_1) + C_{3m}K_m(\kappa a_1) - C_{1m}I_m(\kappa a_1)] = \frac{-e^2 n_0}{m_e \varepsilon_0} \frac{\kappa \frac{\partial I(\kappa r)}{\partial(\kappa r)}\big|_{a_1}}{\omega^2 - \alpha q_1^2 - \beta q_1^4}$$

$$(\omega^2 - \alpha q_1^2 - \beta q_1^4)[C_{2m}I_m(\kappa a_1) + C_{3m}K_m(\kappa a_1) - C_{1m}I_m(\kappa a_1)] = -\frac{e^2 n_0}{m_e \varepsilon_0} \kappa C_{1m}I'_m(\kappa a_1)$$

با استفاده از تعریف دو پارامتر جدید $N, R$ معادله‌ی بالا را به شکل ساده‌تری می‌نویسیم. تعریف پارامتر $N$ با بخش ۴-۱-۱- متفاوت است.

$$R = (\omega^2 - \alpha q_1^2 - \beta q_1^4) \qquad , N = -\frac{e^2 n_0}{m_e \varepsilon_0}\kappa$$

$$RC_{2m}I_m(\kappa a_1) + RC_{3m}K_m(\kappa a_1) - RC_{1m}I_m(\kappa a_1) = NC_{1m}I'_m(\kappa a_1) \tag{۴-۱۳}$$

همان‌طور که انتظار داریم بُعد متغیرهای N,R برابر با توان دو فرکانس است و رابطه‌ی (۴-۱۳) نیز از لحاظ ابعادی درست خواهد بود. (متغیرهای $C_{im}I_{im}, C_{im}K_{im}; i=1,2$ دارای بعد میدان مغناطیسی هستند. در متغیرهای $C_{im}I'_{im}, C_{im}K'_{im}; i=1,2$ نیز چون مشتق‌گیری نسبت به متغیر بی‌بعد $\kappa r$ انجام شده، بعد این متغیرها نیز برابر با بعد میدان مغناطیسی خواهد بود.)

برای ساده‌سازی معادله‌ی (۴-۱۱) نیز مطابق روند بالا عمل کرده و معادله‌ی زیر را به دست می‌آوریم.

$$HC_{4m}I_m(\kappa a_2) - HC_{2m}I_m(\kappa a_2) - HC_{3m}K_m(\kappa a_2) = NC_{4m}K'_m(\kappa a_2) \tag{۴-۱۴}$$

$$H = (\omega^2 - \alpha q_2^2 - \beta q_2^4) \qquad ; q_2 = q^2 + \frac{m^2}{a_2^2}$$

با قرار دادن جواب‌های فرضی میدان در معادلات (۴-۱۰) و (۴-۱۲) مادلات زیر به دست می‌آیند.

$$C_{2m}I'_m(\kappa a_1) + C_{3m}K'_m(\kappa a_1) = C_{1m}I'_m(\kappa a_1) \tag{۴-۱۵}$$

$$C_{2m}I'_m(\kappa a_2) + C_{3m}K'_m(\kappa a_2) = C_{4m}K'_m(\kappa a_2) \tag{۴-۱۶}$$

در پیوست ب، ضرایب $C_{1m}, C_{21m}, C_{3m}, C_{4m}$ را از معادلات (۴-۱۳)، (۴-۱۴)، (۴-۱۵) و (۴-۱۶) حذف می‌کنیم و معادله‌ی زیر را بدست می‌آوریم.

**معادله پاشندگی $\omega(\kappa)$:**

$$\omega^4 - [\omega_1^2 + \omega_2^2 + \alpha(q_1^2 + q_2^2) + \beta(q_1^4 + q_2^4)]\omega^2$$
$$+ \omega_1^2\omega_2^2 - \omega_1^2\omega_2^2 F_{12} + (\alpha q_2^2 + \beta q_2^4)\omega_1^2 + (\alpha q_1^2 + \beta q_1^4)\omega_2^2 + (\alpha q_1^2 + \beta q_1^4)(\alpha q_2^2 + \beta q_2^4) = 0$$

$$q_1^2 = (q^2 + m^2/a_1^2)$$
$$q_2^2 = (q^2 + m^2/a_2^2)$$
$$\omega_1^2 = -\Omega_{p1}^2(\kappa a_1)^2[K'_m(\kappa a_1)I_m(\kappa a_1)]; \Omega_{p1} = (e^2 n_0 / m_e e_0 a_1)^{1/2}$$
$$\omega_2^2 = -\Omega_{p2}^2(\kappa a_2)^2[K'_m(\kappa a_2)I_m(\kappa a_2)]; \Omega_{p2} = (e^2 n_0 / m_e e_0 a_2)^{1/2}$$
$$F_{12} = [I'_m(\kappa a_1) / K'_m(\kappa a_1)][K'_m(\kappa a_2) / I'_m(\kappa a_2)]$$

**توجه**: معادله‌ی پاشندگی فوق کاملا شبیه به معادله‌ی پاشندگی نانوتیوپ دوجداره در مد TM است ولی تعاریف متغیرهای $\omega_1, \omega_2, F_{12}$ در دو مد TE,TM متفاوت هستند.

## بررسی صحت معادله

۱. با توجه به اینکه $\Omega_{p2}^2, \Omega_{p1}^2, \omega_1^2, \omega_2^2$ دارای بعد توان دو فرکانس هستند، تمام جملات معادله پاشندگی دارای بعد توان چهارم فرکانس خواهند بود. در نتیجه معادله‌ی فوق از لحاظ ابعادی درست است.

۲. روش دیگر رسیدن به معادله‌ی پاشندگی مانند بخش ۴-۱-۱، حل معادله‌ی شرط جواب غیر بدیهی دستگاه معادلاتِ مربوط به ضرایب $C_{1m} - C_{4m}$ است.

$$RC_{2m}I_m(\kappa a_1) + RC_{3m}K_m(\kappa a_1) - RC_{1m}I_m(\kappa a_1) = NC_{1m}I'_m(\kappa a_1) \qquad (\text{۴-۱۳})$$

$$HC_{4m}I_m(\kappa a_2) - HC_{2m}I_m(\kappa a_2) - HC_{3m}K_m(\kappa a_2) = NC_{4m}K'_m(\kappa a_2) \qquad (\text{۴-۱۴})$$

$$C_{2m}I'_m(\kappa a_1) + C_{3m}K'_m(\kappa a_1) = C_{1m}I'_m(\kappa a_1) \qquad (\text{۴-۱۵})$$

$$C_{2m}I'_m(\kappa a_2) + C_{3m}K'_m(\kappa a_2) = C_{4m}K'_m(\kappa a_2) \qquad (\text{۴-۱۶})$$

$$\begin{pmatrix} -RI_m(\kappa a_1)-NI'_m(\kappa a_1) & RI_m(\kappa a_1) & RK_m(\kappa a_1) & 0 \\ 0 & HI_m(\kappa a_2) & HK_m(\kappa a_2) & NK'_m(\kappa a_2)-HK_m(\kappa a_2) \\ -I'_m(\kappa a_1) & I'_m(\kappa a_1) & K'_m(\kappa a_1) & 0 \\ 0 & I'_m(\kappa a_2) & K'_m(\kappa a_2) & -K'_m(\kappa a_2) \end{pmatrix} \begin{pmatrix} C_{1m} \\ C_{2m} \\ C_{3m} \\ C_{4m} \end{pmatrix} = \begin{pmatrix} 0 \\ 0 \\ 0 \\ 0 \end{pmatrix}$$

شرط وجود جواب غیربدیهی آن است که رابطه‌ی زیر برقرار باشد.

$$\begin{vmatrix} -RI_m(\kappa a_1)-NI'_m(\kappa a_1) & RI_m(\kappa a_1) & RK_m(\kappa a_1) & 0 \\ 0 & HI_m(\kappa a_2) & HK_m(\kappa a_2) & NK'_m(\kappa a_2)-HK_m(\kappa a_2) \\ -I'_m(\kappa a_1) & I'_m(\kappa a_1) & K'_m(\kappa a_1) & 0 \\ 0 & I'_m(\kappa a_2) & K'_m(\kappa a_2) & -K'_m(\kappa a_2) \end{vmatrix} = 0$$

بسط دترمینان حول ستون اول به صورت زیر است.

$$-RI_m(\kappa a_1)-NI'_m(\kappa a_1) \begin{vmatrix} HI_m(\kappa a_2) & HK_m(\kappa a_2) & NK'_m(\kappa a_2)-HK_m(\kappa a_2) \\ I'_m(\kappa a_1) & K'_m(\kappa a_1) & 0 \\ I'_m(\kappa a_2) & K'_m(\kappa a_2) & -K'_m(\kappa a_2) \end{vmatrix}$$

$$-I'_m(\kappa a_1) \begin{vmatrix} RI_m(\kappa a_1) & RK_m(\kappa a_1) & 0 \\ HI_m(\kappa a_2) & HK_m(\kappa a_2) & NK'_m(\kappa a_2)-HK_m(\kappa a_2) \\ I'_m(\kappa a_2) & K'_m(\kappa a_2) & -K'_m(\kappa a_2) \end{vmatrix} = 0$$

هر دو دترمینان فوق را حول سطرهای اول آن‌ها بسط می‌دهیم.

$$[-RI_m(\kappa a_1) - NI'_m(\kappa a_1)][HI_m(\kappa a_2)][-K'_m(\kappa a_1)K'_m(\kappa a_2))]$$
$$[RI_m(\kappa a_1) + NI'_m(\kappa a_1)][HK_m(\kappa a_2)][-I'_m(\kappa a_1)K'_m(\kappa a_2)]$$
$$[-RI_m(\kappa a_1) - NI'_m(\kappa a_1)][NK'_m(\kappa a_2) - HK_m(\kappa a_2)][I'_m(\kappa a_1)K'_m(\kappa a_2) - K'_m(\kappa a_1)I'_m(\kappa a_2)]$$
$$-RI_m(\kappa a_1)I'_m(\kappa a_1)[-HK'_m(\kappa a_2)K_m(\kappa a_2) + K'_m(\kappa a_2)[HK_m(\kappa a_2) - NK'_m(\kappa a_2)]]$$
$$+RK_m(\kappa a_1)I'_m(\kappa a_1)[-HI_m(\kappa a_2)K'_m(\kappa a_2) + I'_m(\kappa a_2)[HK_m(\kappa a_2) - NK'_m(\kappa a_2)]]$$
$$= 0$$

$$[-RI_m(\kappa a_1)][HI_m(\kappa a_2)][-K'_m(\kappa a_1)K'_m(\kappa a_2))]$$
$$[-NI'_m(\kappa a_1)][HI_m(\kappa a_2)][-K'_m(\kappa a_1)K'_m(\kappa a_2))]$$
$$[RI_m(\kappa a_1)][HK_m(\kappa a_2)][-I'_m(\kappa a_1)K'_m(\kappa a_2)]$$
$$[NI'_m(\kappa a_1)][HK_m(\kappa a_2)][-I'_m(\kappa a_1)K'_m(\kappa a_2)]$$
$$[-RI_m(\kappa a_1)][NK'_m(\kappa a_2) - HK_m(\kappa a_2)][I'_m(\kappa a_1)K'_m(\kappa a_2) - K'_m(\kappa a_1)I'_m(\kappa a_2)]$$
$$[-NI'_m(\kappa a_1)][NK'_m(\kappa a_2) - HK_m(\kappa a_2)][I'_m(\kappa a_1)K'_m(\kappa a_2) - K'_m(\kappa a_1)I'_m(\kappa a_2)]$$
$$-RI_m(\kappa a_1)I'_m(\kappa a_1)[-HK'_m(\kappa a_2)K_m(\kappa a_2)]$$
$$-RI_m(\kappa a_1)I'_m(\kappa a_1)K'_m(\kappa a_2)[HK_m(\kappa a_2) - NK'_m(\kappa a_2)]]$$
$$+RK_m(\kappa a_1)I'_m(\kappa a_1)[-HI_m(\kappa a_2)K'_m(\kappa a_2)]$$
$$+RK_m(\kappa a_1)I'_m(\kappa a_1)I'_m(\kappa a_2)[HK_m(\kappa a_2) - NK'_m(\kappa a_2)]$$
$$= 0$$

۶۹

$+ RHI_m(\kappa a_1)I_m(\kappa a_2)K'_m(\kappa a_1)K'_m(\kappa a_2)$

$+ NHI'_m(\kappa a_1)I_m(\kappa a_2)K'_m(\kappa a_1)K'_m(\kappa a_2)$

$- RHI_m(\kappa a_1)K_m(\kappa a_2)I'_m(\kappa a_1)K'_m(\kappa a_2)$

$- NHI'_m(\kappa a_1)K_m(\kappa a_2)I'_m(\kappa a_1)K'_m(\kappa a_2)$

$- RNI_m(\kappa a_1)K'_m(\kappa a_2)I'_m(\kappa a_1)K'_m(\kappa a_2)$

$+ RNI_m(\kappa a_1)K'_m(\kappa a_2)K'_m(\kappa a_1)I'_m(\kappa a_2)$

$+ RHI_m(\kappa a_1)K_m(\kappa a_2)I'_m(\kappa a_1)K'_m(\kappa a_2)$

$- RHI_m(\kappa a_1)K_m(\kappa a_2)K'_m(\kappa a_1)I'_m(\kappa a_2)$

$- N^2 I'_m(\kappa a_1)K'_m(\kappa a_2)I'_m(\kappa a_1)K'_m(\kappa a_2)$

$+ N^2 I'_m(\kappa a_1)K'_m(\kappa a_2)K'_m(\kappa a_1)I'_m(\kappa a_2)$

$+ NHI'_m(\kappa a_1)K_m(\kappa a_2)I'_m(\kappa a_1)K'_m(\kappa a_2)$

$- NHI'_m(\kappa a_1)K_m(\kappa a_2)K'_m(\kappa a_1)I'_m(\kappa a_2)$

$+ RHI_m(\kappa a_1)I'_m(\kappa a_1)K'_m(\kappa a_2)K_m(\kappa a_2)$

$- RHI_m(\kappa a_1)I'_m(\kappa a_1)K'_m(\kappa a_2)K_m(\kappa a_2)$

$+ RNI_m(\kappa a_1)I'_m(\kappa a_1)K'_m(\kappa a_2)K'_m(\kappa a_2)$

$- RHK_m(\kappa a_1)I'_m(\kappa a_1)I_m(\kappa a_2)K'_m(\kappa a_2)$

$+ RHK_m(\kappa a_1)I'_m(\kappa a_1)I'_m(\kappa a_2)K_m(\kappa a_2)$

$- RNK_m(\kappa a_1)I'_m(\kappa a_1)I'_m(\kappa a_2)K'_m(\kappa a_2)$

$= 0$

با ساده کردن بعضی جملات مشابه و با علامت مخالف، معادله‌ی زیر به دست می‌آید.

$+ RHI_m(\kappa a_1)I_m(\kappa a_2)K'_m(\kappa a_1)K'_m(\kappa a_2)$

$+ NHI'_m(\kappa a_1)I_m(\kappa a_2)K'_m(\kappa a_1)K'_m(\kappa a_2)$

$+ RNI_m(\kappa a_1)K'_m(\kappa a_2)K'_m(\kappa a_1)I'_m(\kappa a_2)$

$- RHI_m(\kappa a_1)K_m(\kappa a_2)K'_m(\kappa a_1)I'_m(\kappa a_2)$

$- N^2 I'_m(\kappa a_1)K'_m(\kappa a_2)I'_m(\kappa a_1)K'_m(\kappa a_2)$

$+ N^2 I'_m(\kappa a_1)K'_m(\kappa a_2)K'_m(\kappa a_1)I'_m(\kappa a_2)$

$- NHI'_m(\kappa a_1)K_m(\kappa a_2)K'_m(\kappa a_1)I'_m(\kappa a_2)$

$- RHK_m(\kappa a_1)I'_m(\kappa a_1)I_m(\kappa a_2)K'_m(\kappa a_2)$

$+ RHK_m(\kappa a_1)I'_m(\kappa a_1)I'_m(\kappa a_2)K_m(\kappa a_2)$

$- RNK_m(\kappa a_1)I'_m(\kappa a_1)I'_m(\kappa a_2)K'_m(\kappa a_2)$

$= 0$

فرمول (B) از پیوست ب:

-R* H* K (ka۲) * I'(ka۲)/ K'(ka۲) *K (ka۱) * I'(ka۱)/ K'(ka۱)

+ R* H* K (ka۲) * I'(ka۲)/ K'(ka۲) * I (ka۱)

+ H*N * K (ka۲) * I'(ka۲)/ K'(ka۲) *I'(ka۱)

+ H*R *I (ka۲) *K (ka۱) * I'(ka۱)/ K'(ka۱)

- H*R *I (ka۲) * I (ka۱)                           (B)

- H* N *I (ka۲) * I'(ka۱)

+N*R * K'(ka۲) * I'(ka۲)/ K'(ka۲) *K (ka۱) * I'(ka۱)/ K'(ka۱)

-N* R * K'(ka۲) * I'(ka۲)/ K'(ka۲) * I (ka۱)

-N*N * K'(ka۲) * I'(ka۲)/ K'(ka۲) *I'(ka۱)

+ N* N* K'(ka۲)*I'(ka۱)/ K'(ka۱)* I'(ka۱)

=۰

طرفین فرمول (B) را در (a۲) 'K *(a۱) 'K ضرب می‌کنیم.

-R* H* K (ka۲) * I'(ka۲) *K (ka۱) * I'(ka۱)

+ R* H* K (ka۲) * I'(ka۲)* K'(ka۱) * I (ka۱)

+ H*N * K (ka۲) * I'(ka۲)*K'(ka۱) *I'(ka۱)

+ H*R *I (ka۲) *K (ka۱) * I'(ka۱)* K'(ka۲)

- H*R *I (ka۲) * I (ka۱)* K' (a۱)* K' (a۲)

- H* N *I (ka۲) * I'(ka۱)* K' (a۱)* K' (a۲)

+N*R * K'(ka۲) * I'(ka۲) *K (ka۱) * I'(ka۱)

-N* R * K'(ka۲) * I'(ka۲)* K'(ka۱) * I (ka۱)

-N*N * K'(ka۲) * I'(ka۲)* K'(ka۱) *I'(ka۱)

+ N* N* K'(ka۲)*I'(ka۱)* K'(ka۲)* I'(ka۱)

=۰

معادله‌ی فوق که از پیوست (B) به دست آمده، منفی معادله‌ی حاصل از حل دترمینان است. این امر درستی محاسبات پیوست ب تا معادله‌ی (B) را نشان می‌دهد.

## ۴-۲-۲- نمودارهای معادله‌ی پاشندگی در شرایط مختلف

معادله‌ی پاشندگی نانوتیوپ کربنی دوجداره در مد TE به صورت زیر است.

$$\omega^4 - [\omega_1^2 + \omega_2^2 + \alpha(q_1^2 + q_2^2) + \beta(q_1^4 + q_2^4)]\omega^2$$
$$+ \omega_1^2\omega_2^2 - \omega_1^2\omega_2^2 F_{12} + (\alpha q_2^2 + \beta q_2^4)\omega_1^2 + (\alpha q_1^2 + \beta q_1^4)\omega_2^2 + (\alpha q_1^2 + \beta q_1^4)(\alpha q_2^2 + \beta q_2^4) = 0$$

$$q_1^2 = (q^2 + m^2/a_1^2)$$
$$q_2^2 = (q^2 + m^2/a_2^2)$$
$$\omega_1^2 = -\Omega_{p1}^2(\kappa a_1)^2[K'_m(\kappa a_1)I'_m(\kappa a_1)]; \Omega_{p1} = (e^2 n_0 / m_e e_0 a_1)^{1/2}$$
$$\omega_2^2 = -\Omega_{p2}^2(\kappa a_2)^2[K'_m(\kappa a_2)I'_m(\kappa a_2)]; \Omega_{p2} = (e^2 n_0 / m_e e_0 a_2)^{1/2}$$
$$F_{12} = [I'_m(\kappa a_1)/K'_m(\kappa a_1)][K'_m(\kappa a_2)/I'_m(\kappa a_2)]$$

قبل از رسم معادله‌ی پاشندگی به بی‌بعد کردن آن می‌پردازیم.

کمیت‌های بی‌بعد را همان‌طور که در فصل ۴-۱-۲ توضیح داده شد به شکل زیر تعریف می‌کنیم.

$$x = \kappa\sqrt{a_1 a_2}, \quad y = \frac{\omega}{\sqrt{\Omega_{p1}\Omega_{p2}}}$$

که در آن فرکانس‌های ویژه‌ی سیستم به شکل زیر هستند.

$$[\Omega_{p1} = (e^2 n_0 / \varepsilon_0 m_e a_1)^{1/2}, \Omega_{p2} = (e^2 n_0 / \varepsilon_0 m_e a_2)^{1/2}]$$

به منظور بی‌بعد کردن معادله‌ی پاشندگی طرفین آن‌را بر $\Omega_{p1}^2\Omega_{p2}^2$ تقسیم می‌کنیم.

$$\frac{\omega^4}{\Omega_{p1}^2\Omega_{p2}^2} - [\frac{\omega_1^2}{\Omega_{p1}\Omega_{p2}} + \frac{\omega_2^2}{\Omega_{p1}\Omega_{p2}} + \frac{\alpha}{\Omega_{p1}\Omega_{p2}}(q_1^2 + q_2^2) + \frac{\beta}{\Omega_{p1}\Omega_{p2}}(q_1^4 + q_2^4)]\frac{\omega^2}{\Omega_{p1}\Omega_{p2}}$$
$$+ \frac{\omega_1^2\omega_2^2}{\Omega_{p1}^2\Omega_{p2}^2} - \frac{\omega_1^2\omega_2^2 F_{12}}{\Omega_{p1}^2\Omega_{p2}^2} + \frac{(\alpha q_2^2 + \beta q_2^4)\omega_1^2}{\Omega_{p1}^2\Omega_{p2}^2} + \frac{(\alpha q_1^2 + \beta q_1^4)\omega_2^2}{\Omega_{p1}^2\Omega_{p2}^2} + \frac{(\alpha q_1^2 + \beta q_1^4)(\alpha q_2^2 + \beta q_2^4)}{\Omega_{p1}^2\Omega_{p2}^2} = 0$$

برای تبدیل پارامترهای $q_1, q_2$ به $\kappa$ و بعد از آن به x ، عبارت $\sqrt{a_1 a_2}$ را در $q_1, q_2$ های معادله ضرب و تقسیم می‌کنیم.

$$y^4 - [\frac{\omega_1^2}{\Omega_{p1}\Omega_{p2}} + \frac{\omega_2^2}{\Omega_{p1}\Omega_{p2}} + \frac{\alpha}{\Omega_{p1}\Omega_{p2}a_1a_2}(a_1a_2q_1^2 + a_1a_2q_2^2)$$

$$+ \frac{\beta}{\Omega_{p1}\Omega_{p2}a_1^2a_2^2}(a_1^2a_2^2q_1^4 + a_1^2a_2^2q_2^4)]y^2$$

$$+ \frac{\omega_1^2\omega_2^2}{\Omega_{p1}^2\Omega_{p2}^2} - \frac{\omega_1^2\omega_2^2 F_{12}}{\Omega_{p1}^2\Omega_{p2}^2}$$

$$+ (\frac{\alpha}{\Omega_{p1}\Omega_{p2}a_1a_2}a_1a_2q_2^2)\frac{\omega_1^2}{\Omega_{p1}\Omega_{p2}} + (\frac{\beta}{\Omega_{p1}\Omega_{p2}a_1^2a_2^2}a_1^2a_2^2q_2^4)\frac{\omega_1^2}{\Omega_{p1}\Omega_{p2}}$$

$$+ (\frac{\alpha}{\Omega_{p1}\Omega_{p2}a_1a_2}a_1a_2q_1^2)\frac{\omega_2^2}{\Omega_{p1}\Omega_{p2}} + (\frac{\beta}{\Omega_{p1}\Omega_{p2}a_1^2a_2^2}a_1^2a_2^2q_1^4)\frac{\omega_2^2}{\Omega_{p1}\Omega_{p2}}$$

$$+ (\frac{\alpha}{\Omega_{p1}\Omega_{p2}a_1a_2}a_1a_2q_1^2 + \frac{\beta}{\Omega_{p1}\Omega_{p2}a_1^2a_2^2}a_1^2a_2^2q_1^4)(\frac{\alpha}{\Omega_{p1}\Omega_{p2}a_1a_2}a_1a_2q_2^2 + \frac{\beta}{\Omega_{p1}\Omega_{p2}a_1^2a_2^2}a_1^2a_2^2q_2^4)$$

$$= 0$$

متغیرهای بی‌بعد $\sigma, \alpha_1, \beta_1$ را به صورت زیر تعریف کرده‌ایم.

$$\sigma = \Omega_p^2 a^2/c^2 : SingleWalled \qquad , \sigma = \Omega_{p1}\Omega_{p2}a_1a_2/c^2 : DoubleWalled$$

$$\alpha_1 = \frac{\alpha}{\Omega_p^2 a^2} : SingleWalled \qquad , \alpha_1 = \frac{\alpha}{\Omega_{p1}\Omega_{p2}a_1a_2} : DoubleWalled$$

$$\beta_1 = \frac{\beta}{\Omega_p^2 a^2}; SingleWalled \qquad , \beta_1 = \frac{\beta}{\Omega_{p1}\Omega_{p2}a_1^2a_2^2}; DoubleWalled$$

معادله‌ی پاشندگی با استفاده از متغیرهای بالا به صورت زیر تبدیل می‌شود.

$$y^4 - [\frac{\omega_1^2}{\Omega_{p1}\Omega_{p2}} + \frac{\omega_2^2}{\Omega_{p1}\Omega_{p2}} + \alpha_1(a_1a_2q_1^2 + a_1a_2q_2^2) + \beta_1(a_1^2a_2^2q_1^4 + a_1^2a_2^2q_2^4)]y^2$$

$$+ \frac{\omega_1^2\omega_2^2}{\Omega_{p1}^2\Omega_{p2}^2} - \frac{\omega_1^2\omega_2^2 F_{12}}{\Omega_{p1}^2\Omega_{p2}^2}$$

$$+ \alpha_1(a_1a_2q_2^2)\frac{\omega_1^2}{\Omega_{p1}\Omega_{p2}} + \beta_1(a_1^2a_2^2q_2^4)\frac{\omega_1^2}{\Omega_{p1}\Omega_{p2}}$$

$$+ \alpha_1(a_1a_2q_1^2)\frac{\omega_2^2}{\Omega_{p1}\Omega_{p2}} + \beta_1(a_1^2a_2^2q_1^4)\frac{\omega_2^2}{\Omega_{p1}\Omega_{p2}}$$

$$+ (\alpha_1 a_1a_2q_1^2 + \beta_1 a_1^2a_2^2q_1^4)(\alpha_1 a_1a_2q_2^2 + \beta_1 a_1^2a_2^2q_2^4)$$

$$= 0$$

بقیه‌ی پارامترها در معادله پاشندگی به صورت زیر ساده می‌شوند.

$$q_1^2 = (q^2 + m^2/a_1^2) = [\kappa^2 + \frac{\omega^2}{c^2} + m^2/a_1^2]$$

$$a_1 a_2 q_1^2 = [\kappa^2 a_1 a_2 + \frac{\Omega_{P1}\Omega_{P2} a_1 a_2}{c^2} \frac{\omega^2}{\Omega_{P1}\Omega_{P2}} + m^2 a_1 a_2/a_1^2] = [x^2 + \sigma y^2 + m^2 a_2/a_1]$$

$$q_2^2 = (q^2 + m^2/a_2^2) = [(\kappa^2 + \frac{\omega^2}{c^2}) + m^2/a_2^2]$$

$$a_1 a_2 q_2^2 = [a_1 a_2 \kappa^2 + \frac{\Omega_{P1}\Omega_{P2} a_1 a_2}{c^2} \frac{\omega^2}{\Omega_{P1}\Omega_{P2}} + m^2 a_1 a_2/a_2^2] = [x^2 + \sigma y^2 + m^2 a_1/a_2]$$

$$\omega_1^2 = -\Omega_{p1}^2 (\kappa a_1)^2 [K'_m(\kappa a_1) I'_m(\kappa a_1)]; \Omega_{p1} = (e^2 n_0 / m_e e_0 a_1)^{1/2}$$
$$\kappa a_1 = \kappa \sqrt{a_1/a_2} \sqrt{a_1 a_2}, (\kappa a_1)^2 = (\kappa^2 a_1 a_2)(a_1/a_2) \rightarrow$$
$$\omega_1^2 = -\Omega_{P1}^2 (\kappa^2 a_1 a_2)(a_1/a_2) K'_m(\kappa \sqrt{a_1 a_2} \sqrt{a_1/a_2}) I'_m(\kappa \sqrt{a_1 a_2} \sqrt{a_1/a_2})$$
$$= -\Omega_{P1}^2 (a_1/a_2) x^2 K'_m(\sqrt{a_1/a_2}\, x) I'_m(\sqrt{a_1/a_2}\, x)$$

$$\omega_2^2 = -\Omega_{p2}^2 (\kappa a_2)^2 [K'_m(\kappa a_2) I'_m(\kappa a_2)]; \Omega_{p2} = (e^2 n_0 / m_e e_0 a_2)^{1/2}$$
$$(\kappa a_2)^2 = (\kappa^2 a_1 a_2)(a_2/a_1), \kappa a_2 = \kappa \sqrt{a_1 a_2} \sqrt{a_2/a_1} \rightarrow$$
$$\omega_2^2 = -\Omega_{P2}^2 (\kappa \sqrt{a_1 a_2})^2 (a_2/a_1) K'_m(\kappa \sqrt{a_1 a_2} \sqrt{a_2/a_1}) I'_m(\kappa \sqrt{a_1 a_2} \sqrt{a_2/a_1})$$
$$= -\Omega_{P2}^2 (a_2/a_1) x^2 K'_m(\sqrt{a_2/a_1}\, x) I'_m(\sqrt{a_2/a_1}\, x)$$

$$\Omega_{p1} = (e^2 n_0 / \varepsilon_0 m_e a_1)^{1/2}, \Omega_{p2} = (e^2 n_0 / \varepsilon_0 m_e a_2)^{1/2} \rightarrow \Omega_{p1}/\Omega_{p2} = \sqrt{a_2/a_1}$$

$$\frac{\omega_1^2}{\Omega_{p1}\Omega_{p2}} = -\frac{\Omega_{p1}}{\Omega_{p2}}(a_1/a_2) x^2 K'_m(\sqrt{a_1/a_2}\, x) I'_m(\sqrt{a_1/a_2}\, x)$$
$$= -x^2 \sqrt{a_1/a_2}\, K'_m(\sqrt{a_1/a_2}\, x) I'_m(\sqrt{a_1/a_2}\, x)$$

$$\frac{\omega_2^2}{\Omega_{p1}\Omega_{p2}} = -\frac{\Omega_{p2}}{\Omega_{p1}}(a_2/a_1)x^2 K'_m(\sqrt{a_2/a_1}x)I'_m(\sqrt{a_2/a_1}x)$$
$$= -x^2\sqrt{a_2/a_1}K'_m(\sqrt{a_2/a_1}x)I'_m(\sqrt{a_2/a_1}x)$$

$$\frac{\omega_1^2\omega_2^2}{\Omega_{p1}^2\Omega_{p2}^2} = x^4 I'_m(\sqrt{a_1/a_2}x)I'_m(\sqrt{a_2/a_1}x)K'_m(\sqrt{a_1/a_2}x)K'_m(\sqrt{a_2/a_1}x)$$

$$\frac{\omega_1^2\omega_2^2 F_{12}}{\Omega_{p1}^2\Omega_{p2}^2} = x^4 K'_m(\sqrt{a_1/a_2}x)I'_m(\sqrt{a_1/a_2}x)K'_m(\sqrt{a_2/a_1}x)I'_m(\sqrt{a_2/a_1}x)$$
$$\times I'_m(\sqrt{a_1/a_2}x)K'_m(\sqrt{a_2/a_1}x)/[K'_m(\sqrt{a_1/a_2}x)I'_m(\sqrt{a_2/a_1}x)]$$
$$= x^4 I'^2_m(\sqrt{a_1/a_2}x)K'^2_m(\sqrt{a_2/a_1}x)$$

با جایگذاری مقادیر ثابت، معادله‌ی پاشندگی به شکل زیر تبدیل می‌شود.

$$y^4 - [-x^2\sqrt{a_1/a_2}K'_m(\sqrt{a_1/a_2}x)I'_m(\sqrt{a_1/a_2}x) - x^2\sqrt{a_2/a_1}K'_m(\sqrt{a_2/a_1}x)I'_m(\sqrt{a_2/a_1}x)$$
$$+\alpha_1([x^2+\sigma y^2+m^2 a_2/a_1]+[x^2+\sigma y^2+m^2 a_1/a_2])$$
$$+\beta_1([x^2+\sigma y^2+m^2 a_2/a_1]^2+[x^2+\sigma y^2+m^2 a_1/a_2]^2)]y^2$$
$$+x^4 I'_m(\sqrt{a_1/a_2}x)I'_m(\sqrt{a_2/a_1}x)K'_m(\sqrt{a_1/a_2}x)K'_m(\sqrt{a_2/a_1}x)$$
$$-x^4 I'^2_m(\sqrt{a_1/a_2}x)K'^2_m(\sqrt{a_2/a_1}x)$$
$$-\alpha_1[x^2+\sigma y^2+m^2 a_1/a_2]x^2\sqrt{a_1/a_2}K'_m(\sqrt{a_1/a_2}x)I'_m(\sqrt{a_1/a_2}x)$$
$$-\beta_1[x^2+\sigma y^2+m^2 a_1/a_2]^2 x^2\sqrt{a_1/a_2}K'_m(\sqrt{a_1/a_2}x)I'_m(\sqrt{a_1/a_2}x)$$
$$-\alpha_1[x^2+\sigma y^2+m^2 a_2/a_1]x^2\sqrt{a_2/a_1}K'_m(\sqrt{a_2/a_1}x)I'_m(\sqrt{a_2/a_1}x)$$
$$-\beta_1[x^2+\sigma y^2+m^2 a_2/a_1]^2 x^2\sqrt{a_2/a_1}K'_m(\sqrt{a_2/a_1}x)I'_m(\sqrt{a_2/a_1}x)$$
$$+(\alpha_1[x^2+\sigma y^2+m^2 a_2/a_1]+\beta_1[x^2+\sigma y^2+m^2 a_2/a_1]^2)\times$$
$$(\alpha_1[x^2+\sigma y^2+m^2 a_1/a_2]+\beta_1[x^2+\sigma y^2+m^2 a_1/a_2]^2)$$
$$=0$$

جملات $\sigma y^2$ معادله پاشندگی بی‌بعد شده را مانند قبل حذف می‌کنیم.

معادله‌ی زیر، معادله‌ی پاشندگی بی‌بعد شده‌ی نانوتیوپ کربنی دوجداره در مد TE است.

$$y^4 - [-x^2\sqrt{a_1/a_2}K'_m(\sqrt{a_1/a_2}x)I'_m(\sqrt{a_1/a_2}x) - x^2\sqrt{a_2/a_1}K'_m(\sqrt{a_2/a_1}x)I'_m(\sqrt{a_2/a_1}x)$$
$$+ \alpha_1([x^2+m^2a_2/a_1]+[x^2+m^2a_1/a_2])$$
$$+ \beta_1([x^2+m^2a_2/a_1]^2+[x^2+m^2a_1/a_2]^2)]y^2$$
$$+ x^4 I'_m(\sqrt{a_1/a_2}x)I'_m(\sqrt{a_2/a_1}x)K'_m(\sqrt{a_1/a_2}x)K'_m(\sqrt{a_2/a_1}x)$$
$$- x^4 I'^2_m(\sqrt{a_1/a_2}x)K'^2_m(\sqrt{a_2/a_1}x)$$
$$- \alpha_1[x^2+m^2a_1/a_2]x^2\sqrt{a_1/a_2}K'_m(\sqrt{a_1/a_2}x)I'_m(\sqrt{a_1/a_2}x)$$
$$- \beta_1[x^2+m^2a_1/a_2]^2 x^2\sqrt{a_1/a_2}K'_m(\sqrt{a_1/a_2}x)I'_m(\sqrt{a_1/a_2}x)$$
$$- \alpha_1[x^2+m^2a_2/a_1]x^2\sqrt{a_2/a_1}K'_m(\sqrt{a_2/a_1}x)I'_m(\sqrt{a_2/a_1}x)$$
$$- \beta_1[x^2+m^2a_2/a_1]^2 x^2\sqrt{a_2/a_1}K'_m(\sqrt{a_2/a_1}x)I'_m(\sqrt{a_2/a_1}x)$$
$$+ (\alpha_1[x^2+m^2a_2/a_1]+\beta_1[x^2+m^2a_2/a_1]^2)\times$$
$$(\alpha_1[x^2+m^2a_1/a_2]+\beta_1[x^2+m^2a_1/a_2]^2)$$
$$= 0$$

به منظور رسم این معادله‌ی درجه چهار، به حل آن می‌پردازیم.

$$Ay^4 + By^2 + C = 0$$
$$y^2 = \frac{-B \pm \sqrt{B^2-4AC}}{2A} = \frac{-B}{2A} \pm \sqrt{\frac{B^2}{4A^2} - \frac{C}{A}}$$
$$A = 1$$
$$B = -[-x^2\sqrt{a_1/a_2}K'_m(\sqrt{a_1/a_2}x)I'_m(\sqrt{a_1/a_2}x) - x^2\sqrt{a_2/a_1}K'_m(\sqrt{a_2/a_1}x)I'_m(\sqrt{a_2/a_1}x)$$
$$\quad + \alpha_1([x^2+m^2a_2/a_1]+[x^2+m^2a_1/a_2])$$
$$\quad + \beta_1([x^2+m^2a_2/a_1]^2+[x^2+m^2a_1/a_2]^2)]$$
$$C = +x^4 I'_m(\sqrt{a_1/a_2}x)I'_m(\sqrt{a_2/a_1}x)K'_m(\sqrt{a_1/a_2}x)K'_m(\sqrt{a_2/a_1}x)$$
$$\quad - x^4 I'^2_m(\sqrt{a_1/a_2}x)K'^2_m(\sqrt{a_2/a_1}x)$$
$$\quad - \alpha_1[x^2+m^2a_1/a_2]x^2\sqrt{a_1/a_2}K'_m(\sqrt{a_1/a_2}x)I'_m(\sqrt{a_1/a_2}x)$$
$$\quad - \beta_1[x^2+m^2a_1/a_2]^2 x^2\sqrt{a_1/a_2}K'_m(\sqrt{a_1/a_2}x)I'_m(\sqrt{a_1/a_2}x)$$
$$\quad - \alpha_1[x^2+m^2a_2/a_1]x^2\sqrt{a_2/a_1}K'_m(\sqrt{a_2/a_1}x)I'_m(\sqrt{a_2/a_1}x)$$
$$\quad - \beta_1[x^2+m^2a_2/a_1]^2 x^2\sqrt{a_2/a_1}K'_m(\sqrt{a_2/a_1}x)I'_m(\sqrt{a_2/a_1}x)$$
$$\quad + (\alpha_1[x^2+m^2a_2/a_1]+\beta_1[x^2+m^2a_2/a_1]^2)\times(\alpha_1[x^2+m^2a_1/a_2]+\beta_1[x^2+m^2a_1/a_2]^2)$$

$$y = \sqrt{\frac{-B}{2} \pm \sqrt{\frac{B^2}{4} - C}} \quad ; (\frac{B^2}{4} - C) \geq 0 \quad , (\frac{-B}{2} \pm \sqrt{\frac{B^2}{4} - C}) \geq 0$$

ضرایب $\alpha_1, \beta_1$ درمعادله را برحسب شعاع‌های $a_1, a_2$ مشخص می‌کنیم.

$$\alpha_1 = \alpha/(\Omega_{p1}\Omega_{p2}a_1 a_2) \quad ; \alpha = \frac{e^2 n_0 a_B}{4\varepsilon_0 m_e} \quad , a_B = 5.29 \times 10^{-11} m$$

$$, \Omega_{P1} = (\frac{e^2 n_0}{\varepsilon_0 m_e a_1})^{1/2}, \Omega_{P2} = (\frac{e^2 n_0}{\varepsilon_0 m_e a_2})^{1/2}$$

$$\beta_1 = \beta/(\Omega_{p1}\Omega_{p2}a_1^2 a_2^2) \quad ; \beta = \frac{e^2 a_B}{16\pi\varepsilon_0 m_e}$$

$$\alpha_1 = \frac{\alpha}{\Omega_{p1}\Omega_{p2}a_1 a_2} = \frac{e^2 n_0 a_B}{4\varepsilon_0 m_e} \cdot \frac{1}{(\frac{e^2 n_0}{\varepsilon_0 m_e a_1})^{1/2}(\frac{e^2 n_0}{\varepsilon_0 m_e a_2})^{1/2} a_1 a_2} = \frac{a_B}{4\sqrt{a_1 a_2}}$$

$$= \frac{1.32 \times 10^{-11} m}{\sqrt{a_1(m)a_2(m)}} = \frac{1.32 \times 10^{-2} nm}{\sqrt{a_1(nm)a_2(nm)}}$$

با توجه به مقدار $n_0 = 152 nm^{-2}$ که در فصل ۳ محاسبه شد، داریم:

$$\beta_1 = \frac{\beta}{\Omega_{p1}\Omega_{p2}a_1^2 a_2^2} = \frac{e^2 a_B}{16\pi\varepsilon_0 m_e} \cdot \frac{1}{(\frac{e^2 n_0}{\varepsilon_0 m_e a_1})^{1/2}(\frac{e^2 n_0}{\varepsilon_0 m_e a_2})^{1/2} a_1^2 a_2^2} = \frac{a_B}{16\pi n_0 \sqrt{(a_1 a_2)^3}}$$

$$= \frac{5.29 \times 10^{-11} m}{16 \times 3.14 \times 152 \times 10^{18} m^{-2} \times \sqrt{[a_1(m)a_2(m)]^3}} = \frac{6.92 \times 10^{-4} \times 10^{-29}}{\sqrt{[a_1(m)a_2(m)]^3}}$$

$$= \frac{6.92 \times 10^{-6}}{\sqrt{[a_1(nm)a_2(nm)]^3}}$$

قبل از رسم نمودارهای پاشندگی برای این مد یادآور می‌شویم که معادلات این بخش با معادلات مد TM تنها در ویژه فرکانس‌های $\omega_1, \omega_2$ متفاوت بودند. به این ترتیب نمودارهای پاشندگی برای دو مد TM,TE تفاوت چندانی با یکدیگر نخواهند داشت.

نانوتیوپی به شعاع‌های داخلی و خارجی $\begin{cases} a_1 = 1nm \\ a_2 = 1.35nm \end{cases}$ را در نظر می‌گیریم. برای مدهای عرضی ۰،۱،۲،۳،۴=m ، نمودارهای پاشندگی مربوط به آن را رسم می‌کنیم.

ضرایب به‌کار رفته در معادله‌ی پاشندگی فوق به صورت زیر هستند.

۷۷

$$\alpha_1 = \frac{1.32 \times 10^{-2} \, nm}{\sqrt{a_1(nm) a_2(nm)}} = \frac{1.32 \times 10^{-2} \, nm}{\sqrt{1(nm) 1.35(nm)}} = 1.14 \times 10^{-3}$$

$$\beta_1 = \frac{6.92 \times 10^{-6}}{\sqrt{[a_1(nm) a_2(nm)]^3}} = \frac{6.92 \times 10^{-6}}{\sqrt{[1(nm) 1.35(nm)]^3}} = 4.41 \times 10^{-6}$$

$$a_1/a_2 = 0.74, \sqrt{a_1/a_2} = 0.86, a_2/a_1 = 1.35, \sqrt{a_2/a_1} = 1.16$$

در شکل (۴-۶) نمودارهای پاشندگی برای مدهای عرضی m=۰،۱،۲،۳،۴ درنانوتیوپ کربنی فوق به طور جداگانه نشان داده شده‌اند.

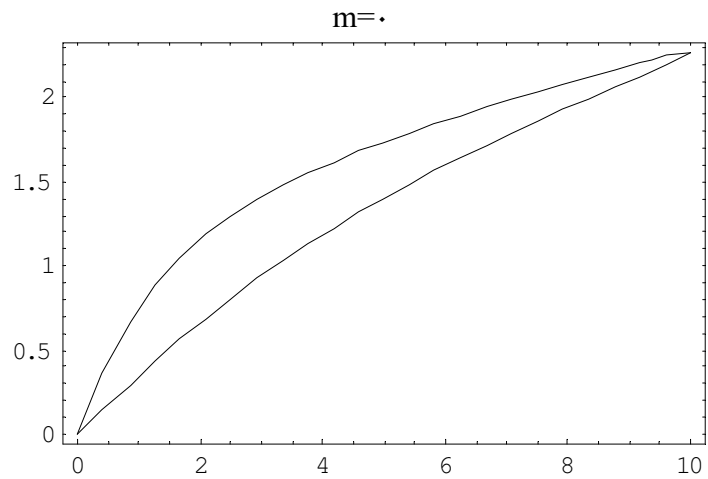

m=۰

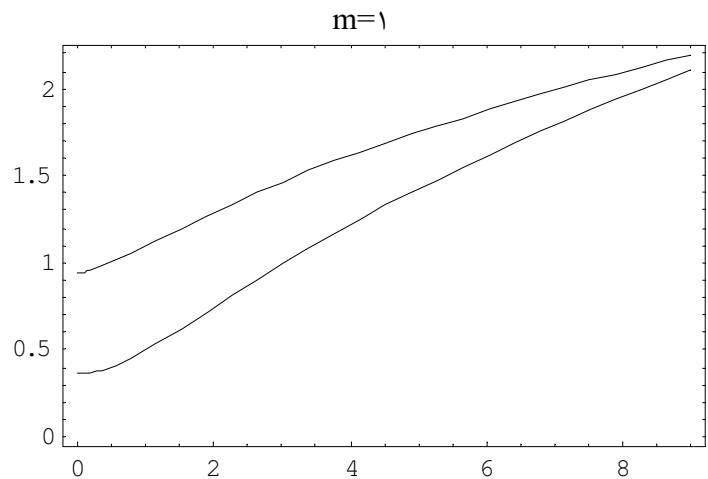

m=۱

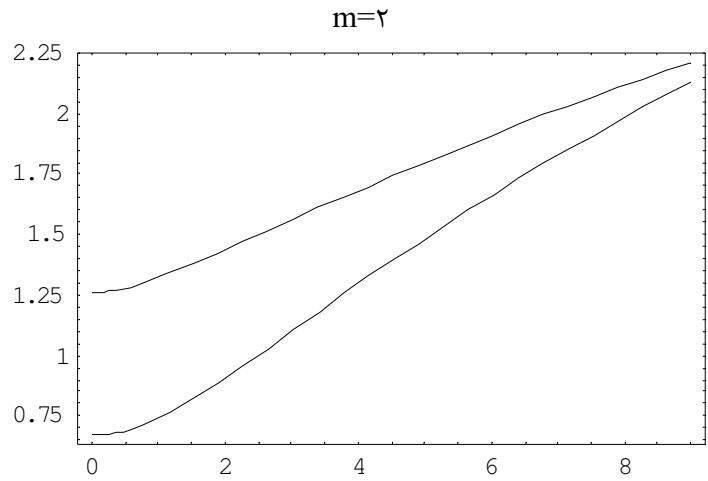

m=۲

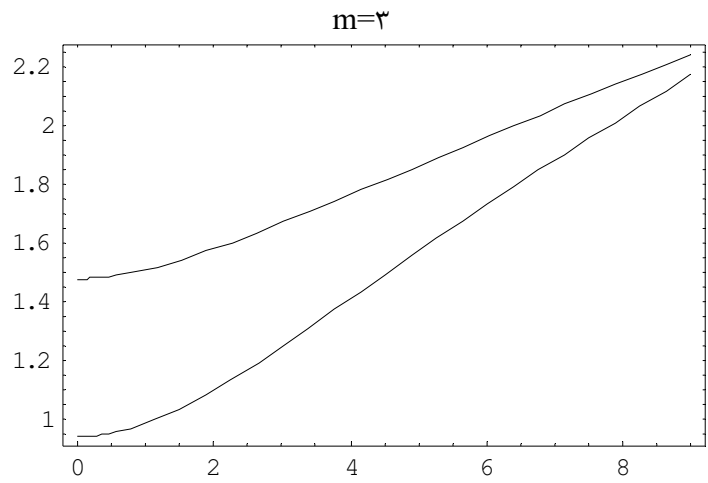

m=۳

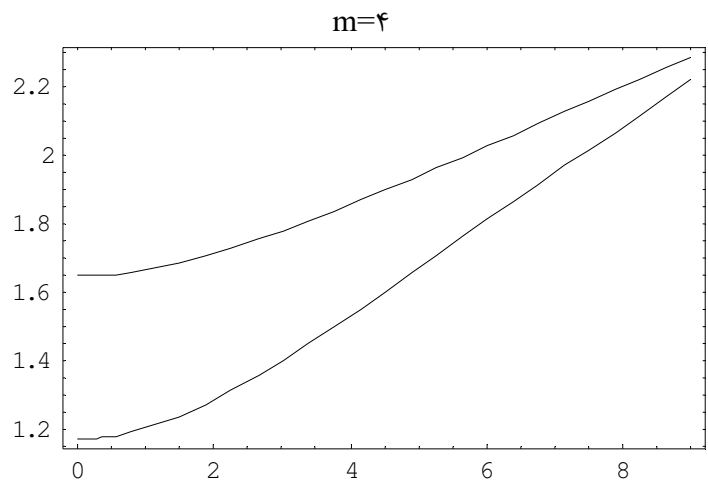

m=۴

شکل (۴-۶)- نمودار پاشندگی با مدهای عرضی مختلف نانوتیوپ دوجداره. محورهای عمودی فرکانس بی‌بعد شده‌ی

$\omega/\sqrt{\Omega_{p1}\Omega_{p2}}$ و محورهای افقی عدد موج بی‌بعد شده‌ی $q\sqrt{a_1 a_2}$ هستند.

در شکل (۴-۷) نمودارهای مربوط به مدهای مختلف عرضی همگی در یک نمودار آورده شده‌اند.

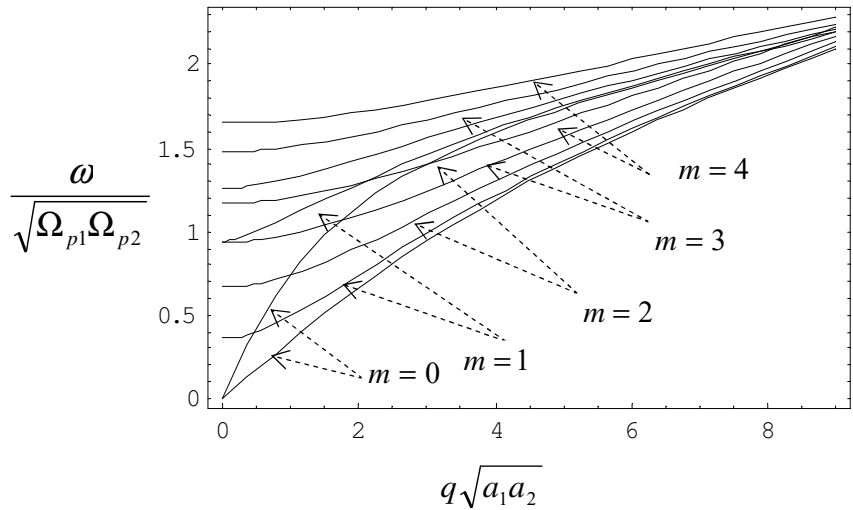

شکل (۴-۷)- نمودار پاشندگی با مدهای عرضی مختلف نانوتیوپ کربنی دوجداره

همان‌طور که در شکل‌های (۴-۶) و (۴-۷) مشخص است، تنها تفاوت نمودارهای مد TE,TM آن است که در مد TE شاخه‌ها کمی به یکدیگر نزدیک‌تر هستند. تفسیرهای مربوط به این مد (غیر از مواردی که ذکر خواهند شد) شبیه به تفسیرهای نمودارهای مربوطه در مد TM هستند.

حال به رسم معادلات پاشندگی با شعاع‌های داخلی و خارجی متفاوت و فاصله‌ی بین جدارهای ثابت $0.35nm$ می‌پردازیم. ضرایب موجود در معادلات پاشندگی با شعاع‌های مختلف در بخش مربوط به مد TM آورده شده‌اند. شکل (۴-۸) نمودارهای پاشندگی نانوتیوپ‌هایی با شعاع‌های مختلف را در مد عرضی صفر نشان می‌دهد.

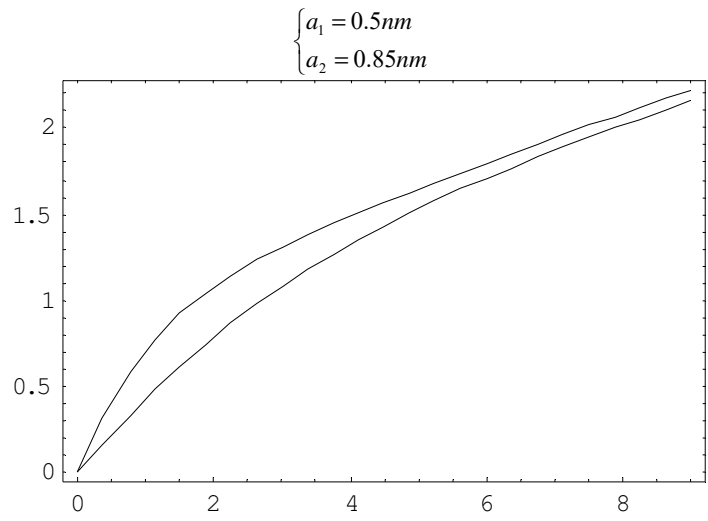

$$\begin{cases} a_1 = 0.5nm \\ a_2 = 0.85nm \end{cases}$$

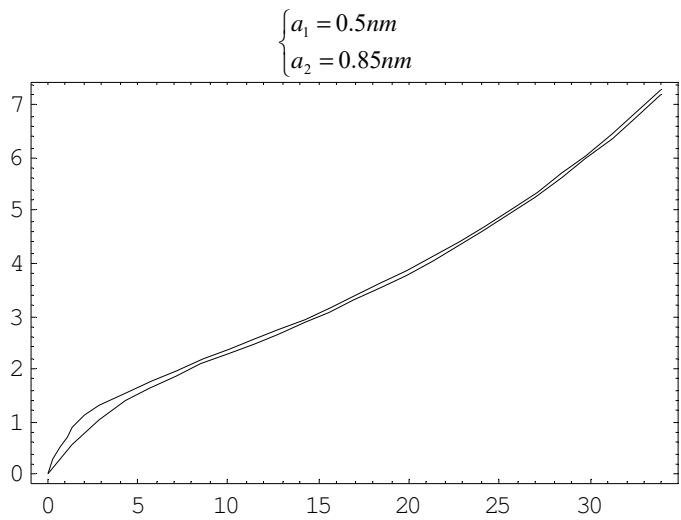

$$\begin{cases} a_1 = 0.5nm \\ a_2 = 0.85nm \end{cases}$$

$$\begin{cases} a_1 = 1nm \\ a_2 = 1.35nm \end{cases}$$

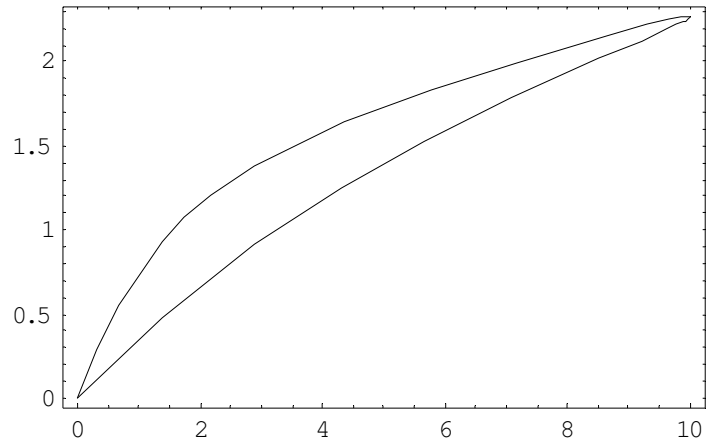

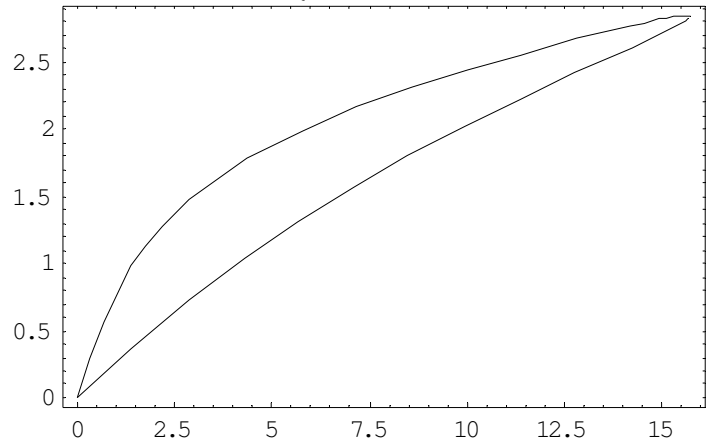

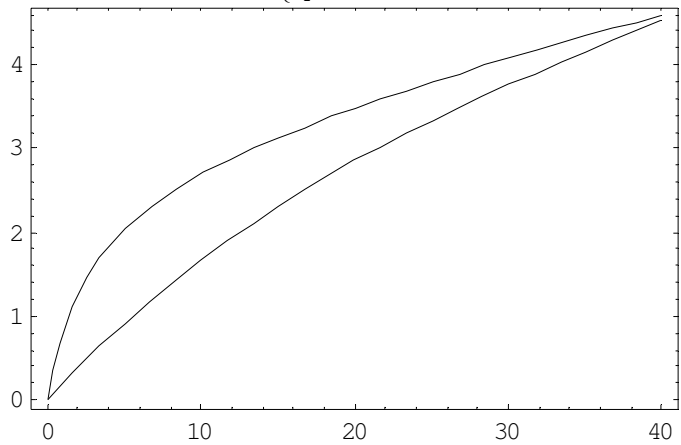

شکل (۴-۸)- نمودار پاشندگی با شعاع‌های مختلف داخلی و خارجی در نانوتیوپ کربنی دوجداره.

محورهای عمودی فرکانس بی‌بعد شده‌ی $y = \omega/\sqrt{\Omega_{p1}\Omega_{p2}}$ ومحورهای افقی عدد موج بی‌بعد شده‌ی $x = q\sqrt{a_1 a_2}$ هستند.

در شکل (۴-۸) معادله‌ی پاشندگی برای نانوتیوپی با شعاع‌های داخلی و خارجی $\begin{cases} a_1 = 0.5nm \\ a_2 = 0.85nm \end{cases}$ در دو بازه‌ی $0 < x < 9$ و $0 < x < 35$ آورده شده است. این نمودار در بازه‌ی $0 < x < 9$ شبیه به نمودار مد TM بوده ولی دو سر شاخه‌های بالایی و پایینی در آن به هم نرسیده و مانند آنچه که در بازه‌ی $0 < x < 35$ نشان داده شده، با فاصله از یکدیگر قرار می‌گیرند.

در شکل (۴-۹) نمودارهای پاشندگی نانوتیوپ‌های دوجداره با شعاع‌های داخلی و خارجی مختلف همگی در یک نمودار آورده شده‌اند.

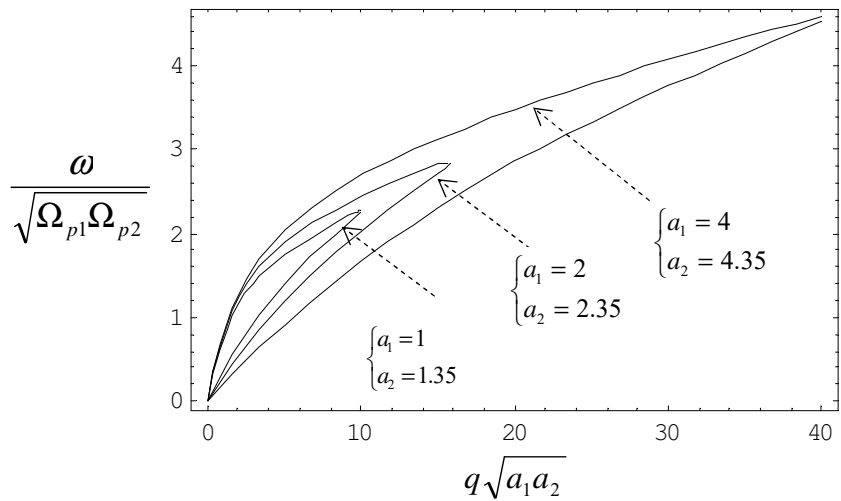

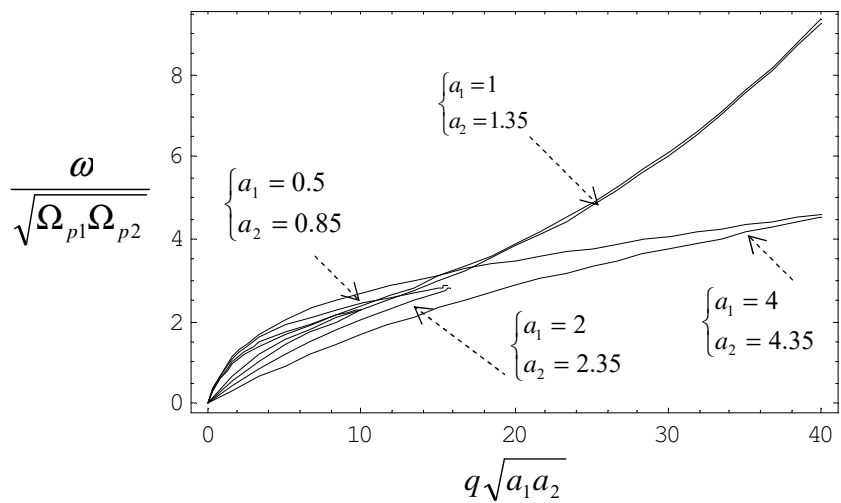

شکل (۴-۹)- نمودار پاشندگی با شعاع‌های مختلف داخلی و خارجی در نانوتیوپ کربنی دوجداره

شکل بالایی در (۴-۹) نمودارهای پاشندگی را برای شعاع‌های مختلف (به جز مورد $\begin{cases} a_1 = 0.5nm \\ a_2 = 0.85nm \end{cases}$ که با مد TM متفاوت بود) نشان می‌دهد. همان‌طور که انتظار داشتیم نمودارها کاملا شبیه به نمودارهای مد TM هستند و تنها اندکی در پهنای منحنی‌ها اختلاف وجود دارد. در شکل پایینی (۴-۹) همه‌ی شعاع‌های داخلی و خارجی مختلف در کنار یکدیگر آورده شده‌اند.

حال اثر نیروهای برهم‌کنش داخلی در دو پوسته‌ی گاز الکترونی دوبعدی را بر روی معادله پاشندگی نانوتیوپ دوجداره در مد TE بررسی می‌نماییم.

با صفر در نظر گرفتن ضرایب $\alpha, \beta$ می‌توان اثر نیروهای برهم‌کنش داخلی گاز الکترونی را حذف کرد. معادلات پاشندگی حاصل برای مد عرضی صفر و شعاع‌های داخلی و خارجی $\begin{cases} a_1 = 1nm \\ a_2 = 1.35nm \end{cases}$ در شکل (۴-۱۰) رسم شده‌اند.

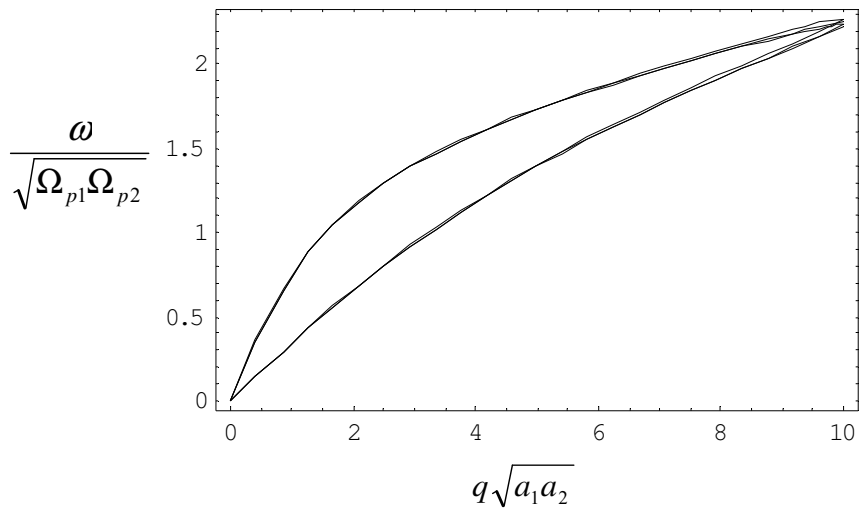

شکل (۴-۱۰)- نمودار پاشندگی با شعاع‌های مختلف داخلی و خارجی در نانوتیوپ کربنی دوجداره

همان‌طور که انتظار داشته‌ایم نمودارهای شکل (۴-۱۰) شبیه به نمودارهای معادل خود در مد TM (شکل (۴-۵)) هستند که قبلا به آن‌ها پرداختیم.

## ۴-۳- حالت حدی معادله پاشندگی نانوتیوپ های کربنی دوجداره

در بخش‌های ۴-۱-۱ و ۴-۱-۲ دیدیم که معادلات پاشندگی نانوتیوپ دوجداره برای دو مد TE,TM فرم یکسانی دارند و تنها تعاریف متغیرهای $\omega_1, \omega_2, F_{12}$ در آن‌ها متفاوت هستند. پس برای فرم مشترک، حالت حدی را می‌یابیم.

$$\omega^4 - [\omega_1^2 + \omega_2^2 + \alpha(q_1^2 + q_2^2) + \beta(q_1^4 + q_2^4)]\omega^2$$
$$+ \omega_1^2\omega_2^2 - \omega_1^2\omega_2^2 F_{12} + (\alpha q_2^2 + \beta q_2^4)\omega_1^2 + (\alpha q_1^2 + \beta q_1^4)\omega_2^2 + (\alpha q_1^2 + \beta q_1^4)(\alpha q_2^2 + \beta q_2^4) = 0$$

$$q_1^2 = (q^2 + m^2/a_1^2)$$
$$q_2^2 = (q^2 + m^2/a_2^2)$$

تعاریف متغیرها در مد TM :

$$\omega_1^2 = \Omega_{p1}^2(\kappa^2 a_1^2 + m^2) K_m(\kappa a_1) I_m(\kappa a_1); \Omega_{p1} = (e^2 n_0 / m_e \varepsilon_0 a_1)^{1/2}$$
$$\omega_2^2 = \Omega_{p2}^2(\kappa^2 a_2^2 + m^2) K_m(\kappa a_2) I_m(\kappa a_2); \Omega_{p2} = (e^2 n_0 / m_e \varepsilon_0 a_2)^{1/2}$$
$$F_{12} = [I_m(\kappa a_1)/K_m(\kappa a_1)][K_m(\kappa a_2)/I_m(\kappa a_2)]$$

تعاریف متغیرها در مد TE :

$$\omega_1^2 = -\Omega_{p1}^2(\kappa a_1)^2 [K'_m(\kappa a_1) I'_m(\kappa a_1)]; \Omega_{p1} = (e^2 n_0 / m_e e_0 a_1)^{1/2}$$
$$\omega_2^2 = -\Omega_{p2}^2(\kappa a_2)^2 [K'_m(\kappa a_2) I'_m(\kappa a_2)]; \Omega_{p2} = (e^2 n_0 / m_e e_0 a_2)^{1/2}$$
$$F_{12} = [I'_m(\kappa a_1)/K'_m(\kappa a_1)][K'_m(\kappa a_2)/I'_m(\kappa a_2)]$$

اگر شعاع‌های دو دیواره‌ی داخلی و خارجی در نانوتیوپ دوجداره را برابر فرض کنیم، انتظار داریم که معادله‌ی پاشندگی نانوتیوپ کربنی دوجداره در هر مد، به معادله‌ی پاشندگی نانوتیوپ کربنی تک‌جداره‌ای تبدیل شود که چگالی الکترونی در آن دو برابر شده است.

$$a_1 = a_2 = a \Rightarrow \omega_1 = \omega_2 = \omega', q_1 = q_2 = q', F_{12} = 1 \tag{۴-۱۷}$$

$\omega'$ برای دو مد مختلف به صورت زیر تعریف می‌شود.

$$\omega'^2 = \begin{cases} \omega'^2_{TE} = -\Omega_p^2(\kappa a)^2 K'_m(\kappa a) I'_m(\kappa a) \\ \omega'^2_{TM} = \Omega_p^2(\kappa^2 a^2 + m^2) K_m(\kappa a) I_m(\kappa a) \end{cases}$$

با استفاده از ساده‌سازی به کمک رابطه‌ی (۴-۱۷) معادله‌ی پاشندگی مشترک به صورت زیر خواهد بود.

$$\omega^4 - 2[\omega'^2 + \alpha q'^2 + \beta q'^4]\omega^2 + 2(\alpha q'^2 + \beta q'^4)\omega'^2 + (\alpha q'^2 + \beta q'^4)^2 = 0$$

$$\omega^4 + B\omega^2 + C = 0$$

$$\Rightarrow \omega^2 = \frac{-B}{2} \pm \sqrt{\frac{B^2}{4} - C}$$

$$\omega^2 = [\omega'^2 + \alpha q'^2 + \beta q'^4] \pm \sqrt{[\omega'^2 + \alpha q'^2 + \beta q'^4]^2 - 2(\alpha q'^2 + \beta q'^4)\omega'^2 + (\alpha q'^2 + \beta q'^4)^2}$$

$$= [\omega'^2 + \alpha q'^2 + \beta q'^4] \pm \sqrt{\begin{array}{l}\omega'^4 + (\alpha q'^2 + \beta q'^4)^2 + 2(\alpha q'^2 + \beta q'^4)\omega'^2 \\ -2(\alpha q'^2 + \beta q'^4)\omega'^2 - (\alpha q'^2 + \beta q'^4)^2\end{array}}$$

$$= \omega'^2 + \alpha q'^2 + \beta q'^4 \pm \omega'^2$$

$$\omega^2 - \alpha q'^2 - \beta q'^4 = \begin{cases} 0 \\ 2\omega'^2 \end{cases}$$

**مد TE :**

$$\omega^2 - \alpha q'^2 - \beta q'^4 = \begin{cases} 0 \\ 2\omega'^2 = -2\Omega_p^2 (\kappa a)^2 K_m'(\kappa a) I_m'(\kappa a) = -\frac{e^2(2n_0)}{\varepsilon_0 m_e a}(\kappa a)^2 K_m'(\kappa a) I_m'(\kappa a) \\ \qquad = \Omega_{p(2n_0)}^2 (\kappa a)^2 K_m'(\kappa a) I_m'(\kappa a) \end{cases}$$

قسمت دوم آکولاد، مشابه معادله پاشندگی نانوتیوپ کربنی تک‌جداره در مد TE است.

**مد TM :**

$$\omega^2 - \alpha q'^2 - \beta q'^4 = \begin{cases} 0 \\ 2\omega'^2 = 2\Omega_p^2(\kappa^2 a^2 + m^2) K_m(\kappa a) I_m(\kappa a) = \frac{e^2(2n_0)}{\varepsilon_0 m_e a}(\kappa^2 a^2 + m^2) K_m(\kappa a) I_m(\kappa a) \\ \qquad = \Omega_{p(2n_0)}^2 (\kappa^2 a^2 + m^2) K_m(\kappa a) I_m(\kappa a) \end{cases}$$

قسمت دوم آکولاد، مشابه معادله پاشندگی نانوتیوپ کربنی تک‌جداره در مد TM است.

# فصل پنجم- برانگیزش پلزمون های سطحی در نانوتیوپ های کربنی

به طور کلی برای برانگیزش پلزمون‌های سطحی در نانوتیوپ‌های کربنی و استفاده از نانوتیوپ‌هـای کربنـی بـه عنـوان موجبرهای نوری نانومتری دو راه وجود دارد: ۱- تابش نور لیزر و ۲- تابش باریکه‌ی الکترونی به موجبر کربنی می‌دانیم اگر یک فوتون یا الکترون بخواهد انرژی خود را به پلزمون‌های سطحی در نـانوتیوپ کربنـی منتقـل کنـد، بایـد دقیقا دارای همان انرژی باشد که پلزمون‌های سطحی مربوط به آن نانوتیوپ کربنی خاص در حالت تشدیدی دارا هستند. پس به منظور بررسی استفاده از هر یک از دو روش بالا برای برانگیزش پلزمون‌های سطحی، خطـوط پاشـندگی نورهـای لیزر در خلا و باریکه‌های الکترونی مختلف را در کنار نمودارهای پاشندگی نانوتیوپ‌های کربنی در خلا رسم کرده و نقـاط تلاقی آن‌ها را مورد توجه قرار دهیم. زیرا فرکانس‌های مربوط به نقاط تلاقی نمودارهای پاشـندگی، همـان فرکـانس‌هـایی هستند که پلزمون‌های سطحی تولید شده در نانوتیوپ‌های کربنی با آن‌ها نوسان خواهند کرد. با توجه به آن‌که بـا تـابش باریکه‌های الکترونی، محیط نانوتیوپ‌های کربنی تغییر می‌کند، برای تعیین طول موج پلزمونی تولیـد شـده توسـط باریکه‌های الکترونی، معادلات پاشندگی نانوتیوپ‌های کربنی را در حضور باریکه‌های الکترونی به دست آورده و طول موج مربوط به فرکانس نقطه‌ی تلاقی نمودارهای پاشندگی را از روی نمودارهای پاشندگی فـوق تعیـین مـی‌کنـیم. همچنـین سرعت گروه امواج پلزمونی از روی شیب نمودارهای فوق تعیین می‌شوند. در بخش‌های بعد بـه بررسـی فرکـانس و طـول موج امواج پلزمون‌های سطحی تشکیل شده در نانوتیوپ‌های کربنی تک‌جداره و دوجداره می‌پردازیم.

## ۵-۱- نانوتیوپ کربنی تک جداره

## ۵-۱-۱- فرکانس امواج پلزمونی در نانوتیوپ کربنی تک جداره

در این بخش به یافتن فرکانس‌های امواج پلزمون‌های سطحی که قابلیت تشکیل شدن در نانوتیوپ‌های کربنی *تک‌جـداره* را دارند، می‌پردازیم.

بدین منظور، نمودارهای پاشندگی نانوتیوپ‌های کربنی تک‌جداره در خلا را که برای مدهای عرضی مختلـف و در دو مـد TE,TM در فصل ۳ به دست آمده‌اند (شکل‌های (۳-۲) و(۳-۷))، در کنار نمودارهای پاشندگی نور در خلا و دو باریکه‌ی الکترونی با سرعت‌های متفاوت، قرار می‌دهیم.

معادله‌ی پاشندگی یک باریکه‌ی نوری و یا یک باریکه‌ی الکترونی که با سرعت $v$ در راستای z حرکت می‌کنند به صورت زیر است.

$$\omega = v k_z$$

متغیرهای $\omega, k_z$ را به صورت زیر بی‌بعد می‌کنیم.

$$\begin{cases} y = \omega/\Omega_p \\ x = k_z a \end{cases} \rightarrow y = vx/(\Omega_p a)$$

مقادیر در نظر گرفته‌شده به منظور بی‌بعد کردن متغیرها به شکل زیر هستند.

$$\Omega_p = 1.08 \times 10^{16} (1/s)$$
$$a = 4.15 \times 10^{-9} m$$

معادلات پاشندگی بی‌بعد شده‌ی برای خط نور در خلا و باریکه‌های الکترونی با دو سرعت متفاوت مطابق زیر خواهند بود.

$$v = c \rightarrow y = 3\times 10^8 \times 2.23\times 10^{-8} x \rightarrow y = 6.7x$$
$$v = 2\times 10^7 (m/s) \rightarrow y = 2\times 10^7 \times 2.23\times 10^{-8} x \rightarrow y = 0.45x$$
$$v = 2\times 10^6 (m/s) \rightarrow y = 2\times 10^6 \times 2.23\times 10^{-8} x \rightarrow y = 0.045x$$

شکل (۵-۱) نمودارهای پاشندگی برای مدهای عرضی مختلف در نانوتیوپ کربنی تک‌جداره به شعاع $5nm$ را در کنار خطوط پاشندگی نور در خلا و باریکه‌های الکترونی فوق برای دو مد TE,TM نشان می‌دهد.

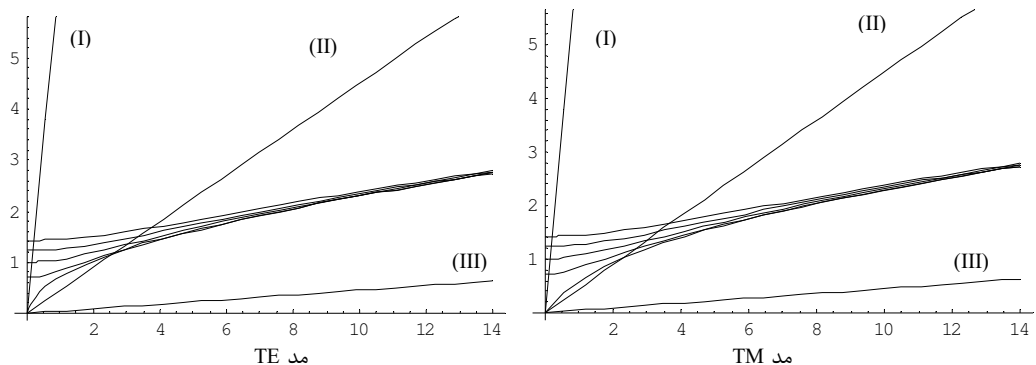

شکل (۵-۱)- نمودارها از پایین به بالا مربوط به معادلات پاشندگی نانوتیوپ کربنی تک‌جداره در مدهای عرضی صفر تا چهار هستند. نمودار (I) : نمودار پاشندگی نور در خلا. (II) : نمودار پاشندگی یک باریکه‌ی الکترونی با سرعت $2\times 10^7 m/s$. (III) : نمودار پاشندگی یک باریکه‌ی الکترونی با سرعت $2\times 10^6 m/s$. محورهای عمودی نمایانگر متغیر $\omega/\Omega_P$ و محورهای افقی متغیر $k_z a$ هستند.

همان‌طور که در شکل (۵-۱) مشخص است، خط نور در خلا و نمودار پاشندگی مربوط به مد عرضی صفر در نانوتیوپ کربنی تک‌جداره، تقاطعی ندارند و با تاباندن یک نور لیزر به یک نانوتیوپ کربنی تک‌جداره نمی‌توان در مد عرضی صفر موج پلزمونی ایجاد نمود. تلاقی خط نور با نمودارهای پاشندگی برای مدهای بزرگتر از صفر نیز در نزدیکی فرکانس‌های قطع واقع شده‌اند. این امر نشان می‌دهد که برای مدهای عرضی بزرگتر از یک، تنها می‌توان فرکانس‌های فوق که طول موج‌های بی‌نهایت دارند را ایجاد نمود.

ولی یک نانوتیوپ کربنی تک‌جداره، مطابق نمودارهای پاشندگی آن، قابلیت انتقال امواج پلزمونی با فرکانس‌های مختلف و طول‌موج‌های کمتر را نیز داراست. برای برانگیزش پلزمون‌های سطحی در فرکانس‌هایی دورتر از فرکانس قطع و در مدهای عرضی مختلف باید از باریکه‌های الکترونی به جای پرتوهای نوری استفاده نمود. البته برای کم‌کردن شیب خط نور یا کاهش سرعت نور می‌توان محیط نانوتیوپ کربنی را غلیظ‌تر کرده و به عبارتی ضریب شکست آن‌را بالا برد. ولی این‌کار علاوه بر مشکلات عملی برای ساخت، بازهم تغییر محسوسی در شیب خط نور ایجاد نخواهد کرد.

به این ترتیب طبق نمودارهای (I),(II),(III) در شکل (۵-۱) تنها باریکه‌های الکترونی با سرعت‌های در حدود $10^7 m/s$ قابلیت برانگیزش پلزمون‌های سطحی با فرکانس‌های مختلف را در نانوتیوپ‌های کربنی تک‌جداره دارا هستند.

فرکانس موج پلزمون سطحی تشکیل شده در موجبر کربنی تک‌جداره به شعاع $5nm$، توسط باریکه‌ی الکترونی با سرعت $2\times 10^7 m/s$، از محل تقاطع نمودارهای پاشندگی شکل (۵-۱) قابل دستیابی است.

برای به دست آوردن فرکانس امواج پلزمون سطحی برای نانوتیوپ‌های کربنی با شعاع‌های دیگر و با سرعت‌های متفاوت باریکه‌ی الکترونی، می‌بایست نمودارهای بالا را برای آن‌ها مجددا رسم کنیم.

## ۵-۱-۲- معادله پاشندگی نانوتیوپ کربنی تک جداره در حضور باریکه‌ی الکترونی گذرنده از داخــل و خارج جداره ها

برای تعیین طول موج امواج پلزمونی در نانوتیوپ کربنی تک‌جداره، با توجه به آن که با تابش باریکـه‌ی الکترونـی، محـیط نانوتیوپ کربنی تغییر می‌کند، معادله‌ی پاشندگی نانوتیوپ کربنی را در حضور باریکه‌ی الکترونی به دسـت آورده و طـول موج مربوط به فرکانس موج پلزمون سطحی را از روی نمودار پاشندگی فوق تعیین می‌کنیم.

قابل ذکر است که تنها یک نقطه از نمودار پاشندگی نانوتیوپ کربنی در حضور باریکه‌ی الکترونی مورد استفاده قرار مـی‌گیرد. زیرا برای امواج پلزمونی با فرکانس‌های دیگر که از سرعت‌های متفاوت و چگالی‌های الکترونـی دیگـری در باریکـه‌ی الکترونی به دست آمده‌اند، نمودار پاشندگی در حضور باریکه‌ی الکترونی تغییر کرده و مجددا باید رسم شود.

اگر یک نانوتیوپ کربنی تک‌جداره را در حضور یک باریکه‌ی الکترونی قرار دهیم، در معادلات بسل تعمیم‌یافته‌ی فصل دوم (معادلات (۲-۱۰))، متغیر $\kappa$ در $\sqrt{\varepsilon_{ll}}$ ضرب خواهد شد.

$\varepsilon_{ll}$ مولفه‌ای از تانسور گذردهی الکتریکی نسبت داده شده به باریکه‌ی الکترونی است که به صورت زیر تعریف می‌شود.

$$\varepsilon = \varepsilon_0 \begin{pmatrix} 1 & 0 & 0 \\ 0 & 1 & 0 \\ 0 & 0 & \varepsilon_{ll} \end{pmatrix}$$

$$\varepsilon_{ll} = 1 - \frac{\omega_{beam}^2}{(\omega - qu)^2} \quad ; \omega_{beam}^2 = \frac{n_0 e^2}{\varepsilon_0 m_e}$$

$u$ – سرعت باریکه‌ی الکترونی
$n_0$ – چگالی الکترونی باریکه
$e$ – بار الکتریکی الکترون
$m_e$ – جرم الکترون

تعریف متغیر $\kappa$ در معادلات بسل تعمیم‌یافته به صورت $\sqrt{q^2 - \omega^2/c^2}$ است. که در آن $\omega$ فرکانس و $q$ عدد موجِ در راستای z ($k_z$) برای موج الکترومغناطیسی پیشرونده در باریکه‌ی الکترونی است.

به این ترتیب برای به دست آوردن معادله‌ی پاشندگی در حضـور باریکـه، تمـامی متغیرهـای $\kappa$ در معـادلات پاشـندگی نانوتیوپ کربنی تک‌جداره را در متغیر $\sqrt{\varepsilon_{ll}}$ که از این پس برای سادگی $\sqrt{\varepsilon}$ نوشته می‌شود، ضرب می‌کنیم.

قبل از ورود این ضریب به معادلات پاشندگی آن را بی‌بعد می‌کنیم.

$$\sqrt{\varepsilon} = (1 - \frac{\omega_b^2}{(\omega - uq)^2})^{1/2} = (1 - \frac{\omega_b^2}{(\sqrt{\Omega_{p1}\Omega_{p2}}\frac{\omega}{\sqrt{\Omega_{p1}\Omega_{p2}}} - \frac{u}{\sqrt{a_1 a_2}}q\sqrt{a_1 a_2})^2})^{1/2}$$

در محاسبات مربوط به معادلات پاشندگی بخش‌های قبل، به علت تقریب به کار رفته (حذف متغیـر $\sigma y^2$)، معـادلات پاشندگی بر حسب $q, \kappa$ صورت‌های یکسانی داشته‌اند. درنتیجه در این بخش نیز متغیر بی‌بعد x را به صورت زیر در نظر می‌گیریم.

$$x = q\sqrt{a_1 a_2}$$

$$\rightarrow \sqrt{\varepsilon} = (1 - \frac{a}{(\sqrt{\Omega_{p1}\Omega_{p2}}\, y - \frac{u}{\sqrt{a_1 a_2}} x)^2})^{1/2}$$

تصحیح دیگری که باید در نظر داشت، تغییر گذردهی الکتریکی محیط نانوتیوپ کربنی در حضور باریکه‌ی الکترونی است که در شرایط مرزی باید اعمال شود. به راحتی می‌توان نشان داد، در حالتی که باریکه‌ی الکترونی از درون یا بیرون یا کل (درون و بیرون) نانوتیوپ کربنی تک‌جداره عبور کند، تصحیح فوق در معادله‌ی پاشندگی اثری نخواهد داشت.

به این ترتیب معادلات پاشندگی نانوتیوپ کربنی تک‌جداره در حضور باریکه‌ی الکترونی برای دو مد TE,TM همان معادلات پاشندگی نانوتیوپ کربنی تک‌جداره در خلا بوده که تمامی متغیرهای $x$ در آن به $x\sqrt{\varepsilon}$ تبدیل شده‌اند.

به علت وجود متغیرهای x,y در ضریب $\sqrt{\varepsilon}$، این معادلات مانند قبل دیگر به طور تحلیلی قابل حل نبوده و باید با روش‌های عددی، جواب‌های آن‌ها را به دست آورده و نمودارهای مربوطه را رسم کنیم.

در رسم نمودارها از روش "تکراری" (Iterative) در MATHEMATICA استفاده شده است. در این روش یک نقطه‌ی شروع به عنوان حدس اولیه به برنامه داده می‌شود و در هر بار راه‌اندازی، یک دسته جواب برای بازه‌ی $1 \leq x \leq 8$ با گام یک، به دست می‌آید. برای هر نقطه‌ی شروع در روش فوق، یک دسته جواب برای مسئله به دست می‌آید، در نتیجه برای داشتن تمامی جواب‌های مسئله در بازه‌ی فوق، نقاط شروع در روش "تکراری" را تمامی اعداد از ۱ تا ۶ با فواصل ۰٫۲ انتخاب کرده‌ایم. برای نقاط شروع با اعداد بیشتر از ۶ جواب‌های تکراری به دست می‌آمدند.

در شکل‌های (۵-۳) و (۵-۴) جواب‌های معادلات پاشندگی نانوتیوپ کربنی تک‌جداره در حضور یک باریکه‌ی الکترونی با مشخصات زیر برای دو مد TE,TM که به طور عددی محاسبه شد، رسم گردیده‌اند. قابل ذکر است که در این نمودارها تنها بخش‌های حقیقی و مثبت فرکانس‌ها را در نظر گرفته‌ایم.

سرعت باریکه‌ی الکترونی: $2 \times 10^7 (m/s)$

چگالی الکترونی باریکه: $n_0 = 1.26 \times 10^{27} (1/m^3)$

شعاع نانوتیوپ کربنی تک‌جداره: $a = 5nm$

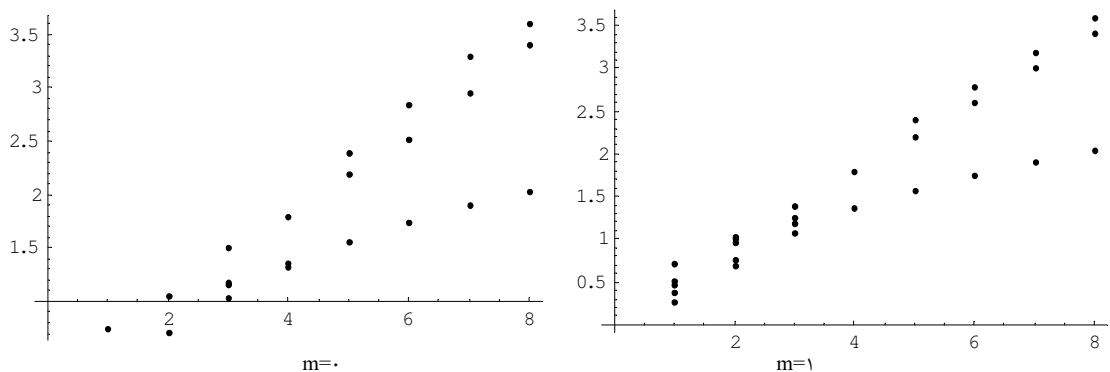

شکل (۵-۳) - نمودار پاشندگی نانوتیوپ کربنی تک‌جداره در حضور باریکه‌ی الکترونی برای مد TM و در دو مد عرضی صفر و یک – محورهای عمودی نمایانگر متغیر $\omega/\Omega_P$ و محورهای افقی متغیر $k_z a$ هستند.

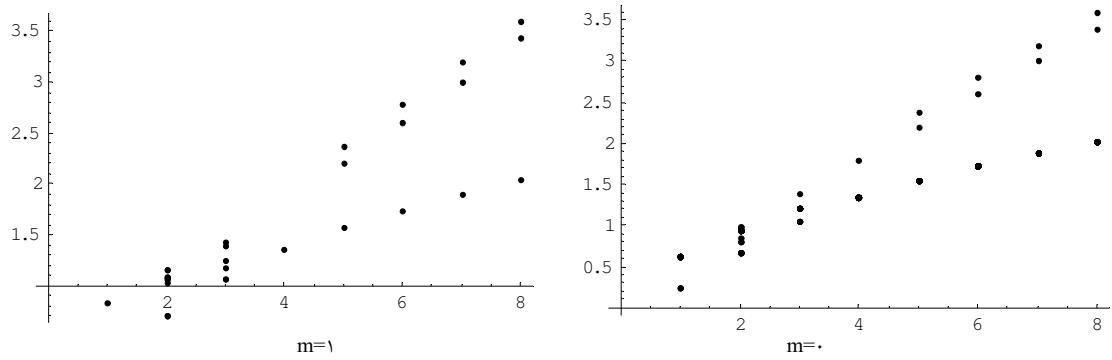

شکل (۵-۴) - نمودار پاشندگی نانوتیوپ کربنی تک‌جداره در حضور باریکه‌ی الکترونی برای مد TE و در دو مد عرضی صفر و یک- محورهای عمودی نمایان‌گر متغیر $\omega/\Omega_P$ و محورهای افقی متغیر $k_z a$ هستند.

نمودارهای پاشندگی نانوتیوپ کربنی تک‌جداره، برای هر مد عرضی در خلا دارای یک شاخه هستند (شکل‌هـای (۲-۳) و (۳-۷) ). ولی با توجه به شکل‌های (۵-۳) و (۵-۴)، این نمودارها در حضور باریکه‌های الکترونی دارای تعـداد شـاخه‌هـای بیشتری می‌شوند.

مطابق اشکال فوق با افزایش مدهای عرضی (مقایسه‌ی مد صفر و یک)، فاصله‌ی بین شاخه‌ی پایینی از دو شاخه‌ی بالایی نیز افزایش می‌یابد.

در شکل (۵-۲) مقدار $\sqrt{\varepsilon}$ برای x,y های مختلف نشان داده شده است. در این شکل میزان تاثیر ضریب فوق در معادلات پاشندگی نمایان می‌شود. مطابق این شکل در بعضی نواحی تاثیر این ضریب بیشتر از نقاط دیگر است.

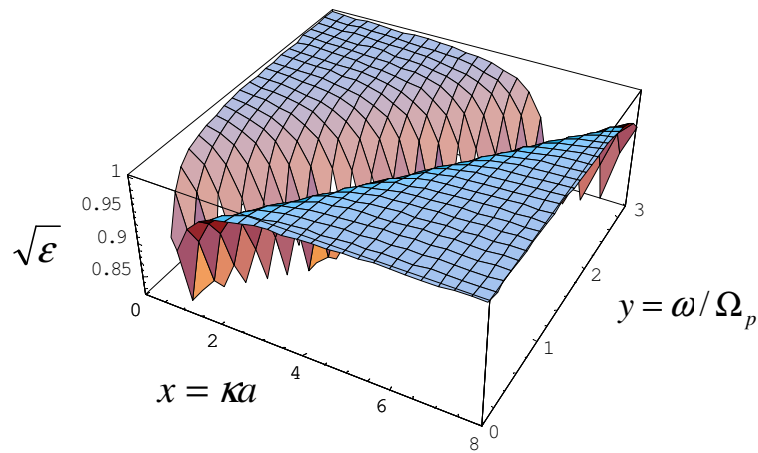

شکل (۵-۲) – مقدار عددی $\sqrt{\varepsilon}$ نسبت به متغیرهای x,y

## ۵-۲- نانوتیوپ کربنی دوجداره

### ۵-۲-۱- فرکانس امواج پلزمونی در نانوتیوپ کربنی دو جداره

در این بخش به یافتن فرکانس‌های امواج پلزمون‌های سطحی که قابلیت تشکیل شدن در نانوتیوپ‌های کربنی دوجـداره را دارند، می‌پردازیم.

نمودارهای پاشندگی نانوتیوپ‌های کربنی دوجداره در خلا را که بـرای مـدهای عرضـی مختلـف و در دو مـد TE,TM در فصل ۴ به دست آمدند (شکل‌های (۴-۲) و(۴-۷) )، در کنار نمودارهای پاشندگی نور در خلا و دو باریکـه‌ی الکترونـی بـا سرعت‌های متفاوت، قرار می‌دهیم.

معادله‌ی پاشندگی یک خط نور یا باریکه‌ی الکترونی که با سرعت u در راستای z حرکت می‌کند به صورت زیر است.

$$\omega = v k_z$$

معادله‌ی فوق را برای یک نانوتیوپ کربنی دو جداره با فرکانس‌های ویژه‌ی $\Omega_{p1}, \Omega_{p2}$ و شعاع‌های داخلی و خارجی $a_1, a_2$ به صورت زیر بی‌بعد می‌کنیم.

$$\begin{cases} y = \omega / \sqrt{\Omega_{p1} \Omega_{p2}} \\ x = \sqrt{a_1 a_2} k_z \end{cases} \rightarrow y = vx / (\sqrt{\Omega_{p1} \Omega_{p2}} \sqrt{a_1 a_2})$$

میانگین هندسی شعاع‌های داخلی و خارجی نانوتیوپ دوجداره و فرکانس‌های ویژه‌ی هر یک از جداره‌های آن که ضریبی از چگالی الکترون‌های آزاد جداره‌ها هستند، به صورت زیر در نظر گرفته می‌شوند.

$$\sqrt{\Omega_{p1} \Omega_{p2}} = 1.08 \times 10^{16} (1/s)$$
$$\sqrt{a_1 a_2} = 4.15 \times 10^{-9} m$$

معادلات پاشندگی بی‌بعد شده‌ی مربوط به خط نور و باریکه‌های الکترونی با سرعت‌های مختلف به صورت‌های زیر خواهند بود.

$$v = c \rightarrow y = 3 \times 10^8 \times 2.23 \times 10^{-8} x \rightarrow y = 6.7 x$$
$$v = 2 \times 10^7 (m/s) \rightarrow y = 2 \times 10^7 \times 2.23 \times 10^{-8} x \rightarrow y = 0.45 x$$
$$v = 2 \times 10^6 (m/s) \rightarrow y = 2 \times 10^6 \times 2.23 \times 10^{-8} x \rightarrow y = 0.045 x$$

شکل‌های (۵-۶) و (۵-۷) معادلات پاشندگی نانوتیوپ‌های کربنی دو جداره را بـرای مـدهای عرضـی مختلـف در دو مـد TE,TM و درحضور معادلات پاشندگی نور در خلا و دو باریکه‌ی الکترونی با سرعت‌های متفاوت، نشان می‌دهند.

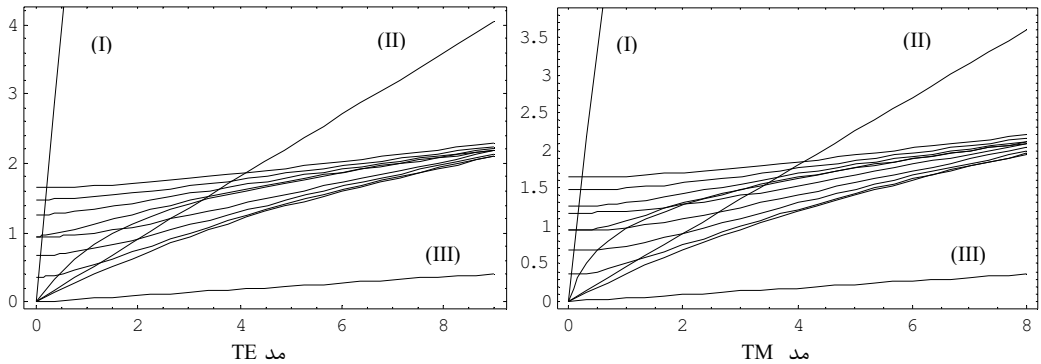

شکل (۵-۶)- خطوط پهلوی هم، مربوط به نمودارهای معادلات پاشندگی، در دو مد TM,TE برای نانوتیوپ کربنی دوجداره با مدهای عرضی مختلف و با شعاع‌های داخلی و خارجی $(a_1 = 1nm, a_2 = 1.35nm)$ هستند که از اشکال (۴-۲)و(۴-۷) گرفته شده‌اند. نمودار (I) : خط پاشندگی نور در خلا. نمودار (II) : خط پاشندگی یک باریکه‌ی الکترونی با سرعت $2\times10^7 m/s$ و نمودار (III) : خط پاشندگی یک باریکه‌ی الکترونی با سرعت $2\times10^6 m/s$ هستند. محورهای عمودی متغیر $\omega/\sqrt{\Omega_{p1}\Omega_{p2}}$ و محورهای افقی نمایان‌گر متغیر $q\sqrt{a_1 a_2}$ هستند.

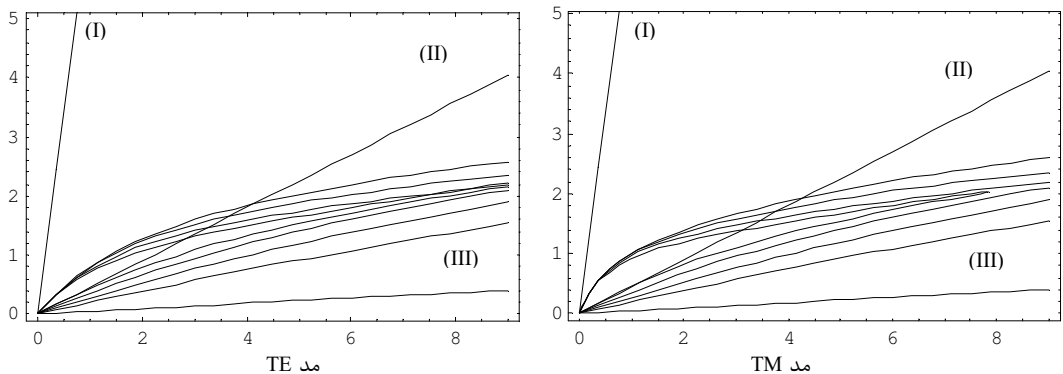

شکل (۵-۷)-خطوط پهلوی هم، مربوط به نمودارهای معادلات پاشندگی، در دو مد TM,TE برای نانوتیوپ کربنی دوجداره با شعاع‌های داخلی و خارجی مختلف و مد عرضی صفر هستند که از اشکال (۴-۴)و(۴-۹) گرفته شده‌اند. نمودار (I) : خط پاشندگی نور در خلا. نمودار (II) : خط پاشندگی یک باریکه‌ی الکترونی با سرعت $2\times10^7 m/s$ و نمودار (III) : خط پاشندگی یک باریکه‌ی الکترونی با سرعت $2\times10^6 m/s$ - محورهای عمودی متغیر $\omega/\sqrt{\Omega_{p1}\Omega_{p2}}$ و محورهای افقی نمایان‌گر متغیر $q\sqrt{a_1 a_2}$ هستند.

همان‌طور که در شکل‌های (۵-۶) و (۵-۷) مشخص است، خط نور در خلا و نمودارهای پاشندگی مربوط به مد عرضی صفر در نانوتیوپ کربنی دوجداره، تقاطعی ندارند و با تاباندن یک نور لیزر از خلا به یک نانوتیوپ کربنی دوجداره نمی‌توان در مد عرضی صفر در نانوتیوپ، موج پلازمونی ایجاد نمود. تلاقی خط نور با نمودارهای پاشندگی برای مدهای عرضی بزرگ‌تر از صفر نیز در نزدیکی فرکانس‌های قطع واقع شده‌اند. این امر نشان می‌دهد که برای مدهای عرضی بزرگ‌تر از یک، تنها می‌توان فرکانس‌های فوق که طول موج‌های بی‌نهایت دارند را ایجاد نمود.

ولی یک نانوتیوپ کربنی دوجداره، مطابق نمودارهای پاشندگی آن‌ها، قابلیت انتقال فرکانس‌های مختلف با طول‌موج‌های کمتر را نیز داراست. برای برانگیزش پلازمون‌های سطحی در مد عرضی صفر و در مدهای عرضی بالاتر با فرکانس‌هایی دورتر از فرکانس قطع، باید از باریکه‌های الکترونی به جای پرتوهای نوری استفاده نمود. البته برای کاهش سرعت نور می‌توان محیط نانوتیوپ کربنی را غلیظ‌تر کرده و به عبارتی ضریب شکست آن‌را بالا برد. ولی این‌کار علاوه بر مشکلات عملی برای ساخت، بازهم تغییر محسوسی در شیب خط نور ایجاد نخواهد کرد.

به این ترتیب طبق نمودارهای (I),(II),(III) در شکل‌های (۵-۶) و (۵-۷) تنها باریکه‌های الکترونی با سرعت‌هـای در حـدود $10^7 m/s$ قابلیت برانگیزش پلزمون‌های سطحی با فرکانس‌های مختلف را در نانوتیوپ‌های کربنی دوجداره دارا هستند. به این ترتیب فرکانس‌های امواج پلزمون‌های سطحی تشکیل شده در موجبرهای کربنی دوجداره با شعاع‌های مختلف کـه توسط باریکه‌های الکترونی متفاوت تولید شده‌اند، از محـل تقـاطع نمودارهـای پاشـندگی اشـکال (۵-۶) و (۵-۷) قابـل دستیابی است.

## ۵-۲-۲- معادله پاشندگی نانوتیوپ کربنی دوجداره در حضور باریکه‌ی الکترونی گذرنده از داخل و خارج جداره‌ها

برای تعیین طول موج امواج پلزمونی در نانوتیوپ کربنی دوجداره نیز مانند بخش ۵-۲-۱، با توجه به آن‌که با تابش باریکه‌ی الکترونی، محیط نانوتیوپ تغییر می‌کند، معادله‌ی پاشندگی نانوتیوپ کربنی را در حضور باریکه‌ی الکترونی به دست آورده و طول موج مربوط به فرکانس موج پلزمون سطحی را از روی نمودار پاشندگی فوق تعیین می‌کنیم. تنها یک نقطه از نمودارهای پاشندگی نانوتیوپ کربنی در حضور باریکه‌ی الکترونی مورد استفاده قرار می‌گیرد. زیرا برای امواج پلزمونی با فرکانس‌های دیگر که از سرعت‌های متفاوت و چگالی‌های الکترونی دیگری در باریکه‌ی الکترونی نتیجـه شده‌اند، نمودار پاشندگی در حضور باریکه‌ی الکترونی تغییر کرده و مجددا باید رسم شود.

اگر یک نانوتیوپ کربنی دوجداره را در حضور یک باریکه‌ی الکترونی قرار دهیم، در معادلات بسل تعمیم‌یافته‌ی فصل دوم (معادلات (۲-۱۰) )، متغیر $K$ در $\sqrt{\varepsilon}$ ضرب خواهد شد. تعریف این متغیر در بخش ۵-۱-۲ آورده شده است.

تصحیح دیگری که باید در نظر داشت، تغییر گذردهی الکتریکی محیط نانوتیوپ کربنی در حضور باریکه‌ی الکترونی است که در شرایط مرزی باید اعمال شود. به راحتی می‌توان نشان داد، در حالتی که باریکه‌ی الکترونی از کل (داخل و خارج جداره‌های) نانوتیوپ کربنی دوجداره عبور کند، تصحیح فوق در معادله‌ی پاشندگی اثری نخواهد داشت.

به این ترتیب معادلات پاشندگی نانوتیوپ کربنی دوجداره در حضور باریکه‌ی الکترونی برای دو مد TE,TM همان معادلات پاشندگی نانوتیوپ کربنی دوجداره در خلا بوده که تمامی متغیرهای $x$ در آن به $x\sqrt{\varepsilon}$ تبدیل شده‌اند. به علت وجود متغیرهای x,y در ضریب $\sqrt{\varepsilon}$، این معادلات مانند قبل دیگر به طور تحلیلی قابل حل نبوده و باید با روش‌های عددی، جواب‌های آن‌ها را به دست آورده و نمودارهای مربوطه را رسم نمود.

در رسم نمودارها با یک روش "تکراری" (Iterative) در MATHEMATICA نقاط شـروع را تمـامی اعـداد بـین ۱ تـا ۶ و بـا فواصل ۰٫۵ انتخاب کرده‌ایم. برای نقاط شروع با اعداد بیشتر از ۶ جواب‌ها، تکراری به دست می‌آمدند.

در شکل‌های (۵-۸)و(۵-۹) جواب‌های معادلات پاشندگی نانوتیوپ کربنی دوجداره در حضور یک باریکه‌ی الکترونی با مشخصات زیر برای دو مد TE,TM که به طور عددی محاسبه شد، رسم گردیده‌اند. قابل ذکر است که در این نمودارها تنها بخش‌های حقیقی و مثبت فرکانس‌ها را در نظر گرفته‌ایم.

سرعت باریکه‌ی الکترونی: $2\times10^7 (m/s)$

چگالی الکترونی باریکه: $n_0 = 1.26\times10^{27} (1/m^3)$

شعاع‌های داخلی و خارجی نانوتیوپ کربنی دوجداره : $(a_1 = 4nm, a_2 = 4.35nm)$

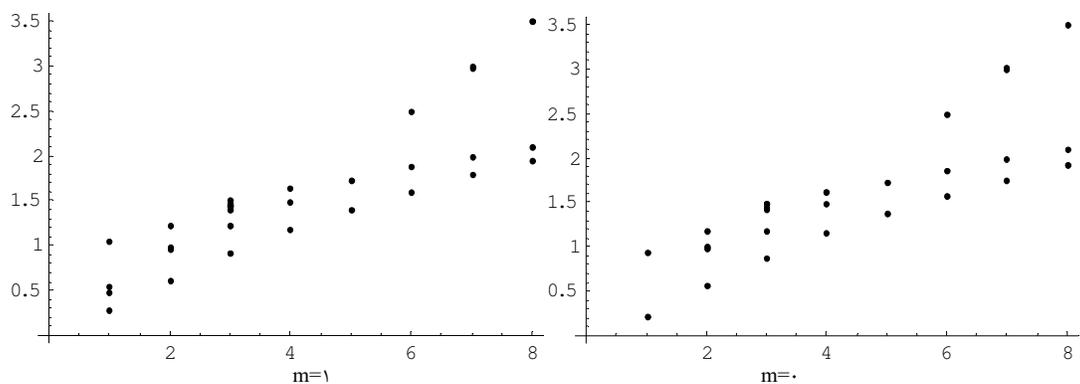

شکل (5-8) – نمودارهای پاشندگی مد TM نانوتیوپ کربنی دوجداره در حضور باریکه‌ی الکترونی گذرنده از داخل و خارج جدارهای نانوتیوپ. محورهای عمودی در نمودارها نمایان‌گر متغیر $\omega/\sqrt{\Omega_{p1}\Omega_{p2}}$ و محورهای افقی نمایان‌گر متغیر $q\sqrt{a_1 a_2}$ است.

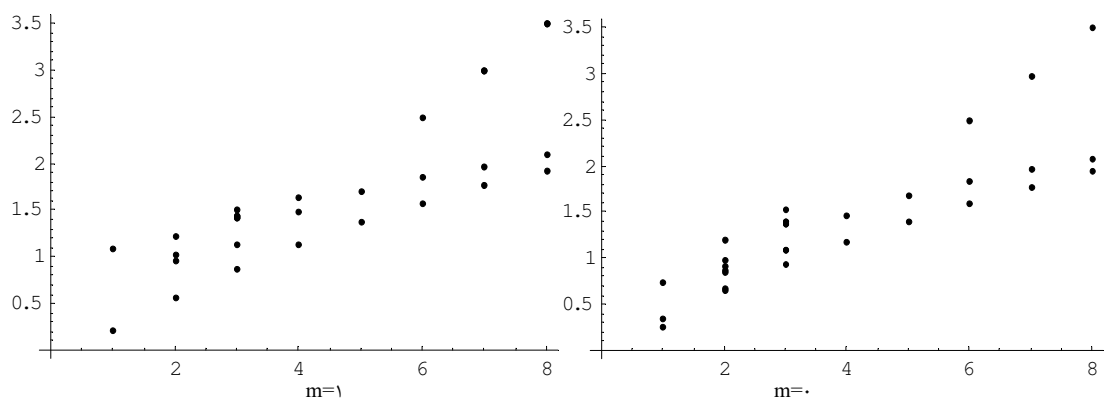

شکل (5-9) – نمودارهای پاشندگی مد TE نانوتیوپ کربنی دوجداره در حضور باریکه‌ی الکترونی گذرنده از داخل و خارج جدارهای نانوتیوپ. محورهای عمودی نمایان‌گر متغیر $\omega/\sqrt{\Omega_{p1}\Omega_{p2}}$ و محورهای افقی نمایان‌گر متغیر $q\sqrt{a_1 a_2}$ هستند.

نمودارهای پاشندگی نانوتیوپ کربنی دوجداره، برای هر مد عرضی در خلا دارای دو شاخه هستند (شکل‌های (3-2) و (3-7) ) ولی با توجه به شکل‌های (5-8) و (5-9)، نمودارهای پاشندگی در حضور باریکه‌های الکترونی دارای تعداد شاخه‌های بیشتری می‌شوند. همچنین مدهای TE,TM مانند نمودارهای پاشندگی نانوتیوپ تک‌جداره تفاوت ناچیزی با یکدیگر دارند.

## ۵-۲-۳- معادله پاشندگی نانوتیوپ کربنی دوجداره در حضور باریکه‌ی الکترونی گذرنده از بین جداره‌ها- مد TM

در این بخش به محاسبه‌ی معادله‌ی پاشندگی نانوتیوپ کربنی دوجداره در مد TM و در حضور باریکه‌ی الکترونی می‌پردازیم که از بین دوجداره‌ی نانوتیوپ کربنی دوجداره عبور می‌کند.

در این حالت به دلیل آن‌که معادلات شرط مرزی با اعمال تغییرات ناشی از گذردهی محیط تغییر می‌کنند، محاسبات را به طور کلی وارد می‌کنیم.

معادلات شرط مرزی برای میدان الکتریکی در اطراف یک نانولوله‌ی کربنی دوجداره به شعاع‌های درونی و بیرونی $a_1$ و $a_2$، با در نظر گرفتن بسط موج تخت برای میدان و چگالی بار الکترونی نانولوله (معادلات (۲-۱) و (۲-۲) ) به صورت زیر هستند.

پوسته‌ی درونی:

$$\varepsilon E_{rm}(a_1)\big|_{r>a_1} - E_{rm}(a_1)\big|_{r<a_1} = -\frac{eN_{m1}}{\varepsilon_0}$$

$$\varepsilon E_{llm}(a_1)\big|_{r>a_1} - E_{llm}(a_1)\big|_{r<a_1} = 0 \rightarrow \begin{cases} \varepsilon E_{zm}(a_1)\big|_{r>a_1} - E_{zm}(a_1)\big|_{r<a_1} = 0 \\ \varepsilon E_{\varphi m}(a_1)\big|_{r>a_1} - E_{\varphi m}(a_1)\big|_{r<a_1} = 0 \end{cases}$$

$$N_{m1} = -i\frac{en_0}{m_e}\frac{(qE_{zm} + \frac{m}{a_1}E_{\varphi m})}{\omega^2 - \alpha(q^2 + \frac{m^2}{a_1^2}) - \beta(q^2 + \frac{m^2}{a_1^2})^2}$$

پوسته‌ی بیرونی:

$$E_{rm}(a_2)\big|_{r>a_2} - \varepsilon E_{rm}(a_2)\big|_{r<a_2} = -\frac{eN_{m2}}{\varepsilon_0}$$

$$E_{llm}(a_2)\big|_{r>a_2} - \varepsilon E_{llm}(a_2)\big|_{r<a_2} = 0 \rightarrow \begin{cases} E_{zm}(a_2)\big|_{r>a_2} - \varepsilon E_{zm}(a_2)\big|_{r<a_2} = 0 \\ E_{\varphi m}(a_2)\big|_{r>a_2} - \varepsilon E_{\varphi m}(a_2)\big|_{r<a_2} = 0 \end{cases}$$

$$N_{m2} = -i\frac{en_0}{m_e}\frac{(qE_{zm} + \frac{m}{a_2}E_{\varphi m})}{\omega^2 - \alpha(q^2 + \frac{m^2}{a_2^2}) - \beta(q^2 + \frac{m^2}{a_2^2})^2}$$

با توجه به معادله‌ی (۲-۹) مولفه‌های $z, \varphi$ میدان الکتریکی در مد TM ($B_z = 0$) و با در نظر گرفتن تصحیح $\kappa \rightarrow \kappa\sqrt{\varepsilon}$ به شکل زیر خواهند بود.

$$E_{\varphi m} = \frac{qm}{\kappa^2 \varepsilon r}E_{zm} \qquad ; \kappa^2 = q^2 - k^2 \quad , k = \frac{\omega}{c}$$

$$E_{rm} = -\frac{iq}{\kappa^2 \varepsilon}\frac{\partial E_{zm}}{\partial r}$$

روابط بالا را در معادلات شرط مرزی دو پوسته‌ی بیرونی و درونی قرار می‌دهیم.

$$-\frac{iq}{\kappa^2\varepsilon}(\varepsilon\frac{\partial E_{zm}(a_1)}{\partial r}\Big|_{r>a_1} - \frac{\partial E_{zm}(a_1)}{\partial r}\Big|_{r<a_1}) = -\frac{e}{\varepsilon_0}N_{m1} \qquad (5-1)$$

$$\begin{cases} E_{zm}(a_1)\Big|_{r>a_1} - \varepsilon E_{zm}(a_1)\Big|_{r<a_1} = 0 \\ \frac{qm}{\kappa^2\varepsilon r}(E_{zm}(a_1)\Big|_{r>a_1} - \varepsilon E_{zm}(a_1)\Big|_{r<a_1}) = 0 \end{cases} \qquad (5-2)$$

$$-\frac{iq}{\kappa^2\varepsilon}(\frac{\partial E_{zm}(a_2)}{\partial r}\Big|_{r>a_2} - \varepsilon\frac{\partial E_{zm}(a_2)}{\partial r}\Big|_{r<a_2}) = -\frac{e}{\varepsilon_0}N_{m2} \qquad (5-3)$$

$$\begin{cases} E_{zm}(a_2)\Big|_{r>a_2} - \varepsilon E_{zm}(a_2)\Big|_{r<a_2} = 0 \\ \frac{qm}{\kappa^2\varepsilon r}(E_{zm}(a_2)\Big|_{r>a_2} - \varepsilon E_{zm}(a_2)\Big|_{r<a_2}) = 0 \end{cases} \qquad (5-4)$$

معادلات شرط مرزی در کروشه‌ها یکسان هستند.
مولفه‌های z میدان الکتریکی در معادلات بسل تعمیم‌یافته صدق می‌کنند (معادلات (۹-۲)) و جواب فرضی میدان در نواحی مختلف نانوتیوپ هنگامی که یک باریکه‌ی الکترونی از بین دوجداره‌ی آن عبور می‌کند به شکل زیر است.

$$E_{zm}(r) = C_{1m}I_m(\kappa r) \qquad\qquad r<a_1$$
$$E_{zm}(r) = C_{2m}I_m(\sqrt{\varepsilon}\kappa r) + C_{3m}K_m(\sqrt{\varepsilon}\kappa r) \qquad a_1<r<a_2$$
$$E_{zm}(r) = C_{4m}K_m(\kappa r) \qquad\qquad r>a_2$$

مقدار $N_{m1}$ را در معادله‌ی (۵-۱) قرار می‌دهیم.

$$-\frac{iq}{\kappa^2\varepsilon}(\varepsilon\frac{\partial E_{zm}(a_1)}{\partial r}\Big|_{r>a_1} - \frac{\partial E_{zm}(a_1)}{\partial r}\Big|_{r<a_1}) = -\frac{e}{\varepsilon_0}(-i\frac{en_0}{m_e}\frac{(qE_{zm}+\frac{m}{a_1}E_{\varphi m})}{\omega^2 - \alpha(q^2+\frac{m^2}{a_1^2}) - \beta(q^2+\frac{m^2}{a_1^2})^2})$$

$$\frac{q}{\kappa^2\varepsilon}(\varepsilon\frac{\partial E_{zm}(a_1)}{\partial r}\Big|_{r>a_1} - \frac{\partial E_{zm}(a_1)}{\partial r}\Big|_{r<a_1}) = -\frac{e^2n_0}{m_e\varepsilon_0}\frac{(qE_{zm}+\frac{m}{a_1}E_{\varphi m})}{\omega^2 - \alpha q_1^2 - \beta q_1^4} \qquad ; q_1^2 = q^2+\frac{m^2}{a_1^2}$$

$$E_{\varphi m} = \frac{qm}{\kappa^2 \varepsilon r} E_{zm} \Rightarrow$$

$$\frac{q}{\kappa^2 \varepsilon}(\varepsilon \frac{\partial E_{zm}(a_1)}{\partial r}\Big|_{r>a_1} - \frac{\partial E_{zm}(a_1)}{\partial r}\Big|_{r<a_1}) = -\frac{e^2 n_0}{m_e \varepsilon_0} \frac{(qE_{zm} + \frac{qm^2}{\kappa^2 \varepsilon a_1^2} E_{zm})}{\omega^2 - \alpha q_1^2 - \beta q_1^4}$$

با استفاده از جواب‌های فرضی مولفه‌ی $z$ میدان الکتریکی، معادله‌ی بالا به شکل زیر در می‌آید.

$$\frac{1}{\kappa^2 \varepsilon}(\varepsilon \frac{\partial E_{zm}(a_1)}{\partial r}\Big|_{r>a_1} - \frac{\partial E_{zm}(a_1)}{\partial r}\Big|_{r<a_1})(\omega^2 - \alpha q_1^2 - \beta q_1^4) = -\frac{e^2 n_0}{m_e \varepsilon_0}(1 + \frac{m^2}{\kappa^2 \varepsilon a_1^2}) E_{zm}$$

$$\frac{1}{\kappa \varepsilon}(C_{2m}\varepsilon\sqrt{\varepsilon}\frac{\partial I(\sqrt{\varepsilon}\kappa r)}{\partial(\sqrt{\varepsilon}\kappa r)}\Big|_{r>a_1} + C_{3m}\varepsilon\sqrt{\varepsilon}\frac{\partial K(\sqrt{\varepsilon}\kappa r)}{\partial(\sqrt{\varepsilon}\kappa r)}\Big|_{r>a_1} - C_{1m}\frac{\partial I(\kappa r)}{\partial(\kappa r)}\Big|_{r<a_1})(\omega^2 - \alpha q_1^2 - \beta q_1^4)$$

$$= -\frac{e^2 n_0}{m_e \varepsilon_0}(1 + \frac{m^2}{\kappa^2 \varepsilon a_1^2}) C_{1m} I(\kappa a_1)$$

$$(\omega^2 - \alpha q_1^2 - \beta q_1^4)[C_{2m}\varepsilon\sqrt{\varepsilon}I'_m(\sqrt{\varepsilon}\kappa a_1) + C_{3m}\varepsilon\sqrt{\varepsilon}K'_m(\sqrt{\varepsilon}\kappa a_1) - C_{1m}I'_m(\kappa a_1)]$$

$$= -\frac{e^2 n_0}{m_e \varepsilon_0}\kappa\varepsilon(1 + \frac{m^2}{\kappa^2 \varepsilon a_1^2}) C_{1m} I_m(\kappa a_1)$$

$$(\omega^2 - \alpha q_1^2 - \beta q_1^4)[C_{2m}\varepsilon\sqrt{\varepsilon}I'_m(\sqrt{\varepsilon}\kappa a_1) + C_{3m}\varepsilon\sqrt{\varepsilon}K'_m(\sqrt{\varepsilon}\kappa a_1) - C_{1m}I'_m(\kappa a_1)]$$

$$= -\frac{e^2 n_0}{m_e \varepsilon_0}\frac{1}{\kappa}(\kappa^2\varepsilon + m^2/a_1^2) C_{1m} I_m(\kappa a_1)$$

با استفاده از تعریف دو پارامتر جدید $N, R$ معادله‌ی بالا را به شکل ساده‌تری می‌نویسیم.

$$R = (\omega^2 - \alpha q_1^2 - \beta q_1^4) \quad , N = -\frac{e^2 n_0}{m_e \varepsilon_0}\frac{1}{\kappa}(\kappa^2\varepsilon + m^2/a_1^2)$$

$$RC_{2m}\varepsilon\sqrt{\varepsilon}I'_m(\sqrt{\varepsilon}\kappa a_1) + RC_{3m}\varepsilon\sqrt{\varepsilon}K'_m(\sqrt{\varepsilon}\kappa a_1) - RC_{1m}I'_m(\kappa a_1) = NC_{1m}I_m(\kappa a_1) \qquad (۵-۵)$$

برای ساده‌سازی معادله‌ی(۴-۳) نیز مطابق روند بالا عمل کرده و معادله‌ی زیر را به دست می‌آوریم.

$$HC_{4m}I'_m(\kappa a_2) - HC_{2m}\varepsilon\sqrt{\varepsilon}I'_m(\sqrt{\varepsilon}\kappa a_2) - HC_{3m}\varepsilon\sqrt{\varepsilon}K'_m(\sqrt{\varepsilon}\kappa a_2) = FC_{4m}K_m(\kappa a_2) \qquad (۶-۵)$$

$$H = (\omega^2 - \alpha q_2^2 - \beta q_2^4), F = -\frac{e^2 n_0}{m_e \varepsilon_0} \frac{1}{\kappa}(\kappa^2 \varepsilon + m^2/a_2^2)$$

جواب‌های فرضی میدان را در معادلات (۴-۲) و (۴-۴) قرار می‌دهیم.

$$C_{2m}\varepsilon I_m(\sqrt{\varepsilon}\kappa a_1) + C_{3m}\varepsilon K_m(\sqrt{\varepsilon}\kappa a_1) = C_{1m} I_m(\kappa a_1) \tag{۵-۷}$$

$$C_{2m}\varepsilon I_m(\sqrt{\varepsilon}\kappa a_2) + C_{3m}\varepsilon K_m(\sqrt{\varepsilon}\kappa a_2) = C_{4m} K_m(\kappa a_2) \tag{۵-۸}$$

به کمک معادلات (۵-۵)، (۵-۶)، (۵-۷) و (۵-۸) دستگاهی با مجهولات $C_{1m}, C_{2m}, C_{3m}, C_{4m}$ را به صورت زیر می‌نویسیم.

$$\begin{pmatrix} -RI'_m(\kappa a_1) - NI_m(\kappa a_1) & R\varepsilon\sqrt{\varepsilon}I'_m(\sqrt{\varepsilon}\kappa a_1) & R\varepsilon\sqrt{\varepsilon}K'_m(\sqrt{\varepsilon}\kappa a_1) & 0 \\ 0 & H\varepsilon\sqrt{\varepsilon}I'_m(\sqrt{\varepsilon}\kappa a_2) & H\varepsilon\sqrt{\varepsilon}K'_m(\sqrt{\varepsilon}\kappa a_2) & FK_m(\kappa a_2) - HK'_m(\kappa a_2) \\ -I_m(\kappa a_1) & \varepsilon I_m(\sqrt{\varepsilon}\kappa a_1) & K_m\varepsilon(\sqrt{\varepsilon}\kappa a_1) & 0 \\ 0 & I_m\varepsilon(\sqrt{\varepsilon}\kappa a_2) & K_m\varepsilon(\sqrt{\varepsilon}\kappa a_2) & -K_m(\kappa a_2) \end{pmatrix} \begin{pmatrix} C_{1m} \\ C_{2m} \\ C_{3m} \\ C_{4m} \end{pmatrix} = \begin{pmatrix} 0 \\ 0 \\ 0 \\ 0 \end{pmatrix}$$

شرط وجود جواب غیربدیهی آن است که رابطه‌ی زیر برقرار باشد.

$$\begin{vmatrix} -RI'_m(\kappa a_1) - NI_m(\kappa a_1) & R\varepsilon\sqrt{\varepsilon}I'_m(\sqrt{\varepsilon}\kappa a_1) & R\varepsilon\sqrt{\varepsilon}K'_m(\sqrt{\varepsilon}\kappa a_1) & 0 \\ 0 & H\varepsilon\sqrt{\varepsilon}I'_m(\sqrt{\varepsilon}\kappa a_2) & H\varepsilon\sqrt{\varepsilon}K'_m(\sqrt{\varepsilon}\kappa a_2) & FK_m(\kappa a_2) - HK'_m(\kappa a_2) \\ -I_m(\kappa a_1) & \varepsilon I_m(\sqrt{\varepsilon}\kappa a_1) & \varepsilon K_m(\sqrt{\varepsilon}\kappa a_1) & 0 \\ 0 & \varepsilon I_m(\sqrt{\varepsilon}\kappa a_2) & \varepsilon K_m(\sqrt{\varepsilon}\kappa a_2) & -K_m(\kappa a_2) \end{vmatrix} = 0$$

$$\varepsilon^2 \begin{vmatrix} -RI'_m(\kappa a_1) - NI_m(\kappa a_1) & R\sqrt{\varepsilon}I'_m(\sqrt{\varepsilon}\kappa a_1) & R\sqrt{\varepsilon}K'_m(\sqrt{\varepsilon}\kappa a_1) & 0 \\ 0 & H\sqrt{\varepsilon}I'_m(\sqrt{\varepsilon}\kappa a_2) & H\sqrt{\varepsilon}K'_m(\sqrt{\varepsilon}\kappa a_2) & FK_m(\kappa a_2) - HK'_m(\kappa a_2) \\ -I_m(\kappa a_1) & I_m(\sqrt{\varepsilon}\kappa a_1) & K_m(\sqrt{\varepsilon}\kappa a_1) & 0 \\ 0 & I_m(\sqrt{\varepsilon}\kappa a_2) & K_m(\sqrt{\varepsilon}\kappa a_2) & -K_m(\kappa a_2) \end{vmatrix} = 0$$

دترمینان را حول ستون اول بسط می‌دهیم.

$$-RI'_m(\kappa a_1) - NI_m(\kappa a_1) \begin{vmatrix} H\sqrt{\varepsilon}I'_m(\sqrt{\varepsilon}\kappa a_2) & H\sqrt{\varepsilon}K'_m(\sqrt{\varepsilon}\kappa a_2) & FK_m(\kappa a_2) - HK'_m(\kappa a_2) \\ I_m(\sqrt{\varepsilon}\kappa a_1) & K_m(\sqrt{\varepsilon}\kappa a_1) & 0 \\ I_m(\sqrt{\varepsilon}\kappa a_2) & K_m(\sqrt{\varepsilon}\kappa a_2) & -K_m(\kappa a_2) \end{vmatrix}$$

$$-I_m(\kappa a_1) \begin{vmatrix} R\sqrt{\varepsilon}I'_m(\sqrt{\varepsilon}\kappa a_1) & R\sqrt{\varepsilon}K'_m(\sqrt{\varepsilon}\kappa a_1) & 0 \\ H\sqrt{\varepsilon}I'_m(\sqrt{\varepsilon}\kappa a_2) & H\sqrt{\varepsilon}K'_m(\sqrt{\varepsilon}\kappa a_2) & FK_m(\kappa a_2) - HK'_m(\kappa a_2) \\ I_m(\sqrt{\varepsilon}\kappa a_2) & K_m(\sqrt{\varepsilon}\kappa a_2) & -K_m(\kappa a_2) \end{vmatrix} = 0$$

هر دو دترمینان را حول سطر اول بسط می‌دهیم.

$$[-RI'_m(\kappa a_1) - NI_m(\kappa a_1)][H\sqrt{\varepsilon}I'_m(\sqrt{\varepsilon}\kappa a_2)][-K_m(\sqrt{\varepsilon}\kappa a_1)K_m(\kappa a_2))]$$
$$[RI'_m(\kappa a_1) + NI_m(\kappa a_1)][H\sqrt{\varepsilon}K'_m(\sqrt{\varepsilon}\kappa a_2)][-I_m(\sqrt{\varepsilon}\kappa a_1)K_m(\kappa a_2)]$$
$$[-RI'_m(\kappa a_1) - NI_m(\kappa a_1)][FK_m(\kappa a_2) - HK'_m(\kappa a_2)][I_m(\sqrt{\varepsilon}\kappa a_1)K_m(\sqrt{\varepsilon}\kappa a_2) - K_m(\sqrt{\varepsilon}\kappa a_1)I_m(\sqrt{\varepsilon}\kappa a_2)]$$
$$-R\sqrt{\varepsilon}I'_m(\sqrt{\varepsilon}\kappa a_1)I_m(\kappa a_1)[-HK_m(\kappa a_2)\sqrt{\varepsilon}K'_m(\sqrt{\varepsilon}\kappa a_2) + K_m(\sqrt{\varepsilon}\kappa a_2)[HK'_m(\kappa a_2) - FK_m(\kappa a_2)]]$$
$$+R\sqrt{\varepsilon}K'_m(\sqrt{\varepsilon}\kappa a_1)I_m(\kappa a_1)[-H\sqrt{\varepsilon}I'_m(\sqrt{\varepsilon}\kappa a_2)K_m(\kappa a_2) + I_m(\sqrt{\varepsilon}\kappa a_2)[HK'_m(\kappa a_2) - FK_m(\kappa a_2)]]$$
$$= 0$$

$$[-RI'_m(\kappa a_1)][H\sqrt{\varepsilon}I'_m(\sqrt{\varepsilon}\kappa a_2)][-K_m(\sqrt{\varepsilon}\kappa a_1)K_m(\kappa a_2))]$$
$$[-NI_m(\kappa a_1)][H\sqrt{\varepsilon}I'_m(\sqrt{\varepsilon}\kappa a_2)][-K_m(\sqrt{\varepsilon}\kappa a_1)K_m(\kappa a_2))]$$
$$[RI'_m(\kappa a_1)][H\sqrt{\varepsilon}K'_m(\sqrt{\varepsilon}\kappa a_2)][-I_m(\sqrt{\varepsilon}\kappa a_1)K_m(\kappa a_2)]$$
$$[NI_m(\kappa a_1)][H\sqrt{\varepsilon}K'_m(\sqrt{\varepsilon}\kappa a_2)][-I_m(\sqrt{\varepsilon}\kappa a_1)K_m(\kappa a_2)]$$
$$[-RI'_m(\kappa a_1)][FK_m(\kappa a_2)-HK'_m(\kappa a_2)][I_m(\sqrt{\varepsilon}\kappa a_1)K_m(\sqrt{\varepsilon}\kappa a_2)-K_m(\sqrt{\varepsilon}\kappa a_1)I_m(\sqrt{\varepsilon}\kappa a_2)]$$
$$[-NI_m(\kappa a_1)][FK_m(\kappa a_2)-HK'_m(\kappa a_2)][I_m(\sqrt{\varepsilon}\kappa a_1)K_m(\sqrt{\varepsilon}\kappa a_2)-K_m(\sqrt{\varepsilon}\kappa a_1)I_m(\sqrt{\varepsilon}\kappa a_2)]$$
$$-R\sqrt{\varepsilon}I'_m(\sqrt{\varepsilon}\kappa a_1)I_m(\kappa a_1)[-H\sqrt{\varepsilon}K_m(\kappa a_2)K'_m(\sqrt{\varepsilon}\kappa a_2)]$$
$$-R\sqrt{\varepsilon}I'_m(\sqrt{\varepsilon}\kappa a_1)I_m(\kappa a_1)K_m(\sqrt{\varepsilon}\kappa a_2)[HK'_m(\kappa a_2)-FK_m(\kappa a_2)]]$$
$$+R\sqrt{\varepsilon}K'_m(\sqrt{\varepsilon}\kappa a_1)I_m(\kappa a_1)[-H\sqrt{\varepsilon}I'_m(\sqrt{\varepsilon}\kappa a_2)K_m(\kappa a_2)]$$
$$+R\sqrt{\varepsilon}K'_m(\sqrt{\varepsilon}\kappa a_1)I_m(\kappa a_1)I_m(\sqrt{\varepsilon}\kappa a_2)[HK'_m(\kappa a_2)-FK_m(\kappa a_2)]$$
$$=0$$
$$+RH\sqrt{\varepsilon}I'_m(\kappa a_1)I'_m(\sqrt{\varepsilon}\kappa a_2)K_m(\sqrt{\varepsilon}\kappa a_1)K_m(\kappa a_2)$$
$$+NH\sqrt{\varepsilon}I_m(\kappa a_1)I'_m(\sqrt{\varepsilon}\kappa a_2)K_m(\sqrt{\varepsilon}\kappa a_1)K_m(\kappa a_2)$$
$$-RH\sqrt{\varepsilon}I'_m(\kappa a_1)K'_m(\sqrt{\varepsilon}\kappa a_2)I_m(\sqrt{\varepsilon}\kappa a_1)K_m(\kappa a_2)$$
$$-NH\sqrt{\varepsilon}I_m(\kappa a_1)K'_m(\sqrt{\varepsilon}\kappa a_2)I_m(\sqrt{\varepsilon}\kappa a_1)K_m(\kappa a_2)$$
$$-RFI'_m(\kappa a_1)K_m(\kappa a_2)I_m(\sqrt{\varepsilon}\kappa a_1)K_m(\sqrt{\varepsilon}\kappa a_2)$$
$$+RFI'_m(\kappa a_1)K_m(\kappa a_2)K_m(\sqrt{\varepsilon}\kappa a_1)I_m(\sqrt{\varepsilon}\kappa a_2)$$
$$+RHI'_m(\kappa a_1)K'_m(\kappa a_2)I_m(\sqrt{\varepsilon}\kappa a_1)K_m(\sqrt{\varepsilon}\kappa a_2)$$
$$-RHI'_m(\kappa a_1)K'_m(\kappa a_2)K_m(\sqrt{\varepsilon}\kappa a_1)I_m(\sqrt{\varepsilon}\kappa a_2)$$
$$-NFI_m(\kappa a_1)K_m(\kappa a_2)I_m(\sqrt{\varepsilon}\kappa a_1)K_m(\sqrt{\varepsilon}\kappa a_2)$$
$$+NFI_m(\kappa a_1)K_m(\kappa a_2)K_m(\sqrt{\varepsilon}\kappa a_1)I_m(\sqrt{\varepsilon}\kappa a_2)$$
$$+NHI_m(\kappa a_1)K'_m(\kappa a_2)I_m(\sqrt{\varepsilon}\kappa a_1)K_m(\sqrt{\varepsilon}\kappa a_2)$$
$$-NHI_m(\kappa a_1)K'_m(\kappa a_2)K_m(\sqrt{\varepsilon}\kappa a_1)I_m(\sqrt{\varepsilon}\kappa a_2)$$
$$+RH\varepsilon I'_m(\sqrt{\varepsilon}\kappa a_1)I_m(\kappa a_1)K_m(\kappa a_2)K'_m(\sqrt{\varepsilon}\kappa a_2)$$
$$-RH\sqrt{\varepsilon}I'_m(\sqrt{\varepsilon}\kappa a_1)I_m(\kappa a_1)K_m(\sqrt{\varepsilon}\kappa a_2)K'_m(\kappa a_2)$$
$$+RF\sqrt{\varepsilon}I'_m(\sqrt{\varepsilon}\kappa a_1)I_m(\kappa a_1)K_m(\sqrt{\varepsilon}\kappa a_2)K_m(\kappa a_2)$$
$$-RH\varepsilon K'_m(\sqrt{\varepsilon}\kappa a_1)I_m(\kappa a_1)I'_m(\sqrt{\varepsilon}\kappa a_2)K_m(\kappa a_2)$$
$$+RH\sqrt{\varepsilon}K'_m(\sqrt{\varepsilon}\kappa a_1)I_m(\kappa a_1)I_m(\sqrt{\varepsilon}\kappa a_2)K'_m(\kappa a_2)$$
$$-RF\sqrt{\varepsilon}K'_m(\sqrt{\varepsilon}\kappa a_1)I_m(\kappa a_1)I_m(\sqrt{\varepsilon}\kappa a_2)K_m(\kappa a_2)$$
$$=0$$

$$+ RH(\sqrt{\varepsilon}I'_m(\kappa a_1)I'_m(\sqrt{\varepsilon}\kappa a_2)K_m(\sqrt{\varepsilon}\kappa a_1)K_m(\kappa a_2)$$
$$-\sqrt{\varepsilon}I'_m(\kappa a_1)K'_m(\sqrt{\varepsilon}\kappa a_2)I_m(\sqrt{\varepsilon}\kappa a_1)K_m(\kappa a_2)$$
$$+I'_m(\kappa a_1)K'_m(\kappa a_2)I_m(\sqrt{\varepsilon}\kappa a_1)K_m(\sqrt{\varepsilon}\kappa a_2)$$
$$-I'_m(\kappa a_1)K'_m(\kappa a_2)K_m(\sqrt{\varepsilon}\kappa a_1)I_m(\sqrt{\varepsilon}\kappa a_2)$$
$$+\varepsilon I'_m(\sqrt{\varepsilon}\kappa a_1)I_m(\kappa a_1)K_m(\kappa a_2)K'_m(\sqrt{\varepsilon}\kappa a_2)$$
$$-\sqrt{\varepsilon}I'_m(\sqrt{\varepsilon}\kappa a_1)I_m(\kappa a_1)K_m(\sqrt{\varepsilon}\kappa a_2)K'_m(\kappa a_2)$$
$$-\varepsilon K'_m(\sqrt{\varepsilon}\kappa a_1)I_m(\kappa a_1)I'_m(\sqrt{\varepsilon}\kappa a_2)K_m(\kappa a_2)$$
$$+\sqrt{\varepsilon}K'_m(\sqrt{\varepsilon}\kappa a_1)I_m(\kappa a_1)I_m(\sqrt{\varepsilon}\kappa a_2)K'_m(\kappa a_2))$$

$$+ NH(\sqrt{\varepsilon}I_m(\kappa a_1)I'_m(\sqrt{\varepsilon}\kappa a_2)K_m(\sqrt{\varepsilon}\kappa a_1)K_m(\kappa a_2)$$
$$-\sqrt{\varepsilon}I_m(\kappa a_1)K'_m(\sqrt{\varepsilon}\kappa a_2)I_m(\sqrt{\varepsilon}\kappa a_1)K_m(\kappa a_2)$$
$$+I_m(\kappa a_1)K'_m(\kappa a_2)I_m(\sqrt{\varepsilon}\kappa a_1)K_m(\sqrt{\varepsilon}\kappa a_2)$$
$$-I_m(\kappa a_1)K'_m(\kappa a_2)K_m(\sqrt{\varepsilon}\kappa a_1)I_m(\sqrt{\varepsilon}\kappa a_2))$$

$$RF(-I'_m(\kappa a_1)K_m(\kappa a_2)I_m(\sqrt{\varepsilon}\kappa a_1)K_m(\sqrt{\varepsilon}\kappa a_2)$$
$$+I'_m(\kappa a_1)K_m(\kappa a_2)K_m(\sqrt{\varepsilon}\kappa a_1)I_m(\sqrt{\varepsilon}\kappa a_2)$$
$$+\sqrt{\varepsilon}I'_m(\sqrt{\varepsilon}\kappa a_1)I_m(\kappa a_1)K_m(\sqrt{\varepsilon}\kappa a_2)K_m(\kappa a_2)$$
$$-\sqrt{\varepsilon}K'_m(\sqrt{\varepsilon}\kappa a_1)I_m(\kappa a_1)I_m(\sqrt{\varepsilon}\kappa a_2)K_m(\kappa a_2))$$

$$NF(-I_m(\kappa a_1)K_m(\kappa a_2)I_m(\sqrt{\varepsilon}\kappa a_1)K_m(\sqrt{\varepsilon}\kappa a_2)$$
$$+I_m(\kappa a_1)K_m(\kappa a_2)K_m(\sqrt{\varepsilon}\kappa a_1)I_m(\sqrt{\varepsilon}\kappa a_2))$$
$$=0$$

$$+ RH(\sqrt{\varepsilon}I'_m(\kappa a_1)I'_m(\sqrt{\varepsilon}\kappa a_2)K_m(\sqrt{\varepsilon}\kappa a_1)K_m(\kappa a_2)$$
$$- \sqrt{\varepsilon}I'_m(\kappa a_1)K'_m(\sqrt{\varepsilon}\kappa a_2)I_m(\sqrt{\varepsilon}\kappa a_1)K_m(\kappa a_2)$$
$$+ I'_m(\kappa a_1)K'_m(\kappa a_2)I_m(\sqrt{\varepsilon}\kappa a_1)K_m(\sqrt{\varepsilon}\kappa a_2)$$
$$- I'_m(\kappa a_1)K'_m(\kappa a_2)K_m(\sqrt{\varepsilon}\kappa a_1)I_m(\sqrt{\varepsilon}\kappa a_2)$$
$$+ \varepsilon I'_m(\sqrt{\varepsilon}\kappa a_1)I_m(\kappa a_1)K_m(\kappa a_2)K'_m(\sqrt{\varepsilon}\kappa a_2)$$
$$- \varepsilon K'_m(\sqrt{\varepsilon}\kappa a_1)I_m(\kappa a_1)I'_m(\sqrt{\varepsilon}\kappa a_2)K_m(\kappa a_2)$$
$$+ \sqrt{\varepsilon}K'_m(\sqrt{\varepsilon}\kappa a_1)I_m(\kappa a_1)I_m(\sqrt{\varepsilon}\kappa a_2)K'_m(\kappa a_2))$$
$$- \sqrt{\varepsilon}I'_m(\sqrt{\varepsilon}\kappa a_1)I_m(\kappa a_1)K_m(\sqrt{\varepsilon}\kappa a_2)K'_m(\kappa a_2)$$

$$+ NH(\sqrt{\varepsilon}I_m(\kappa a_1)I'_m(\sqrt{\varepsilon}\kappa a_2)K_m(\sqrt{\varepsilon}\kappa a_1)K_m(\kappa a_2)$$
$$- \sqrt{\varepsilon}I_m(\kappa a_1)K'_m(\sqrt{\varepsilon}\kappa a_2)I_m(\sqrt{\varepsilon}\kappa a_1)K_m(\kappa a_2)$$
$$+ I_m(\kappa a_1)K'_m(\kappa a_2)I_m(\sqrt{\varepsilon}\kappa a_1)K_m(\sqrt{\varepsilon}\kappa a_2)$$
$$- I_m(\kappa a_1)K'_m(\kappa a_2)K_m(\sqrt{\varepsilon}\kappa a_1)I_m(\sqrt{\varepsilon}\kappa a_2))$$

$$RF(-I'_m(\kappa a_1)K_m(\kappa a_2)I_m(\sqrt{\varepsilon}\kappa a_1)K_m(\sqrt{\varepsilon}\kappa a_2)$$
$$+ I'_m(\kappa a_1)K_m(\kappa a_2)K_m(\sqrt{\varepsilon}\kappa a_1)I_m(\sqrt{\varepsilon}\kappa a_2)$$
$$+ \sqrt{\varepsilon}I'_m(\sqrt{\varepsilon}\kappa a_1)I_m(\kappa a_1)K_m(\sqrt{\varepsilon}\kappa a_2)K_m(\kappa a_2)$$
$$- \sqrt{\varepsilon}K'_m(\sqrt{\varepsilon}\kappa a_1)I_m(\kappa a_1)I_m(\sqrt{\varepsilon}\kappa a_2)K_m(\kappa a_2))$$

$$NF(-I_m(\kappa a_1)K_m(\kappa a_2)I_m(\sqrt{\varepsilon}\kappa a_1)K_m(\sqrt{\varepsilon}\kappa a_2)$$
$$+ I_m(\kappa a_1)K_m(\kappa a_2)K_m(\sqrt{\varepsilon}\kappa a_1)I_m(\sqrt{\varepsilon}\kappa a_2))$$
$$= 0$$

$$N = -\frac{e^2 n_0}{m_e \varepsilon_0}\frac{1}{\kappa}(\kappa^2\varepsilon + m^2/a_1^2), F = -\frac{e^2 n_0}{m_e \varepsilon_0}\frac{1}{\kappa}(\kappa^2\varepsilon + m^2/a_2^2)$$

در تعریف پارامترهای N,F، دیگر نباید تبدیل $\kappa \to \kappa\sqrt{\varepsilon}$ را انجام دهیم. زیرا این تبدیل در $\kappa$ های شرط مرزی وارد شده.

$$RH = \omega^4 + \omega^2(-\alpha(q_1^2 + q_2^2) - \beta(q_1^4 + q_2^4)) + (\alpha q_1^2 + \beta q_1^4)(\alpha q_2^2 + \beta q_2^4)$$
$$NH = \frac{-e^2 n_0}{m_e \varepsilon_0 a_1}\frac{1}{\kappa a_1}(\varepsilon\kappa^2 a_1^2 + m^2)(\omega^2 - \alpha q_2^2 - \beta q_2^4) = -\Omega_{p1}^2\frac{1}{\kappa a_1}(\varepsilon\kappa^2 a_1^2 + m^2)(\omega^2 - \alpha q_2^2 - \beta q_2^4)$$

$$FR = \frac{-e^2 n_0}{m_e \varepsilon_0 a_2} \frac{1}{\kappa a_2} (\varepsilon \kappa^2 a_2^2 + m^2)(\omega^2 - \alpha q_1^2 - \beta q_1^4)$$

$$= -\Omega_{p2}^2 \frac{1}{\kappa a_2} (\varepsilon \kappa^2 a_2^2 + m^2)(\omega^2 - \alpha q_1^2 - \beta q_1^4)$$

$$NF = (\frac{e^2 n_0}{m_e \varepsilon_0 a_1 a_2})^2 \frac{1}{\kappa^2} (\varepsilon \kappa^2 a_1^2 + m^2)(\varepsilon \kappa^2 a_2^2 + m^2)$$

$$= \Omega_{p1}^2 \Omega_{p2}^2 \frac{1}{(\kappa a_1)(\kappa a_2)} (\varepsilon \kappa^2 a_1^2 + m^2)(\varepsilon \kappa^2 a_2^2 + m^2)$$

$$RH/(\Omega_{p1}^2 \Omega_{p2}^2) = \omega^4/(\Omega_{p1}^2 \Omega_{p2}^2) - \frac{\omega^2}{\Omega_{p1}\Omega_{p2}}(\frac{\alpha}{\Omega_{p1}\Omega_{p2}}(q_1^2 + q_2^2) + \frac{\beta}{\Omega_{p1}\Omega_{p2}}(q_1^4 + q_2^4))$$

$$+ \frac{(\alpha q_1^2 + \beta q_1^4)(\alpha q_2^2 + \beta q_2^4)}{\Omega_{p1}^2 \Omega_{p2}^2}$$

$$NH/(\Omega_{p1}^2 \Omega_{p2}^2) = -\frac{\Omega_{p1}}{\Omega_{p2}} \frac{1}{\kappa a_1} (\varepsilon \kappa^2 a_1^2 + m^2)(\frac{\omega^2}{\Omega_{p1}\Omega_{p2}} - \frac{\alpha}{\Omega_{p1}\Omega_{p2}} q_2^2 - \frac{\beta}{\Omega_{p1}\Omega_{p2}} q_2^4)$$

$$= -\sqrt{\frac{a_2}{a_1}} \frac{1}{\kappa a_1} (\varepsilon \kappa^2 a_1^2 + m^2)(\frac{\omega^2}{\Omega_{p1}\Omega_{p2}} - \frac{\alpha}{\Omega_{p1}\Omega_{p2}} q_2^2 - \frac{\beta}{\Omega_{p1}\Omega_{p2}} q_2^4)$$

$$FR/(\Omega_{p1}^2 \Omega_{p2}^2) = -\frac{\Omega_{p2}}{\Omega_{p1}} \frac{1}{\kappa a_2} (\varepsilon \kappa^2 a_2^2 + m^2)(\frac{\omega^2}{\Omega_{p1}\Omega_{p2}} - \frac{\alpha}{\Omega_{p1}\Omega_{p2}} q_1^2 - \frac{\beta}{\Omega_{p1}\Omega_{p2}} q_1^4)$$

$$= -\sqrt{\frac{a_1}{a_2}} \frac{1}{\kappa a_2} (\varepsilon \kappa^2 a_2^2 + m^2)(\frac{\omega^2}{\Omega_{p1}\Omega_{p2}} - \frac{\alpha}{\Omega_{p1}\Omega_{p2}} q_1^2 - \frac{\beta}{\Omega_{p1}\Omega_{p2}} q_1^4)$$

$$NF/(\Omega_{p1}^2 \Omega_{p2}^2) = \frac{1}{(\kappa a_1)(\kappa a_2)} (\varepsilon \kappa^2 a_1^2 + m^2)(\varepsilon \kappa^2 a_2^2 + m^2)$$

$$RH/(\Omega_{p1}^2 \Omega_{p2}^2) = \omega^4/(\Omega_{p1}^2 \Omega_{p2}^2) - \frac{\omega^2}{\Omega_{p1}\Omega_{p2}}(\frac{\alpha}{\Omega_{p1}\Omega_{p2} a_1 a_2}(a_1 a_2 q_1^2 + a_1 a_2 q_2^2)$$

$$+ \frac{\beta}{\Omega_{p1}\Omega_{p2} a_1^2 a_2^2}(a_1^2 a_2^2 q_1^4 + a_1^2 a_2^2 q_2^4)) + (\frac{\alpha}{\Omega_{p1}\Omega_{p2} a_1 a_2} a_1 a_2 q_1^2 +$$

$$\frac{\beta}{\Omega_{p1}\Omega_{p2} a_1^2 a_2^2} a_1^2 a_2^2 q_1^4)(\frac{\alpha}{\Omega_{p1}\Omega_{p2} a_1 a_2} a_1 a_2 q_2^2 + \frac{\beta}{\Omega_{p1}\Omega_{p2} a_1^2 a_2^2} a_1^2 a_2^2 q_2^4)$$

$$NH/(\Omega_{p1}^2 \Omega_{p2}^2) = -\sqrt{\frac{a_2}{a_1}} \frac{1}{\kappa \sqrt{a_1 a_2}} \sqrt{\frac{a_2}{a_1}} (\varepsilon \kappa^2 a_1 a_2 \frac{a_1}{a_2} + m^2)(\frac{\omega^2}{\Omega_{p1}\Omega_{p2}} - \frac{\alpha}{\Omega_{p1}\Omega_{p2} a_1 a_2} a_1 a_2 q_2^2$$

$$- \frac{\beta}{\Omega_{p1}\Omega_{p2} a_1^2 a_2^2} a_1^2 a_2^2 q_2^4)$$

$$= -\frac{a_2}{a_1} \frac{1}{\kappa \sqrt{a_1 a_2}} (\varepsilon \kappa^2 a_1 a_2 \frac{a_1}{a_2} + m^2)(\frac{\omega^2}{\Omega_{p1}\Omega_{p2}} - \frac{\alpha}{\Omega_{p1}\Omega_{p2} a_1 a_2} a_1 a_2 q_2^2 - \frac{\beta}{\Omega_{p1}\Omega_{p2} a_1^2 a_2^2} a_1^2 a_2^2 q_2^4)$$

$$FR/(\Omega_{p1}^2\Omega_{p2}^2) = -\sqrt{\frac{a_1}{a_2}}\frac{1}{\kappa\sqrt{a_1a_2}}\sqrt{\frac{a_1}{a_2}}(\kappa^2 a_1 a_2 \frac{a_2}{a_1}+m^2)$$

$$(\frac{\omega^2}{\Omega_{p1}\Omega_{p2}}-\frac{\alpha}{\Omega_{p1}\Omega_{p2}a_1a_2}a_1a_2q_1^2-\frac{\beta}{\Omega_{p1}\Omega_{p2}a_1^2a_2^2}a_1^2a_2^2q_1^4)$$

$$=-\frac{a_1}{a_2}\frac{1}{\kappa\sqrt{a_1a_2}}(\kappa^2a_1a_2\frac{a_2}{a_1}+m^2)(\frac{\omega^2}{\Omega_{p1}\Omega_{p2}}-\frac{\alpha}{\Omega_{p1}\Omega_{p2}a_1a_2}a_1a_2q_1^2-\frac{\beta}{\Omega_{p1}\Omega_{p2}a_1^2a_2^2}a_1^2a_2^2q_1^4)$$

$$NF/(\Omega_{p1}^2\Omega_{p2}^2) = \frac{1}{(\kappa\sqrt{a_1a_2})(\kappa\sqrt{a_1a_2})}(\varepsilon\kappa^2 a_1 a_2 \frac{a_1}{a_2}+m^2)(\varepsilon\kappa^2 a_1 a_2 \frac{a_2}{a_1}+m^2)$$

$$q_1^2 = (q^2+m^2/a_1^2) = [\kappa^2+\frac{\omega^2}{c^2}+m^2/a_1^2]$$

$$a_1a_2q_1^2 = [\kappa^2 a_1 a_2 + \frac{\Omega_{P1}\Omega_{P2}a_1a_2}{c^2}\frac{\omega^2}{\Omega_{P1}\Omega_{P2}}+m^2 a_1 a_2/a_1^2]=[x^2+\sigma y^2+m^2 a_2/a_1]\approx[x^2+m^2 a_2/a_1]$$

$$\kappa\to\kappa\sqrt{\varepsilon}\Rightarrow a_1a_2q_1^2=[x^2\varepsilon+m^2 a_2/a_1]$$

متغیرهای q چون در ابتدا از فرمول ابتدایی $N_m = -i\frac{en_0}{m_e}\frac{(qE_{zm}+m/aE_{\varphi m})}{\omega^2-\alpha(q^2+m^2/a^2)-\beta(q^2+m^2/a^2)^2}$ که در فصل دوم محاسبه شده، وارد معادلات پاشندگی می‌شوند، وقتی بر حسب $\kappa$ نوشته می‌شوند باید تبدیل $\kappa\to\kappa\sqrt{\varepsilon}$ روی آن‌ها صورت گیرد.

$$q_2^2 = (q^2+m^2/a_2^2) = [(\kappa^2+\frac{\omega^2}{c^2})+m^2/a_2^2]$$

$$a_1a_2q_2^2 = [a_1a_2\kappa^2+\frac{\Omega_{P1}\Omega_{P2}a_1a_2}{c^2}\frac{\omega^2}{\Omega_{P1}\Omega_{P2}}+m^2 a_1 a_2/a_2^2]=[x^2+\sigma y^2+m^2 a_1/a_2]\approx[x^2+m^2 a_1/a_2]$$

$$\kappa\to\kappa\sqrt{\varepsilon}\Rightarrow a_1a_2q_2^2=[x^2\varepsilon+m^2 a_1/a_2]$$

$$RH/(\Omega_{p1}^2\Omega_{p2}^2) = y^4-y^2(\alpha_1([x^2\varepsilon+m^2 a_2/a_1]+[x^2\varepsilon+m^2 a_1/a_2])$$
$$+\beta_1([x^2\varepsilon+m^2 a_2/a_1]^2+[x^2\varepsilon+m^2 a_1/a_2]^2))$$
$$+(\alpha_1[x^2\varepsilon+m^2 a_2/a_1]+\beta_1[x^2\varepsilon+m^2 a_2/a_1]^2)(\alpha_1[x^2\varepsilon+m^2 a_1/a_2]+\beta_1[x^2\varepsilon+m^2 a_1/a_2]^2)$$

$$NH/(\Omega_{p1}^2\Omega_{p2}^2) = -\frac{a_2}{a_1}\frac{1}{x}(\frac{a_1}{a_2}x^2\varepsilon+m^2)(y^2-\alpha_1[x^2\varepsilon+m^2 a_1/a_2]-\beta_1[x^2\varepsilon+m^2 a_1/a_2]^2)$$

$$FR/(\Omega_{p1}^2\Omega_{p2}^2) = -\frac{a_1}{a_2}\frac{1}{x}(\frac{a_2}{a_1}x^2\varepsilon+m^2)(y^2-\alpha_1[x^2\varepsilon+m^2 a_2/a_1]-\beta_1[x^2\varepsilon+m^2 a_2/a_1]^2)$$

$$NF/(\Omega_{p1}^2\Omega_{p2}^2) = \frac{1}{x^2}(\frac{a_1}{a_2}x^2\varepsilon+m^2)(\frac{a_2}{a_1}x^2\varepsilon+m^2)$$

$$[y^4 - y^2(\alpha_1([(x\sqrt{\varepsilon})^2 + m^2 a_2/a_1] + [(x\sqrt{\varepsilon})^2 + m^2 a_1/a_2]) + \beta_1([(x\sqrt{\varepsilon})^2 + m^2 a_2/a_1]^2$$
$$+ [(x\sqrt{\varepsilon})^2 + m^2 a_1/a_2]^2)) + (\alpha_1[(x\sqrt{\varepsilon})^2 + m^2 a_2/a_1] + \beta_1[(x\sqrt{\varepsilon})^2 + m^2 a_2/a_1]^2) \times$$
$$(\alpha_1[(x\sqrt{\varepsilon})^2 + m^2 a_1/a_2] + \beta_1[(x\sqrt{\varepsilon})^2 + m^2 a_1/a_2]^2)] \times$$
$$(\sqrt{\varepsilon} I'_m(\kappa a_1) I'_m(\sqrt{\varepsilon}\kappa a_2) K_m(\sqrt{\varepsilon}\kappa a_1) K_m(\kappa a_2)$$
$$-\sqrt{\varepsilon} I'_m(\kappa a_1) K'_m(\sqrt{\varepsilon}\kappa a_2) I_m(\sqrt{\varepsilon}\kappa a_1) K_m(\kappa a_2)$$
$$+ I'_m(\kappa a_1) K'_m(\kappa a_2) I_m(\sqrt{\varepsilon}\kappa a_1) K_m(\sqrt{\varepsilon}\kappa a_2)$$
$$- I'_m(\kappa a_1) K'_m(\kappa a_2) K_m(\sqrt{\varepsilon}\kappa a_1) I_m(\sqrt{\varepsilon}\kappa a_2)$$
$$+ \varepsilon I'_m(\sqrt{\varepsilon}\kappa a_1) I_m(\kappa a_1) K_m(\kappa a_2) K'_m(\sqrt{\varepsilon}\kappa a_2)$$
$$- \varepsilon K'_m(\sqrt{\varepsilon}\kappa a_1) I_m(\kappa a_1) I'_m(\sqrt{\varepsilon}\kappa a_2) K_m(\kappa a_2)$$
$$+ \sqrt{\varepsilon} K'_m(\sqrt{\varepsilon}\kappa a_1) I_m(\kappa a_1) I_m(\sqrt{\varepsilon}\kappa a_2) K'_m(\kappa a_2))$$
$$- \sqrt{\varepsilon} I'_m(\sqrt{\varepsilon}\kappa a_1) I_m(\kappa a_1) K_m(\sqrt{\varepsilon}\kappa a_2) K'_m(\kappa a_2)$$
$$- \frac{a_2}{a_1}\frac{1}{x}(\frac{a_1}{a_2}x^2\varepsilon + m^2)(y^2 - \alpha_1[(x\sqrt{\varepsilon})^2 + m^2 a_1/a_2] - \beta_1[(x\sqrt{\varepsilon})^2 + m^2 a_1/a_2]^2) \times$$
$$(\sqrt{\varepsilon} I_m(\kappa a_1) I'_m(\sqrt{\varepsilon}\kappa a_2) K_m(\sqrt{\varepsilon}\kappa a_1) K_m(\kappa a_2)$$
$$- \sqrt{\varepsilon} I_m(\kappa a_1) K'_m(\sqrt{\varepsilon}\kappa a_2) I_m(\sqrt{\varepsilon}\kappa a_1) K_m(\kappa a_2)$$
$$+ I_m(\kappa a_1) K'_m(\kappa a_2) I_m(\sqrt{\varepsilon}\kappa a_1) K_m(\sqrt{\varepsilon}\kappa a_2)$$
$$- I_m(\kappa a_1) K'_m(\kappa a_2) K_m(\sqrt{\varepsilon}\kappa a_1) I_m(\sqrt{\varepsilon}\kappa a_2))$$
$$- \frac{a_1}{a_2}\frac{1}{x}(\frac{a_2}{a_1}x^2\varepsilon + m^2)(y^2 - \alpha_1[(x\sqrt{\varepsilon})^2 + m^2 a_2/a_1] - \beta_1[(x\sqrt{\varepsilon})^2 + m^2 a_2/a_1]^2) \times$$
$$(-I'_m(\kappa a_1) K_m(\kappa a_2) I_m(\sqrt{\varepsilon}\kappa a_1) K_m(\sqrt{\varepsilon}\kappa a_2)$$
$$+ I'_m(\kappa a_1) K_m(\kappa a_2) K_m(\sqrt{\varepsilon}\kappa a_1) I_m(\sqrt{\varepsilon}\kappa a_2)$$
$$+ \sqrt{\varepsilon} I'_m(\sqrt{\varepsilon}\kappa a_1) I_m(\kappa a_1) K_m(\sqrt{\varepsilon}\kappa a_2) K_m(\kappa a_2)$$
$$- \sqrt{\varepsilon} K'_m(\sqrt{\varepsilon}\kappa a_1) I_m(\kappa a_1) I_m(\sqrt{\varepsilon}\kappa a_2) K_m(\kappa a_2))$$
$$+ \frac{1}{x^2}(\frac{a_1}{a_2}x^2\varepsilon + m^2)(\frac{a_2}{a_1}x^2\varepsilon + m^2) \times$$
$$(-I_m(\kappa a_1) K_m(\kappa a_2) I_m(\sqrt{\varepsilon}\kappa a_1) K_m(\sqrt{\varepsilon}\kappa a_2)$$
$$+ I_m(\kappa a_1) K_m(\kappa a_2) K_m(\sqrt{\varepsilon}\kappa a_1) I_m(\sqrt{\varepsilon}\kappa a_2))$$
$$= 0$$

فصل ۵- برانگیزش پلزمون‌های سطحی در نانوتیوپ‌های کربنی دوجداره

در رسم نمودارها با یک روش "تکراری" (Iterative) در MATHEMATICA نقاط شروع، تمامی اعداد از ۱ تا ۶ با فواصل ۰٫۲ انتخاب شدند. برای نقاط شروع با اعداد بیشتر از ۶ جواب‌ها تکراری به دست می‌آمدند.

در شکل (۵-۱۰) جواب‌های معادله‌ی پاشندگی بی‌بعد شده‌ی نانوتیوپ کربنی دوجداره در حضور یک باریکه‌ی الکترونی گذرنده از بین جداره‌ها و با مشخصات زیر برای دو مد عرضی صفر و یک به طور عددی محاسبه شده و رسم گردیده‌اند. قابل ذکر است که در این نمودارها تنها بخش‌های حقیقی و مثبت فرکانس‌ها را در نظر گرفته‌ایم.

سرعت باریکه‌ی الکترونی:

$$2\times 10^7 (m/s)$$

چگالی الکترونی و فرکانس ویژه‌ی باریکه:

$$n_0 = 1.26\times 10^{29}(1/m^3), \omega_b^2 = 4\times 10^{32}(1/s^2)$$

شعاع نانوتیوپ کربنی تک‌جداره:

$$(a_1 = 4nm, a_2 = 4.35nm)$$

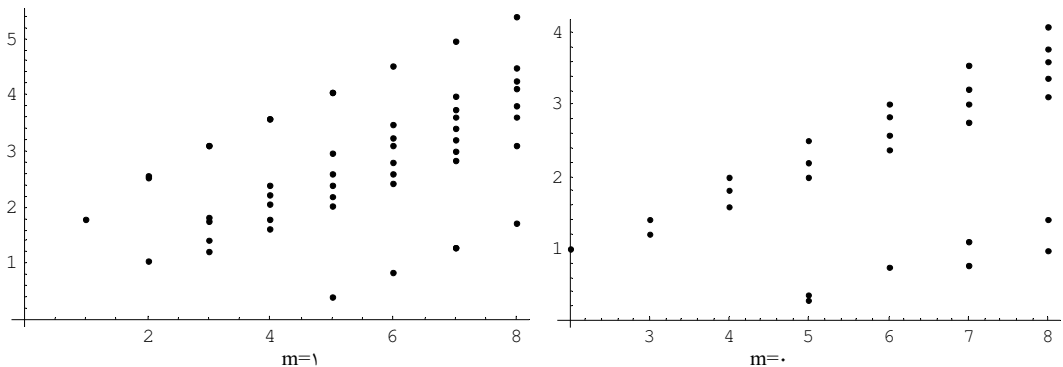

شکل (۵-۱۰)- نمودارهای پاشندگی نانوتیوپ کربنی دوجداره در مد TM در حضور باریکه‌ی الکترونی گذرنده از بین دوجداره - محور عمودی متغیر $\omega/\sqrt{\Omega_{p1}\Omega_{p2}}$ و محور افقی نمایان‌گر متغیر $q\sqrt{a_1 a_2}$ است.

نمودارهای پاشندگی نانوتیوپ کربنی دوجداره، برای هر مد عرضی در خلا دارای دو شاخه هستند (شکل (۴-۲)). ولی با توجه به شکل (۵-۱۰)، این نمودارها در حضور باریکه‌ی الکترونی گذرنده از بین جداره‌ها، دارای تعداد شاخه‌های بیشتری می‌شوند. همچنین تعداد شاخه‌های مد یک نسبت به مد صفر بیشتر بوده و محل آن‌ها کمی بالاتر از شاخه‌های معادل خود در مد صفر است.

در حالتی که باریکه‌ی الکترونی رقیق باشد و مرتبه‌ی چگالی الکترونی آن از مرتبه‌ی چگالی الکترونی پلاسما (نانوتیوپ کربنی فلزی که با گاز الکترون آزاد مدل‌سازی شده است) نباشد، باریکه اثری روی معادله‌ی پاشندگی نانوتیوپ کربنی نمی‌گذارد. زیرا در این حالت مقدار ضریب $\sqrt{\varepsilon}$، یک خواهد بود. حال اگر جواب‌های معادلات پاشندگی در حضور یک باریکه‌ی رقیق با جواب‌های تحلیلی به دست آمده از معادلات پاشندگی نانوتیوپ در خلا بر هم منطبق باشند، درستی بدنه‌ی معادلات پاشندگی نوشته شده در حضور باریکه اثبات می‌شود. منظور از بدنه‌ی معادلات پاشندگی نوشته شده در حضور باریکه، همان معادلات پاشندگی بدون حضور $\sqrt{\varepsilon}$ است. زیرا وقتی مقدار $\sqrt{\varepsilon}$ در معادله برابر با ۱ شود، درستی مکان و توان $\varepsilon$ در معادلات بررسی نخواهد شد و تنها درستی عناصر باقی‌مانده در معادله‌ی پاشندگی بررسی می‌شود که باید مانند حالت بدون حضور باریکه باشد.

همان‌طور که در شکل (۵-۱۱) مشخص است وقتی باریکه‌ی الکترونی رقیق می‌شود، برای هر x در بازه‌ی مورد نظر تنها دو جواب به دست می‌آید که منطبق بر جواب‌های معادلات پاشندگی بدون حضور باریکه هستند.

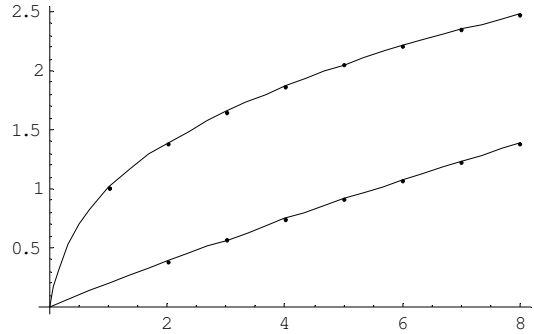

شکل (۵-۱۱) – نقاط نشان‌گر جواب‌های معادله‌ی پاشندگی نانوتیوپ کربنی در مد TM در حضور باریکه‌ی الکترونی رقیق بوده و خطوط پر، جواب‌های معادلات پاشندگی نانوتیوپ کربنی دوجداره در خلا و بدون حضور باریکه‌ی الکترونی را نشان می‌دهند. شعاع‌های داخلی و خارجی نانوتیوپ کربنی فوق $a_1 = 4nm, a_2 = 4.35nm$ بوده و مد عرضی، صفر در نظر گرفته شده است. ویژه فرکانس باریکه‌ی الکترونی رقیق نیز $\omega_b^2 = 4 \times 10^9 (1/s^2)$ می‌باشد.

## 5-2-4- معادله پاشندگی نانوتیوپ کربنی دوجداره در حضور باریکه‌ی الکترونی گذرنده از بین جداره‌ها- مد TE

در این بخش به محاسبه‌ی معادله‌ی پاشندگی نانوتیوپ کربنی دوجداره در مد TE و در حضور باریکه‌ی الکترونی می‌پردازیم که از بین دوجداره‌ی نانوتیوپ کربنی دوجداره عبور می‌کند.

در این حالت نیز مانند مد TM به دلیل آن‌که معادلات شرط مرزی با اعمال تغییرات ناشی از گذردهی محیط تغییر می‌کنند، تمامی محاسبات را انجام می‌دهیم.

معادلات شرط مرزی برای میدان الکتریکی در اطراف یک نانولوله‌ی کربنی دوجداره به شعاع‌های درونی و بیرونی $a_1$ و $a_2$، با در نظر گرفتن بسط موج تخت برای میدان و چگالی بار الکترونی نانولوله (معادلات (2-1) و (2-2) ) به صورت زیر هستند.

پوسته‌ی درونی:

$$\varepsilon E_{rm}(a_1)\big|_{r>a_1} - E_{rm}(a_1)\big|_{r<a_1} = -\frac{eN_{m1}}{\varepsilon_0}$$

$$\varepsilon E_{llm}(a_1)\big|_{r>a_1} - E_{llm}(a_1)\big|_{r<a_1} = 0 \to \begin{cases} \varepsilon E_{zm}(a_1)\big|_{r>a_1} - E_{zm}(a_1)\big|_{r<a_1} = 0 \\ \varepsilon E_{\varphi m}(a_1)\big|_{r>a_1} - E_{\varphi m}(a_1)\big|_{r<a_1} = 0 \end{cases}$$

$$N_{m1} = -i\frac{en_0}{m_e}\frac{(qE_{zm}+\frac{m}{a_1}E_{\varphi m})}{\omega^2 - \alpha(q^2+\frac{m^2}{a_1^2})-\beta(q^2+\frac{m^2}{a_1^2})^2}$$

پوسته‌ی بیرونی:

$$E_{rm}(a_2)\big|_{r>a_2} - \varepsilon E_{rm}(a_2)\big|_{r<a_2} = -\frac{eN_{m2}}{\varepsilon_0}$$

$$E_{llm}(a_2)\big|_{r>a_2} - \varepsilon E_{llm}(a_2)\big|_{r<a_2} = 0 \to \begin{cases} E_{zm}(a_2)\big|_{r>a_2} - \varepsilon E_{zm}(a_2)\big|_{r<a_2} = 0 \\ E_{\varphi m}(a_2)\big|_{r>a_2} - \varepsilon E_{\varphi m}(a_2)\big|_{r<a_2} = 0 \end{cases}$$

$$N_{m2} = -i\frac{en_0}{m_e}\frac{(qE_{zm}+\frac{m}{a_2}E_{\varphi m})}{\omega^2 - \alpha(q^2+\frac{m^2}{a_2^2})-\beta(q^2+\frac{m^2}{a_2^2})^2}$$

با توجه به معادله‌ی (2-9) مولفه‌های $z,\varphi$ میدان الکتریکی در مد TE ($E_z = 0$) و با در نظر گرفتن تصحیح $\kappa \to \kappa\sqrt{\varepsilon}$ به شکل زیر خواهند بود.

$$E_{\varphi m} = \frac{i\omega}{\kappa^2\varepsilon}\frac{\partial B_{zm}}{\partial r} \qquad \kappa^2 = q^2 - k^2 \quad, k = \frac{\omega}{c}$$

$$E_{rm} = \frac{m\omega}{\kappa^2\varepsilon r}B_{zm}$$

روابط بالا را در معادلات شرط مرزی دو پوسته‌ی بیرونی و درونی قرار می‌دهیم.

$$\frac{m\omega}{\kappa^2 \varepsilon r}(\varepsilon B_{zm}(a_1)|_{r>a_1} - B_{zm}(a_1)|_{r<a_1}) = -\frac{e}{\varepsilon_0} N_{m1} \qquad (5-9)$$

$$\begin{cases} \varepsilon E_{zm}(a_1)|_{r>a_1} - E_{zm}(a_1)|_{r<a_1} = 0 \\ \frac{i\omega}{\kappa^2 \varepsilon}(\varepsilon \frac{\partial B_{zm}}{\partial r}(a_1)|_{r>a_1} - \varepsilon \frac{\partial B_{zm}}{\partial r}(a_1)|_{r<a_1}) = 0 \end{cases} \qquad (5-10)$$

$$\frac{m\omega}{\kappa^2 \varepsilon r}(B_{zm}(a_2)|_{r>a_2} - \varepsilon B_{zm}(a_2)|_{r<a_2}) = -\frac{e}{\varepsilon_0} N_{m2} \qquad (5-11)$$

$$\begin{cases} E_{zm}(a_2)|_{r>a_2} - \varepsilon E_{zm}(a_2)|_{r<a_2} = 0 \\ \frac{i\omega}{\kappa^2 \varepsilon}(\frac{\partial B_{zm}}{\partial r}(a_2)|_{r>a_2} - \varepsilon \frac{\partial B_{zm}}{\partial r}(a_2)|_{r<a_2}) = 0 \end{cases} \qquad (5-12)$$

معادلات شرط مرزی در کروشه‌ها یکسان هستند.
مولفه‌های z میدان مغناطیسی در معادلات بسل تعمیم‌یافته صدق می‌کنند (معادلات (۹-۲) ) و جواب فرضی میدان در نواحی مختلف نانوتیوپ هنگامی که یک باریکه‌ی الکترونی از بین دوجداره‌ی آن عبور می‌کند به شکل زیر است.

$$B_{zm}(r) = C_{1m} I_m(\kappa r) \qquad r < a_1$$
$$B_{zm}(r) = C_{2m} I_m(\sqrt{\varepsilon}\kappa r) + C_{3m} K_m(\sqrt{\varepsilon}\kappa r) \qquad a_1 < r < a_2$$
$$B_{zm}(r) = C_{4m} K_m(\kappa r) \qquad r > a_2$$

مقدار $N_{m1}$ را در معادله‌ی (۵-۱) قرار می‌دهیم.

$$\frac{m\omega}{\kappa^2 \varepsilon r}(\varepsilon B_{zm}(a_1)|_{r>a_1} - B_{zm}(a_1)|_{r<a_1}) = -\frac{e}{\varepsilon_0}(-i\frac{en_0}{m_e}\frac{(qE_{zm} + \frac{m}{a_1}E_{\varphi m})}{\omega^2 - \alpha(q^2 + \frac{m^2}{a_1^2}) - \beta(q^2 + \frac{m^2}{a_1^2})^2})$$

$$E_{zm} = 0, E_{\varphi m} = \frac{i\omega}{\kappa^2 \varepsilon}\frac{\partial B_{zm}}{\partial r} \Rightarrow$$

$$\frac{m\omega}{\kappa^2 \varepsilon r}(\varepsilon B_{zm}(a_1)|_{r>a_1} - B_{zm}(a_1)|_{r<a_1}) = \frac{ie^2 n_0}{m_e \varepsilon_0}\frac{(\frac{im\omega}{\kappa^2 \varepsilon a_1}\frac{\partial B_{zm}}{\partial r}(a_1))}{\omega^2 - \alpha q_1^2 - \beta q_1^4} \qquad ; q_1^2 = q^2 + \frac{m^2}{a_1^2}$$

$$(\varepsilon B_{zm}(a_1)\big|_{r>a_1} - B_{zm}(a_1)\big|_{r<a_1})(\omega^2 - \alpha q_1^2 - \beta q_1^4) = \frac{-e^2 n_0}{m_e \varepsilon_0}\frac{\partial B_{zm}}{\partial r}(a_1)$$

با استفاده از جواب‌های فرضی مولفه‌ی $z$ میدان مغناطیسی، معادله‌ی بالا به شکل زیر در می‌آید.

$$(C_{2m}\varepsilon I(\sqrt{\varepsilon}\kappa a_1) + C_{3m}\varepsilon K(\sqrt{\varepsilon}\kappa a_1) - C_{1m}I(\kappa a_1))(\omega^2 - \alpha q_1^2 - \beta q_1^4) = \frac{-e^2 n_0}{m_e \varepsilon_0}\kappa\frac{\partial I(\kappa a_1)}{\partial(\kappa r)}$$

با استفاده از جواب‌های فرضی برای مولفه‌ی $z$ میدان در نواحی مختلف معادله‌ی بالا به شکل زیر در می‌آید.

$$(C_{2m}\varepsilon I(\sqrt{\varepsilon}\kappa a_1) + C_{3m}\varepsilon K(\sqrt{\varepsilon}\kappa a_1) - C_{1m}I(\kappa a_1))(\omega^2 - \alpha q_1^2 - \beta q_1^4) = \frac{-e^2 n_0}{m_e \varepsilon_0}\kappa C_{1m}I'(\kappa a_1)$$

با استفاده از تعریف دو پارامتر جدید $N, R$ معادله‌ی بالا را به شکل ساده‌تری می‌نویسیم. تعریف این پارامتر با بخش ۵-۲-۲- متفاوت است.

$$R = (\omega^2 - \alpha q_1^2 - \beta q_1^4) \qquad , N = -\frac{e^2 n_0}{m_e \varepsilon_0}\kappa$$

$$RC_{2m}\varepsilon I_m(\sqrt{\varepsilon}\kappa a_1) + RC_{3m}\varepsilon K_m(\sqrt{\varepsilon}\kappa a_1) - RC_{1m}I_m(\kappa a_1) = NC_{1m}K'_m(\kappa a_1) \qquad (۵-۱۳)$$

برای ساده‌سازی معادله‌ی (۵-۱۱) نیز مطابق روند بالا عمل کرده و معادله‌ی زیر را به دست می‌آوریم.

$$HC_{4m}I_m(\sqrt{\varepsilon}\kappa a_2) - HC_{2m}\varepsilon I_m(\sqrt{\varepsilon}\kappa a_2) - HC_{3m}\varepsilon K_m(\kappa a_2) = NC_{4m}K'_m(\kappa a_2) \qquad (۵-۱۴)$$

$$H = (\omega^2 - \alpha q_2^2 - \beta q_2^4) \qquad ; q_2 = q^2 + \frac{m^2}{a_2^2}, \qquad , F = N = -\frac{e^2 n_0}{m_e \varepsilon_0}\kappa$$

جواب‌های فرضی میدان را در معادلات (۵-۱۰) و (۵-۱۲) قرار می‌دهیم.

$$C_{2m}\varepsilon\sqrt{\varepsilon}I'_m(\sqrt{\varepsilon}\kappa a_1) + C_{3m}\varepsilon\sqrt{\varepsilon}K'_m(\sqrt{\varepsilon}\kappa a_1) = C_{1m}I'_m(\kappa a_1) \qquad (۵-۱۵)$$

$$C_{2m}\varepsilon\sqrt{\varepsilon}I'_m(\sqrt{\varepsilon}\kappa a_2) + C_{3m}\varepsilon\sqrt{\varepsilon}K'_m(\sqrt{\varepsilon}\kappa a_2) = C_{4m}K'_m(\kappa a_2) \qquad (۵-۱۶)$$

$$\begin{pmatrix} -RI_m(\kappa a_1) - NI'_m(\kappa a_1) & R\varepsilon I_m(\sqrt{\varepsilon}\kappa a_1) & R\varepsilon K_m(\sqrt{\varepsilon}\kappa a_1) & 0 \\ 0 & H\varepsilon I_m(\sqrt{\varepsilon}\kappa a_2) & H\varepsilon K_m(\sqrt{\varepsilon}\kappa a_2) & FK'_m(\kappa a_2) - HK_m(\kappa a_2) \\ -I'_m(\kappa a_1) & \varepsilon\sqrt{\varepsilon}I'_m(\sqrt{\varepsilon}\kappa a_1) & \varepsilon\sqrt{\varepsilon}K'_m(\sqrt{\varepsilon}\kappa a_1) & 0 \\ 0 & \varepsilon\sqrt{\varepsilon}I'_m(\sqrt{\varepsilon}\kappa a_2) & \varepsilon\sqrt{\varepsilon}K'_m(\sqrt{\varepsilon}\kappa a_2) & -K'_m(\kappa a_2) \end{pmatrix} \begin{pmatrix} C_{1m} \\ C_{2m} \\ C_{3m} \\ C_{4m} \end{pmatrix} = \begin{pmatrix} 0 \\ 0 \\ 0 \\ 0 \end{pmatrix}$$

شرط وجود جواب غیربدیهی آن است که رابطه‌ی زیر برقرار باشد.

$$\begin{vmatrix} -RI_m(\kappa a_1) - NI'_m(\kappa a_1) & R\varepsilon I_m(\sqrt{\varepsilon}\kappa a_1) & R\varepsilon K_m(\sqrt{\varepsilon}\kappa a_1) & 0 \\ 0 & H\varepsilon I_m(\sqrt{\varepsilon}\kappa a_2) & H\varepsilon K_m(\sqrt{\varepsilon}\kappa a_2) & FK'_m(\kappa a_2) - HK_m(\kappa a_2) \\ -I'_m(\kappa a_1) & \varepsilon\sqrt{\varepsilon}I'_m(\sqrt{\varepsilon}\kappa a_1) & \varepsilon\sqrt{\varepsilon}K'_m(\sqrt{\varepsilon}\kappa a_1) & 0 \\ 0 & \varepsilon\sqrt{\varepsilon}I'_m(\sqrt{\varepsilon}\kappa a_2) & \varepsilon\sqrt{\varepsilon}K'_m(\sqrt{\varepsilon}\kappa a_2) & -K'_m(\kappa a_2) \end{vmatrix} = 0$$

$$\varepsilon^2 \begin{vmatrix} -RI_m(\kappa a_1) - NI'_m(\kappa a_1) & RI_m(\sqrt{\varepsilon}\kappa a_1) & RK_m(\sqrt{\varepsilon}\kappa a_1) & 0 \\ 0 & HI_m(\sqrt{\varepsilon}\kappa a_2) & HK_m(\sqrt{\varepsilon}\kappa a_2) & FK'_m(\kappa a_2) - HK_m(\kappa a_2) \\ -I'_m(\kappa a_1) & \sqrt{\varepsilon}I'_m(\sqrt{\varepsilon}\kappa a_1) & \sqrt{\varepsilon}K'_m(\sqrt{\varepsilon}\kappa a_1) & 0 \\ 0 & \sqrt{\varepsilon}I'_m(\sqrt{\varepsilon}\kappa a_2) & \sqrt{\varepsilon}K'_m(\sqrt{\varepsilon}\kappa a_2) & -K'_m(\kappa a_2) \end{vmatrix} = 0$$

دترمینان را حول ستون اول بسط می‌دهیم.

$$-RI_m(\kappa a_1) - NI'_m(\kappa a_1) \begin{vmatrix} HI_m(\sqrt{\varepsilon}\kappa a_2) & HK_m(\sqrt{\varepsilon}\kappa a_2) & FK'_m(\kappa a_2) - HK_m(\kappa a_2) \\ \sqrt{\varepsilon}I'_m(\sqrt{\varepsilon}\kappa a_1) & \sqrt{\varepsilon}K'_m(\sqrt{\varepsilon}\kappa a_1) & 0 \\ \sqrt{\varepsilon}I'_m(\sqrt{\varepsilon}\kappa a_2) & \sqrt{\varepsilon}K'_m(\sqrt{\varepsilon}\kappa a_2) & -K'_m(\kappa a_2) \end{vmatrix}$$

$$-I'_m(\kappa a_1) \begin{vmatrix} RI_m(\sqrt{\varepsilon}\kappa a_1) & RK_m(\sqrt{\varepsilon}\kappa a_1) & 0 \\ HI_m(\sqrt{\varepsilon}\kappa a_2) & HK_m(\sqrt{\varepsilon}\kappa a_2) & FK'_m(\kappa a_2) - HK_m(\kappa a_2) \\ \sqrt{\varepsilon}I'_m(\sqrt{\varepsilon}\kappa a_2) & \sqrt{\varepsilon}K'_m(\sqrt{\varepsilon}\kappa a_2) & -K'_m(\kappa a_2) \end{vmatrix} = 0$$

هر دو دترمینان را حول سطر اول بسط می‌دهیم.

$$[-RI_m(\kappa a_1) - NI'_m(\kappa a_1)][HI_m(\sqrt{\varepsilon}\kappa a_2)][-\sqrt{\varepsilon}K'_m(\sqrt{\varepsilon}\kappa a_1)K'_m(\kappa a_2))]$$
$$[RI_m(\kappa a_1) + NI'_m(\kappa a_1)][HK_m(\sqrt{\varepsilon}\kappa a_2)][-\sqrt{\varepsilon}I'_m(\sqrt{\varepsilon}\kappa a_1)K'_m(\kappa a_2)]$$
$$[-RI_m(\kappa a_1) - NI'_m(\kappa a_1)][FK'_m(\kappa a_2) - HK_m(\kappa a_2)]\varepsilon[I'_m(\sqrt{\varepsilon}\kappa a_1)K'_m(\sqrt{\varepsilon}\kappa a_2) - K'_m(\sqrt{\varepsilon}\kappa a_1)I'_m(\sqrt{\varepsilon}\kappa a_2)]$$
$$-RI_m(\sqrt{\varepsilon}\kappa a_1)I'_m(\kappa a_1)[-HK'_m(\kappa a_2)K_m(\sqrt{\varepsilon}\kappa a_2) + \sqrt{\varepsilon}K'_m(\sqrt{\varepsilon}\kappa a_2)[HK_m(\kappa a_2) - FK'_m(\kappa a_2)]]$$
$$+RK_m(\sqrt{\varepsilon}\kappa a_1)I'_m(\kappa a_1)[-HI_m(\sqrt{\varepsilon}\kappa a_2)K'_m(\kappa a_2) + \sqrt{\varepsilon}I'_m(\sqrt{\varepsilon}\kappa a_2)[HK_m(\kappa a_2) - FK'_m(\kappa a_2)]]$$
$$= 0$$

$$[-RI_m(\kappa a_1)][HI_m(\sqrt{\varepsilon}\kappa a_2)][-\sqrt{\varepsilon}K'_m(\sqrt{\varepsilon}\kappa a_1)K'_m(\kappa a_2))]$$
$$[-NI'_m(\kappa a_1)][HI_m(\sqrt{\varepsilon}\kappa a_2)][-\sqrt{\varepsilon}K'_m(\sqrt{\varepsilon}\kappa a_1)K'_m(\kappa a_2))]$$
$$[RI_m(\kappa a_1)][HK_m(\sqrt{\varepsilon}\kappa a_2)][-\sqrt{\varepsilon}I'_m(\sqrt{\varepsilon}\kappa a_1)K'_m(\kappa a_2)]$$
$$[NI'_m(\kappa a_1)][HK_m(\sqrt{\varepsilon}\kappa a_2)][-\sqrt{\varepsilon}I'_m(\sqrt{\varepsilon}\kappa a_1)K'_m(\kappa a_2)]$$
$$[-RI_m(\kappa a_1)][FK'_m(\kappa a_2) - HK_m(\kappa a_2)]\varepsilon[I'_m(\sqrt{\varepsilon}\kappa a_1)K'_m(\sqrt{\varepsilon}\kappa a_2) - K'_m(\sqrt{\varepsilon}\kappa a_1)I'_m(\sqrt{\varepsilon}\kappa a_2)]$$
$$[-NI'_m(\kappa a_1)][FK'_m(\kappa a_2) - HK_m(\kappa a_2)]\varepsilon[I'_m(\sqrt{\varepsilon}\kappa a_1)K'_m(\sqrt{\varepsilon}\kappa a_2) - K'_m(\sqrt{\varepsilon}\kappa a_1)I'_m(\sqrt{\varepsilon}\kappa a_2)]$$
$$-RI_m(\sqrt{\varepsilon}\kappa a_1)I'_m(\kappa a_1)[-HK'_m(\kappa a_2)K_m(\sqrt{\varepsilon}\kappa a_2)]$$
$$-RI_m(\sqrt{\varepsilon}\kappa a_1)I'_m(\kappa a_1)\sqrt{\varepsilon}K'_m(\sqrt{\varepsilon}\kappa a_2)[HK_m(\kappa a_2) - FK'_m(\kappa a_2)]]$$
$$+RK_m(\sqrt{\varepsilon}\kappa a_1)I'_m(\kappa a_1)[-HI_m(\sqrt{\varepsilon}\kappa a_2)K'_m(\kappa a_2)]$$
$$+RK_m(\sqrt{\varepsilon}\kappa a_1)I'_m(\kappa a_1)\sqrt{\varepsilon}I'_m(\sqrt{\varepsilon}\kappa a_2)[HK_m(\kappa a_2) - FK'_m(\kappa a_2)]$$
$$= 0$$

$$+ RH\sqrt{\varepsilon} I_m(\kappa a_1) I_m(\sqrt{\varepsilon}\kappa a_2) K'_m(\sqrt{\varepsilon}\kappa a_1) K'_m(\kappa a_2)$$
$$+ NH\sqrt{\varepsilon} I'_m(\kappa a_1) I_m(\sqrt{\varepsilon}\kappa a_2) K'_m(\sqrt{\varepsilon}\kappa a_1) K'_m(\kappa a_2)$$
$$- RH\sqrt{\varepsilon} I_m(\kappa a_1) K_m(\sqrt{\varepsilon}\kappa a_2) I'_m(\sqrt{\varepsilon}\kappa a_1) K'_m(\kappa a_2)$$
$$- NH\sqrt{\varepsilon} I'_m(\kappa a_1) K_m(\sqrt{\varepsilon}\kappa a_2) I'_m(\sqrt{\varepsilon}\kappa a_1) K'_m(\kappa a_2)$$
$$- RF\varepsilon I_m(\kappa a_1) K'_m(\kappa a_2) I'_m(\sqrt{\varepsilon}\kappa a_1) K'_m(\sqrt{\varepsilon}\kappa a_2)$$
$$+ RF\varepsilon I_m(\kappa a_1) K'_m(\kappa a_2) K'_m(\sqrt{\varepsilon}\kappa a_1) I'_m(\sqrt{\varepsilon}\kappa a_2)$$
$$+ RH\varepsilon I_m(\kappa a_1) K_m(\kappa a_2) I'_m(\sqrt{\varepsilon}\kappa a_1) K'_m(\sqrt{\varepsilon}\kappa a_2)$$
$$- RH\varepsilon I_m(\kappa a_1) K_m(\kappa a_2) K'_m(\sqrt{\varepsilon}\kappa a_1) I'_m(\sqrt{\varepsilon}\kappa a_2)$$
$$- NF\varepsilon I'_m(\kappa a_1) K'_m(\kappa a_2) I'_m(\sqrt{\varepsilon}\kappa a_1) K'_m(\sqrt{\varepsilon}\kappa a_2)$$
$$+ NF\varepsilon I'_m(\kappa a_1) K'_m(\kappa a_2) K'_m(\sqrt{\varepsilon}\kappa a_1) I'_m(\sqrt{\varepsilon}\kappa a_2)$$
$$+ NH\varepsilon I'_m(\kappa a_1) K_m(\kappa a_2) I'_m(\sqrt{\varepsilon}\kappa a_1) K'_m(\sqrt{\varepsilon}\kappa a_2)$$
$$- NH\varepsilon I'_m(\kappa a_1) K_m(\kappa a_2) K'_m(\sqrt{\varepsilon}\kappa a_1) I'_m(\sqrt{\varepsilon}\kappa a_2)$$
$$+ RH I_m(\sqrt{\varepsilon}\kappa a_1) I'_m(\kappa a_1) K'_m(\kappa a_2) K_m(\sqrt{\varepsilon}\kappa a_2)$$
$$- RH\sqrt{\varepsilon} I_m(\sqrt{\varepsilon}\kappa a_1) I'_m(\kappa a_1) K'_m(\sqrt{\varepsilon}\kappa a_2) K_m(\kappa a_2)$$
$$+ RF\sqrt{\varepsilon} I_m(\sqrt{\varepsilon}\kappa a_1) I'_m(\kappa a_1) K'_m(\sqrt{\varepsilon}\kappa a_2) K'_m(\kappa a_2)$$
$$- RH K_m(\sqrt{\varepsilon}\kappa a_1) I'_m(\kappa a_1) I_m(\sqrt{\varepsilon}\kappa a_2) K'_m(\kappa a_2)$$
$$+ RH\sqrt{\varepsilon} K_m(\sqrt{\varepsilon}\kappa a_1) I'_m(\kappa a_1) I'_m(\sqrt{\varepsilon}\kappa a_2) K_m(\kappa a_2)$$
$$- RF\sqrt{\varepsilon} K_m(\sqrt{\varepsilon}\kappa a_1) I'_m(\kappa a_1) I'_m(\sqrt{\varepsilon}\kappa a_2) K'_m(\kappa a_2)$$
$$= 0$$

$$+ RH(\sqrt{\varepsilon}I_m(\kappa a_1)I_m(\sqrt{\varepsilon}\kappa a_2)K'_m(\sqrt{\varepsilon}\kappa a_1)K'_m(\kappa a_2)$$
$$-\sqrt{\varepsilon}I_m(\kappa a_1)K_m(\sqrt{\varepsilon}\kappa a_2)I'_m(\sqrt{\varepsilon}\kappa a_1)K'_m(\kappa a_2)$$
$$+\varepsilon I_m(\kappa a_1)K_m(\kappa a_2)I'_m(\sqrt{\varepsilon}\kappa a_1)K'_m(\sqrt{\varepsilon}\kappa a_2)$$
$$-\varepsilon I_m(\kappa a_1)K_m(\kappa a_2)K'_m(\sqrt{\varepsilon}\kappa a_1)I'_m(\sqrt{\varepsilon}\kappa a_2)$$
$$+ I_m(\sqrt{\varepsilon}\kappa a_1)I'_m(\kappa a_1)K'_m(\kappa a_2)K_m(\sqrt{\varepsilon}\kappa a_2)$$
$$-\sqrt{\varepsilon}I_m(\sqrt{\varepsilon}\kappa a_1)I'_m(\kappa a_1)K'_m(\sqrt{\varepsilon}\kappa a_2)K_m(\kappa a_2)$$
$$- K_m(\sqrt{\varepsilon}\kappa a_1)I'_m(\kappa a_1)I_m(\sqrt{\varepsilon}\kappa a_2)K'_m(\kappa a_2)$$
$$+\sqrt{\varepsilon}K_m(\sqrt{\varepsilon}\kappa a_1)I'_m(\kappa a_1)I'_m(\sqrt{\varepsilon}\kappa a_2)K_m(\kappa a_2))$$

$$+ NH(\sqrt{\varepsilon}I'_m(\kappa a_1)I_m(\sqrt{\varepsilon}\kappa a_2)K'_m(\sqrt{\varepsilon}\kappa a_1)K'_m(\kappa a_2)$$
$$-\sqrt{\varepsilon}I'_m(\kappa a_1)K_m(\sqrt{\varepsilon}\kappa a_2)I'_m(\sqrt{\varepsilon}\kappa a_1)K'_m(\kappa a_2)$$
$$+\varepsilon I'_m(\kappa a_1)K_m(\kappa a_2)I'_m(\sqrt{\varepsilon}\kappa a_1)K'_m(\sqrt{\varepsilon}\kappa a_2)$$
$$-\varepsilon I'_m(\kappa a_1)K_m(\kappa a_2)K'_m(\sqrt{\varepsilon}\kappa a_1)I'_m(\sqrt{\varepsilon}\kappa a_2))$$

$$- RF(\varepsilon I_m(\kappa a_1)K'_m(\kappa a_2)I'_m(\sqrt{\varepsilon}\kappa a_1)K'_m(\sqrt{\varepsilon}\kappa a_2)$$
$$+\varepsilon I_m(\kappa a_1)K'_m(\kappa a_2)K'_m(\sqrt{\varepsilon}\kappa a_1)I'_m(\sqrt{\varepsilon}\kappa a_2)$$
$$+\sqrt{\varepsilon}I_m(\sqrt{\varepsilon}\kappa a_1)I'_m(\kappa a_1)K'_m(\sqrt{\varepsilon}\kappa a_2)K'_m(\kappa a_2)$$
$$-\sqrt{\varepsilon}K_m(\sqrt{\varepsilon}\kappa a_1)I'_m(\kappa a_1)I'_m(\sqrt{\varepsilon}\kappa a_2)K'_m(\kappa a_2))$$

$$- NF(-\varepsilon I'_m(\kappa a_1)K'_m(\kappa a_2)I'_m(\sqrt{\varepsilon}\kappa a_1)K'_m(\sqrt{\varepsilon}\kappa a_2)$$
$$+\varepsilon I'_m(\kappa a_1)K'_m(\kappa a_2)K'_m(\sqrt{\varepsilon}\kappa a_1)I'_m(\sqrt{\varepsilon}\kappa a_2))$$

$$= 0$$

$$R = (\omega^2 - \alpha q_1^2 - \beta q_1^4) \qquad , F = N = -\frac{e^2 n_0}{m_e \varepsilon_0}\kappa$$

در تعریف پارامتر N، دیگر نباید تبدیل $\kappa \to \kappa\sqrt{\varepsilon}$ را انجام دهیم. زیرا این تبدیل در $\kappa$‌های شرط مرزی وارد شده است.

$$NH = -\frac{e^2 n_0}{m_e \varepsilon_0}\kappa(\omega^2 - \alpha q_2^2 - \beta q_2^4) = -\Omega_{p1}^2(\kappa a_1)(\omega^2 - \alpha q_2^2 - \beta q_2^4)$$
$$NR = -\frac{e^2 n_0}{m_e \varepsilon_0}\kappa(\omega^2 - \alpha q_1^2 - \beta q_1^4) = -\Omega_{p2}^2(\kappa a_2)(\omega^2 - \alpha q_1^2 - \beta q_1^4)$$

$$N^2 = (-\frac{e^2 n_0}{m_e \varepsilon_0}\kappa)^2 = \Omega_{p1}^2 \Omega_{p2}^2 (\kappa a_1)(\kappa a_2)$$

$$RH = \omega^4 + \omega^2(-\alpha(q_1^2 + q_2^2) - \beta(q_1^4 + q_2^4)) + (\alpha q_1^2 + \beta q_1^4)(\alpha q_2^2 + \beta q_2^4)$$

$$NH = -\frac{e^2 n_0}{m_e \varepsilon_0}\kappa(\omega^2 - \alpha q_2^2 - \beta q_2^4) = -\Omega_{p1}^2(\kappa a_1)(\omega^2 - \alpha q_2^2 - \beta q_2^4)$$

$$NR = -\frac{e^2 n_0}{m_e \varepsilon_0}\kappa(\omega^2 - \alpha q_1^2 - \beta q_1^4) = -\Omega_{p2}^2(\kappa a_2)(\omega^2 - \alpha q_1^2 - \beta q_1^4)$$

$$N^2 = (-\frac{e^2 n_0}{m_e \varepsilon_0}\kappa)^2 = \Omega_{p1}^2 \Omega_{p2}^2 (\kappa a_1)(\kappa a_2)$$

$$RH/(\Omega_{p1}^2 \Omega_{p2}^2) = \omega^4/(\Omega_{p1}^2 \Omega_{p2}^2) - \frac{\omega^2}{\Omega_{p1}\Omega_{p2}}(\frac{\alpha}{\Omega_{p1}\Omega_{p2}}(q_1^2 + q_2^2) + \frac{\beta}{\Omega_{p1}\Omega_{p2}}(q_1^4 + q_2^4))$$

$$+ \frac{(\alpha q_1^2 + \beta q_1^4)(\alpha q_2^2 + \beta q_2^4)}{\Omega_{p1}^2 \Omega_{p2}^2}$$

$$NH/(\Omega_{p1}^2 \Omega_{p2}^2) = -\frac{\Omega_{p1}}{\Omega_{p2}}(\kappa a_1)(\frac{\omega^2}{\Omega_{p1}\Omega_{p2}} - \frac{\alpha}{\Omega_{p1}\Omega_{p2}}q_2^2 - \frac{\beta}{\Omega_{p1}\Omega_{p2}}q_2^4)$$

$$= -\sqrt{\frac{a_2}{a_1}}(\kappa a_1)(\frac{\omega^2}{\Omega_{p1}\Omega_{p2}} - \frac{\alpha}{\Omega_{p1}\Omega_{p2}}q_2^2 - \frac{\beta}{\Omega_{p1}\Omega_{p2}}q_2^4)$$

$$FR/(\Omega_{p1}^2 \Omega_{p2}^2) = -\frac{\Omega_{p2}}{\Omega_{p1}}(\kappa a_2)(\frac{\omega^2}{\Omega_{p1}\Omega_{p2}} - \frac{\alpha}{\Omega_{p1}\Omega_{p2}}q_1^2 - \frac{\beta}{\Omega_{p1}\Omega_{p2}}q_1^4)$$

$$= -\sqrt{\frac{a_1}{a_2}}(\kappa a_2)(\frac{\omega^2}{\Omega_{p1}\Omega_{p2}} - \frac{\alpha}{\Omega_{p1}\Omega_{p2}}q_1^2 - \frac{\beta}{\Omega_{p1}\Omega_{p2}}q_1^4)$$

$$NF/(\Omega_{p1}^2 \Omega_{p2}^2) = (\kappa a_1)(\kappa a_2)$$

$$RH/(\Omega_{p1}^2\Omega_{p2}^2) = \omega^4/(\Omega_{p1}^2\Omega_{p2}^2) - \frac{\omega^2}{\Omega_{p1}\Omega_{p2}}(\frac{\alpha}{\Omega_{p1}\Omega_{p2}a_1a_2}(a_1a_2q_1^2+a_1a_2q_2^2)+\frac{\beta}{\Omega_{p1}\Omega_{p2}a_1^2a_2^2}(a_1^2a_2^2q_1^4+a_1^2a_2^2q_2^4))$$

$$+(\frac{\alpha}{\Omega_{p1}\Omega_{p2}a_1a_2}a_1a_2q_1^2+\frac{\beta}{\Omega_{p1}\Omega_{p2}a_1^2a_2^2}a_1^2a_2^2q_1^4)(\frac{\alpha}{\Omega_{p1}\Omega_{p2}a_1a_2}a_1a_2q_2^2+\frac{\beta}{\Omega_{p1}\Omega_{p2}a_1^2a_2^2}a_1^2a_2^2q_2^4)$$

$$NH/(\Omega_{p1}^2\Omega_{p2}^2) = -\sqrt{\frac{a_2}{a_1}}(\kappa a_1)(\frac{\omega^2}{\Omega_{p1}\Omega_{p2}}-\frac{\alpha}{\Omega_{p1}\Omega_{p2}}q_2^2-\frac{\beta}{\Omega_{p1}\Omega_{p2}}q_2^4)$$

$$= -(\kappa\sqrt{a_1a_2})(\frac{\omega^2}{\Omega_{p1}\Omega_{p2}}-\frac{\alpha}{\Omega_{p1}\Omega_{p2}a_1a_2}a_1a_2q_2^2-\frac{\beta}{\Omega_{p1}\Omega_{p2}a_1^2a_2^2}a_1^2a_2^2q_2^4)$$

$$FR/(\Omega_{p1}^2\Omega_{p2}^2) = -\sqrt{\frac{a_1}{a_2}}(\kappa a_2)(\frac{\omega^2}{\Omega_{p1}\Omega_{p2}}-\frac{\alpha}{\Omega_{p1}\Omega_{p2}}q_1^2-\frac{\beta}{\Omega_{p1}\Omega_{p2}}q_1^4)$$

$$= -(\kappa\sqrt{a_1a_2})(\frac{\omega^2}{\Omega_{p1}\Omega_{p2}}-\frac{\alpha}{\Omega_{p1}\Omega_{p2}a_1a_2}a_1a_2q_1^2-\frac{\beta}{\Omega_{p1}\Omega_{p2}a_1^2a_2^2}a_1^2a_2^2q_1^4)$$

$$NF/(\Omega_{p1}^2\Omega_{p2}^2) = (\kappa\sqrt{a_1a_2})(\kappa\sqrt{a_1a_2})$$

$$q_1^2 = (q^2+m^2/a_1^2) = [\kappa^2+\frac{\omega^2}{c^2}+m^2/a_1^2]$$

$$a_1a_2q_1^2 = [\kappa^2a_1a_2+\frac{\Omega_{P1}\Omega_{P2}a_1a_2}{c^2}\frac{\omega^2}{\Omega_{P1}\Omega_{P2}}+m^2a_1a_2/a_1^2] = [x^2+\sigma y^2+m^2a_2/a_1] \approx [x^2+m^2a_2/a_1]$$

$$\kappa \to \kappa\sqrt{\varepsilon} \Rightarrow a_1a_2q_1^2 = [x^2\varepsilon+m^2a_2/a_1]$$

متغیرهای q چون در ابتدا از فرمول ابتدایی $N_m = -i\frac{en_0}{m_e}\frac{(qE_{zm}+m/aE_{\varphi m})}{\omega^2-\alpha(q^2+m^2/a^2)-\beta(q^2+m^2/a^2)^2}$ که در فصل دوم محاسبه شده، وارد معادلات پاشندگی می‌شوند، وقتی بر حسب $\kappa$ نوشته می‌شوند باید تبدیل $\kappa \to \kappa\sqrt{\varepsilon}$ روی آن‌ها صورت گیرد.

$$q_2^2 = (q^2+m^2/a_2^2) = [(\kappa^2+\frac{\omega^2}{c^2})+m^2/a_2^2]$$

$$a_1a_2q_2^2 = [a_1a_2\kappa^2+\frac{\Omega_{P1}\Omega_{P2}a_1a_2}{c^2}\frac{\omega^2}{\Omega_{P1}\Omega_{P2}}+m^2a_1a_2/a_2^2] = [x^2+\sigma y^2+m^2a_1/a_2] \approx [x^2+m^2a_1/a_2]$$

$$\kappa \to \kappa\sqrt{\varepsilon} \Rightarrow a_1a_2q_2^2 = [x^2\varepsilon+m^2a_1/a_2]$$

$$RH/(\Omega_{p1}^2\Omega_{p2}^2) = y^4 - y^2(\alpha_1([x^2\varepsilon + m^2 a_2/a_1] + [x^2\varepsilon + m^2 a_1/a_2]) + \beta_1([x^2\varepsilon + m^2 a_2/a_1]^2 + [x^2\varepsilon + m^2 a_1/a_2]^2))$$
$$+ (\alpha_1[x^2\varepsilon + m^2 a_2/a_1] + \beta_1[x^2\varepsilon + m^2 a_2/a_1]^2)(\alpha_1[x^2\varepsilon + m^2 a_1/a_2] + \beta_1[x^2\varepsilon + m^2 a_1/a_2]^2)$$

$$NH/(\Omega_{p1}^2\Omega_{p2}^2) = -(x)(y^2 - \alpha_1[x^2\varepsilon + m^2 a_1/a_2] - \beta_1[x^2\varepsilon + m^2 a_1/a_2]^2)$$
$$FR/(\Omega_{p1}^2\Omega_{p2}^2) = -(x)(y^2 - \alpha_1[x^2\varepsilon + m^2 a_2/a_1] - \beta_1[x^2\varepsilon + m^2 a_2/a_1]^2)$$
$$NF/(\Omega_{p1}^2\Omega_{p2}^2) = x^2$$

$$[y^4 - y^2(\alpha_1([x^2\varepsilon + m^2 a_2/a_1] + [x^2\varepsilon + m^2 a_1/a_2]) + \beta_1([x^2\varepsilon + m^2 a_2/a_1]^2 + [x^2\varepsilon + m^2 a_1/a_2]^2))$$
$$+ (\alpha_1[x^2\varepsilon + m^2 a_2/a_1] + \beta_1[x^2\varepsilon + m^2 a_2/a_1]^2)(\alpha_1[x^2\varepsilon + m^2 a_1/a_2] + \beta_1[x^2\varepsilon + m^2 a_1/a_2]^2)] \times$$
$$(\sqrt{\varepsilon} I_m(\kappa a_1) I_m(\sqrt{\varepsilon}\kappa a_2) K'_m(\sqrt{\varepsilon}\kappa a_1) K'_m(\kappa a_2)$$
$$- \sqrt{\varepsilon} I_m(\kappa a_1) K_m(\sqrt{\varepsilon}\kappa a_2) I'_m(\sqrt{\varepsilon}\kappa a_1) K'_m(\kappa a_2)$$
$$+ \varepsilon I_m(\kappa a_1) K_m(\kappa a_2) I'_m(\sqrt{\varepsilon}\kappa a_1) K'_m(\sqrt{\varepsilon}\kappa a_2)$$
$$- \varepsilon I_m(\kappa a_1) K_m(\kappa a_2) K'_m(\sqrt{\varepsilon}\kappa a_1) I'_m(\sqrt{\varepsilon}\kappa a_2)$$
$$+ I_m(\sqrt{\varepsilon}\kappa a_1) I'_m(\kappa a_1) K'_m(\kappa a_2) K_m(\sqrt{\varepsilon}\kappa a_2)$$
$$- \sqrt{\varepsilon} I_m(\sqrt{\varepsilon}\kappa a_1) I'_m(\kappa a_1) K'_m(\sqrt{\varepsilon}\kappa a_2) K_m(\kappa a_2)$$
$$- K_m(\sqrt{\varepsilon}\kappa a_1) I'_m(\kappa a_1) I_m(\sqrt{\varepsilon}\kappa a_2) K'_m(\kappa a_2)$$
$$+ \sqrt{\varepsilon} K_m(\sqrt{\varepsilon}\kappa a_1) I'_m(\kappa a_1) I'_m(\sqrt{\varepsilon}\kappa a_2) K_m(\kappa a_2))$$
$$- (x)(y^2 - \alpha_1[x^2\varepsilon + m^2 a_1/a_2] - \beta_1[x^2\varepsilon + m^2 a_1/a_2]^2) \times$$
$$(\sqrt{\varepsilon} I'_m(\kappa a_1) I_m(\sqrt{\varepsilon}\kappa a_2) K'_m(\sqrt{\varepsilon}\kappa a_1) K'_m(\kappa a_2)$$
$$- \sqrt{\varepsilon} I'_m(\kappa a_1) K_m(\sqrt{\varepsilon}\kappa a_2) I'_m(\sqrt{\varepsilon}\kappa a_1) K'_m(\kappa a_2)$$
$$+ \varepsilon I'_m(\kappa a_1) K_m(\kappa a_2) I'_m(\sqrt{\varepsilon}\kappa a_1) K'_m(\sqrt{\varepsilon}\kappa a_2)$$
$$- \varepsilon I'_m(\kappa a_1) K_m(\kappa a_2) K'_m(\sqrt{\varepsilon}\kappa a_1) I'_m(\sqrt{\varepsilon}\kappa a_2))$$
$$+ (x)(y^2 - \alpha_1[x^2\varepsilon + m^2 a_2/a_1] - \beta_1[x^2\varepsilon + m^2 a_2/a_1]^2) \times$$
$$(\varepsilon I_m(\kappa a_1) K'_m(\kappa a_2) I'_m(\sqrt{\varepsilon}\kappa a_1) K'_m(\sqrt{\varepsilon}\kappa a_2)$$
$$+ \varepsilon I_m(\kappa a_1) K'_m(\kappa a_2) K'_m(\sqrt{\varepsilon}\kappa a_1) I'_m(\sqrt{\varepsilon}\kappa a_2)$$
$$+ \sqrt{\varepsilon} I_m(\sqrt{\varepsilon}\kappa a_1) I'_m(\kappa a_1) K'_m(\sqrt{\varepsilon}\kappa a_2) K'_m(\kappa a_2)$$
$$- \sqrt{\varepsilon} K_m(\sqrt{\varepsilon}\kappa a_1) I'_m(\kappa a_1) I'_m(\sqrt{\varepsilon}\kappa a_2) K'_m(\kappa a_2))$$
$$- x^2 \times$$
$$(-\varepsilon I'_m(\kappa a_1) K'_m(\kappa a_2) I'_m(\sqrt{\varepsilon}\kappa a_1) K'_m(\sqrt{\varepsilon}\kappa a_2)$$
$$+ \varepsilon I'_m(\kappa a_1) K'_m(\kappa a_2) K'_m(\sqrt{\varepsilon}\kappa a_1) I'_m(\sqrt{\varepsilon}\kappa a_2))$$
$$= 0$$

در رسم نمودارها با یک روش "تکراری" (Iterative) در MATHEMATICA نقاط شروع، تمامی اعداد از ۱ تا ۶ با فواصـل ۰،۲ انتخاب شدند. برای نقاط شروع با اعداد بیشتر از ۶ جواب‌ها تکراری به دست می‌آمدند.

در شکل (۵-۱۲) جواب‌های معادله‌ی پاشندگی بی‌بعد شده‌ی نانوتیوپ کربنی دوجداره در حضور یک باریکه‌ی الکترونی گذرنده از بین جداره‌ها و با مشخصات زیر برای دو مد عرضی صفر و یک به طور عددی محاسبه شده و رسم گردیده‌اند. قابل ذکر است که در این نمودارها تنها بخش‌های حقیقی و مثبت فرکانس‌ها را در نظر گرفته‌ایم.

سرعت باریکه‌ی الکترونی:

$2 \times 10^7 (m/s)$

چگالی الکترونی و فرکانس ویژه‌ی باریکه:

$n_0 = 1.26 \times 10^{29} (1/m^3), \omega_b^2 = 4 \times 10^{32} (1/s^2)$

شعاع نانوتیوپ کربنی تک‌جداره :

$(a_1 = 4nm, a_2 = 4.35nm)$

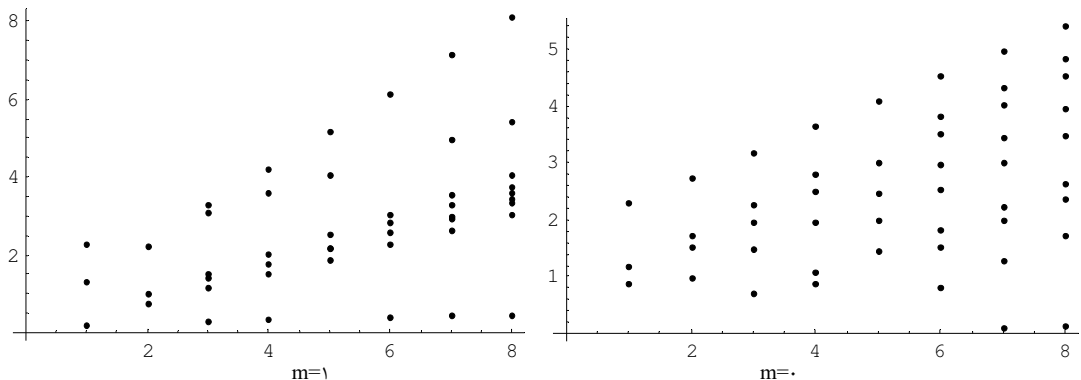

شکل (۵-۱۲)- نمودارهای پاشندگی نانوتیوپ کربنی دوجداره در مد TE در حضور باریکه‌ی الکترونی گذرنده از بین دوجداره - محور عمودی متغیر $\omega/\sqrt{\Omega_{p1}\Omega_{p2}}$ و محور افقی نمایانگر متغیر $q\sqrt{a_1 a_2}$ است.

نمودارهای پاشندگی نانوتیوپ کربنی دوجداره، برای هر مد عرضی در خلا دارای دو شاخه هستند (شکل (۴-۷) ). ولی بـا توجه به شکل‌های (۵-۱۲) ، این نمودارها در حضور باریکه‌ی الکترونی گذرنده از بین جداره‌ها، دارای تعـداد شـاخه‌هـای بیشتری می‌شوند. همچنین شاخه‌های نمودارهای پاشندگی در مد صفر به طور کلی فواصل بیشتری از یکدیگر دارند.

# نتیجه گیری

در این پایان‌نامه معادلات پاشندگی مدهای TE,TM برای نانوتیوپ‌های کربنی تک‌جداره و دوجداره محاسبه شده است. همچنین معادلات پاشندگی در حضور یک باریکه‌ی الکترونی نیز به دست آورده شده‌اند. نمودارهای مربوط به معادلات پاشندگی برای مد TM در نانوتیوپ تک‌جداره (بدون حضور باریکه‌ی الکترونی) و در مدهای عرضی مختلف نشان می‌دهند که با افزایش مدهای عرضی، فرکانس‌های قطع برای نانوتیوپ کربنی تک‌جداره افزایش می‌یابند. همچنین برای شعاع‌های مختلفِ نانوتیوپ‌های کربنی تک‌جداره، با افزایش شعاع، شیب منحنی پاشندگی که نمایان‌گر سرعت گروه امواج پلزمونی منتشر شده در موجبر است کاهش می‌یابد. با رسم نمودارهای پاشندگی اثر نیروهای برهم‌کنش داخلی در گاز الکترون آزاد را بررسی کردیم و نشان دادیم که اثر محسوسی روی معادلات پاشندگی نگذاشته و تنها شیب آن‌ها را اندکی کاهش می‌دهند. در نانوتیوپ تک‌جداره با مد TE نمودارهای معادلات پاشندگی به دلایل مطرح شده‌ی ریاضی در پایان‌نامه، شبیه به نمودارهای معادلات پاشندگی مد TM، به دست آمدند و تفسیرهای مشابهی برای آن‌ها وجود خواهد داشت. نمودارهای پاشندگی در موجبرهای کربنی دوجداره، برای مدهای TM, TE (بدون حضور باریکه) رسم شدند. نمودارهای پاشندگی مدهای TM, TE مانند حالت تک‌جداره تا حدود زیادی شبیه به هم شدند. نمودارها دارای دو شاخه‌ی بالایی و پایینی بوده و برای مدهای بالاتر از صفر، در نقاط مینیمم ِ منحنی‌ها، سرعت گروه موج در نانویتوپ‌های کربنی دوجداره صفر خواهد بود. با رسم نمودارهای پاشندگی برای شعاع‌های داخلی و خارجی متفاوت، به این نتیجه رسیدیم که نقطه‌ی ماگزیمم در شاخه‌ی پایینی با افزایش شعاع‌های داخلی و خارجی بیشتر شده و همچنین دو شاخه معادله‌ی پاشندگی به یکدیگر نزدیکتر می‌شوند. نمودارهای پاشندگی نانوتیوپ‌های کربنی تک‌جداره و دوجداره در حضور باریکه‌های الکترونی همگی تعداد شاخه‌های بیشتری نسبت به حالات بدون حضور باریکه در فصل‌های ۳ و ۴ داشته‌اند. کاربرد این نوع نمودارها در این پایان‌نامه

شرح داده شده است. بدین ترتیب با رسم نمودارهای پاشندگی در حضور باریکه می‌توان به سرعت‌های گروه امواج پلزمونی دست یافت و با بینش درستی از نانوتیوپ‌های کربنی تک‌-جداره و دوجداره به عنوان موجبر در مدارات نوری نانومتری استفاده نمود. با استفاده از نمودارهای به دست آمده در این پایان‌نامه طراحی‌های بهینه‌ای برای مدارات مجتمع نوری نانومتری که در آن‌ها موجبرهای کربنی به کار رفته امکان‌پذیر می‌شود.

# پیوست ها

## پیوست الف- بخشی از محاسبات مربوط به معادله پاشندگی نانوتیوپ کربنی دوجداره در مد TM

در این پیوست ادامه‌ی محاسبات مربوط به معادله پاشندگی مد TM در نانوتیوپ دوجداره آورده شده است. حروفی که در محاسبات این پیوست با فونت انگلیسی آورده شده‌اند، در بقیه‌ی گزارش به شکل زیر هستند.

( w = $\omega$ ),( a = $\alpha$ ),( b = $\beta$ ),( a۱, a۲ = $a_1, a_2$ ),( E۰ = $\varepsilon_0$ ),( n۰ = $n_0$ ),( Me = $m_e$ ),( q۱^۲, q۲^۲ = $q_{1m}^2, q_{2m}^2$ ),( k = $\kappa$ ),( C۱,C۲,C۳,C۴ = $C_{1m}, C_{2m}, C_{3m}, C_{4m}$ )

[ I (ka۱), I (ka۲) = $I_m(\kappa a_1), I_m(\kappa a_2)$ ],[ I'(ka۱), I'(ka۲) = $I'_m(\kappa a_1), I'_m(\kappa a_2)$ ]
[ K (ka۱), K (ka۲) = $K_m(\kappa a_1), K_m(\kappa a_2)$ ],[ K'(ka۱), K'(ka۲) = $K'_m(\kappa a_1), K'_m(\kappa a_2)$ ]

معادله‌ی شرط مرزی (۴-۷) :

C۱*I (ka۱) = C۲*I (ka۱) + C۳*K (ka۱)

با حذف ضریب C۳ از آن و قرار دادن این ضریب در معادله‌ی (۴-۵) ضریب C۲ بر حسب ضریب C۱ بدست می‌آمد.

C۳ = I (ka۱)/ K (ka۱) *C۱ – I (ka۱)/ K (ka۱)*C۲

R* C۲*I'(ka۱) + R*C۳ *K'(ka۱) – R*C۱* I'(ka۱) = N*C۱* I (ka۱)             (۴-۵)

R* I'(ka۱) *C۲ + R*K'(ka۱) * I (ka۱)/ K (ka۱) *C۱ – R*K'(ka۱) *I (ka۱)/ K (ka۱)*C۲
 – R* I'(ka۱) *C۱ = N* I (ka۱)*C۱

$\Big($R* I'(ka۱) - R*K'(ka۱) *I (ka۱)/ K (ka۱)$\Big)$*C۲ = $\Big($-R*K'(ka۱) * I (ka۱)/ K (ka۱) + R* I'(ka۱)+N* I (ka۱)$\Big)$ *C۱

=> C۲ = $\Big($-R*K'(ka۱) * I (ka۱)/ K (ka۱) + R* I'(ka۱)+N* I (ka۱)$\Big) \Big/ \Big($R* I'(ka۱) - R*K'(ka۱) *I (ka۱)/ K (ka۱)$\Big)$ *C۱

حال که ضریب C۲ بر حسب C۱ به دست آمده، می‌توان ضریب C۳ را تنها برحسب C۱ نوشت.

C۳ = I (ka۱)/ K (ka۱) *C۱ – $\Big($-I (ka۱)/ K (ka۱)*R*K'(ka۱) * I (ka۱)/ K (ka۱) + I (ka۱)/ K (ka۱)*R* I'(ka۱) + I (ka۱)/ K (ka۱)*N* I (ka۱)$\Big)$/$\Big($R* I'(ka۱) - R*K'(ka۱) *I (ka۱)/ K (ka۱)$\Big)$ *C۱

با استفاده از معادله‌ی شرط مرزی (۴-۸) در بخش ۴-۱-۱ ، ضریب C۴ بر حسب سه ضریب دیگر نوشته می‌شود.

C۴*K (ka۲) = C۲*I (ka۲) + C۳*K (ka۲)
=> C۴ = I (ka۲)/ K (ka۲) *C۲ + C۳

با جایگذاری ضرایب C۲,C۳ برحسب C۱ ، ضریب C۴ نیز تنها بر حسب C۱ بدست می‌آید.

C۴ = $\Big($-R* I (ka۲)/ K (ka۲) *K'(ka۱) * I (ka۱)/ K (ka۱) + R* I (ka۲)/ K (ka۲) * I'(ka۱)+N* I (ka۲)/ K (ka۲) *I (ka۱)$\Big)$/ $\Big($R* I'(ka۱) - R*K'(ka۱) *I (ka۱)/ K (ka۱)$\Big)$*C۱+ I (ka۱)/ K (ka۱) *C۱ + $\Big($I (ka۱)/ K (ka۱)*R*K'(ka۱) * I (ka۱)/ K (ka۱) – I (ka۱)/ K (ka۱)*R* I'(ka۱) – I (ka۱)/ K (ka۱)*N* I (ka۱)$\Big)$/$\Big($R* I'(ka۱) - R*K'(ka۱) *I (ka۱)/ K (ka۱)$\Big)$*C۱

با جایگذاری ضرایب C۲,C۳,C۴ بر حسب ضریب C۱ در معادله (۴-۶) تنها ضریب C۱ باقی می‌ماند که از طرفین معادله حذف شده و معادله‌ی پاشندگی به دست می‌آید.

H* K'(ka۲) *C۴ – H*I'(ka۲) *C۲ – H* K'(ka۲) *C۳ = F* K (ka۲)*C۴     (۴-۶)

$\Big($-R* QH* K'(ka۲) *I (ka۲)/ K (ka۲) *K'(ka۱) * I (ka۱)/ K (ka۱) + R* H* K'(ka۲) * I (ka۲)/ K (ka۲) * I'(ka۱)+N* H* K'(ka۲) *I (ka۲)/ K (ka۲) *I (ka۱)$\Big)$/ $\Big($R* I'(ka۱) - R*K'(ka۱) *I (ka۱)/ K (ka۱)$\Big)$*C۱ + H* K'(ka۲) * I (ka۱)/ K (ka۱) *C۱ + $\Big($H* K'(ka۲) *I (ka۱)/ K (ka۱)*R*K'(ka۱) * I (ka۱)/ K (ka۱) – H* K'(ka۲) *I (ka۱)/ K (ka۱)*R* I'(ka۱) – H* K'(ka۲) *I (ka۱)/ K (ka۱)*N* I (ka۱)$\Big)$/ $\Big($R* I'(ka۱) - R*K'(ka۱) *I (ka۱)/ K (ka۱)$\Big)$*C۱

$$+ \Big(R* H*I'(ka_2) *K'(ka_1) * I(ka_1)/ K(ka_1) - R* H*I'(ka_2) *I'(ka_1) - N* H*I'(ka_2) * I(ka_1)\Big)/ \Big(R* I'(ka_1) - R*K'(ka_1) *I(ka_1)/ K(ka_1)\Big)*C_1$$

- $H* K'(ka_2)* I(ka_1)/ K(ka_1) *C_1 + \Big(-H* K'(ka_2)* I(ka_1)/ K(ka_1)*R*K'(ka_1) * I(ka_1)/ K(ka_1) + H* K'(ka_2)* I(ka_1)/ K(ka_1)*R* I'(ka_1) + H* K'(ka_2)* I(ka_1)/ K(ka_1)*N* I(ka_1)\Big)/ \Big(R* I'(ka_1) - R*K'(ka_1) *I(ka_1)/ K(ka_1)\Big)*C_1$

$+ \Big(R* F* K(ka_2)* I(ka_2)/ K(ka_2) *K'(ka_1) * I(ka_1)/ K(ka_1) - R* F* K(ka_2)*I(ka_2)/ K(ka_2) * I'(ka_1) - N* F* K(ka_2)*I(ka_2)/ K(ka_2) *I(ka_1)\Big)/ \Big(R* I'(ka_1) - R*K'(ka_1) *I(ka_1)/ K(ka_1)\Big)*C_1 - F* K(ka_2)*I(ka_1)/ K(ka_1) *C_1 + \Big(-I(ka_1)/ K(ka_1)*R* F* K(ka_2)*K'(ka_1) * I(ka_1)/ K(ka_1) + F* K(ka_2)*I(ka_1)/ K(ka_1)*R* I'(ka_1) + F* K(ka_2)*I(ka_1)/ K(ka_1)*N* I(ka_1)\Big)/ \Big(R* I'(ka_1) - R*K'(ka_1) *I(ka_1)/ K(ka_1)\Big)*C_1$

$= .$

هر جمله در معادله‌ی بالا را در یک سطر جداگانه می‌نویسیم.

-R* H* K'(ka۲) *I (ka۲)/ K (ka۲) *K'(ka۱) * I (ka۱)/ K (ka۱)

+ R* H* K'(ka۲) * I (ka۲)/ K (ka۲) * I'(ka۱)

+N* H* K'(ka۲) *I (ka۲)/ K (ka۲) *I (ka۱)

+ R* H* I'(ka۱) * K'(ka۲) * I (ka۱)/ K (ka۱)

- R* H* K'(ka۲) * I (ka۱)/ K (ka۱)* K'(ka۱) *I (ka۱)/ K (ka۱)

+ H* K'(ka۲) *I (ka۱)/ K (ka۱)*QR*K'(ka۱) * I (ka۱)/ K (ka۱)

- R* H* K'(ka۲) *I (ka۱)/ K (ka۱)* I'(ka۱)

- N* H* K'(ka۲) *I (ka۱)/ K (ka۱)* I (ka۱)

+ R* H*I'(ka۲) *K'(ka۱) * I (ka۱)/ K (ka۱)

- R*H*I'(ka۲) *I'(ka۱)

- N* H*I'(ka۲) * I (ka۱)

-R* H* I'(ka۱) * K'(ka۲)* I (ka۱)/ K (ka۱)

+ R* H* K'(ka۲)* I (ka۱)/ K (ka۱)*K'(ka۱) *I (ka۱)/ K (ka۱)

-H* K'(ka۲)* I (ka۱)/ K (ka۱)*QR*K'(ka۱) * I (ka۱)/ K (ka۱)

+ R*H* I'(ka۱) * K'(ka۲)* I (ka۱)/ K (ka۱)

+ N* H* K'(ka۲)* I (ka۱)/ K (ka۱)* I (ka۱)

+ R* F* K (ka۲)* I (ka۲)/ K (ka۲) *K'(ka۱) * I (ka۱)/ K (ka۱)

- R* F* K (ka۲)*I (ka۲)/ K (ka۲) * I'(ka۱)

- N* F* K (ka۲) *I (ka۱) *I (ka۲)/ K (ka۲)

-R* F* K (ka۲)*I (ka۱)/ K (ka۱)* I'(ka۱)

+ R* F* K (ka۲)*I (ka۱)/ K (ka۱)* K'(ka۱) *I (ka۱)/ K (ka۱)

- R* F* K (ka۲)*I (ka۱)/ K (ka۱)* K'(ka۱) * I (ka۱)/ K (ka۱)

+ R*F* K (ka۲)*I (ka۱)/ K (ka۱)* I'(ka۱)

+ N* F* K (ka۲) * I (ka۱)*I (ka۱)/ K (ka۱)

= ۰

پس از ساده کردن جملات یکسان و با علامت مخالف به معادله‌ی زیر می‌رسیم.

-R* H*K'(ka۱) * K'(ka۲) *I (ka۲)/ K (ka۲) * I (ka۱)/ K (ka۱)
+ R* H* I'(ka۱)* K'(ka۲) * I (ka۲)/ K (ka۲)
+N* H* K'(ka۲) *I (ka۲)/ K (ka۲) *I (ka۱)
+ R* H*I'(ka۲) *K'(ka۱) * I (ka۱)/ K (ka۱)
 - R*H*I'(ka۲) *I'(ka۱)                                    (A)
 - N* H*I'(ka۲) * I (ka۱)
+ R* F*K'(ka۱)* I (ka۲) * I (ka۱)/ K (ka۱)
- R* F* I'(ka۱) *I (ka۲)
- N* F* K (ka۲) *I (ka۱) *I (ka۲)/ K (ka۲)
+ N* F* K (ka۲) * I (ka۱)*I (ka۱)/ K (ka۱)
= ۰

R* H* $\Big($ -K'(ka۱) * K'(ka۲) *I (ka۲)/ K (ka۲) * I (ka۱)/ K (ka۱)

+ I'(ka۱)* K'(ka۲) * I (ka۲)/ K (ka۲)

+ I'(ka۲) *K'(ka۱) * I (ka۱)/ K (ka۱)

- I'(ka۲) *I'(ka۱)$\Big)$

+N* H*$\Big($K'(ka۲) *I (ka۲)/ K (ka۲) *I (ka۱)

- I'(ka۲) * I (ka۱)$\Big)$

+ R* F*$\Big($K'(ka۱)* I (ka۲) * I (ka۱)/ K (ka۱)

- I'(ka۱) *I (ka۲)$\Big)$

+ N* F* $\Big($K (ka۲) * I (ka۱)*I (ka۱)/ K (ka۱)

- K (ka۲) *I (ka۱) *I (ka۲)/ K (ka۲)$\Big)$

= ۰

فرمول رونسکین:

I'(X)* K(X) – I(X)* K'(X) = ۱/X
I'(ka۱)* K(ka۱) – I(ka۱)* K'(ka۱) = ۱/(ka۱)

با ضرب بعضی جملات در عبارات واحد می‌توان جملات رونسکین را تشکیل داد.

R* H* $\Big($ -K'(ka۱) * K'(ka۲) *I (ka۲)/ K (ka۲) * I (ka۱)/ K (ka۱)

+ I'(ka۱)* K'(ka۲) * I (ka۲)/ K (ka۲)*( K (ka۱)/ K (ka۱))

+ I'(ka۲) *K'(ka۱) * I (ka۱)/ K (ka۱)

- I'(ka۲) *I'(ka۱)$\Big)$

+N* H*$\Big($K'(ka۲) *I (ka۲)/ K (ka۲) *I (ka۱)

- I'(ka۲) * I (ka۱)$\Big)$

+ R* F*$\Big($K'(ka۱)* I (ka۲) * I (ka۱)/ K (ka۱)

- I'(ka۱) *I (ka۲)$\Big)$

+ N* F* $\Big($K (ka۲) * I (ka۱)*I (ka۱)/ K (ka۱)

- K (ka۲) *I (ka۱)  *I (ka۲)/ K (ka۲)$\Big)$
= ۰

R* H* $\Big($ K'(ka۲) *I (ka۲)/ K (ka۲)/ K (ka۱) * (۱/(ka۱))

+ I'(ka۲) *K'(ka۱) * I (ka۱)/ K (ka۱)

- I'(ka۲) *I'(ka۱)*( K (ka۱)/ K (ka۱))$\Big)$

+N* H*$\Big($K'(ka۲) *I (ka۲)/ K (ka۲) *I (ka۱)

- I'(ka۲) * I (ka۱)$\Big)$

۱۲۹

+ R* F*$\big($K'(ka₁)* I (ka₂) * I (ka₁)/ K (ka₁)

- I'(ka₁) *I (ka₂)$\big)$

+ N* F* $\big($K (ka₂) * I (ka₁)*I (ka₁)/ K (ka₁)

- K (ka₂) *I (ka₁) *I (ka₂)/ K (ka₂)$\big)$

= ۰

R* H* $\big($ K'(ka₂) *I (ka₂)/ K (ka₂)/ K (ka₁) (۱/(ka₁))

- I'(ka₂) / K (ka₁) *(K (ka₂)/K(ka₂))* (۱/(ka₁))$\big)$

+N* H*$\big($K'(ka₂) *I (ka₂)/ K (ka₂) *I (ka₁)

- I'(ka₂) * I (ka₁)$\big)$

+ R* F*$\big($K'(ka₁)* I (ka₂) * I (ka₁)/ K (ka₁)

- I'(ka₁) *I (ka₂)$\big)$

+ N* F* $\big($K (ka₂) * I (ka₁)*I (ka₁)/ K (ka₁)

- K (ka₂) *I (ka₁) *I (ka₂)/ K (ka₂)$\big)$

= ۰

R* H* $\big($-۱/ K (ka₂)/ K (ka₁) (۱/(ka₁)) (۱/(ka₂))$\big)$

۱۳۰

$+ N^* H^* \big( K'(ka_2) * I(ka_2) / K(ka_2) * I(ka_1)$

$- I'(ka_2) * I(ka_1) \big)$

$+ R^* F^* \big( K'(ka_1) * I(ka_2) * I(ka_1) / K(ka_1)$

$- I'(ka_1) * I(ka_2) \big)$

$+ N^* F^* \big( K(ka_2) * I(ka_1) * I(ka_1) / K(ka_1)$

$- K(ka_2) * I(ka_1) * I(ka_2) / K(ka_2) \big)$

$= 0$

$- R^* H / \big( ka_1 * ka_2 * K(ka_2) * K(ka_1) \big)$

$+ N^* H^* \big( K'(ka_2) * I(ka_2) / K(ka_2) * I(ka_1)$

$- I'(ka_2) * I(ka_1) * (K(ka_2) / K(ka_2)) \big)$

$+ R^* F^* \big( K'(ka_1) * I(ka_2) * I(ka_1) / K(ka_1)$

$- I'(ka_1) * I(ka_2) \big)$

$+ N^* F^* \big( K(ka_2) * I(ka_1) * I(ka_1) / K(ka_1)$

$- K(ka_2) * I(ka_1) * I(ka_2) / K(ka_2) \big)$

$= 0$

$- R^* H / \big( ka_1 * ka_2 * K(ka_2) * K(ka_1) \big)$

$- N^* H^* \big( I(ka_1) / (ka_2 * K(ka_2)) \big)$

$+ R^* F^* \big( K'(ka_1) * I(ka_2) * I(ka_1) / K(ka_1)$

$- I'(ka_1) * I(ka_2) * (K(ka_1) / K(ka_1)) \big)$

$+ N^* F^* \big( K(ka_2) * I(ka_1) * I(ka_1) / K(ka_1)$

$- K(ka_2) * I(ka_1) * I(ka_2) / K(ka_2) \big)$

$= 0$

۱۳۱

<dir="rtl">

<dir="rtl">پیوست الف- بخشی از محاسبه‌ی معادله‌ی پاشندگی نانوتیوپ دوجداره در مد TM</dir>

- R* H / $\left(ka_1*ka_2*K(ka_2)* K(ka_1)\right)$

- N* H* $\left(I(ka_1)/(ka_2*K(ka_2))\right)$

- F*R* $\left(I(ka_2)/(ka_1*K(ka_1))\right)$

+ N* F* $\left(K(ka_2)*I(ka_1)^2/K(ka_1) – I(ka_1)*I(ka_2)\right)$

= 0

<dir="rtl">ضرب طرفین در</dir> $\left(ka_1*ka_2*K(ka_2)*K(ka_1)\right)$ :

R* H

+ N * H* $\left(ka_1*I(ka_1)\times K(ka_1)\right)$

+ F * R* $\left(ka_2*I(ka_2)*K(ka_2)\right)$

- N* F* $\left(ka_1*ka_2*I(ka_1)^2*K(ka_2)^2 – ka_1*ka_2*I(ka_1)*K(ka_1)*I(ka_2)*K(ka_2)\right)$

= 0

R = $w^2-a* q_1^2- b*q_1^4$

H = $w^2-a* q_2^2- b*q_2^4$

$q_1^2=(q^2+m^2/a_1^2)$

$q_2^2=(q^2+m^2/a_2^2)$

N= $-e^2 n_0/(Me*E_0) *(k^2 + m^2/a_1^2)/k$

F = $-e^2 n_0/(Me*E_0) *(k^2 + m^2/a_2^2)/k$

<dir="rtl">پرانتزهایی با مقدار واحد با نمای ایتالیک در عبارات زیر برای ساده‌تر کردن شکل پارامترها اضافه می‌شوند.</dir>

N= $-e^2 n_0/(Me*E_0) *(k^2 + m^2/a_1^2)/k *(a_1^2/a_1^2)$

= $- e^2 n_0/(Me*E_0*a_1^2) * (k^2 *a_1^2+ m^2)/k$

F=-e^2n۰/(Me*E۰) *( k^2 + m^2/a2^2)/k * (a۳۲/ a۳۲)

   = - e^2n۰/(Me*E۰* a2^2) *( k^2 *a2^2+ m^2)/k

N*F= e^4n۰^2/(Me*E۰)^2*(k^2 + m^2*a1^2)/k *( k^2 + m^2*a2^2)/k

   = e^4n۰^2/(Me*E۰* a1*a2)^2*(k^2 *a1^2+ m^2) *( k^2 *a2^2+ m^2)/k^2

RH= (w^2-a* q1^2- b * q1^4)*( w^2-a* q2^2- b * q2^4)

   = w^4 + w^2 * [-a*( q1^2+ q2^2)- b * (q1^4 + q2^4)]

   + (a* q1^2+ b * q1^4)*( a* q2^2+ b * q2^4)

N*H= - e^2n۰/(Me*E۰*a1^2) * (k^2 *a1^2+ m^2)/k * (w^2-a* q2^2- b * q2^4)

F*R= - e^2n۰/(Me*E۰* a2^2) *( k^2 *a2^2+ m^2)/k *( w^2-a* q1^2- b * q1^4)

R* H

+ N * H*(ka1*I (ka1)×K(ka1))

+ F * R* (ka2*I (ka2)*K (ka2))

- N* F* (ka1*ka2* I (ka1)^2*K (ka2)^2– ka1*ka2* I (ka1)* K (ka1)* I (ka2)*K (ka2))

= ۰

با جایگذاری ثابت‌های تعریف شده در معادله‌ی پاشندگی داریم:

w^4 + w^2 * [-a*( q1^2+ q2^2)- b * (q1^4 + q2^4)] + (a* q1^2+ b * q1^4)

*( a* q2^2+ b * q2^4)

- e^2n۰/(Me*E۰*a1^2) * (k^2 *a1^2+ m^2)/k * (w^2-a* q2^2- b * q2^4) *(ka1*I

(a1)×K(a1))

- e^2n۰/(Me*E۰* a2^2) *( k^2 *a2^2+ m^2)/k *( w^2-a* q1^2- b * q1^4) * (ka2*I (a2)*K

(a2))

- e^4n0^2/(Me*E0* a1*a2)^2*(k^2 *a1^2+ m^2) *( k^2 *a2^2+ m^2)/k^2* $\Big($ka1*ka2* I (a1)^2*K (a2)^2− ka1*ka2* I (a1)* K (a1)* I (a2)*K (a2)$\Big)$
= 0

اعداد ثابت و ضرایب $w^2$ را در معادله از هم تفکیک کرده و معادله‌ی زیر را بدست می‌آوریم.

w^4 + w^2 *[ - *a*( q1^2+ q2^2) - *b* * (q1^4 + q2^4)+ (- e^2n0)/(Me*E0*a1^2) * (k^2 *a1^2+ m^2)*$\big($a1*I (a1)×K(a1)$\big)$* (- e^2n0)/(Me*E0* a2^2) *( k^2 *a2^2+ m^2) * $\big($a2*I (a2)*K (a2)$\big)$]
+ {(*a*\*q1^2+*b*\*q1^4)*(*a*\*q2^2+ *b* * q2^4) + ( *a** q2^2+ *b* * q2^4)* e^2n0/(Me*E0*a1^2) * (k^2 *a1^2+ m^2) *$\big($a1*I (a1)×K(a1)$\big)$ +( *a** q1^2 + *b* * q1^4)* e^2n0/(Me*E0* a2^2) *( k^2 *a2^2+ m^2) * $\big($a2*I (a2)*K (a2)$\big)$
- e^4n0^2/(Me^2*E0^2*a1*a2)*(k^2*a1^2+m^2)*(k^2*a2^2+m^2)*$\big($I(a1)^2*K (a2)^2− I (a1)* K (a1)* I (a2)*K (a2)$\big)$}
= 0

در ساده کردن جمله‌ی آخر از رابطه‌ی زیر استفاده می‌کنیم.

I (a1)^2* K (a2)^2= $\big($[I (a1) /K (a1)] *[ K (a2)/I (a2)]$\big)$*$\big($I (a1)* I (a2) * K (a1)*K (a2)$\big)$
= F12*$\big($I (a1) * K (a1)* I (a2) *K (a2)$\big)$

پارامتر F12 در زیر تعریف شده است. معادله نهایی پاشندگی به شکل زیر در می‌آید.

معادله‌ی پاشندگی w(k) :

w^4 + w^2 *[ - *a*( q1^2+ q2^2) - *b* * (q1^4 + q2^4)- w1^2 - w2^2 ]
+{(*a** q1^2+ *b* * q1^4)*( *a** q2^2+ *b* * q2^4) + ( *a** q2^2+ *b* * q2^4)* w1^2+( *a** q1^2 + *b* * q1^4)* w2^2- w1^2 * w2^2*F12 + w1^2 * w2^2}
=0

$q_1^2 = (q^2 + m^2/a_1^2)$

$q_2^2 = (q^2 + m^2/a_2^2)$

$w_1^2 = e^2 n_0 /(M_e * E_0 * a_1) * (k^2 * a_1^2 + m^2) * \bigl( I(a_1) \times K(a_1) \bigr)$

$w_2^2 = e^2 n_0 /(M_e * E_0 * a_2) * (k^2 * a_2^2 + m^2) * \bigl( I(a_2) * K(a_2) \bigr)$

$F_{12} = \bigl( [I(a_1)/K(a_1)] * [K(a_2)/I(a_2)] \bigr)$

# پیوست ب- بخشی از محاسبات مربوط به معادله پاشندگی نانوتیوپ کربنی دوجداره در مـد TE

در این پیوست ادامه‌ی محاسبات مربوط به معادله پاشندگی مد TE در نانوتیوپ دوجداره آورده شده است. حروفی را که در محاسبات با فونت انگلیسی نوشته شده‌اند، در بقیه‌ی گزارش به شکل زیر مطرح شده‌اند.

( w= $\omega$ ),( a= $\alpha$ ),( b= $\beta$ ),( a۱, a۲ = $a_1, a_2$ ),( E۰ = $\varepsilon_0$ ),( n۰ = $n_0$ ),( Me = $m_e$ ),( q۱^۲, q۲^۲ = $q_{1m}^2, q_{2m}^2$ ),( k= $\kappa$ ),( C۱,C۲,C۳,C۴ = $C_{1m}, C_{2m}, C_{3m}, C_{4m}$ )

[ I (ka۱), I (ka۲) = $I_m(\kappa a_1), I_m(\kappa a_2)$ ],[ I'(ka۱), I'(ka۲) = $I'_m(\kappa a_1), I'_m(\kappa a_2)$ ]
[ K (ka۱), K (ka۲) = $K_m(\kappa a_1), K_m(\kappa a_2)$ ],[ K'(ka۱), K'(ka۲) = $K'_m(\kappa a_1), K'_m(\kappa a_2)$ ]

معادله‌ی شرط مرزی (۴-۱۵) :

C۱*I'(ka۱) = C۲* I'(ka۱)+ C۳* K'(ka۱)

با حذف ضریب C۳ از آن و قرار دادن این ضریب در معادله‌ی (۴-۱۳) ضریب C۲ بر حسب ضریب C۱ بدست می‌آمد.

=> C۳ = I'(ka۱)/ K'(ka۱) *C۱ - I '(ka۱)/ K'(ka۱)*C۲

R* I (ka۱) *C۲ + R*K (ka۱) *C۳ – R* I (ka۱) *C۱ = N* I'(ka۱)*C۱         (4–13)

R* I (ka۱) *C۲ + R*K (ka۱) * I'(ka۱)/ K'(ka۱) *C۱ - R*K (ka۱) *I'(ka۱)/ K'(ka۱)*C۲
– R* I (ka۱) *C۱ = N* I (ka۱)*C۱

$\Big($R* I (ka۱)  - R*K (ka۱) *I'(ka۱)/ K'(ka۱)$\Big)$*C۲ = $\Big($-R*K (ka۱) * I'(ka۱)/ K'(ka۱) + R* I (ka۱) + N* I'(ka۱)$\Big)$ *C۱

=> C۲ = $\Big($-R*K (ka۱) * I'(ka۱)/ K'(ka۱) + R* I (ka۱) + N* I'(ka۱)$\Big)\Big/\Big($R* I (ka۱)  - R*K (ka۱) *I'(ka۱)/ K'(ka۱)$\Big)$ *C۱

حال که ضریب C۲ بر حسب C۱ به دست آمده، می‌توان ضریب C۳ را تنها برحسب C۱ نوشت.

$$\Rightarrow C_3 = I'(ka_1)/ K'(ka_1) *C_1 + \Big(I'(ka_1)/ K'(ka_1) *R*K(ka_1) * I'(ka_1)/ K'(ka_1) - I'(ka_1)/ K'(ka_1) *R* I(ka_1) - I'(ka_1)/ K'(ka_1) * N* I'(ka_1)\Big)\Big/\Big(R* I(ka_1) - R*K(ka_1) *I'(ka_1)/ K'(ka_1)\Big) *C_1$$

با استفاده از معادله‌ی شرط مرزی (۴-۱۶) در بخش ۴-۲-۱، ضریب $C_4$ بر حسب سه ضریب دیگر نوشته می‌شود.

$$C_4*K'(ka_2) = C_2*I'(ka_2) + C_3*K'(ka_2)$$
$$\Rightarrow C_4 = I'(ka_2)/ K'(ka_2) *C_2 + C_3$$

با جایگذاری ضرایب $C_2, C_3$ برحسب $C_1$، ضریب $C_4$ نیز تنها بر حسب $C_1$ به‌دست می‌آید.

$$\Rightarrow C_4 = \Big(-R* I'(ka_2)/ K'(ka_2) *K(ka_1) * I'(ka_1)/ K'(ka_1) + R* I'(ka_2)/ K'(ka_2) * I(ka_1) + N* I'(ka_2)/ K'(ka_2) *I'(ka_1)\Big)\Big/\Big(R* I(ka_1) - R*K(ka_1) *I'(ka_1)/ K'(ka_1)\Big) *C_1 + I'(ka_1)/ K'(ka_1) *C_1 + \Big(I'(ka_1)/ K'(ka_1)*R*K(ka_1) * I'(ka_1)/ K'(ka_1) - I'(ka_1)/ K'(ka_1)*R* I(ka_1) - I'(ka_1)/ K'(ka_1)*N* I'(ka_1)\Big)\Big/\Big(R* I(ka_1) - R*K(ka_1) *I'(ka_1)/ K'(ka_1)\Big) *C_1$$

با جایگذاری ضرایب $C_2, C_3, C_4$ بر حسب $C_1$ در معادله (۴-۱۲) تنها ضریب $C_1$ باقی می‌ماند که از طرفین معادله حذف شده و معادله‌ی پاشندگی به‌دست می‌آید.

$$H* K(ka_2) *C_4 - H*I(ka_2) *C_2 – H* K(ka_2) *C_3 = N* K'(ka_2)*C_4 \qquad (4\text{-}14)$$

$\Big($-R* H* K (ka۲) * I'(ka۲)/ K'(ka۲) *K (ka۱) * I'(ka۱)/ K'(ka۱) + R* H* K (ka۲) * I'(ka۲)/ K'(ka۲) * I (ka۱)+ H* K (ka۲) *N* I'(ka۲)/ K'(ka۲) *I'(ka۱)$\Big)$/$\Big($R* I (ka۱) - R*K (ka۱) *I'(ka۱)/ K'(ka۱)$\Big)$ *C۱+ H* K (ka۲) * I'(ka۱)/ K'(ka۱) *C۱ + $\Big($H* K (ka۲) *I'(ka۱)/ K'(ka۱)*R*K (ka۱) * I'(ka۱)/ K'(ka۱) – H* K (ka۲) * I'(ka۱)/ K'(ka۱)*R* I (ka۱) – H* K (ka۲) * I'(ka۱)/ K'(ka۱)*N* I'(ka۱)$\Big)$/$\Big($R* I (ka۱) - R*K (ka۱) *I'(ka۱)/ K'(ka۱)$\Big)$ *C۱

+$\Big($H*I (ka۲) *R*K (ka۱) * I'(ka۱)/ K'(ka۱) - H*I (ka۲) *R* I (ka۱) - H*I (ka۲) * N* I'(ka۱)$\Big)$/$\Big($R* I (ka۱) - R*K (ka۱) *I'(ka۱)/ K'(ka۱)$\Big)$ *C۱

– H* K (ka۲) *I'(ka۱)/ K'(ka۱) *C۱ + $\Big($– H* K (ka۲) *I'(ka۱)/ K'(ka۱) *R*K (ka۱) * I'(ka۱)/ K'(ka۱) + H* K (ka۲) * I'(ka۱)/ K'(ka۱) *R* I (ka۱) + H* K (ka۲) *I'(ka۱)/ K'(ka۱) * N* I'(ka۱)$\Big)$/ $\Big($R* I (ka۱) - R*K (ka۱) *I'(ka۱)/ K'(ka۱)$\Big)$ *C۱

+$\Big($N* K'(ka۲)*R* I'(ka۲)/ K'(ka۲) *K (ka۱) * I'(ka۱)/ K'(ka۱) -N* K'(ka۲)* R* I'(ka۲)/ K'(ka۲) * I (ka۱) -N* K'(ka۲)*N* I'(ka۲)/ K'(ka۲) *I'(ka۱)$\Big)$/$\Big($R* I (ka۱) - R*K (ka۱) *I'(ka۱)/ K'(ka۱)$\Big)$ *C۱ -N* K'(ka۲)* I'(ka۱)/ K'(ka۱) *C۱ + $\Big($-N* K'(ka۲)*I'(ka۱)/ K'(ka۱)*R*K (ka۱) * I'(ka۱)/ K'(ka۱) + N* K'(ka۲)* I'(ka۱)/ K'(ka۱)*R* I (ka۱) + N* K'(ka۲)*I'(ka۱)/ K'(ka۱)*N* I'(ka۱)$\Big)$/$\Big($R* I (ka۱) - R*K (ka۱) *I'(ka۱)/ K'(ka۱)$\Big)$ *C۱ = ۰

$\bigl($-R* H* K (ka۲) * I'(ka۲)/ K'(ka۲) *K (ka۱) * I'(ka۱)/ K'(ka۱) + R* H* K (ka۲) * I'(ka۲)/ K'(ka۲) * I (ka۱)+ H* K (ka۲) *N* I'(ka۲)/ K'(ka۲) *I'(ka۱)$\bigr)$ +$\bigl($R* I (ka۱) - R*K (ka۱) *I'(ka۱)/ K'(ka۱)$\bigr)$ *H* K (ka۲) * I'(ka۱)/ K'(ka۱) + $\bigl($H* K (ka۲) *I'(ka۱)/ K'(ka۱)*R*K (ka۱) * I'(ka۱)/ K'(ka۱) – H* K (ka۲) * I'(ka۱)/ K'(ka۱)*R* I (ka۱) – H* K (ka۲) * I'(ka۱)/ K'(ka۱)*N* I'(ka۱)$\bigr)$

+ $\bigl($H*I (ka۲) *R*K (ka۱) * I'(ka۱)/ K'(ka۱) - H*I (ka۲) *R* I (ka۱) - H*I (ka۲) * N* I'(ka۱)$\bigr)$

– H* K (ka۲) *I'(ka۱)/ K'(ka۱)* $\bigl($R* I (ka۱) - R*K (ka۱) *I'(ka۱)/ K'(ka۱)$\bigr)$ + $\bigl($– H* K (ka۲) *I'(ka۱)/ K'(ka۱) *R*K (ka۱) * I'(ka۱)/ K'(ka۱) + H* K (ka۲) * I'(ka۱)/ K'(ka۱) *R* I (ka۱) + H* K (ka۲) *I'(ka۱)/ K'(ka۱) * N* I'(ka۱)$\bigr)$

+$\bigl($+ N* K'(ka۲)*R* I'(ka۲)/ K'(ka۲) *K (ka۱) * I'(ka۱)/ K'(ka۱) -N* K'(ka۲)* R* I'(ka۲)/ K'(ka۲) * I (ka۱) -N* K'(ka۲)*N* I'(ka۲)/ K'(ka۲) *I'(ka۱)$\bigr)$ -N* K'(ka۲)* I'(ka۱)/ K'(ka۱)* $\bigl($R* I (ka۱) - R*K (ka۱) *I'(ka۱)/ K'(ka۱)$\bigr)$ + $\bigl($-N* K'(ka۲)*I'(ka۱)/ K'(ka۱)*R*K (ka۱) * I'(ka۱)/ K'(ka۱) + N* K'(ka۲)* I'(ka۱)/ K'(ka۱)*R* I (ka۱) + N* K'(ka۲)*I'(ka۱)/ K'(ka۱)*N* I'(ka۱)$\bigr)$

= ۰

-R* H* K (ka۲) * I'(ka۲)/ K'(ka۲) *K (ka۱) * I'(ka۱)/ K'(ka۱) + R* H* K (ka۲) * I'(ka۲)/ K'(ka۲) * I (ka۱)+ H* K (ka۲) *N* I'(ka۲)/ K'(ka۲) *I'(ka۱) +R* I (ka۱) *H* K (ka۲) * I'(ka۱)/ K'(ka۱) - R*K (ka۱) *I'(ka۱)/ K'(ka۱) *H* K (ka۲) * I'(ka۱)/ K'(ka۱) + H* K (ka۲) *I'(ka۱)/ K'(ka۱)*R*K (ka۱) * I'(ka۱)/ K'(ka۱) – H* K (ka۲) * I'(ka۱)/ K'(ka۱)*R* I (ka۱) – H* K (ka۲) * I'(ka۱)/ K'(ka۱)*N* I'(ka۱) + H*I (ka۲) *R*K (ka۱) * I'(ka۱)/ K'(ka۱) - H*I (ka۲) *R* I (ka۱) - H*I (ka۲) * N* I'(ka۱)– H* K (ka۲) *I'(ka۱)/ K'(ka۱)* R* I (ka۱) + H* K (ka۲) *I'(ka۱)/ K'(ka۱)* R*K (ka۱) *I'(ka۱)/ K'(ka۱)– H* K (ka۲) *I'(ka۱)/ K'(ka۱) *R*K (ka۱) * I'(ka۱)/ K'(ka۱) + H* K (ka۲) * I'(ka۱)/ K'(ka۱) *R* I (ka۱) + H* K (ka۲) *I'(ka۱)/ K'(ka۱) * N* I'(ka۱) +N* K'(ka۲)*R* I'(ka۲)/ K'(ka۲) *K (ka۱) * I'(ka۱)/ K'(ka۱) -N* K'(ka۲)* R* I'(ka۲)/ K'(ka۲) * I (ka۱) -N* K'(ka۲)*N* I'(ka۲)/ K'(ka۲) *I'(ka۱)-N* K'(ka۲)* I'(ka۱)/ K'(ka۱)* R* I (ka۱) + N* K'(ka۲)* I'(ka۱)/ K'(ka۱)* R*K (ka۱) *I'(ka۱)/ K'(ka۱) -N* K'(ka۲)*I'(ka۱)/ K'(ka۱)*R*K (ka۱) * I'(ka۱)/ K'(ka۱) + N* K'(ka۲)* I'(ka۱)/ K'(ka۱)*R* I (ka۱) + N* K'(ka۲)*I'(ka۱)/ K'(ka۱)*N* I'(ka۱)

= ۰

هر جمله در معادله‌ی بالا را در یک سطر جداگانه می‌نویسیم.

-R* H* K (ka۲) * I'(ka۲)/ K'(ka۲) *K (ka۱) * I'(ka۱)/ K'(ka۱)
+ R* H* K (ka۲) * I'(ka۲)/ K'(ka۲) * I (ka۱)
+ H*N * K (ka۲) * I'(ka۲)/ K'(ka۲) *I'(ka۱)
+R*H * I (ka۱) * K (ka۲) * I'(ka۱)/ K'(ka۱)
- R*H *K (ka۱) *I'(ka۱)/ K'(ka۱) * K (ka۲) * I'(ka۱)/ K'(ka۱)
+H*R * K (ka۲) *I'(ka۱)/ K'(ka۱) *K (ka۱) * I'(ka۱)/ K'(ka۱)
– H*R * K (ka۲) * I'(ka۱)/ K'(ka۱) * I (ka۱)
– H*N * K (ka۲) * I'(ka۱)/ K'(ka۱) * I'(ka۱)
+ H*R *I (ka۲) *K (ka۱) * I'(ka۱)/ K'(ka۱)

- H*R *I (ka۲) * I (ka۱)
- H* N *I (ka۲) * I'(ka۱)
– H*R * K (ka۲) *I'(ka۱)/ K'(ka۱) * I (ka۱)
+ H* R * K (ka۲) *I'(ka۱)/ K'(ka۱) *K (ka۱) *I'(ka۱)/ K'(ka۱)
– H*R * K (ka۲) *I'(ka۱)/ K'(ka۱) *K (ka۱) * I'(ka۱)/ K'(ka۱)
+ H*R * K (ka۲) * I'(ka۱)/ K'(ka۱) * I (ka۱)
+ H* N * K (ka۲) *I'(ka۱)/ K'(ka۱) * I'(ka۱)
+N*R * K'(ka۲) * I'(ka۲)/ K'(ka۲) *K (ka۱) * I'(ka۱)/ K'(ka۱)
-N* R * K'(ka۲) * I'(ka۲)/ K'(ka۲) * I (ka۱)
-N*N * K'(ka۲) * I'(ka۲)/ K'(ka۲) *I'(ka۱)
-N*R * K'(ka۲)* I'(ka۱)/ K'(ka۱) * I (ka۱)
+N*R * K'(ka۲)* I'(ka۱)/ K'(ka۱) *K (ka۱)*I'(ka۱)/ K'(ka۱)
-N*R * K'(ka۲)*I'(ka۱)/ K'(ka۱) *K (ka۱) * I'(ka۱)/ K'(ka۱)
+ N*R * K'(ka۲)* I'(ka۱)/ K'(ka۱) * I (ka۱)
+ N* N* K'(ka۲)*I'(ka۱)/ K'(ka۱)* I'(ka۱)
= ۰

پس از ساده کردن جملات یکسان و با علامت مخالف، معادله‌ی پاشندگی به صورت زیر تبدیل می‌شود.

-R* H* K (ka۲) * I'(ka۲)/ K'(ka۲) *K (ka۱) * I'(ka۱)/ K'(ka۱)
+ R* H* K (ka۲) * I'(ka۲)/ K'(ka۲) * I (ka۱)
+ H*N * K (ka۲) * I'(ka۲)/ K'(ka۲) *I'(ka۱)
+ H*R *I (ka۲) *K (ka۱) * I'(ka۱)/ K'(ka۱)
- H*R *I (ka۲) * I (ka۱)     (B)
- H* N *I (ka۲) * I'(ka۱)
+N*R * K'(ka۲) * I'(ka۲)/ K'(ka۲) *K (ka۱) * I'(ka۱)/ K'(ka۱)
-N* R * K'(ka۲) * I'(ka۲)/ K'(ka۲) * I (ka۱)
-N*N * K'(ka۲) * I'(ka۲)/ K'(ka۲) *I'(ka۱)
+ N* N* K'(ka۲)*I'(ka۱)/ K'(ka۱)* I'(ka۱)
=۰

**نکته :**

با توجه به معادلات شرط مرزی اولیه برای دو مد (معادلات (۴-۵) تا (۴-۸) ) در پیوست الف و معادلات (۴-۱۳) تا (۴-۱۶) در پیوست ب )، مشاهده می‌شودکه تمامی محاسبات در این دو پیوست تا معادله‌ی (B) بالا

در این پیوست و معادله‌ی (A) در پیوست الف، شبیه به هم بوده و تنها توابع بسل تعمیم‌یافته‌ی بدون پریم به پریم‌دار تبدیل شده‌اند و بالعکس.

R* H* ( -K (ka۲) * I'(ka۲)/ K'(ka۲) *K (ka۱) * I'(ka۱)/ K'(ka۱)

+ K (ka۲) * I (ka۱)* I'(ka۲)/ K'(ka۲)

+ I (ka۲) *K (ka۱) * I'(ka۱)/ K'(ka۱)

-I (ka۲) * I (ka۱) )

+ H*N * ( K (ka۲) * I'(ka۲)/ K'(ka۲) *I'(ka۱)

- I (ka۲) * I'(ka۱) )

+N*R *( I'(ka۲)*K (ka۱) * I'(ka۱)/ K'(ka۱)

- K'(a۲) * I'(ka۲)/ K'(ka۲) * I (ka۱) )

+ N* N*( K'(ka۲)*I'(ka۱)/ K'(ka۱)* I'(ka۱)

- K'(ka۲) * I'(ka۲)/ K'(ka۲) *I'(ka۱) )

=۰.

فرمول رونسکین:

I'(X)* K(X) – I(X)* K'(X) = ۱/X

I'(ka۱)* K(ka۱) – I(ka۱)* K'(ka۱) = ۱/(ka۱)

با ضرب بعضی جملات در عبارات واحد، جملات رونسکین را تشکیل می‌دهیم.

R* H* ( -K (ka۲) * I'(ka۲)/ K'(ka۲) *K (ka۱) * I'(ka۱)/ K'(ka۱)

+ K (ka۲) * I (ka۱)* I'(ka۲)/ K'(ka۲) )*( K'(ka۱)/ K'(ka۱)

+ I (ka۲) *K (ka۱) * I'(ka۱)/ K'(ka۱)

-I (ka۲) * I (ka۱) )

+ H*N * ( K (ka۲) * I'(ka۲)/ K'(ka۲) *I'(ka۱)

- I (ka۲) * I'(ka۱) )

+N*R *( I'(ka۲)*K (ka۱) * I'(ka۱)/ K'(ka۱)

- K'(ka۲) * I'(ka۲)/ K'(ka۲) * I (ka۱))

+ N* N*( K'(ka۲)*I'(ka۱)/ K'(ka۱)* I'(ka۱)

- K'(ka۲) * I'(ka۲)/ K'(ka۲) *I'(ka۱))

=۰

R* H* ( K (ka۲) * I'(ka۲)/ K'(ka۲) / K'(ka۱) (-۱/(ka۱))

+ I (ka۲) *K (ka۱) * I'(ka۱)/ K'(ka۱)

-I (ka۲) * I (ka۱) *( K'(ka۱)/ K'(ka۱))

+ H*N * ( K (ka۲) * I'(ka۲)/ K'(ka۲) *I'(ka۱)

- I (ka۲) * I'(ka۱))

+N*R *( I'(ka۲)*K (ka۱) * I'(ka۱)/ K'(ka۱)

- K'(ka۲) * I'(ka۲)/ K'(ka۲) * I (ka۱))

+ N* N*( K'(ka۲)*I'(ka۱)/ K'(ka۱)* I'(ka۱)

- K'(ka۲) * I'(ka۲)/ K'(ka۲) *I'(ka۱))

=۰

R* H* ( -K (ka۲) * I'(ka۲)/ K'(ka۲) / K'(ka۱) (۱/(ka۱))

+ I (ka۲) / K'(ka۱) * (۱/(ka۱)) *( K'(ka۲)/ K'(ka۲))

+ H*N * ( K (ka۲) * I'(ka۲)/ K'(ka۲) *I'(ka۱)

- I (ka۲) * I'(ka۱))

+N*R *( I'(ka۲)*K (ka۱) * I'(ka۱)/ K'(ka۱)

- K'(ka۲) * I'(ka۲)/ K'(ka۲) * I (ka۱))

+ N* N*( K'(ka۲)*I'(ka۱)/ K'(ka۱)* I'(ka۱)

- K'(ka۲) * I'(ka۲)/ K'(ka۲) *I'(ka۱))

۱۴۳

$= \cdot$

$R* H* \left( \, \backslash / \, K'(ka\Upsilon) \, / \, K'(ka\backslash) \, \underline{(\backslash/(ka\backslash))(-\backslash/(ka\Upsilon))} \right)$

$+ H*N * \big( K(ka\Upsilon) * I'(ka\Upsilon)/ K'(ka\Upsilon) *I'(ka\backslash)$

$- I(ka\Upsilon) * I'(ka\backslash) *(K'(ka\Upsilon)/ K'(ka\Upsilon)) \big)$

$+N*R *\big( I'(ka\Upsilon)*K(ka\backslash) * I'(ka\backslash)/ K'(ka\backslash)$

$- K'(ka\Upsilon) * I'(ka\Upsilon)/ K'(ka\Upsilon) * I(ka\backslash) \big)$

$+ N* N*\big( K'(ka\Upsilon)*I'(ka\backslash)/ K'(ka\backslash)* I'(ka\backslash)$

$- K'(ka\Upsilon) * I'(ka\Upsilon)/ K'(ka\Upsilon) *I'(ka\backslash) \big)$

$= \cdot$

$R* H* \left( \backslash / K'(ka\Upsilon) / K'(ka\backslash) (\backslash/(ka\backslash))(-\backslash/(ka\Upsilon)) \right)$

$+ H*N * \big( \backslash / K'(ka\Upsilon) *I'(ka\backslash) *(\backslash/(ka\Upsilon)) \big)$

$+N*R *\big( I'(ka\Upsilon)*K(ka\backslash) * I'(ka\backslash)/ K'(ka\backslash)$

$- K'(ka\Upsilon) * I'(ka\Upsilon)/ K'(ka\Upsilon) * I(ka\backslash) *( K'(ka\backslash)/ K'(ka\backslash)) \big)$

$+ N* N*\big( K'(ka\Upsilon)*I'(ka\backslash)/ K'(ka\backslash)* I'(ka\backslash)$

$- K'(ka\Upsilon) * I'(ka\Upsilon)/ K'(ka\Upsilon) *I'(ka\backslash) \big)$

$= \cdot$

$R* H* \left( \backslash / K'(ka\Upsilon) / K'(ka\backslash) (\backslash/(ka\backslash))(-\backslash/(ka\Upsilon)) \right)$

$+ H*N * \big( \backslash / K'(ka\Upsilon) *I'(ka\backslash) *(\backslash/(ka\Upsilon)) \big)$

$+N*R *\big( I'(ka\Upsilon)/ K'(ka\backslash)*(\backslash/(ka\backslash)) \big)$

$+ N* N*\big( K'(ka\Upsilon) * I'(ka\backslash)*[I'(ka\backslash)/ K'(ka\backslash)- I'(ka\Upsilon)/ K'(ka\Upsilon)] \big)$

$= \cdot$

ضرب طرفین در $\left(-ka_1*ka_2*K'(ka_2)*K'(ka_1)\right)$ :

$+R*H$

$-H*N*\left(K'(ka_1)*I'(ka_1)*ka_1\right)$

$-N*R*\left(I'(ka_2)*K'(ka_2)*ka_2\right)$

$-N*N*\left(K'(ka_2)*I'(ka_1)*[I'(ka_1)/K'(ka_1)-I'(ka_2)/K'(ka_2)]\right)*\left(ka_1*ka_2*K'(ka_2)*K'(ka_1)\right)$

$=0$

$R = w^2 - a*q_1^2 - b*q_1^2$
$H = w^2 - a*q_2^2 - b*q_2^2$
$q_1^2 = q^2 + m^2/a_1^2$
$q_2^2 = q^2 + m^2/a_2^2$
$N = -e^2 n_0/(Me*E_0)*k$

متغیر N در این‌جا با مقدار آن در بخش TM فرق دارد.

$N^2 = e^4 n_0^2/(Me*E_0)^2 *k^2$
$= + e^4 n_0^2/(Me^2*E_0^2*a_1^4)*(a_1*a_2)^2*k^2$

$RH = (w^2 - a*q_1^2 - b*q_1^4)*(w^2 - a*q_2^2 - b*q_2^4)$
$= w^4 - w^2*(a*(q_1^2 + q_2^2) + b*(q_1^4 + q_2^4)) + (a*q_1^2 + b*q_1^4)*(a*q_2^2 + b*q_2^4)$

$N*H = -e^{2}n_0/(M_e*E_0)*k *(w^2 - a* q_2^2 - b*q_2^4)$

$N*R = -e^{2}n_0/(M_e*E_0)*k *(w^2 - a*q_1^2 - b* q_1^4)$

$+R*H$

$- N*H*\left(K'(ka_1)*I'(ka_1)*ka_1\right)$

$-N*R*\left(I'(ka_2)*K'(ka_2)*ka_2\right)$

$- N^2*\left(K'(ka_2)*I'(ka_1)*[I'(ka_1)/K'(ka_1) - I'(ka_2)/K'(ka_2)]\right)*\left(ka_1*ka_2*K'(ka_2)*K'(ka_1)\right)$

$= 0$

با جایگذاری ثابت‌های تعریف شده در معادله‌ی پاشندگی داریم:

$w^4 - (a*(q_1^2 + q_2^2) + b*(q_1^4 + q_2^4))*w^2 + (a*q_1^2 + b*q_1^4)*(a*q_2^2 + b*q_2^4)$

$+ e^{2}n_0/(M_e*E_0)*k *(w^2 - a* q_2^2 - b*q_2^4) *\left(K'(ka_1)*I'(ka_1)*ka_1\right)$

$+ e^{2}n_0/(M_e*E_0)*k *(w^2 - a*q_1^2 - b* q_1^4) *\left(I'(ka_2)*K'(ka_2)*ka_2\right)$

$- e^{4}n_0^2/(M_e*E_0)^2*k^4* a_1*a_2*\left(K'(ka_2)^2 * I'(ka_1)^2 - K'(ka_2)*K'(ka_1)*I'(ka_1)*I'(ka_2)\right)$

$= 0$

اعداد ثابت و ضرایب توان‌های مختلف w را در معادله از هم تفکیک می‌کنیم و معادله‌ی زیر را به‌دست می‌آوریم.

$w^4 + w^2 \left[- a*(q_1^2 + q_2^2) - b*(q_1^4 + q_2^4) + e^{2}n_0/(M_e*E_0)*k*\left(K'(ka_1)*I'(ka_1)*ka_2\right) + e^{2}n_0/(M_e*E_0)*k *\left(I'(ka_2)*K'(ka_2)*ka_2\right)\right]$

$+ \{(a*q_1^2 + b*q_1^4)*(a*q_2^2 + b*q_2^4)$

$+ e^{2}n_0/(M_e*E_0)*k *(- a*q_2^2 - b*q_2^4) *\left(K'(ka_1)*I'(ka_1)*ka_1\right)$

$+ e^{2}n_0/(M_e*E_0)*k *(-a*q_1^2 - b* q_1^4) *\left(I'(ka_2)*K'(ka_2)*ka_2\right)$

- e^۴n۰^۲/(Me^۲*E۰^۲)*k^۴ *a۱*a۲* $\big($K'(ka۲)^۲*I'(ka۱)^۲ - I'(ka۱) K'(ka۱)*I'(ka۲) K'(ka۲)$\big)$}
=۰

در ساده کردن جمله‌ی آخر از رابطه‌ی زیر استفاده می‌کنیم.

I(ka۱)^۲*K(ka۲)^۲=$\big($[I'(ka۱)/K'(ka۱)]*[K'(ka۲)/I'(ka۲)]$\big)$*$\big($I'(ka۱)*I'(ka۲)*K'(ka۱)*K'(ka۲)$\big)$

= F'۱۲*$\big($I'(ka۱) * K'(ka۱)* I'(ka۲) *K'(ka۲)$\big)$

معادله‌ی پاشندگی w(k) :

w^۴ + w^۲ [- $a$* (q۱^۲ + q۲^۲) - $b$ *(q۱^۴ + q۲^۴) + w'۱^۲ + w'۲^۲]
+{($a$* q۱^۲ + $b$ * q۱^۴)*( $a$* q۲^۲ + $b$ * q۲^۴)
- ( $a$* q۲^۲ + $b$*q۲^۴) w'۱^۲
- ( $a$* q۱^۲ + $b$*q۱^۴) w'۲^۲
+ w'۱^۲ * w'۲^۲ - w'۱^۲ * w'۲^۲*F'۱۲}
=۰

w۱^۲= e^۲n۰/(Me*E۰)*k^۲ * a۱* $\big($I'(ka۱)×K'(ka۱)$\big)$

w۲^۲= e^۲n۰/(Me*E۰)*k^۲ * a۲ $\big($I'(ka۲)*K'(ka۲)$\big)$

F۱۲ = $\big($[I'(ka۱) /K'(ka۱)] *[ K'(ka۲)/I'(ka۲)]$\big)$

# کارهای بعدی

❖ در نظر گرفتن نیروی حاصل از برهم‌کنش جداره‌های مجاور در نانوتیوپ کربنی دوجداره، به صورت اضافه کردن جمله‌ای در معادله‌ی حرکت الکترون در گاز الکترونی دوبعدی.

❖ استفاده از مدل الکترون نیمه‌آزاد به جای مدل الکترون آزاد در به دست آوردن معادلات پاشندگی نانوتیوپ‌های تک‌جداره و دوجداره

❖ بررسی میزان تلفات نوری در نانوتیوپ‌های کربنی که به عنوان موجبر در مدارات نوری مورد استفاده قرار می‌گیرند.

❖ محاسبه‌ی ضریب گذردهی نوری در نانوتیوپ‌های خمیده به عنوان موجبرهایی در مدارات مجتمع نوری نانومتری.